\documentclass[12pt,a4paper,Times New Roman]{book}
\pdfoutput=1
 
\pdfoutput=1
\usepackage[12hr]{datetime} 

\RequirePackage{ifpdf}
\usepackage{amsmath,amsfonts,amssymb,amsthm,mathtools} 
\usepackage{braket}
\usepackage{pstricks}
\usepackage[final]{pdfpages} 
\usepackage{slashed,cancel}
\usepackage{color,xcolor} 
\usepackage{fixmath}
\usepackage{caption} 
\usepackage{subcaption} 
\usepackage{cite}
\usepackage{float}
\usepackage{afterpage}
\usepackage{lscape,rotating}
\usepackage{makeidx} 
\usepackage{bbold}
\usepackage{cuted,afterpage}
\usepackage{calrsfs}
\RequirePackage{graphicx}
\RequirePackage{mathptmx}      
\RequirePackage{flushend}
\usepackage{multirow}
\usepackage{enumitem}
\allowdisplaybreaks[4]

\usepackage{graphicx}
\usepackage{setspace}
\usepackage[T1]{fontenc}
\usepackage[utf8x]{inputenc}
\usepackage{txfonts}

\usepackage[margin=2.0cm]{geometry}
\usepackage[toc,page]{appendix}

\usepackage{tikz}
\usetikzlibrary{positioning,arrows}
\usetikzlibrary{decorations.pathmorphing}
\usetikzlibrary{decorations.markings}
\usetikzlibrary{shapes.geometric}
\usepackage{pgflibraryarrows}
\usepackage{pgflibrarysnakes}
\tikzset{
    vector/.style={decorate, decoration={snake}, draw},
    provector/.style={decorate, decoration={snake,amplitude=2.5pt}, draw},
    antivector/.style={decorate, decoration={snake,amplitude=-2.5pt}, draw},
    fermion/.style={thick,draw=blue, postaction={decorate},decoration={markings,mark=at position .55 with {\arrow[draw=black]{>}}}},
    fermionbar/.style={draw=black, postaction={decorate},
                       decoration={markings,mark=at position .55 with {\arrow[draw=black]{<}}}},
    fermionnoarrow/.style={draw=black},
    gluon/.style={decorate, draw=black,decoration={coil,amplitude=4pt, segment length=5pt}},
    scalar/.style={dashed,draw=black, postaction={decorate},decoration={markings,mark=at position .55 with {\arrow[draw=black]{>}}}},
    scalarbar/.style={dashed,draw=black, postaction={decorate},decoration={markings,mark=at position .55 with {\arrow[draw=black]{<}}}},
    scalarnoarrow/.style={dashed,draw=black},
    electron/.style={draw=black, postaction={decorate},decoration={markings,mark=at position .55 with {\arrow[draw=black]{>}}}},
    bigvector/.style={decorate, decoration={snake,amplitude=4pt}, draw},
}

\tikzset{
  photon/.style={decorate, decoration={snake}, draw=red},
  fermion/.style={thick,draw=blue, postaction={decorate},decoration={markings,mark=at position .55 with {\arrow{>}}}},
  vertex/.style={draw,shape=circle,fill=black,minimum size=3pt,inner sep=0pt},
 particle/.style={thick,draw=black, postaction={decorate}},
 gluon/.style={decorate, draw=magenta,
        decoration={coil,amplitude=4pt, segment length=5pt}} 
}

\usepackage[font=small,labelfont=bf,margin=3mm,labelsep=period,tableposition=top]{caption}
\usepackage{booktabs}

\newcommand{\mytitle}{Study of anomalous  gauge boson self-couplings and the role of spin-$1$ polarizations}
\newcommand{\myauthor}{Rafiqul Rahaman}

\definecolor{urlblue}{rgb}{0.2,0.4,0.7}
\definecolor{citegreen}{rgb}{0,0.6,0.2}
\definecolor{linkred}{rgb}{0.9,0.2,0.1}
\usepackage[pdftex,pdfpagelabels,bookmarks,hyperindex,hyperfigures]{hyperref}\hypersetup{
colorlinks,citecolor=citegreen,linkcolor=blue,urlcolor=urlblue,linktocpage,linktoc=all%
pdfauthor={\myauthor},pdftitle={\mytitle}}

\definecolor{headercolor}{gray}{0.65} 

\usepackage{fancyhdr}
\fancypagestyle{plain}{%
\fancyhf{} 
\fancyfoot[C]{\fontfamily{ppl}\selectfont \small\bfseries \thepage} 
 
}
\usepackage{emptypage}

\definecolor{halfgray}{gray}{0.55} 
\newfont{\chapNumFont}{eurb10 scaled 7000}
\newfont{\chapTitFont}{pplr9d}
\usepackage[explicit]{titlesec}
\titleformat{\section}[block]{\bfseries\Large}{\fontfamily{ppl}\selectfont \thesection}{15pt}{\fontfamily{ppl}\selectfont #1}
\titleformat{\subsection}[block]{\bfseries\large}{\fontfamily{ppl}\selectfont \thesubsection}{14pt}{\fontfamily{ppl}\selectfont #1}
\titleformat{\subsubsection}[block]{\bfseries}{\fontfamily{ppl}\selectfont \thesubsubsection}{13pt}{\fontfamily{ppl}\selectfont #1}
\titleformat{\chapter}[block]%
{\Huge}{\raggedleft{\color{halfgray}\chapNumFont\thechapter}}{20pt}%
{\raggedright{\fontfamily{ppl}\selectfont #1}}

\usepackage{titletoc}
\titlecontents{chapter}[1.4em]{\vspace{8pt}}{\fontfamily{ppl}\selectfont\bfseries \contentslabel{1.4em}}%
{\fontfamily{ppl}\selectfont\bfseries \hspace*{-1.4em}}{\titlerule*[1pc]{.}\fontfamily{ppl}\selectfont\bfseries\contentspage}
\titlecontents{section}[4em]{}{\fontfamily{ppl}\selectfont\small \contentslabel{2.2em}}{}{\titlerule*[1pc]{.}\fontfamily{ppl}\selectfont\small\contentspage}
\titlecontents{subsection}[7em]{}{\fontfamily{ppl}\selectfont\footnotesize \contentslabel{2.6em}}{}{\hfill\contentspage}

\usepackage[notoccite,undotted]{minitoc}
\let\minitocORIG\minitoc
\renewcommand{\minitoc}{\minitocORIG \vspace{1.5em}}

\numberwithin{equation}{section}

\usepackage{titlesec}

\usepackage{mdframed,lipsum}
\usepackage{soul}
\setstcolor{red}

\newmdenv[
linewidth = 0.5pt,
topline = false,
bottomline = false
]{leftbar}
\makeatletter
\def\true@true@true{\fi\fi\iftrue\iftrue\iftrue}
\makeatother

\begin{document}

\unitlength1cm

\def\wtil#1{\widetilde{#1}}
\def\p{\partial}
\def\s{\slashed}
\def\ol{\overline}
\def\tc{\textcolor}
\def\Lag{{\cal L}}
\def\MGvATNLO{{\tt {\sc MadGraph5}\_aMC@NLO}}
\renewcommand\thesubsubsection{\thesubsection.\alph{subsubsection}}

\def\zo{\overline{z}_1}
\def\zt{\overline{z}_2}
\def\C{\overline{C}}
\def\D{{\cal D}}
\def\DD{\overline{\cal D}}
\def\g{\overline{\cal G}}
\def\gm{\gamma}
\def\M{{\cal M}}
\def\ep{\epsilon}
\def\epm1{\frac{1}{\epsilon}}
\def\epm2{\frac{1}{\epsilon^{2}}}
\def\epm3{\frac{1}{\epsilon^{3}}}
\def\epm4{\frac{1}{\epsilon^{4}}}
\def\unM{\hat{\cal M}}
\def\ashat{\hat{a}_{s}}
\def\asmur{a_{s}^{2}(\mu_{R}^{2})}
\def\sigbar{{{\overline {\sigma}}}\left(a_{s}(\mu_{R}^{2}), L\left(\mu_{R}^{2}, m_{H}^{2}\right)\right)}
\def\sigbarn{{{{\overline \sigma}}_{n}\left(a_{s}(\mu_{R}^{2}) L\left(\mu_{R}^{2}, m_{H}^{2}\right)\right)}}
\def\unas{ \left( \frac{\hat{a}_s}{\mu_0^{\epsilon}} S_{\epsilon} \right) }
\def\rnM{{\cal M}}
\def\bt{\beta}
\def\cD{{\cal D}}
\def\cC{{\cal C}}
\def\ca{\text{\tiny C}_\text{\tiny A}}
\def\cf{\text{\tiny C}_\text{\tiny F}}
\def\ct{{\red []}}
\def\sv{\text{SV}}
\def\murOmu{\left( \frac{\mu_{R}^{2}}{\mu^{2}} \right)}
\def\bb{b{\bar{b}}}
\def\bt0{\beta_{0}}
\def\bt1{\beta_{1}}
\def\bt2{\beta_{2}}
\def\bt3{\beta_{3}}
\def\gm0{\gamma_{0}}
\def\gm1{\gamma_{1}}
\def\gm2{\gamma_{2}}
\def\gm3{\gamma_{3}}
\def\nn{\nonumber}
\def\l{\left}
\def\r{\right}
\def\CA{\mathbf{C_A}}
\def\F{{\cal F}}
\newcommand{\dis}{}
\newcommand{\overbar}[1]{mkern-1.5mu\overline{\mkern-1.5mu#1\mkern-1.5mu}\mkern
1.5mu}
\newcommand{\iu}{{i\mkern1mu}}
\newcommand{\ConferenceEntry}[7]{
		\noindent #1 \href{#3}{\textbf{\textit{#2}}}, #4,
                \href{#6}{#5} #7 }
\newcommand{\TalkEntry}[4]{
		\noindent #1 \textbf{\textit{#2}}, #3 #4}

\frontmatter 
\thispagestyle{empty}

\begin{titlepage}
\newlength{\centeroffset}
\setlength{\centeroffset}{0cm}
\setlength{\centeroffset}{-0.5\oddsidemargin}
\addtolength{\centeroffset}{0.5\evensidemargin}



\vskip 3cm

\noindent\hspace*{\centeroffset}\makebox[0pt][l]{%
\begin{minipage}{\textwidth}
\begin{center}
{\setstretch{3}
\noindent{\fontfamily{ppl}\fontsize{24.88pt}{0pt}\selectfont\bfseries \mytitle}\\[1.5cm]
}

\noindent{\Large\bfseries\fontfamily{ppl}\selectfont \textsc{Rafiqul Rahaman}\\ Roll. no. 13RS033}\\
\vspace*{0.4cm}
Supervisor: Dr. Ritesh K. Singh

\end{center}

\end{minipage}}
\vspace{0.5cm}

\noindent\hspace*{\centeroffset}\makebox[0pt][l]{%
\begin{minipage}{\textwidth}
\begin{center}
\begin{tabular}{rr}
\end{tabular}
\end{center}
\end{minipage}}

\vspace{0.5cm}

\noindent\hspace*{\centeroffset}\makebox[0pt][l]{%
\begin{minipage}{\textwidth}
\begin{center}
\includegraphics[width=0.25\textwidth]{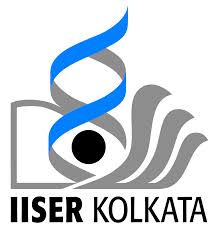}\\
\textsc{\fontfamily{ppl}\selectfont \small Department of Physical Sciences\\ Indian Institute of Science Education and Research, Kolkata\\ Mohanpur, Nadia, WB-741246, India}
\\[1.5cm]
\end{center}
\begin{center}
\noindent\textit{\bfseries\fontfamily{ppl}\selectfont A dissertation submitted to IISER Kolkata\\ for the partial fulfilment of the requirements for the Degree of\\ Doctor of Philosophy}
\end{center}
\vspace{0.5cm}
\begin{center}
July  2020
\end{center}
\end{minipage}
}

\end{titlepage}

\newpage\null\thispagestyle{empty}\newpage

\newpage
\thispagestyle{empty}


\newpage
\thispagestyle{empty}
\vspace*{\fill}
\centerline{{\bf \textit{{\Large I dedicate this thesis to}}}}
\vskip 0.5cm

\hspace{2cm}\centerline{{\bf \textit{{\huge My Parents}}}}
\vspace*{\fill}

\newpage\null\thispagestyle{empty}\newpage
\thispagestyle{empty}
\centerline{{\bf{\large DECLARATION OF AUTHORSHIP}}}
\vskip 1cm
\begin{flushright}
Date: \underline{03/07/2020}
\end{flushright}
I,  \textbf{\myauthor} with Registration No. \textbf{13RS033} dated $26/12/2013$, a student of Department of Physical Sciences of the PhD Program of IISER Kolkata, hereby declare that this thesis is my own work and, to the best of my knowledge, it neither contains materials previously published or written  by  any  other  person, nor it has been submitted for  any  degree/diploma  or  any other academic award anywhere before. \\

I also declare that all copyrighted material incorporated into this thesis is in compliance  with the Indian Copyright (Amendment) Act, 2012 and that I have received written permission  from the copyright owners for my use of their work.

\vspace{2.0cm}
\hspace{-0.7cm}
\vspace{0.2cm}
\textbf{Rafiqul Rahaman} \\
Department of Physical Sciences \\
Indian Institute of Science Education and Research Kolkata\\
Mohanpur 741246, West Bengal, India

\newpage\null\thispagestyle{empty}\newpage
\thispagestyle{empty}
\centerline{{\bf{\large CERTIFICATE FROM  SUPERVISOR}}}
\vskip 1cm
\begin{flushright}
	Date: \underline{03/07/2020}
\end{flushright}
This is to certify that the thesis entitled \textbf{``\mytitle "} submitted by Mr. \textbf{\myauthor} with Registration No. \textbf{13RS033} dated $26/12/2013$, a student of Department of Physical Sciences of the PhD  Program of IISER  Kolkata, is based upon his own research work under my supervision. This is also to certify  that neither the thesis nor any part of it has been submitted for any degree/diploma or any  other academic award anywhere before. In my opinion, the thesis fulfils the requirement for  the award of the degree of Doctor of Philosophy.

\vspace{2cm}
{\textbf{Dr. Ritesh K.  Singh}} \\
Associate  Professor \\
Department of Physical Sciences \\
Indian Institute of Science Education and Research Kolkata\\
Mohanpur 741246, West Bengal, India

\newpage\null\thispagestyle{empty}\newpage

\newpage
\thispagestyle{empty}
~
\vskip 1.0cm
\centerline{{\bf{\large ACKNOWLEDGMENTS}}}
\vskip 1cm
%

\textit{
I can not forget the people who have their contributions directly or indirectly in the long journey of my Ph.D. Without their support, I can not imagine myself writing my Ph.D. thesis now. I will try to acknowledge their contribution to my Ph.D. in the best possible way.}

\textit{
First of all,  I would like to thank Maa and Abba for their unconditional love and support. It is impossible to mention their contributions in a few words, but it is worth mentioning the crucial ones. They supported my educational aspirations and passionate cultivation of my curiosities through their persistent emotional, financial aid, even in their empty stomach. They unconditionally support me to fulfil my dream of becoming a scientist, a selfish one based on the fact that we have a poor financial condition from the very beginning.
I could not forget the contribution of my brother Majibul, who had to quit school after his Madhyamik and join work to keep our stomach full. It is
not enough just to thank him for his unconditional love, along with emotional and financial support.
I would like to thank my other brother and sisters, for their love and support.
I especially thank Mabo, my late grandmother, for her words of encouragement and praise.
I thank my uncles, aunts,  and cousins for their love and support.
}

\textit{
Secondly, I would like to acknowledge my Ph.D. supervisor Dr. Ritesh K. Singh, who directly contributed to my Ph.D. In the long journey of my Ph.D., he gave me his valuable times for discussions. He kept his calm down with patience to many of my irritating stupid questions.  He is more like a friend than a guide during any interaction. He is the coolest person I ever met. He continued discussing with me even on his own and his family member's health issues.
For his guidance and suggestions, I could publish articles in peer-reviewed journals. 
My thirst for knowledge has increased gradually through discussions with him. Once again, I thank Dr. Ritesh to boost my quest to unravel the mysteries of the Universe.
}

\textit{
Next, I would like to thank all my teachers throughout my school and college. I would like to thank my school headmaster Dr. Ganguly for his support and encouragement to science.  I would especially like to acknowledge my science and mathematics teachers late Bumba Da, Samsul Sir, Natu Da, who not only encourage my interests through support and praise but also did not take fees from me during my high school.
I especially thank Dr. Mainak Gupta, who seeded the interests of quantum mechanics and relativity in me in college.
I also thank Dr. Kumar Rao, with whom I did my master's thesis in particle physics and later chose to do a Ph.D. in the same area.
I would like to acknowledge Asuthosh College in Kolkata for inviting me and giving me the opportunity to complete my B.Sc. with free food and lodging. I would like to thank the  Department of Science and Technology (DST), Government of India, for support through stipend during my B.Sc. and M.Sc. I would like to express my sincere gratitude to my research collaborators Subhasish Behera,  Dr. Rashidul Islam, Dr. Mukesh Kumar, Prof. Poulose, Dr. Satendra Kumar, and Dr. Adam Falkowski for 
wonderful collaborations.
}

\textit{
I am grateful to the Department of Science and Technology (DST), Government of India, for support through DST-INSPIRE Fellowship for my doctoral program, INSPIRE CODE IF140075, 2014.  I am grateful to IISER Kolkata for financial assistance to attend the APS April meeting in the United States. I am also thankful to CEFIPRA for giving me funds to visit  LPT Orsay for collaborative work.  I would like to thank the taxpayers of India for their contribution to fellowship.
}

\textit{
Lastly, I would like to thank my friends Momidul and Mosayeb for their company during high school. I would like to thank my friends Rakesh, Vivek, Deep, and Anirban, for their support and company. Late-night gossips, tea break gossips in various topics like social, political, science, etc.,
with them has chased away the monotony of life during my Ph.D.
I would especially like to thank Atanu for his company and various discussions.   
I would like to thank my seniors Soumitra, Priyasri, Ipsita, Swati, Biswarup, Soumen, Gopal, Ankan, and  Santanu, for their advice. I also thank my juniors Sayak, Soumik, Anurag, Chiranjib, Subhojit, Avijit, Srijita, Debangana, Dipanjan, Pandey, Souraj, Sajal, Souradip, Arkayan, and Chiranjit for keeping a vibrant and joyful environment in the workplace.  I would like to thank Bimal kaka, in whose shop we take tea, have breakfast and gossip/discuss everything.  
}

\thispagestyle{empty}
\vskip 1cm
\begin{flushright}
\textit{Rafiqul Rahaman}
\end{flushright}


 \newpage
\thispagestyle{empty}


\centerline{{\bf{\large ABSTRACT}}}
\vskip 1cm

The prime goal of this thesis is to study anomalous gauge boson self couplings, triple gauge boson couplings in particular, with the help of spin polarization observables of the gauge bosons $Z$ and $W^\pm$ in the presence of beam polarizations where ever possible. The neutral triple gauge boson couplings, i.e., $ZZZ$, $ZZ\gamma$, $Z\gamma\gamma$, are studied in $ZZ/Z\gamma$ ($2lq\bar{q}/2l\gamma$) production at an $e^+e^-$ collider with and without beam polarization. Some of these anomalous couplings are also studied in $ZZ$ ($4l$) production at the LHC.  In the charge sector the anomalous gauge boson couplings, i.e., $WWZ$, $WW\gamma$ have been studied at an $e^+e^-$ collider in $W^+W^-$ ($l^-\bar{\nu_l}q^\prime\bar{q}$) production. The $WWZ$ anomalous couplings are also studied in $ZW^\pm$ production at LHC in $3l+\cancel{E}_T$ channel. All the analyses at an $e^+$-$e^-$ collider have been performed for center-of-mass (CM) energy of $500$ GeV and integrated luminosity of $100$ fb$^{-1}$. The analyses at the LHC are performed at $13$ TeV CM energy of $pp$ collisions. The cross sections and polarization asymmetries, along with other asymmetries (forward-backward, azimuthal), are used to obtain simultaneous limits on the anomalous couplings using  Markov-Chain--Monte-Carlo (MCMC) method in each process. The polarization asymmetries can distinguish between $CP$-even and $CP$-odd couplings and help to put tighter constraints on the couplings. The polarization of the initial $e^-$  and $e^+$ beam, in case of $e^+e^-$ collider,  are used to increase the signal to background ratio, putting tighter constraints on the anomalous couplings. The polarization asymmetries are instrumental in the measurement of anomalous couplings should a deviation from the SM be observed.

 \newpage
\thispagestyle{empty}
\newpage\null\thispagestyle{empty}\newpage

\centerline{{\bf{\large List of Publications (included in this thesis)}}}

\begin{itemize}
		\item[1.]
		{\textbf{R.~Rahaman} and R.~K. Singh, {``On polarization parameters of spin-1 particles and
				anomalous couplings in $e^+e^-\rightarrow ZZ/Z\gamma $''},
			\href{http://dx.doi.org/10.1140/epjc/s10052-016-4374-4}{{\em Eur. Phys. J.}
				{\bfseries C76} no.~10, (2016) 539},
			\href{http://arxiv.org/abs/1604.06677}{{\ttfamily arXiv:1604.06677 [hep-ph]}}.}
		
		\item[2.]
		{\textbf{R.~Rahaman} and R.~K. Singh, {``On the choice of beam polarization in
				$e^+e^-\rightarrow ZZ/Z\gamma $ and anomalous triple gauge-boson couplings''},
			\href{http://dx.doi.org/10.1140/epjc/s10052-017-5093-1}{{\em Eur. Phys. J.}
				{\bfseries C77} no.~8, (2017) 521},
			\href{http://arxiv.org/abs/1703.06437}{{\ttfamily arXiv:1703.06437 [hep-ph]}}.}
	
		\item[3.]{\textbf{R.~Rahaman} and R.~K. Singh, {``Anomalous triple gauge boson couplings in
				$ZZ$ production at the LHC and the role of $Z$ boson polarizations"},
			\href{http://dx.doi.org/https://doi.org/10.1016/j.nuclphysb.2019.114754}{{\em
					Nuclear Physics B} {\bfseries 948} (2019) 114754},
			\href{http://arxiv.org/abs/1810.11657}{{\ttfamily arXiv:1810.11657
					[hep-ph]}}.} 

\item[4.]{\textbf{R.~Rahaman} and R.~K. Singh, {\it {``Probing the anomalous triple gauge boson
			couplings in $e^+e^-\to W^+W^-$ using $W$ polarizations with polarized
			beams"}},  \href{http://dx.doi.org/10.1103/PhysRevD.101.075044}{{\em Phys.
			Rev. D} {\bfseries 101} no.~7, (2020) 075044},
	\href{http://arxiv.org/abs/1909.05496}{{\ttfamily arXiv:1909.05496
			[hep-ph]}}}
		
\item[5.]{\textbf{R.~Rahaman} and R.~K. Singh, {\it {``Unravelling the anomalous gauge boson
			couplings in $ZW^\pm$ production at the LHC and the role of spin-$1$
			polarizations"}},  \href{http://dx.doi.org/10.1007/JHEP04(2020)075}{{\em JHEP}
		{\bfseries 04} (2020) 075}, \href{http://arxiv.org/abs/1911.03111}{{\ttfamily
			arXiv:1911.03111 [hep-ph]}}.
		}
\end{itemize}

\vskip 1cm
\centerline{{\bf{\large List of Publications (not included in this thesis)}}}
\vskip 1cm

\begin{itemize}
		\item[1.]{Subhasish Behera, Rashidul Islam, Mukesh Kumar and Poulose Poulose and \textbf{Rafiqul~Rahaman}, {``Fingerprinting the
				Top quark FCNC via anomalous $Ztq$ couplings at the LHeC''},   \href{http://dx.doi.org/10.1103/PhysRevD.100.015006}{{\em Phys. Rev.}
				{\bfseries D100} no.~1, (2019) 015006},
			\href{http://arxiv.org/abs/1811.04681}{{\ttfamily arXiv:1811.04681 [hep-ph]}}.}
		
		\item[2.]{S.~Kumar, P.~Poulose, \textbf{R.~Rahaman}, and R.~K. Singh,  {``Measuring Higgs
				self-couplings in the presence of VVH and VVHH at the ILC"},
			\href{http://dx.doi.org/10.1142/S0217751X19500945}{{\em Int. J. Mod. Phys.}
				{\bfseries A34} no.~18, (2019) 1950094},
			\href{http://arxiv.org/abs/1905.06601}{{\ttfamily arXiv:1905.06601 [hep-ph]}}.}
\end{itemize}

\newpage
\newpage\null\thispagestyle{empty}\newpage

\newpage

\centerline{{\bf{\large UNIT,  SYMBOLS AND ABBREVIATIONS}}}
\vskip 1cm
We use the natural unit system, i.e., $  \hslash = c = 1$ throughout.

\begin{tabular}{lll}
\centering
Symbol & Name & Unit \\ 
$\sqrt{s}$ & CME & GeV\\
${\cal L}$  & Luminosity & fb$^{-1}$\\
$\sigma$  & cross section & pb/fb\\
\end{tabular}

\vskip 1cm
\begin{tabular}{ll}
 SM & Standard Model\\
 QFT & Quantum field theory \\ 
  BSM & Beyond the SM \\
QED & Quantum electro dynamics \\
QCD & Quantum chromo dynamics \\
$CP$ & Charge and parity \\
EFT & Effective field theory \\
EM & Electromagnetic \\
ED & Extra dimension \\
UED & Universal extra dimension\\
KK & Kaluza-Klein\\
DM & Dark matter \\
  EW & Electroweak \\
 EWSB & Electroweak symmetry breaking\\
SUSY & Supersymmetry \\ 
2HDM & Two-Higgs-doublet-model\\
C2HDM & Complex 2HDM \\
MSSM & Minimal supersymmetric SM\\
NCSM & Non commutative extension of the SM \\
GUT & Grand unified theory \\

\end{tabular}

\vskip 1cm
\begin{tabular}{ll}
LEP & Large Electron–Positron Collider \\
LHC & Large Hadron Collider\\
ILC & International Linear Collider\\
LHeC & Large Hadron electron Collider\\
CM & Centre-of-mass\\
aTGC & Anomalous triple gauge boson couplings\\
MCMC  & Markov-Chain--Monte-Carlo\\
 SDM & Spin density matrix \\ 
 C.L. & Confidence level\\
 BCI & Bayesian confidence interval\\ 
 LO & Leading order \\
 NLO & Next to LO \\
 NNLO & Next to NLO\\
 pb & Pico barn \\ 
 fb & Femto barn \\
 GeV & Giga electron volt \\
\end{tabular}


\begingroup
\hypersetup{linkcolor=blue}
\dominitoc
\tableofcontents
\endgroup

\newpage
\newpage\null\thispagestyle{empty}\newpage




\phantomsection
\addcontentsline{toc}{chapter}{\listfigurename}
\listoffigures
\adjustmtc

\cleardoublepage

\phantomsection
\addcontentsline{toc}{chapter}{\listtablename}
\listoftables
\adjustmtc

\mainmatter 

\chapter{Introduction}\label{chapter:intro}
\begingroup
\hypersetup{linkcolor=blue}
\minitoc
\endgroup
The Standard Model (SM)~\cite{Glashow:1961tr,Weinberg:1967tq,Salam:1968rm,Gell-Mann:2015noa,Fritzsch:1973pi,Gross:1973id,Gross:1973ju,Politzer:1973fx,Weinberg:1973un} of particle physics is one of the most remarkably successful fundamental theories to describe the governing principle of elementary constituents of
matter and their interactions in the Universe. It explains  almost all the 
phenomena observed in Nature at a small length scale. The theoretical predictions of the SM are being confirmed time and again with a spectacular accuracy with the discovery of many fundamental particles and interactions. The particle spectrum of the SM is complete with the discovery of its last milestone, the Higgs boson  at the Large Hadron Collider (LHC)~\cite{Chatrchyan:2012xdj,Aad:2012tfa}  in 2012. However, the SM is too far from being a final theory (theory of everything) as it has many issues within the theoretical framework, such as the  hierarchy of mass scales, the strong $CP$ problem, and it is unable to address some experimental facts, such as neutrino oscillation, dark matter, baryogenesis, and many more. Many theories, with SM as a subset,  have been postulated to address the unresolved issues predicting new particles and new interactions, i.e., new physics beyond the SM (BSM).  Searches for such new physics  are going on at the LHC with higher energy and higher luminosity, but sadly one has not found anything beyond the SM~\cite{Summary:ATLAS:exotics,Summary:ATLAS:SUSY,Summary:CMS:exotic,Summary:CMS:longlived,Summary:CMS:B2G} till date except few  fluctuations (e.g., see Refs.~\cite{Aaboud:2016tru,Khachatryan:2016hje,Sirunyan:2018wim}).  
One could expect that the new physics scale is too heavy  to be directly 
explored by the current LHC, and they may leave some footprints in the available energy range.
They will modify the structure of the SM interactions or bring some new interactions  
often through higher-dimensional operators with the SM fields.  
These new vertices and/or the  extra contributions to the SM  vertices are 
termed as anomalous in the sense that they are not present in the SM at leading order. 
The electroweak sector will get affected by the anomalous bosonic self couplings, which alter the paradigm of electroweak symmetry breaking (EWSB)~\cite{Higgs:1964pj,Guralnik:1964eu,Englert:1964et,Higgs:1966ev}. 
To test the SM (or BSM) predictions for the EWSB mechanism, precise measurements of the Higgs  couplings with all other 
gauge bosons, Higgs self-couplings, and gauge boson self-couplings are necessary. 

This thesis is focused on the study of gauge boson self couplings, in particular, anomalous triple gauge boson couplings (aTGC). The anomalous gauge boson couplings which carry information of high scale new physics in the
electroweak sector can be modelled through effective field theory (EFT) in a model independent way. In EFT, one can preserve the SM gauge symmetry or just consider the Lorentz invariance.
Both approaches serve as an open-minded method of describing low energy impacts of new physics at much higher energy scales. 
There are lots of studies of the aTGC in the literature on the theoretical side as well as in various experiments in various ways with cross sections and some asymmetries.
Innovative techniques  with more observables are  required to probe many
unknown anomalous couplings parameters in experiments.
Our strategy, here, is to use 
the polarizations of gauge bosons to probe the aTGC and we will see that they 
give significant contributions in pinning down the aTGC.

Before going to the main contents, we give a brief introduction of the SM which is very important
to understand  the aTGC.
 We will close the introduction 
chapter by giving a description of the aTGC in the EFT approach followed  by contributions from some of the BSM scenarios.

\section{The Standard Model of particle physics}\label{sec:intro-SM}
The SM relies on the framework of quantum field theory (QFT), where the fundamental particles are treated as discrete excitations of an underlying dynamical field. The SM is constructed by postulating a set of underlying symmetries and considering the most general renormalizable Lagrangian with the field contents. The symmetry of the SM  is based on the {\em local} gauge transformation, i.e., separately valid at each space-time point $x$, under  $SU(3)_C\otimes SU(2)_L\otimes U(1)_Y$ group, which describe the fundamental forces in Nature: the strong, the weak and the electromagnetic interactions excluding the gravity. Each gauge symmetry manifestly gives rise to gauge boson mediators for each interaction. 
The particle content of the SM in the matter sector, all with spin $1/2$, are the six quarks, endowed with both color ($SU(3)_C$) and electroweak charges ($SU(2)_L\otimes U(1)_Y$), six leptons with no color but with electroweak charges. The matter fermions  (quarks and leptons)
are present in three generations with identical quantum number but different masses with the pattern
\begin{eqnarray}\label{eq:intro-doublet-fermion}
\text{Leptons}~(l)&:& 
\begin{pmatrix}
\nu_l \\ l^-
\end{pmatrix}
=
\begin{pmatrix}
\nu_e && \nu_\mu && \nu_\tau \\
e^- &&  \mu^- && \tau^-
\end{pmatrix} \hspace{1cm} \text{and} \nonumber  \\
\text{Quarks}~(q)&:& 
\begin{pmatrix}
q_u \\ q_d
\end{pmatrix}
=
\begin{pmatrix}
u  && c && t \\
d &&  s && b
\end{pmatrix}.
\end{eqnarray}
The mediators (gauge bosons) of the strong, weak and electromagnetic forces are, of spin $1$; eight massless gluons ($g$) for strong interaction, three massive gauge  boson ($W^+$, $W^-$ and $Z$) for the weak interaction and one massless photon ($A$) for the electromagnetic interaction.
While the color ($SU(3)_C$) symmetry is conserved, the electroweak ($SU(2)_L\otimes U(1)_Y$) symmetry is broken  leaving some of its generators to be massive. The electroweak gauge symmetry
is spontaneously broken by the celebrated Brout-Englert-Higgs-Kibble (BEHK) mechanism, which generates mass terms for the massive vector  bosons by the spin-$0$  Higgs field ($\Phi$). The quantum excitation of the Higgs field gives rise to the particle Higgs boson ($h$).   

\begin{table}\caption{\label{tab:intro-gaug-quantum-number} The gauge quantum numbers of the SM fields in the 
        $SU(3)_C\otimes SU(2)_L\otimes U(1)_Y$ gauge group.}
    \centering
    \renewcommand{\arraystretch}{1.5}
    \begin{tabular}{|c|c|c|c|c|c|}\hline
        Field/ Quantum number & $SU(3)_C$ & $SU(2)_L$ & $U(1)_Y$& $T_3$ & $Q_{EM}=T_3+Y$ \\ \hline
        \multirow{2}{*}{$q_{L}=\begin{pmatrix}q_u \\ q_d \end{pmatrix}_L$}& \multirow{2}{*}{$3 $} & \multirow{2}{*}{$2 $}& \multirow{2}{*}{$+\frac{1}{6}$}&$+\frac{1}{2}$ & $+\frac{2}{3}$ \\
        &&&& $-\frac{1}{2}$ & $-\frac{1}{3}$ \\ \hline
        ${q_u}_{R}$    & $3 $ & $1 $& $+\frac{2}{3} $&$0$& $+\frac{2}{3} $ \\ \hline
        ${q_d}_{ R}$    & $3 $ & $1 $& $-\frac{1}{3} $&$0$& $-\frac{1}{3} $ \\\hline
        \multirow{2}{*}{$l_{L}=\begin{pmatrix}\nu_l \\ l^- \end{pmatrix}_L$}& \multirow{2}{*}{$1 $} & \multirow{2}{*}{$2 $}& \multirow{2}{*}{$-\frac{1}{2} $}&$+\frac{1}{2}$&$0$\\
        &&&& $+\frac{1}{2}$&$-1$\\ \hline
        $l_{R}$        & $1 $ & $1 $& $-1 $ & $0$ & $-1$ \\ \hline
        \multirow{2}{*}{$\Phi=\begin{pmatrix}\Phi^+\\ \Phi^0\end{pmatrix}$} & \multirow{2}{*}{$1$} & \multirow{2}{*}{$2$} & \multirow{2}{*}{$+\frac{1}{2}$} & $+\frac{1}{2}$ & $+1$ \\ 
        &&&& $-\frac{1}{2}$ & $0$ \\ \hline \hline
    \end{tabular}
\end{table}
The weak interaction in the SM is chiral, i.e., the left chiral and the right chiral fermion fields given by
\begin{equation}\label{eq:intro-chiral-Psi}
\psi_{L/R}=\frac{1}{2}(1\mp \gamma_5)\psi=P_{L/R} \psi
\end{equation}
transform differently under $SU(2)_L$ group. Due to the chirality, each of the three fermion generations come with five different representation as $q_L$, $q_{uR}$, $q_{dR}$, $l_L$, and $l_R$, shown in Table~\ref{tab:intro-gaug-quantum-number} along with their quantum numbers.
The left-handed fields transform as doublet, while the right-handed fields transform as singlet
under the $SU(2)_L$ group. All quarks, having three color degrees of freedom, transform as triplet under the $SU(3)_C$ subgroup equally for left-handed and right-handed fields. The leptons remain singlet under the $SU(3)_C$. 

The gauge field mediators arise naturally when the local gauge invariance is imposed on the free Lagrangian. For example, the free quantum electrodynamic (QED) Lagrangian
\begin{equation}\label{eq:ungauged-LEQD}
\Lag_{0}= i\ol{\psi}(x)\left(\s{\p} -m  \right)\psi(x)~~~~~~ (\s{\p}=\gamma^\mu\p_\mu)
\end{equation}
is invariant under the {\em local}\footnote{{\em Local} or space-time dependence is required as the space time dependent  phase of charged fields should not be observable.} $U(1)$ transformations
\begin{equation}
\psi(x)~~\xrightarrow[]{U(1)}~~\psi^\prime(x)\equiv \exp\{iQ\theta(x)\}\psi(x)
\end{equation}
if 
\begin{equation}\label{eq:intro-COD}
\p_\mu ~~\to~~ \D_\mu\equiv \p_\mu - i e Q A_\mu(x),~~A_\mu(x)~~\to~~A_\mu^\prime(x)\equiv A_\mu(x) + \frac{1}{e}\p_\mu\theta(x),
\end{equation}
i.e., introducing a new spin-$1$ field $A_\mu(x)$ which is realized as
photon in Nature. The local gauge invariance, thus, of a Lagrangian  demands a gauge field or gauge mediator. For the $A_\mu(x)$ to be a free propagating field one needs to add the gauge 
invariant kinetic term for it as
\begin{equation}
\Lag_{kin} = -\frac{1}{4} F_{\mu\nu}(x) F^{\mu\nu}(x),
\end{equation}
where $F_{\mu\nu}=\p_\mu A_\nu - \p_\nu A_\mu$ is the usual electro magnetic field strength
tensor.
A possible mass term of  $A_\mu(x)$  of the form $m_A^2 A^\mu A_\mu$ is forbidden as it breaks the gauge invariance.
The total gauge invariant QED Lagrangian
\begin{eqnarray}\label{eq:intro-LagQED}
\Lag_{QED}=i\ol{\psi}(x)\left(\s{\D} -m  \right)\psi(x)-\frac{1}{4} F_{\mu\nu}(x) F^{\mu\nu}(x)
\end{eqnarray}
gives the well known Maxwell equations:
\begin{equation}
\p_\mu F^{\mu\nu}=J^\nu,~~~~~~ j^\nu = -e Q \ol{\psi}\gamma^\nu \psi. 
\end{equation}
The electromagnetic (EM) current $j^\nu$  couple to the photon field $A_\nu$ with the 
interaction term:
\begin{equation}\label{eq:intro-QED-int}
\Lag_{QED(int)} = A_\nu J^\nu = eQA_\nu(x) \ol{\psi}(x)\gamma^\nu\psi(x).
\end{equation}

The electromagnetic interaction and the weak interaction are unified as the electroweak  theory
under the gauge group $SU(2)_L\otimes U(1)_Y$. For the $U(1)_Y$ abelian gauge symmetry a $B_\mu$ field with
generator $Y$ (hypercharge) is introduced (similar to $A_\mu$ in QED case); for the $SU(2)_L$ non abelian gauge symmetry,  three ($2^2-1=3$) gauge fields $W_\mu^a$ with
three generators $T^a=\sigma^a/2$  are required. 
The strong interaction, however, remain unbroken requiring eight ($3^2-1=8$) gauge field $G_\mu^c$ with eight
generators $\mathbb{T}^a=\lambda^a/2$ for the $SU(3)_C$ gauge invariance.
The {\em local} gauge invariance of the full  $SU(3)_C\otimes SU(2)_L\otimes U(1)_Y$ symmetry
require the co-variant derivative $\D_\mu$ in Eq.~(\ref{eq:intro-COD}) to be 
\begin{eqnarray}\label{eq:Dmu}
\D_\mu &=& \p_\mu - i g^\prime Y B_\mu - i g\frac{1}{2}\sigma^a W_\mu^a - i g_s \frac{1}{2} \lambda^a G_\mu^a.
\end{eqnarray}
Here $g^\prime$, $g$ and $g_s$ are the couplings constant in $U(1)_Y$, $SU(2)_L$ and $SU(3)_C$ subgroup, respectively. The $\sigma^a$ and $\lambda^a$ are Pauli matrices (see Eq.~(\ref{eq:app:pauli-sigma})) and eight Gell-Mann matrices, respectively. The non-Lorentz  indices $a$ on the gauge field  $W$ and $G$ run on $SU(2)_L$
flavour space and the color space, respectively.  The generators $Y$, $T^a$, and the $\mathbb{T}^a$ follow the  algebra:
\begin{eqnarray}
\l[\mathbb{T}^a,\mathbb{T}^b\r]=if^{abc}\mathbb{T}^c,~~\l[T^a,T^b \r]=i\ep^{abc}T^c,~~
\l[\mathbb{T}^a,T^b\r]=\l[\mathbb{T}^a,Y\r]=\l[T^a,Y\r]=0.
\end{eqnarray}

The gauge invariant  matter Lagrangian of the SM is given by
\begin{equation}\label{eq:intro-Lmatter}
\Lag_{matter}= i \ol{q}_L \s{\D} q_L + i \ol{l}_L\s{\D}^{(l_L)} l_L + i \ol{q_u}_{R} \s{\D}^{(q_R)} {q_u}_R + i \ol{q_d}_R \s{\D}^{(q_R)} {q_d}_R + i \ol{l}_R \s{\D}^{(l_R)} l_R,
\end{equation}
where $\D^{(l_L)} $, $\D^{(q_R)}$, and  $\D^{(l_R)}$ are given by (according to Table~\ref{tab:intro-gaug-quantum-number}),
\begin{eqnarray}\label{eq:different-covrainat-derivatives}
\D_\mu^{(l_L)} &=& \p_\mu - i g^\prime Y B_\mu - i g\frac{1}{2}\sigma^a W_\mu^a , \nonumber\\
\D_\mu^{(q_R)} &=& \p_\mu - i g^\prime Y B_\mu  - i g_s \frac{1}{2} \lambda^a G_\mu^a , \nonumber\\
\D_\mu^{(l_R)} &=& \p_\mu - i g^\prime Y B_\mu .
\end{eqnarray}

The Lagrangian with kinetic terms for the gauge fields is 
\begin{equation}\label{eq:intro-Lgauge}
\Lag_{gauge} = -\frac{1}{4g^{\prime^2}}B^{\mu\nu}B_{\mu\nu}  -\frac{1}{4g^2}W_a^{\mu\nu}W_{\mu\nu}^a -\frac{1}{4g_s^2}G_a^{\mu\nu}G_{\mu\nu}^a
\end{equation} 
with the field strength tensors given by
\begin{eqnarray}\label{eq:Fmunu}
G_{\mu\nu}^a&=&g_s \left( \p_\mu G_\nu^a -\p_\nu G_\mu^a + g_s f^{abc} G_\mu^b G_\nu^c\right),\\
W_{\mu\nu}^a&=&g\left( \p_\mu W_\nu^a -\p_\nu W_\mu^a + g \ep^{abc} W_\mu^b W_\nu^c\right),\\
B_{\mu\nu} &=&g^\prime \left( \p_\mu B_\nu -\p_\nu B_\mu\right).
\end{eqnarray}
The mass terms for the gauge fields can not be added by hand as they break the gauge invariance. Fermion masses  can also not be added as they would generate left-chiral and right-chiral mixing breaking the gauge invariance explicitly. 
A scalar doublet under $SU(2)$, the Higgs scalar field (quantum number given in Table~\ref{tab:intro-gaug-quantum-number}),
\begin{equation}
\Phi=\begin{pmatrix}
\Phi^+\\ \Phi^0
\end{pmatrix}
\end{equation}
comes to rescue by  generating mass of the heavy gauge bosons  by spontaneous symmetry breaking (SSB). The gauge invariant Higgs Lagrangian is given by,
\begin{eqnarray}\label{eq:intro-LHiggs}
\Lag_{Higgs} = \l( \D_\mu^{(l_L)} \Phi \r)^\dagger \l( \D_\mu^{(l_L)} \Phi \r) - \mu^2\Phi^\dagger\Phi -\lambda\l(\Phi^\dagger\Phi\r)^2 .
\end{eqnarray} 
The potential
\begin{equation}\label{eq:intro-HiggsPotential}
V(\Phi) =  \mu^2\Phi^\dagger\Phi + \lambda\l(\Phi^\dagger\Phi\r)^2
\end{equation}
get a non-vanishing vacuum expectation value (VEV) for $\mu^2<0$ as,
\begin{equation}\label{eq:intro-VEV}
\langle \Phi \rangle= \frac{{\sf v}}{\sqrt{2}}  =\frac{1}{\sqrt{2}} \sqrt{\frac{-\mu^2}{\lambda}},
\end{equation}
which break the $SU(2)_L\otimes U(1)_Y$ symmetry spontaneously down to $U(1)$ electro magnetic
with generator $Q_{EM}=T_3+Y$.
The scalar field $\Phi$, after removing the would-be Goldstone boson, can be expressed as,
\begin{equation}
\Phi=\frac{1}{\sqrt{2}}
\begin{pmatrix}
0 \\ {\sf v}+h 
\end{pmatrix},
\end{equation}
$h$ being the excitation around the minima. The $h$ turn out to be a physical degree of freedom,
 the Higgs boson with mass 
\begin{equation}
m_h=\sqrt{2\lambda}{\sf v}.
\end{equation}
After the SSB the kinetic term  $\l( \D_\mu \Phi \r)^\dagger \l( \D_\mu \Phi \r)$ of the Higgs field gives the mass terms of the physical gauge bosons $W^\pm$ and $Z$ as,
\begin{equation}\label{eq:intro-Lmass}
\Lag_{mass} = \frac{1}{4}g^2{\sf v}^2 {W^\pm}^\mu W_\mu^\mp  + \frac{1}{8}\l(g^2+{g^\prime}^2\r){\sf v}^2 Z^\mu Z_\mu + 0\times (A^\mu A_\mu)
\end{equation}
after rewriting the gauge field in their mass basis as,
\begin{eqnarray}\label{eq:intro-WZA}
W^\pm &=& \frac{1}{\sqrt{2}} \l( W_1 \mp i W_2\r),\nonumber\\
Z^\mu &=& \cos\theta_W W_3^\mu − \sin\theta_W B^\mu,\nonumber\\
A^\mu &=& \sin\theta_W W_3^\mu + \cos\theta_W B^\mu
\end{eqnarray}
with 
\begin{equation}
\tan\theta_W= \frac{g^\prime}{g} .
\end{equation}
The $\theta_W$,  called the weak mixing angle, 
represents a rotation angle from the ``interaction'' basis (where fields have well-
defined transformation properties under the gauge symmetry), $W_3^\mu$ and $B^\mu$, into their mass basis for
the vector bosons, $Z^\mu$ and $A^\mu$. 
The $W$ and $Z$, thus, acquire  masses as,
\begin{eqnarray}
m_W=\frac{1}{2}g {\sf v},~~ m_Z= \frac{1}{2}\sqrt{ \l(g^2+{g^\prime}^2\r)  } {\sf v},
\end{eqnarray}
wile the photon $A^\mu$  remain massless.
The $\theta_W$ provides a relation between the vector boson masses with the parameter
 \begin{equation}\label{eq:rho-param}
\rho= \frac{m_W^2}{m_Z^2\cos^2\theta_W}=1
\end{equation}
at the tree level.
The fermions also get their masses after SSB via  the Yukawa  terms given by,
\begin{eqnarray}\label{eq:intro-LYukawa}
\Lag_{Yukawa}& =& -Y_{ij}^u\ol{q_{Li}}{q_u}_{Rj}\tilde{\Phi} - Y_{ij}^d \ol{q_{Li}}{q_d}_{Rj}\Phi  - Y_{ij}^l\ol{l_{Li}} l_{Rj} \Phi +h.c,~
\tilde{\Phi}=i\sigma_2 \Phi^\star.
\end{eqnarray}
The Yukawa matrix $Y_{ij}$ are in general $3\times 3$ matrix of dimensionless 
couplings, and can be chosen in diagonal basis  as,
\begin{equation}
Y^{u} = diag(y_u,y_c,y_t),~~
Y^{d} = diag(y_d,y_s,y_b),~~
Y^{l} = diag(y_e,y_\mu,y_\tau)
\end{equation}
without any loss of generality.
After SSB, the $\Lag_{Yukawa}$ generates the mass terms
\begin{equation}
\Lag_{Yukawa(mass)} = -\frac{y_l {\sf v}}{\sqrt{2}}\ol{l}_L l_R  -\frac{y_q {\sf v}}{\sqrt{2}}\ol{q}_L q_R
\end{equation}
with the fermion masses given by,
\begin{equation}
m_l = \frac{y_l {\sf v}}{\sqrt{2}},~~ m_q = \frac{y_q {\sf v}}{\sqrt{2}}.
\end{equation}

The total SM Lagrangian is thus given by,
\begin{equation}\label{eq:intro-LSM}
\Lag_{SM} = \Lag_{matter} + \Lag_{gauge} + \Lag_{Higgs} + \Lag_{Yukawa}.
\end{equation}

The SM Lagrangian needs the gauge fixing of the field strength tensors required
to have a finite propagator and also the ghost term for the $s$-matrix to remain unitary. The gauge fixing terms and the ghosts do not appear in the physical observables.

\subsection{The interactions in the SM electroweak theory}
The matter part $\Lag_{matter}$ in Eq.~(\ref{eq:intro-Lmatter}) and gauge part $\Lag_{gauge}$ in Eq.~(\ref{eq:intro-Lgauge}) of the SM Lagrangian  generates interactions between the 
gauge bosons and fermions as well as the gauge bosons themselves, which are discussed in the next subsection. The relevant  Feynman rules for this thesis in the
electroweak sector (excluding quartic gauge couplings) are given in appendix~\ref{appendix:intro} for completeness.

\subsubsection{Charged and neutral current interactions }
The neutral current and the charge current interactions of the fermions with the electroweak gauge bosons
arise from the Lagrangian $\Lag_{matter}$ in Eq.~(\ref{eq:intro-Lmatter}) and they are given by,
\begin{equation}\label{eq:intro-LEWint}
\Lag_{int} = - i g^\prime B_\mu \sum_{j=L,R} Y_j \ol{\psi}_j\gamma^\mu\psi_j 
-i g \ol{\psi}_L\gamma^\mu T^a W_\mu^a  \psi_L,
\end{equation}
where $\psi$ represents the fermion field given in Table~\ref{tab:intro-gaug-quantum-number}.
The terms containing in the $SU(2)_L$ matrix
\begin{equation}
\frac{1}{2}\sigma^a W_\mu^a = \frac{1}{2}
\begin{pmatrix}
 W_\mu^3 & \sqrt{2}W_\mu^+ \\
\sqrt{2}W_\mu^- & -  W_\mu^3
\end{pmatrix}
\end{equation} 
give the charged current interactions
\begin{equation}\label{eq:intro-LEWCC}
\Lag_{CC} = \frac{g}{2\sqrt{2}} \left\{ W_\mu^+\left[ \bar{u}\gamma^\mu\left(1-\gamma_5\right)d
+ \bar{\nu}_e \gamma^\mu \left(1-\gamma_5\right) e  \right] +h.c. \right\}
\end{equation}
for a single family of quarks and leptons, see Eq.~(\ref{eq:intro-doublet-fermion}).
Gauge symmetry thus brings the universality of the quarks and leptons interacting with the charge
gauge bosons $W^\pm$.
For the quarks, the mass basis (diagonal Yukawa matrix) are not the same as the
interaction basis. When one writes the weak eigenstates of the quarks in their mass basis, a unitary matrix $V$,  called the Cabibbo-Kobayashi-Maskawa (CKM) matrix, arises in the quark charged current interaction as,
\begin{equation}\label{eq:intro-LEWCC-CKM}
\Lag_{CC(CKM)} = \frac{g}{2\sqrt{2}} \left\{ W_\mu^+ \sum_{i,j}\bar{q}_{ui}\gamma^\mu\left(1-\gamma_5\right) V_{ij} q_{dj}  + h.c. \right\}.
\end{equation}
 The CKM matrix $V$ couples any up-type quark ($q_u$) to all down-type
 quarks ($q_d$).

The interaction term $\Lag_{int}$ in Eq.~(\ref{eq:intro-LEWint}) also contains  the
interactions of fermions with the neutral gauge fields $W_\mu^3$ and $B_\mu$, which are related
to the physical neutral gauge bosons $Z_\mu$ and $A_\mu$, as  given in Eq.~(\ref{eq:intro-WZA}).
The neutral interactions in terms of the physical gauge bosons are given by,
\begin{equation}\label{eq:LNC}
\Lag_{NC}= - \sum_{j=L,R} \ol{\psi}_j \gamma^\mu \left\{ A_\mu\left[g\frac{\sigma^3}{2}\sin\theta_W + g^\prime Y_j \cos\theta_W  \right] 
+ Z_\mu \left[ g\frac{\sigma^3}{2}\cos\theta_W - g^\prime Y_j \sin\theta_W  \right] \right\}\psi_j.
\end{equation}
To recover the QED interaction in Eq.~(\ref{eq:intro-QED-int}) containing the $A_\mu$ term in the above equation, one needs to impose the conditions
\begin{equation}
 g\sin\theta_W=g^\prime\cos\theta_W = e,~~~  Y=Q-T^3,
\end{equation}
where  $Q$ denote the charge operator as,
\begin{equation}
Q_L = \begin{pmatrix}
Q_{u/\nu} & 0 \\ 0 & Q_{d/e}
\end{pmatrix}
, ~~~~ Q_R = Q_{u/\nu} , ~ Q_{d/e} .
\end{equation}
The first identity relates the observable EM coupling $e$ to the couplings of the unified electroweak theory. The second identity provides the hypercharges of the fermions in terms of 
their electric charges and weak iso-spin, given in Table~\ref{tab:intro-gaug-quantum-number}.
A hypothetical right handed neutrino would not interact with $W^\pm$ boson as both of it's electric chareg and  weak hypercharge  will be zero.
The neutral current interaction containing the $Z$ boson is given by,
\begin{equation}\label{eq:LZNC}
\Lag_{NC(Z)} = \frac{e}{2\sin\theta_W\cos\theta_W} Z_\mu \sum_{j=L,R} \ol{\psi}_j \gamma^\mu (\sigma^3-2\sin^2\theta_W Q_j)\psi_j
\end{equation}
which can be simplified as,
\begin{equation}\label{eq:LZNC-simplified}
\Lag_{NC(Z)} = \frac{g_Z}{2} Z_\mu \sum_{f} \bar{f} \gamma^\mu ({\sf v}_f - a_f\gamma_5) f, ~~~g_Z=\frac{g}{\cos\theta_W},
\end{equation}
where axial coupling $a_f$ and vector couplings $v_f$ are given by,
\begin{equation}\label{eq:intro-afvf}
a_f = T_f^3,~~ {\sf v}_f = T_f^3  - 2 Q_f \sin^2\theta_W .
\end{equation}
The values of $a_f$ and ${\sf v}_f$ of all the fermions can be deduced from the Table~\ref{tab:intro-gaug-quantum-number} and they are given in Table~\ref{tab:intro-vfaf}.
\begin{table}[h]
\caption{ \label{tab:intro-vfaf} Value of axial coupling $a_f$ and vector couplings ${\sf v}_f$ to the $Z$ boson given in Eq.~(\ref{eq:LZNC-simplified}) \&~(\ref{eq:intro-afvf}).}
\centering
\renewcommand{\arraystretch}{1.5}
\begin{tabular}{|c|c|c|c|c|}\hline
& $q_u$ & $q_d$ &$\nu_l$ & $l$ \\ \hline
$a_f$ & $\frac{1}{2}$ &$-\frac{1}{2}$ &$\frac{1}{2}$ & $-\frac{1}{2}$ \\ \hline 
${\sf v}_f$ & $ \frac{1}{2}-\frac{4}{3} \sin^2\theta_W$ & $ -\frac{1}{2}+\frac{2}{3} \sin^2\theta_W$
&$\frac{1}{2} $ &  $ -\frac{1}{2}+2 \sin^2\theta_W$ \\ \hline 
\end{tabular}
\end{table}

\subsubsection{Gauge self couplings}
The gauge part $\Lag_{gauge}$ in Eq.~(\ref{eq:intro-Lgauge}) of the electroweak theory
generates cubic and quartic  interactions among the gauge bosons in the following form:
\begin{equation}\label{eq:intro-SM-WWV}
\Lag_{WWV} = ig_{WWV}\left[\left(W_{\mu\nu}^+W^{-\mu}-W^{+\mu}W_{\mu\nu}^-\right)V^\nu
+ W_\mu^+W_\nu^-V^{\mu\nu}\right],
\end{equation}
\begin{eqnarray}\label{eq:intro-SM-WWVV}
\Lag_{W^4/W^2V^2} &=& -\frac{g^2}{2} \left[ \left( W_\mu^+W^{-\mu} \right)^2 - W_\mu^+W^{+\mu} W_\nu^-W^{-\nu} \right] \nonumber \\ 
&-& g_{WWV_1V_2} \left[  W_\mu^+W^{-\mu} V_{1\nu}V_2^\nu - W_\mu^+ V_1^\mu W_\nu^- V_2^\nu \right],
\end{eqnarray}
with $V =Z/A$ and
\begin{eqnarray}
&W_{\mu\nu}^\pm = \partial_\mu W_\nu^\pm - \partial_\nu W_\mu^\pm,~~
V_{\mu\nu} = \partial_\mu V_\nu - \partial_\nu V_\mu,& \nonumber\\
&g_{WW\gamma}=-g\sin\theta_W,~~  g_{WWZ}=-g\cos\theta_W,&\nonumber\\
&g_{WWZZ} = \left( g\cos\theta_W \right)^2,~~
g_{WWAA} = \left( g\sin\theta_W \right)^2,~~  g_{WWAZ/WWZA}=g^2\sin\theta_W\cos\theta_W.&\nonumber\\
\end{eqnarray}
There are  no gauge couplings   among only the neutral gauge bosons  in the SM at tree level. However, higher order corrections can generate neutral triple gauge
boson vertex and also can contribute to the existing tree level cubic and quartic gauge 
boson self couplings. Higher dimensional effective operators of the SM fields
and new physics effect can produce triple and quartic gauge boson couplings
beyond the SM tree level structures, which we will discuss in section~\ref{sec:intro-EFT}.  

\subsubsection{Higgs couplings}
The  Lagrangian $\Lag_{Higgs}$ includes Higgs self couplings as well as
Higgs to gauge boson couplings in the form:
\begin{equation}\label{eq:Higss-couplings}
\Lag_{Higgs(int)} = -\frac{m_h^2}{2 {\sf v}} h^3 - \frac{m_h^2}{8 {\sf v}^2} h^4 
+m_W^2 W_\mu^-W^{+\mu} \left( \frac{2h}{{\sf v}} + \frac{h^2}{{\sf v}^2} \right)
+\frac{1}{2}m_Z^2 Z_\mu Z^{\mu} \left( \frac{2h}{{\sf v}} + \frac{h^2}{{\sf v}^2} \right).
\end{equation}
The Higgs ($h$) couples to the gauge bosons proportional to their masses. There are no $hAA$ nor $hhAA$ couplings present in the SM, as the $h$ has zero EM charge
and should not couple to EM force carrier. Another way of looking the absence is 
that the photon is massless, and hence it's  coupling to Higgs is zero.
The fermion couples to the Higgs thorough the Lagrangian $\Lag_{Yukawa}$ in the form:
\begin{equation}
\Lag_{Yukawa(int)} = -\frac{h}{{\sf v}} \left[ m_l \ol{l}_L l_R  + m_q \ol{q}_L q_R\right].
\end{equation}
The  Higgs to fermion couplings are also  proportional to   the fermion masses.

\subsection{Summary of the SM}
The SM beautifully accommodates the electroweak and the strong interactions under the 
$SU(3)_C\otimes SU(2)_L\otimes U(1)_Y$ gauge group predicting or explaining almost all the experimental facts.    
All the theoretical predictions of the  SM, such as the weak neutral current; existence and masses of $W$ and $Z$ bosons; the existence of   $\tau$, $c$, $b$, and $t$  were discovered at various colliders
with spectacular accuracy. The particle spectrum of the SM got completed with the discovery of the last milestone,
the Higgs boson,  at the  LHC in 2012. The $\rho$ parameter (in Eq.~(\ref{eq:rho-param})), predicted
to be $1$ by the SM, was confirmed with great accuracy at colliders. 
It receives a small perturbative correction, however,  related to a broken $SU(2)_L×SU(2)_R$ ``custodial”
symmetry of the Higgs potential~\cite{Diaz:2001yz}. 
However, the SM is believed to be not a complete theory for various issues, which are discussed in the next section. 
\section{Problems in the SM with possible solutions}\label{sec:intro-problem-SM}
Although the SM is a highly successful theory, it suffers various problems within the theoretical framework. It is  also unable to explain many observed phenomena. The SM requires a large number of arbitrary free parameters ($19$ free parameters~\cite{Thomson:2013zua}), too much for a fundamental theory believed by most physicists. The other fundamental problems are discussed below.

\paragraph{Gauge problem} 
The SM has three separate gauge couplings; they do not unify at high scale after running. We do not have an explanation for the electroweak part only being chiral. The SM includes but does not explain the charge quantization, i.e., why electric charges are multiple of $e/3$.
One possible solution includes grand unified theory (GUT) which predicts the existence of 
magnetic monopole and decay of the proton (e.g., see Refs.~\cite{Buras:1977yy,Langacker:1980js}) not observed yet.

\paragraph{Fermion problem} 
The SM contains three generations of fermions, 
whereas $e^-$, $u$, $d$ alone from the first generation 
make up all the visible matter in the Universe. There is no suitable explanation of the existence of the other 
heavier fermions. More ever, the mass of the fermions are input by hand or from the experiment; they do not originate from the SM.
Possible solutions are given as an extension of the SM, such as the model of  composite fermions, the model of radiative hierarchies where the fermion masses are generated at the loop-level~\cite{Balakrishna:1988ks,Babu:1990vx}, model of extra dimension~\cite{Fujimoto:2017lln,Scrucca:2003ra,Ahmed:2019zxm}, etc.

\paragraph{Higgs mass hierarchy}
The Higgs mass gets  divergent corrections from the top quark loop as
\begin{equation}
m_H^2=\l(m_H^2\r)_0 + {\cal O}\l(\Lambda_{uv}^2\r),
\end{equation} 
where the $\Lambda_{uv}$ is the energy  scale of a ultraviolet complete theory, e.g., the  Planck scale (or the grand unification scale).
Thus, the natural scale of $m_H$ is   ${\cal O}(\Lambda_{uv})$, which is much larger than the observed values of $m_H=125$ GeV.
There must be an incredibly unnatural fine tuning to cancel the quadratic corrections to the tree level mass or bare mass. One solution could be to forget about the elementary Higgs field and consider
the dynamical generation of mass: technicolor and composite Higgs model are in favour to this~\cite{Hill:2002ap,Bellazzini:2014yua}. Various other models of SM extension (e.g., extra dimension)  also shed  light on it. The most compelling solution is the supersymmetry (SUSY) for the hierarchy problem. In SUSY, each fermion and boson have their bosonic and fermionic superpartner: the quadratic divergent from the fermionic loop of the Higgs mass cancels way  by the bosonic superpartner loop. The superpartners have not been observed yet, but not ruled out. 

\paragraph{Strong $CP$ problem}
In the current mathematical formulation of quantum chromodynamics (QCD), $CP$ can be violated
by including the term $\theta\times G_{\mu\nu}\wtil{G}^{\mu\nu}$ ($\wtil{G}_{\mu\nu}=1/2\epsilon_{\mu\nu\alpha\beta}G^{\alpha\beta}$ is the dual field) in to the Lagrangian. This would induce electric dipole moment to
neutron ($n$), but the very small value of this put a stringent limit on $\theta$ to be ${\cal O}(10^{-10})$~\cite{Langacker:1995hi}. Thus, QCD does not violate $CP$; whereas, there is no explanation for that in the SM.
There are several extensions of the SM  to solve the strong $CP$ problem. The most well-known solution is the  Peccei–Quinn mechanism involving a new pseudo-scalar (imposing a $U(1)$ {\em global} symmetry) particle named axion~\cite{Peccei:1977hh,Weinberg:1977ma}. The axions are being searched at various experiments, but no evidence of its existence has been found yet~\cite{Graham:2015ouw,Irastorza:2018dyq}.  


\paragraph{Neutrino oscillation}
In the SM, the neutrinos are massless. However, compelling evidences~\cite{Fukuda:1998mi,Ahmad:2001an,Eguchi:2002dm,Allison:1999ms,Cleveland:1998nv} are there for neutrino oscillation suggesting small neutrino mass and   mixing. The SM does not 
provide any mechanism for neutrino mass and their mixing.
The most popular mechanism for the small neutrino mass is the seesaw mechanism~\cite{Mohapatra:1979ia,Akhmedov:1999tm,Mohapatra:2004zh}, where two or more right-handed neutrinos with large Majorana mass  are assumed. The right-handed neutrino induces
a very small mass to the left-handed neutrino reciprocal to the heavy mass. Many derivatives of
the  seesaw mechanism exist, such as type I, type II, inverse, etc~\cite{Valle:2006vb} of which
inverse seesaw mechanism has obtained great interest nowadays ~\cite{PhysRevD.34.1642}. 

\paragraph{Dark matter}
To the current belief in accordance with astrophysical and cosmological data, dark matter
(DM) is one of the main ingredients (about $26~\%$) of the Universe~\cite{Hinshaw:2012aka,Aghanim:2018eyx}. The SM does not provide
any fundamental particles that could be good dark matter candidates.
There are plenty of models with extension to the SM postulating  candidates  for  dark matter~\cite{Garrett:2010hd,Morgante:2018tiq,Lin:2019uvt} with extra  dimension~\cite{CoimbraAraujo:2012xw}, supersymmetry, axion, inverse seesaw~\cite{PhysRevD.100.035034}, etc.

\paragraph{Baryogenesis}
The Universe is made out of mostly the matter (baryon).  According to the SM, the matter and anti-matter should be present in equal amounts, resulting in zero baryon number. No sufficient mechanism exists in the SM to explain the matter dominance over anti-matter. One solution for this is to break baryon number symmetry in a GUT theory with mediating a  massing $X$ boson or heavy Higgs ($H^0$) boson~\cite{Sakharov:1967dj}.

\paragraph{}
The various models which address the issues of the problems mentioned above predict
new particles along with new interactions. However, no new resonance beyond the SM has been observed until now at the current reach of energy. We could thus expect that the new physics, undoubtedly necessary, could be standing at a higher energy scale; these new physics
may leave their footprint to the low energy available to us. 
They will modify the interactions among the SM particles or bring new interactions among them.
Precision study of these modified or new interactions, which are called anomalous, in a way, could reveal new physics scenario. The anomalous interactions can be studied considering a given model and also can be modelled by higher dimensional effective operators formed by the SM fields, which goes by the name of effective field theory (EFT). Our aim here is to study the anomalous gauge boson couplings focusing on the EWSB. 
In the next section, we give a brief description of EFT for the anomalous gauge boson self-interactions with contributions from some of the  BSM models.

\section{EFT and anomalous gauge boson couplings}\label{sec:intro-EFT}
\begin{figure}[h]
    \centering
    \includegraphics[width=1\textwidth]{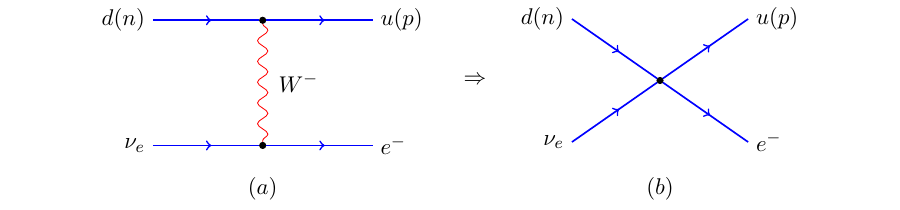}
    \caption{\label{fig:intro-Fermi-theory} Beta decay ($d\to u$ conversion) with $W^-$ (a) exchange and effective four-point interaction (b).}
\end{figure}
To begin with the EFT, let us start with the classic example of Fermi's theory of  weak interaction (beta decay) at low energy.
Fermi was able to describe the beta decay $n(udd) \to p (uud) + e^- +\bar{\nu_e}$
with high level accuracy with a four-point fermion interaction with a coupling constant $G_F$.
The beta decay is proceeded  via the exchange of the   $W^-$ boson at the symmetry breaking scale (VEV). In the low energy, the $W^-$ can not be produced (extreme off-shell $W^-$ is produced) and this interaction effectively boils down to a four-point contact interaction. 
The tree level amplitude for the diagram (a) in Fig.~\ref{fig:intro-Fermi-theory} is
\begin{equation}\label{eq:intro-beta-decay-W}
{\cal M}_W = \left(\frac{-ig}{\sqrt{2}}\right) V_{ud} \left( \bar{u}\gamma^\mu P_L d \right)
\left( \frac{-i g_{\mu\nu}}{p^2-m_W^2} \right)
\left( \bar{e^-}\gamma^\nu P_L \nu_e \right)
\end{equation}
with $V_{ud}$ being the CKM mixing matrix element and $p$ being the momentum transfer of the $W^-$ propagator. In low energy, i.e., $p \ll m_W$, the $W$ propagator can be expanded as,
\begin{equation}
\frac{1}{p^2-m_W^2}=-\frac{1}{m_W^2}\left( 1+ \frac{p^2}{m_W^2} + \frac{p^4}{m_W^4} + \ldots  \right)
\end{equation}
with different order of expansion w.r.t $p/m_W$.  Retaining only the first term, we obtain the
amplitude in Eq.~(\ref{eq:intro-beta-decay-W}) as,
\begin{equation}\label{eq:intro-beta-decay-4point}
{\cal M}_W = \frac{i}{m_W^2}\left(\frac{-ig}{\sqrt{2}}\right) V_{ud} \left( \bar{u}\gamma^\mu P_Ld \right)
\left( \bar{e^-}\gamma_\mu P_L \nu_e \right) + {\cal O}(\frac{1}{m_W^4})
\end{equation}
which resembles the amplitude for the four-point interaction vertex shown in Fig.~\ref{fig:intro-Fermi-theory}(b) as~\cite{aitchison2002gauge},
\begin{equation}\label{eq:intro-beta-decay-4point2}
{\cal M}_{4f} = -\frac{G_F}{\sqrt{2}} V_{ud} \left( \bar{u}\gamma^\mu (1-\gamma_5)d \right)
\left( \bar{e^-}\gamma_\mu (1-\gamma_5) \nu_e \right) .
\end{equation}
One can easily identify the Fermi constant ($G_F$) as,
\begin{equation}
\frac{G_F}{\sqrt{2}} \equiv \frac{g^2}{8 m_W^2}. 
\end{equation}
Thus the effective Lagrangian
\begin{equation}\label{eq:intro-beta-decay-4point-Lag}
{\cal L} = \frac{-4G_F}{\sqrt{2}} V_{ud} \left( \bar{u}\gamma^\mu P_Ld \right)
\left( \bar{e^-}\gamma_\mu P_L \nu_e \right)
\end{equation}
is the low energy limit of the SM. The EFT does not have the dynamical  $W$; however, the effect of $W$
exchange in the SM is incorporated by the  dimension-$6$ four-fermion operator.
New physics beyond the SM can also provide the four-fermion contact interaction
contributing to the Fermi constant $G_F$.
Flavour changing processes such as $\mu \to e$ through  weak current can also be
parametrized by a four fermion contact interaction at low energy.

In EFT, one adds higher dimensional effective operators suppressed by an energy cut-off ($\Lambda$)  with the SM fields and obtain the interactions after symmetry breaking~\cite{Buchmuller:1985jz}. Thus, EFT acts as a bridge between heavy scale new physics and low energy experimental observations.
In EFT, there are two main approaches to follow: top-down and bottom-up.   In the top-down approach, a high energy theory is known (GUT like), and the low energy effective operators are obtained by
integrating out the heavy scale associated with the theory. This approach is thus, model dependent.
The heavy scale gets encoded into the Wilson coefficient or the coupling constants of the remnant interactions. 
In the  bottom-up approach, however, a fundamental theory at high energy is unknown, but one uses the known symmetry and particles of the SM and construct higher dimension operators to model the effects of new physics in a model independent way.   
In the above example, Fermi modelled the weak interaction by four-fermion contact interactions, which falls under the bottom-up approach. We integrated out the $W$ boson
to get the effective four-fermion operator at low energy, which falls under the top-down approach.
In both the way, the effective Lagrangian incorporating the effective higher dimension operators is taken as,
\begin{equation}\label{eq:effective-Lag-SMEFT}
{\cal L}_{eft} = {\cal L}_{SM} +  \sum_{D\ge 4}\sum_i \frac{c_i^{(D)}}{\Lambda^{D-4}} {\cal O}_i^{(D)}  
\end{equation}
encapsulating all the desired operators ($\sum_{i}{\cal O}_i$) at given order ($D$) of $\Lambda$.
Here, $c_i^{(D)}$ are the coefficients of the dimension-$D$ operator ${\cal O}_i^{(D)}$.
Avoiding the baryon number and lepton number violation the effective Lagrangian becomes~\cite{Buchmuller:1985jz}
\begin{equation}\label{eq:effective-Lag-SMEFT-dim6}
{\cal L}_{eft} = {\cal L}_{SM} + \sum_i \frac{c_i^{(6)}}{\Lambda^2}{\cal O}_i^{(6)}
+ \sum_i \frac{c_i^{(8)}}{\Lambda^4}{\cal O}_i^{(8)}  + \dots ,
\end{equation}
which encapsulates only the even order effective operators. As we go to higher and higher order, the effect of them  become lower and lower in the low energy. Almost all of the studies available in literature contain
operators only up to order $\Lambda^{-4}$, of which most of them contain only up to $\Lambda^{-2}$. A complete list  of dimension-$6$ operators (a total of $80$) respecting the SM gauge group has been given
by Buchmuller {\em et} al. in Ref~\cite{Buchmuller:1985jz} in gauge sector, fermion sector, and gauge with fermions.

\begin{figure}
\centering
\includegraphics[width=0.8\textwidth]{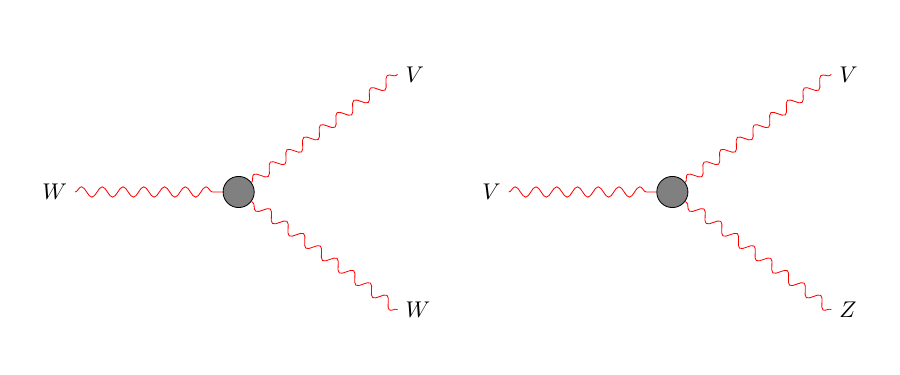}
\caption{\label{fig:intro-TGC-vertex} Triple gauge boson vertex with $V=Z/\gamma$. Anomalous contributions are shown by the shaded blob.}
\end{figure}
The kinetic terms of the gauge fields in the SM Lagrangian (Eq.~(\ref{eq:intro-Lgauge}))
generate interactions of triple gauge bosons $WWV$ ($V=Z/\gamma$) and quartic gauge bosons 
 $WWVV$ due to the  non abelian nature of $SU(2)_L$ symmetry as given in Eqs.~(\ref{eq:intro-SM-WWV}) \&~(\ref{eq:intro-SM-WWVV}).
There are no gauge boson vertices among the neutral gauge bosons, i.e., $ZZZ$, $ZZ\gamma$, $Z\gamma\gamma$, and $\gamma\gamma\gamma$ are not present in the SM.
As we stated earlier, heavy scale new physics may induce new gauge boson vertices and also
modify the SM gauge boson vertices through the effective higher dimension operators.
The new or modified form factors (a function of the momentum of the gauge bosons) are called
anomalous. In the SM, there could be some extra contributions to the gauge boson vertices
through higher order loop corrections. For example, a  triangular diagram with top-quark running in the loop can provide some of the anomalous triple gauge boson couplings. In this thesis, we focus
on the study of the anomalous triple gauge boson vertices (see Fig.~\ref{fig:intro-TGC-vertex})
in the charge sector as well as in the neutral sector with terms up to dimension-$6$ in a model independent way. 

To the lowest order (i.e., upto dimension-$6$), the operators in HISZ basis~\cite{Willenbrock:2014bja} contributing to  $WWV$ couplings,
respecting the SM gauge symmetry, are~\cite{Hagiwara:1993ck,Degrande:2012wf} 
\begin{eqnarray}\label{eq:intro-opertaors-dim6}
   {\cal O}_{WWW}&=&\mbox{Tr}[W_{\mu\nu}W^{\nu\rho}W_{\rho}^{\mu}] ,\nonumber\\
{\cal O}_W&=&(\D_\mu\Phi)^\dagger W^{\mu\nu}(\D_\nu\Phi) ,\nonumber\\
{\cal O}_B&=&(\D_\mu\Phi)^\dagger B^{\mu\nu}(\D_\nu\Phi)  ,\nonumber\\
{\cal O}_{\wtil{WWW}}&=&\mbox{Tr}[\wtil{W}_{\mu\nu}W^{\nu\rho}W_{\rho}^{\mu}]  ,\nonumber\\
{\cal O}_{\wtil W}&=&(\D_\mu\Phi)^\dagger \wtil{ W}^{\mu\nu}(\D_\nu\Phi) ,
\end{eqnarray}
where $\wtil{ W}^{\mu\nu}$ is the dual of  $W^{\mu\nu}$ given by
 $\wtil{W}_{\mu \nu}=1/2 \epsilon_{\mu \nu \rho \sigma}W^{\rho\sigma}$ ($\epsilon^{0123}=+1$).
Among these operators ${\cal O}_{WWW}$,  ${\cal O}_W$ and  ${\cal O}_B$ are
$CP$-even, while ${\cal O}_{\wtil{WWW}}$  and ${\cal O}_{\wtil W}$ are
$CP$-odd. In the neutral sector, however, there are no dimension-$6$ operators contributing to the
neutral triple gauge boson couplings $ZVV$; they appear only at dimension-$8$. The dimension-$8$
operators in HISZ basis under the SM gauge group contributing to the anomalous $ZVV$ are~\cite{Degrande:2013kka},
\begin{eqnarray}\label{eq:intro-dim8-operator-ZVV}
{\cal O}_{BW}&=& i \Phi^\dagger \left(B_{\mu\nu}/g^\prime\right) \left(W^{\mu\rho}/g\right)\{\D_\rho,\D^\nu\}\Phi,\nonumber\\
{\cal O}_{WW}&=& i \Phi^\dagger \left(W_{\mu\nu} W^{\mu\rho}/g^2\right)\{\D_\rho,\D^\nu\}\Phi,\nonumber\\ 
{\cal O}_{BB}&=& i \Phi^\dagger \left(B_{\mu\nu} B^{\mu\rho}/g^{\prime^2}\right)\{\D_\rho,\D^\nu\}\Phi,\nonumber\\ 
{\cal O}_{\wtil{B} W}&=& i \Phi^\dagger \left(\wtil{B}_{\mu\nu}/g^\prime\right) \left(W^{\mu\rho}/g\right)\{\D_\rho,\D^\nu\}\Phi.
\end{eqnarray}
The first three operators are $CP$-odd;  the last one is $CP$-even.

The above operators are invariant under the  $SU(2)\otimes U(1)$ gauge group. In order to
 establish the SM couplings and/or to capture new physics irrespective of any symmetry, one has to go beyond the SM gauge symmetry. 
In this way, one can only consider the Lorentz invariance and $U(1)_{EM}$ symmetry to construct
more general form factors as a function of the momentum involved in a given vertex. In the form factor formalism, the effective $WWV$ and $ZVV$ Lagrangian are given by~\cite{Hagiwara:1986vm},
 \begin{eqnarray} \label{eq:intro-LWWV}
 {\cal L}_{WWV} &=&ig_{WWV}\Bigg[ g_1^V(W_{\mu\nu}^+W^{-\mu}-
 W^{+\mu}W_{\mu\nu}^-)V^\nu
 +ig_4^VW_\mu^+W^-_\nu(\partial^\mu V^\nu+\partial^\nu V^\mu)\nonumber\\
 &-&ig_5^V\epsilon^{\mu\nu\rho\sigma}(W_\mu^+\partial_\rho W^-_
 \nu-\partial_\rho W_\mu^+W^-_\nu)V_\sigma
 +\frac{\lambda^V}{m_W^2}W_\mu^{+\nu}W_\nu^{-\rho}V_\rho^{\mu}\nonumber\\
 &+&\frac{\wtil{\lambda^V}}{m_W^2}W_\mu^{+\nu}W_\nu^{-\rho}\wtil{V}_\rho^{\mu}
 +\kappa^V W_\mu^+W_\nu^-V^{\mu\nu}+\wtil{\kappa^V}W_\mu^+W_\nu^-\wtil{V}^{\mu\nu}
 \Bigg] 
 \end{eqnarray}
 and 
 \begin{eqnarray}\label{eq:intro-LZVV-full}
 {\cal L}_{ZVV} = \frac{e}{m_Z^2} \Bigg [
 &-&\bigg[f_4^\gamma (\partial_\mu F^{\mu \beta})+
 f_4^Z (\partial_\mu Z^{\mu \beta}) \bigg] Z_\alpha
 ( \partial^\alpha Z_\beta)+
 \bigg[f_5^\gamma (\partial^\sigma F_{\sigma \mu})+
 f_5^Z (\partial^\sigma Z_{\sigma \mu}) \bigg] \wtil{Z}^{\mu \beta} Z_\beta
 \nonumber \\
 &-&  \bigg[h_1^\gamma (\partial^\sigma F_{\sigma \mu})
 +h_1^Z (\partial^\sigma Z_{\sigma \mu})\bigg] Z_\beta F^{\mu \beta}
 -\bigg[h_3^\gamma  (\partial_\sigma F^{\sigma \rho})
 + h_3^Z  (\partial_\sigma Z^{\sigma \rho})\bigg] Z^\alpha
 \wtil{F}_{\rho \alpha}
 \nonumber \\
 &- & \left \{\frac{h_2^\gamma}{m_Z^2} \bigg[\partial_\alpha \partial_\beta
 \partial^\rho F_{\rho \mu} \bigg]
 +\frac{h_2^Z}{m_Z^2} \bigg[\partial_\alpha \partial_\beta
 (\square +m_Z^2) Z_\mu\bigg] \right \} Z^\alpha F^{\mu \beta}\nonumber \\
 &+& \left \{
 \frac{h_4^\gamma}{2m_Z^2}\bigg[\square \partial^\sigma
 F^{\rho \alpha}\bigg] +
 \frac{h_4^Z}{2 m_Z^2} \bigg[(\square +m_Z^2) \partial^\sigma
 Z^{\rho \alpha}\bigg] \right \} Z_\sigma \wtil{F}_{\rho \alpha }
 \Bigg ],
 \end{eqnarray}
respectively. 
In the SM $g_1^V=1$, $\kappa^V=1$  and other couplings are zero, see Eq.~(\ref{eq:intro-SM-WWV}).
In the charge sector, the couplings $g_1^V$, $\kappa^V$
and $\lambda^V$  are $CP$-even (both $C$ and $P$-even), 
while $g_4^V$ (odd in $C$, even in $P$), $\wtil{\kappa^V}$
and $\wtil{\lambda^V}$  (even in $C$, odd in $P$) are $CP$-odd.
The $g_5^V$ is, however, both $C$ and $P$-odd making it $CP$-even.
In the neutral sector, the couplings $f_4^V$, $h_1^V$, $h_2^V$
correspond to the CP-odd form factors, while $f_5^V$, $h_3^V$, $h_4^V$
correspond to the CP-even ones. We note that although the Lagrangain in Eq.~(\ref{eq:intro-LZVV-full}) provieds a $Z\gamma\gamma$
couplings as given in Fig.~\ref{fig:intro-TGC-vertex}, this does not appear in a two on-shell photons production process as it is forbidden by Yang-Landau theorem~\cite{Landau:1948kw,Yang:1950rg}.

The couplings of the form factors in Eq.~(\ref{eq:intro-LWWV}) \&  Eq.~(\ref{eq:intro-LZVV-full})
are related to the couplings of the operators in Eq.~(\ref{eq:intro-opertaors-dim6}) 
\& Eq.~(\ref{eq:intro-dim8-operator-ZVV}), respectively when $SU(2)\otimes U(1)$ 
gauge invariance is assumed. The relations between form factor couplings and operator 
couplings  in the charge sector are given by~\cite{Hagiwara:1993ck,Wudka:1994ny,Degrande:2012wf},
\begin{eqnarray}
\Delta g_1^Z & = & c_W\frac{m_Z^2}{2\Lambda^2} ,\nonumber\\
g_4^V &=& g_5^V=\Delta g_1^\gamma=0 ,\nonumber\\
\lambda^\gamma & = & \lambda^Z=\lambda^V = c_{WWW}\frac{3g^2m_W^2}{2\Lambda^2} ,\nonumber\\
\wtil{\lambda^\gamma} & = & \wtil{\lambda^Z}=\wtil{\lambda^V} = c_{\wtil{WWW}}\frac{3g^2m_W^2}{2\Lambda^2} ,\nonumber\\
\Delta\kappa^\gamma & = & (c_W+c_B)\frac{m_W^2}{2\Lambda^2} ,\nonumber\\
\Delta\kappa^Z & = & (c_W-c_B\tan^2\theta_W)\frac{m_W^2}{2\Lambda^2} ,\nonumber\\
\wtil{\kappa^\gamma} & = &
c_{\wtil{W}}\frac{m_W^2}{2\Lambda^2} ,\nonumber\\
\wtil{\kappa^Z} & = &
-c_{\wtil{W}}\tan^2\theta_W\frac{m_W^2}{2\Lambda^2} \ \ .
\label{eq:intro-Operator-to-Lagrangian}
\end{eqnarray}
It is clear from above that some of the vertex factor couplings are dependent on each others in the SM gauge symmetry and 
they are 
\begin{eqnarray}
&&\Delta g_1^Z=\Delta \kappa^Z + \tan^2\theta_W \Delta \kappa^\gamma ,\nonumber\\
&&\wtil {\kappa^Z} + \tan^2\theta_W \wtil{ \kappa^\gamma}=0 \ \ .
\end{eqnarray}
In the neutral sector the relations are~\cite{Degrande:2013kka}
\begin{eqnarray}\label{eq:intro-ntgc-operator-1}
f_5^Z&=&0,\nonumber\\
\frac{f_5^\gamma}{m_Z^2} &=&\frac{  {\sf v}^2  }{4 c_w s_w} \frac{C_{\widetilde{B}W}}{\Lambda^4},\nonumber\\
\frac{f_4^Z}{m_Z^2} &=&  \frac{  {\sf v}^2 \left(c_w{}^2 C_{WW}+2 c_w s_w C_{BW}+4 s_w{}^2
C_{BB}\right)}{2 c_w s_w \Lambda ^4},\nonumber\\
\frac{f_4^\gamma}{m_Z^2} &=&  -\frac{  {\sf v}^2 \left(-c_w s_w
C_{WW}+C_{BW} \left(c_w{}^2-s_w{}^2\right)+4 c_w s_w C_{BB}\right)}{4 c_w s_w\Lambda ^4}
\end{eqnarray}
and 
\begin{eqnarray}\label{eq:intro-ntgc-operator-2}
\frac{h_3^Z}{m_Z^2}&=&\frac{  {\sf v}^2  }{4 c_w s_w} \frac{C_{\widetilde{B}W}}{\Lambda^4},\nonumber\\
h_4^Z&=& 
h_3^\gamma=
h_4^\gamma=h_2^Z=h_2^\gamma=0,\nonumber\\
\frac{h_1^Z}{m_Z^2}&=&\frac{  {\sf v}^2 \left(-c_w s_w C_{WW}+C_{BW} \left(c_w{}^2-s_w{}^2\right)+4 c_w s_w C_{BB}\right)}{4 c_ws_w\Lambda ^4},\nonumber\\
\frac{h_1^\gamma}{m_Z^2}&=&-\frac{ {\sf v}^2 \left(s_w{}^2 C_{WW}-2 c_w s_w C_{BW}+4 c_w{}^2 C_{BB}\right)}{4 c_w s_w\Lambda ^4}
\end{eqnarray}
with the relation
\begin{equation}
f_5^\gamma=h_3^Z \qquad \text{and} \qquad h_1^Z=-f_4^\gamma.
\end{equation}
Here ${\sf v}$ is the VEV and $c_w=\cos\theta_W$, $s_w=\sin\theta_W$.

One can follow both  effective operator approach and effective vertex approach
to study the anomalous gauge boson couplings. We follow both the approach for
studying the charge sector couplings; while, we study only the vertex factors in
the neutral sector.
In this thesis, we restrict ourselves to only dimension-$6$ operators or form factors
with a partial contribution up to $\Lambda^{-4}$. We take the quadratic contribution of dimension-$6$ form factors/ operators to
compare our results with current LHC constraints. 

\subsection{BSM contributions to aTGC}
In the top down approach of the EFT, one has a high scale model which may provide
triple gauge boson couplings in the low energy through loops. 
Some simplified
fermionic models~\cite{Corbett:2017ecn}, the Minimal Supersymmetric SM 
(MSSM)~\cite{Gounaris:2000tb,Choudhury:2000bw} and Little Higgs 
model~\cite{Dutta:2009nf} provide  some of the $CP$-even structure of the neutral
aTGC. Some  $CP$-odd
couplings in the neutral sector can be generated in the MSSM~\cite{Gounaris:2000tb} (at two loops),
in complex two Higgs doublet model (C2HDM)~\cite{Corbett:2017ecn,Grzadkowski:2016lpv,Belusca-Maito:2017iob}.
Besides these, a non-commutative extension of the SM
(NCSM)~\cite{Deshpande:2001mu,Deshpande:2011uk} can also provide an anomalous
coupling  structure in the neutral sector with a possibility of a trilinear 
$\gamma \gamma \gamma$ coupling as well~\cite{Deshpande:2001mu}.
In the charge sector, 
aTGC may be obtained in  MSSM~\cite{Lahanas:1994dv,Arhrib:1996rj,Argyres:1995ib}, extra dimension~\cite{FloresTlalpa:2010rm,Lopez-Osorio:2013xka}, Georgi-Machacek model~\cite{Arroyo-Urena:2016gjt}, etc. by integrating out the heavy degrees of freedom.

\begin{figure}[h]
    \centering
    \includegraphics[width=1\textwidth]{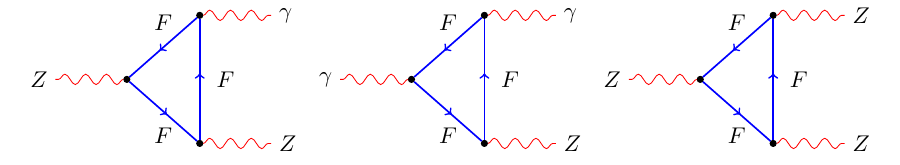}
    \caption{\label{fig:intro-toy-NaTGC} Triangular diagram with a heavy fermion $F$ contributing to neutral aTGC.}
\end{figure}
We discuss some of the  explicit models providing  some of the aTGC.
As a simplest example, a heavy fermion in a toy model can contribute to some of the triple gauge boson couplings in the neutral sector. For example, the  Lagrangian~\cite{Gounaris:1999kf}
\begin{equation}
{\cal L}_{VF\bar{F}} = -e Q_F A^\mu \bar{F} \gamma_\mu F - \frac{e}{2 s_w c_w} Z^\mu \bar{F}
\left( \gamma_\mu g_{VF} -\gamma_\mu\gamma_5 g_{AF} \right) F
\end{equation}
with a heavy fermion $F$ can generate $f^\gamma$, $f^Z$, $h_3^\gamma$ and $h_3^Z$
couplings in the Lagrangian ${\cal L}_{ZVV}$ in Eq.~(\ref{eq:intro-LZVV-full}) through the triangular diagrams shown in Fig.~\ref{fig:intro-toy-NaTGC} with  the $F$  running in loops.
The contributions to aTGC are~\cite{Gounaris:1999kf}
\begin{eqnarray}
&&h_3^Z=-f_5^\gamma = - N_F \dfrac{e^2Q_F g_{VF}g_{AF}}{96 \pi^2 s_w^2 c_w^2} \dfrac{m_Z^2}{m_F^2},\\
&& h_3^\gamma = - N_F \dfrac{e^2Q_F g_{AF}}{48 \pi^2 s_w c_w} \dfrac{m_Z^2}{m_F^2},\\
&& f_5^Z=- N_F \dfrac{e^2g_{AF}\left( 5g_{VF}^2 + g_{AF}^2 \right)}{960 \pi^2 s_w^3 c_w^3} \dfrac{m_Z^2}{m_F^2}.
\end{eqnarray}
Here $Q_F$ is the electric  charge, $m_F$ is the mass of  $F$, $N_F$ is the number of flavour. Other couplings are zero in this model.
The heavy fermion, if associates with a iso-spin partner to form a $SU(2)$ doublet,  will provide $WWZ$ and $WW\gamma$ couplings as well.

\begin{figure}[h]
    \centering
    \includegraphics[width=1\textwidth]{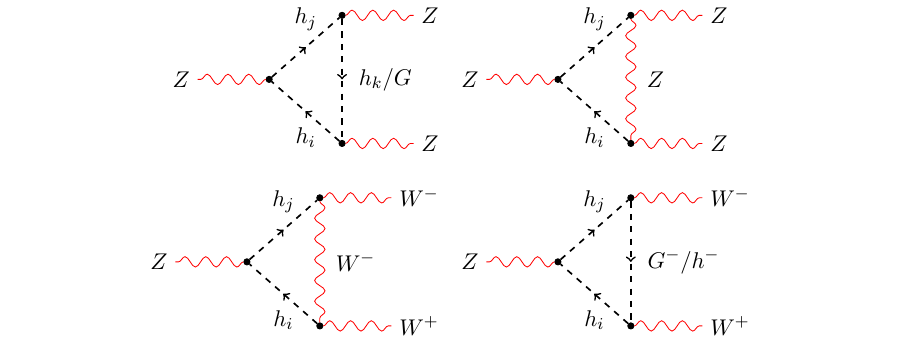}
    \caption{\label{fig:intro-2HDM-aTGC} Contributions to anomalous $ZZZ$ ({\em upper-row}) and $WWZ$ ({\em lower-row}) vertex  from 2HDM with $i\ne j\ne k$ in {\em left-top} and $i\ne j$ for rest.}
\end{figure}
\begin{figure}[h]
    \centering
    \includegraphics[width=1\textwidth]{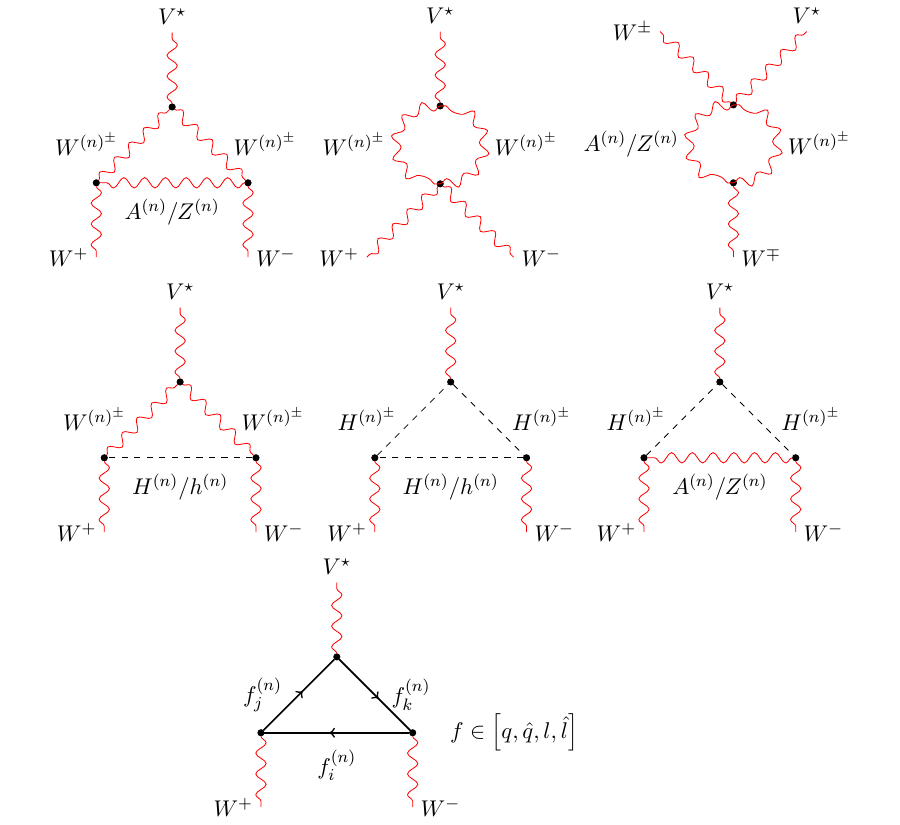}
    \caption{\label{fig:intro-UED-aTGC} Contributions to anomalous $WWV^\star$ ($V=Z/\gamma$)
        vertex with off-shell $V$  in a universal extra dimension. }
\end{figure}
The two-Higgs-doublet model (2HDM) is another example of a detail and renormalizable model containing the SM
where
the anomalous $ZZZ$ vertex  along with $WWZ$ vertex can be generated~\cite{Corbett:2017ecn,Grzadkowski:2016lpv,Belusca-Maito:2017iob}.
The Feynman diagrams in Fig.~\ref{fig:intro-2HDM-aTGC} with the neutral Higgs bosons ($h_i,~i=1,2,3$)
 or Goldstone boson ($G$), after integrated out, will provide 
 $ZZZ$  ($CP$-even/odd, {\em upper-row}) and $WWZ$ vertex
({\em lower-row}).
In another example, anomalous $WWZ$ and $WW\gamma$ vertices ($\Delta\kappa^V$ in particular)  can be generated at loop level 
in a  universal extra dimension (UED) from the Feynman diagrams given in Fig.~\ref{fig:intro-UED-aTGC}
with the  Kaluza–Klein (KK) excited modes ($A^{(n)}/Z^{(n)}/W^{(n)}/H^{(n)}/h^{(n)}/f_i^{(n)}$) running in  loops as discussed in Ref.~\cite{Lopez-Osorio:2013xka}. 

\subsection{Probe of the aTGC}
It is important to probe the aTGC for the precision measurements in a collider. The aTGC predominantly appear in
various di-boson ($ZZ/Z\gamma/WW/WZ$) production processes; a detailed and careful study of these processes 
may reveal new physics through probing the aTGC. 
The aTGC not only change the total cross section in those processes, they affect the distribution of various kinematical variables, such as transverse momentum, invariant mass, etc. They also modify the angular distributions of the daughter particles  of the final state gauge bosons. Some of the asymmetries based on the  angular distribution of the decay products are related to the various polarizations of the gauge bosons. 
So, by studying the cross sections, distribution of kinematical variables, polarizations of gauge bosons, angular asymmetries of the decay products, etc., one could probe the aTGC in a particle collider.
There has been a lot of studies of the aTGC in the neutral sector~\cite{Czyz:1988yt,Baur:1992cd,Choudhury:1994nt,Choi:1994nv,
    Aihara:1995iq,Ellison:1998uy, Gounaris:1999kf,Gounaris:2000dn,Baur:2000ae,
    Rizzo:1999xj,Atag:2003wm,Ananthanarayan:2004eb,Ananthanarayan:2011fr,
    Ananthanarayan:2014sea,Poulose:1998sd,Senol:2013ym,Rahaman:2016pqj,Rahaman:2017qql,Ots:2006dv,
    Ananthanarayan:2003wi,Chiesa:2018lcs,Chiesa:2018chc,Boudjema:Desy1992,Ananthanarayan:2005ib,Rahaman:2018ujg}
as well as in the charge sector~\cite{Gaemers:1978hg,Hagiwara:1986vm,Bilchak:1984ur,Hagiwara:1992eh,
    Wells:2015eba,Buchalla:2013wpa,Zhang:2016zsp,Berthier:2016tkq,Bian:2015zha,Bian:2016umx,
    Choudhury:1996ni,Choudhury:1999fz,Rahaman:2019mnz,Biswal:2014oaa,Cakir:2014swa,Li:2017kfk,Kumar:2015lna,Baur:1987mt,Dixon:1999di,Falkowski:2016cxu,Azatov:2017kzw,Azatov:2019xxn,Bian:2015zha,Campanario:2016jbu,Bian:2016umx,Butter:2016cvz,Baglio:2017bfe,Li:2017esm,Bhatia:2018ndx,Chiesa:2018lcs} in different processes.
The direct measurements of the aTGC have been performed at various colliders for neutral sector in Refs.~\cite{Acciarri:2000yu,Abbiendi:2000cu,
    Abbiendi:2003va,Achard:2004ds,Abdallah:2007ae,Abazov:2007ad,Aaltonen:2011zc,Abazov:2011qp,Chatrchyan:2012sga,Chatrchyan:2013nda,Aad:2013izg,Khachatryan:2015kea,Khachatryan:2016yro,Sirunyan:2017zjc,Aaboud:2017rwm,Aaboud:2018jst} and for the charge sectors in Refs.~~\cite{Abbiendi:2000ei,Abbiendi:2003mk,
    Abdallah:2008sf,Schael:2013ita,Aaltonen:2007sd,Abazov:2012ze,Aaboud:2017cgf,
    Sirunyan:2017bey,Aaboud:2017fye,Khachatryan:2016poo,
    Aad:2016ett,Aad:2016wpd,Chatrchyan:2013yaa,
    RebelloTeles:2013kdy,ATLAS:2012mec,Chatrchyan:2012bd,Aad:2013izg,
    Chatrchyan:2013fya,Sirunyan:2017jej,Sirunyan:2019gkh,Sirunyan:2019dyi,Sirunyan:2019bez,Corbett:2013pja}.
\section{Outline of the thesis}
We organise the rest of the thesis as follows: 
We employ the polarizations of the gauge boson involved in the anomalous couplings
along with the cross sections to study the aTGC. First, we introduce the polarization
observables of a general spin-$1$ particle in chapter~\ref{chap:polarization}.
The thesis then divided into two parts on the basis of neutral aTGC and charged aTGC.
First, we study the neutral aTGC, and then we move on to the charged aTGC.
In chapter~\ref{chap:epjc1}, we study the sensitivities of the polarization observables to the aTGC in the neutral sector and obtain limits
on them in $ZZ/Z\gamma$ productions in a $e^+e^-$ collider. In chapter~\ref{chap:epjc2},
we examine the effect of beam ( $e^+,~e^-$) polarizations on the sensitivity of observables to aTGC in the same processes of $e^+e^-\to ZZ/Z\gamma$.
In chapter~\ref{chap:ZZatLHC}, we  restrict to $ZZ$ production process at the LHC and investigate the role of
$Z$ polarizations. Next in chapter~\ref{chap:eeWW}, we study the $WWV$ anomalous couplings in $e^+e^-\to W^+W^-$ process by employing the 
$W$ polarizations along with the beam polarizations.  In chapter~\ref{chap:WZatLHC}, we study the $WWZ$ anomalous
couplings in $W^\pm Z$ production at the LHC and investigate the role of $Z$ and reconstructed $W$ polarizations.
We conclude in chapter~\ref{chap:conclusion} followed by the outlooks of the thesis.
We keep some important supplementary materials in the appendices for completeness of the main chapters. In appendix~\ref{appendix:intro}, we give some of the Feynman rules   of the electroweak theory   necessary for the chapters. In appendix~\ref{appendix:epjc1}, we give the helicity amplitudes in $e^+e^-\to ZZ/Z\gamma$ process for SM+aTGC along with the expressions for observables. The semi-analytic expressions for the observables in $ZZ$ production are given
in appendix~\ref{appendix:ZZatLHC}. The helicity amplitudes for SM+aTGC in $e^+e^-\to W^+W^-$ process 
are given in appendix~\ref{appendix:eeWW}.  
In appendix~\ref{appendix:WZLHC}, we give the 
 SM values of the asymmetries and corresponding
polarizations along with the numerical fitting procedures for them in $ZW^\pm$ production process. In appendix~\ref{appendix:HEP-packages}, we give some brief descriptions of some HEP packages that we have used in this thesis.

\chapter{\label{chap:polarization} Polarization parameters  of  spin-$1$ particles}

\begingroup
\hypersetup{linkcolor=blue}
\minitoc
\endgroup


The cross section of a process is an important observable to detect new physics through excess rate compared to the SM or through new resonance. Kinematical distributions and cuts may increase the signal to background ratio for a new physics. But the cross section may not be sensitive to some new physics parameters or may not be sufficient when a large number of new  physics parameters have to be measured. In this scenario, one needs as many observables as possible.

One can construct observables related to the polarizations of a particle and use them along with 
total rate and other kinematical observables to study new physics. A spin-$s$ particle offers a 
total of $(2s+1)^2-1=4s(s+1)$ observables related to polarizations of the particle. The polarization 
density matrix of the spin-$s$ particle is a $(2s+1)\times(2s+1)$ hermitian, unit-trace matrix 
that can be parametrized by $4s(s+1)$ real parameters. These parameters are different kinds of 
polarizations. For example, a spin-$1/2$ fermion has three polarization parameters called 
longitudinal, transverse, and normal polarizations (see for 
example~\cite{Godbole:2006tq,Boudjema:2009fz}). Similarly, for a spin-$1$ particle we have a 
total of eight such 
parameters~\cite{Bourrely:1980mr,Abbiendi:2000ei,Ots:2004hk,Boudjema:2009fz,Aguilar-Saavedra:2015yza,Rahaman:2016pqj,Nakamura:2017ihk}; three of them are vectorial like in the spin-$1/2$ case 
and the other five are tensorial~\cite{Boudjema:2009fz,Aguilar-Saavedra:2015yza} as will be 
described in Section~\ref{sect:pol-polarization_matrix} in detail. These polarization
parameters can be calculated analytically from the production process as well as from the angular 
asymmetries of the decayed products of the particle.

The polarization observables of spin-$1$ particles have been used earlier to study new physics. 
The polarization asymmetries of $Z$ and $W$ were used to study the anomalous  gauge boson couplings~\cite{Abbiendi:2000ei,Rahaman:2016pqj,Rahaman:2017qql,Rahaman:2017aab,Rahaman:2018ujg,Rahaman:2019mnz,Rahaman:2019lab},  Higgs-gauge boson 
interaction~\cite{Nakamura:2017ihk,Rao:2018abz},  FCNC interaction~\cite{Behera:2018ryv}, dark matter~\cite{Renard:2018tae}, the top quark mass structure~\cite{Renard:2018bsp,Renard:2018lqv}, 
special interactions of massive particles~\cite{Renard:2018jxe,Renard:2018blr}, and  dark matter
along with heavy resonance~\cite{Aguilar-Saavedra:2017zkn}. In the next sections, we discuss the polarizations of a general spin-$s$ particle with special focus on the spin-$1$ case.

\section{Polarization density matrix}\label{sect:pol-polarization_matrix}
\begin{figure}
    \begin{center}
        \includegraphics[scale=1]{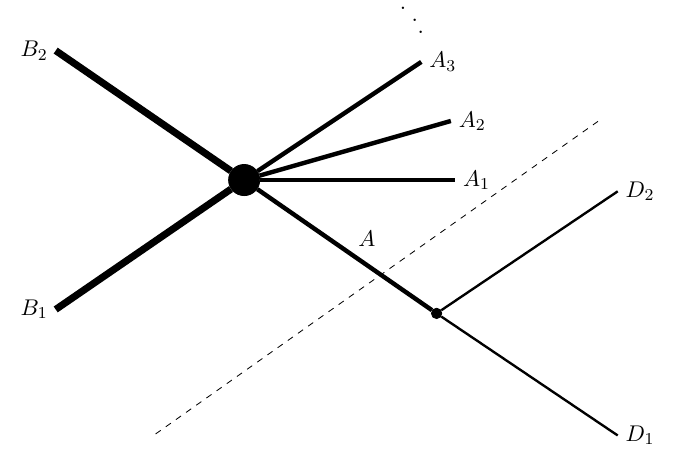}
    \end{center}
    \caption{\label{fig:V_production_decay} Schematic diagram for the production of a particle $A$ and it's decay to $D_1$ and $D_2$. The dashed line separates the production part and the decay part.} 
\end{figure}
To begin with, let us consider the production and decay of an unstable particle $A$ of spin-$s$ in a general
process of $B_1B_2\to A A_1A_2\ldots A_{n-1}$ with $A\to D_1 D_2$, as shown in Fig.~\ref{fig:V_production_decay}. The differential rate for such a process is given by~\cite{Boudjema:2009fz},
\begin{eqnarray}\label{eq:whole-process}
d\sigma &=& \sum_{\lambda,\lambda^\prime}\left[ \frac{1}{I_{B_1B_2}}  \rho^\prime(\lambda,\lambda^\prime) (2\pi)^4\delta^4\l(p_{B_1}+p_{B_1}-p_A-\sum_{i}^{n-1}p_{A_i}\r) \dfrac{d^3p_A}{(2\pi)^3 2E_A}   \prod_i^{n-1} \dfrac{d^3p_{A_i}}{(2\pi)^3 2E_{A_i}}    \right]
\nonumber\\
&\times& \left[\frac{1}{\Gamma_A}\frac{(2\pi)^4}{2m_A} \Gamma^\prime(\lambda,\lambda^\prime)\delta^4\l(p_A-p_{D_1}-p_{D_2} \r) \dfrac{d^3p_{D_1}}{(2\pi)^3 2E_{D_1}} \dfrac{d^3p_{D_2}}{(2\pi)^3 2E_{D_2}}    \right] 
\end{eqnarray} 
after using the narrow width approximation of the unstable particle $A$, allowing the factoring out the production part ($1^{st}$ square-bracket) from the decay ($2^{nd}$ square-bracket). Here, the flux factor $I_{B_1B_2}$ is given by $I_{B_1B_2}=4\sqrt{\l(p_{B_1}\cdot p_{B_2}\r)^2-m_{B_1}^2 m_{B_2}^2}$, $\Gamma_A$ is the total decay width of $A$ with $\Gamma_A \ll m_A$, $m_i$ are the mass of $i^{th}$ particles. The $\lambda$ and $\lambda^\prime$ are the helicities of the particle $A$ and they are given by $\lambda,\lambda^\prime\in\{-s,-s+1,\ldots,s\}$. The helicities of the particles $A_i$ are suppressed, i.e., helicities are summed over the remaining final state particle and 
averaged over the initial particles.
The phase-space integration can be performed in any reference frame without any loss of generality. To get the decay distribution of $A$, we perform the phase-space integration in the rest frame of $A$. We integrate the production part in the first square bracket in Eq.~(\ref{eq:whole-process}) and expresses it as,
\begin{equation}\label{eq:production_density_matrix}
\rho(\lambda,\lambda^\prime) = \frac{1}{I_{B_1B_2}}   \int \rho^\prime(\lambda,\lambda^\prime)  
(2\pi)^4\delta^4\l(p_{B_1}+p_{B_1}-p_A-\sum_{i}^{n-1}p_{A_i}\r) \dfrac{d^3p_A}{(2\pi)^3 2E_A}   \prod_i^{n-1} \dfrac{d^3p_{A_i}}{(2\pi)^3 2E_{A_i}},
\end{equation} 
with $\rho^\prime(\lambda,\lambda^\prime)= \mathcal{M}(\lambda)  
\mathcal{M}^\dagger(\lambda^\prime)$, $\mathcal{M}(\lambda)$ being the helicity amplitude with helicity  $\lambda$.
The $\rho(\lambda,\lambda^\prime)$, which is a $(2s+1)\times (2s+1)$ matrix, is called the production density matrix. 
The total integrated production cross section, without any cuts, will be given by the sum of diagonal terms, i.e.,
\begin{eqnarray}
\sigma_A = \mbox{Tr}[\rho(\lambda,\lambda^\prime)]= \sum_{\lambda} \rho(\lambda,\lambda).
\end{eqnarray} 
 Only the diagonal elements enter into the total rate, while all the 
 elements in certain combinations have the information for the polarizations of the particle. Thus the normalized production density matrix will contain only the information  of polarizations of the particle and can be equated to the spin density matrix (SDM) of spin-$s$ particles. We rewrite $\rho(\lambda,\lambda^\prime)= \sigma_A P_A(\lambda,\lambda^\prime)$,
 $P_A(\lambda,\lambda^\prime)$ is the polarization density matrix. In a similar way, we can integrate the decay part in second square bracket in Eq.~(\ref{eq:whole-process}) and write it as, 
 \begin{eqnarray}\label{eq:general-decay-part}
&&\int\frac{1}{\Gamma_A}\frac{(2\pi)^4}{2m_A} \Gamma^\prime(\lambda,\lambda^\prime)\delta^4\l(p_A-p_{D_1}-p_{D_2} \r) \dfrac{d^3p_{D_1}}{(2\pi)^3 2E_{D_1}} \dfrac{d^3p_{D_2}}{(2\pi)^3 2E_{D_2}} \nonumber \\
&=& \frac{B_{D_1D_2}(2s+1)}{4\pi}\Gamma_A(\lambda,\lambda^\prime) d\Omega_{D_i}, \ \ i=1,2,
 \end{eqnarray}
where $B_{D_1D_2}$ is the branching fraction for $A\to D_1D_2$. The matrix $\Gamma_A(\lambda,\lambda^\prime)$ is the decay density matrix normalized to unit trace;   $d\Omega_{D_i}=\sin\theta_{D_i} d\theta_{D_i} d\phi_{D_i}$ is the measure of solid angle of the daughter $D_1/D_2$.
Now, combining the production density matrix in Eq.~(\ref{eq:production_density_matrix}) and the decay density matrix 
in Eq.~(\ref{eq:general-decay-part}), the decay angular  distribution of either of the daughter $D_1/D_2$ becomes 
\begin{equation}\label{eq:norm_dist}
\dfrac{1}{\sigma}\dfrac{d\sigma}{d\Omega_{D_i}}=\frac{2s+1}{4\pi}
\sum_{\lambda,\lambda^\prime}^{}P_A(\lambda,\lambda^\prime)
\Gamma_A(\lambda,\lambda^\prime),
\end{equation}
where $\sigma=\sigma_A B_{D_1D_2}$ is the total cross section for the whole process. Below, we discuss the polarization density matrix in terms of SDM.

An SDM represents a state of an ensemble of particles.
To understand a SDM, let us begin with a 
pure quantum mechanical spin state of a spin-$s$ particle which  can be expressed as,
\begin{equation}\label{eq:spin-state}
\ket{\Psi_s} = \sum_{\lambda=-s}^{s} c_\lambda \ket{s,s_z=\lambda}.
\end{equation}
In this spin state, the mean value of an arbitrary operator $\hat{A}$ will be given by,
\begin{equation}
\braket{\hat{A}}_{\Psi_s} = \bra{\Psi_s}\hat{A}\ket{\Psi_s} 
= \sum_{\lambda,\lambda^\prime}^{} c_{\lambda^\prime}^\star A_{\lambda,\lambda^\prime} c_{\lambda},
\end{equation}
where $A_{\lambda,\lambda^\prime}$ is the matrix element in the given helicity basis. For a non-pure state (e.g., state in  scattering processes)
with incoherent mixture of some pure state $\ket{\Psi^{(i)}}$ each with a probability $p^{(i)}$
 ($\sum_i p^{(i)}=1$), the mean value of $\hat{A}$ will be given by,
\begin{equation}
\braket{\hat{A}}_{\Psi_s} = \sum_{\lambda,\lambda^\prime}^{} A_{\lambda,\lambda^\prime} \sum_{i} p^{(i)} c_{\lambda^\prime}^{(i)\star} c_{\lambda}^{(i)}.
\end{equation}
The SDM of the non-pure ensemble is thus
\begin{equation}
\rho_s(\lambda,\lambda^\prime) = \sum_{i} p^{(i)} c_{\lambda^\prime}^{(i)\star} c_{\lambda}^{(i)}.
\end{equation}
This density matrix has unit trace, i.e. 
\begin{equation}
\sum_\lambda\rho_s(\lambda,\lambda)=1; 
\end{equation} 
it is a hermitian matrix, i.e.
\begin{equation}
\rho_s^\star(\lambda,\lambda^\prime)= \rho_s(\lambda^\prime,\lambda);
\end{equation} 
the diagonal elements are positive semi-definite, i.e.
\begin{equation}
\rho_s(\lambda,\lambda) \ge 0 .
\end{equation}
A number of $(2s+1)^2-1$ real parameters can completely specify the unit traced hermitian
density matrix $\rho_s$.

The SDM can be expressed  with irreducible spin tensors up to rank $2s$, i.e., identity matrix, linear (spin-$1/2$), bilinear (spin-$1$), trilinear (spin-$3/2$)  combinations of standard spin matrices. The SDM can be represented  in terms of spherical tensor operators or can be given in Cartesian form. The properties of the density matrix will then be specified by the expansion coefficients. In the spherical tensor form, the SDM can be expanded as~\cite{Bourrely:1980mr},
\begin{equation}\label{eq:pol-spherical-form}
\rho_s = \frac{1}{2s+1}\sum_{L,M}(2L+1) (t_M^L)^\star T_M^L,
\end{equation}
where $T_L^M$ is the spherical tensor operator with rank $L$ satisfying $0\le L\le 2s$, $−L\le M\le L$; $t_L^M$ are multi-pole parameters which are generalization of the vector polarizations. We, however, are interested in the Cartesian form of SDM. In Cartesian system, the SDM of spin-$1/2$ particles, which is  
$2\times 2$, can always be expressed as,
\begin{equation}\label{eq:spin-density-half}
\rho_{1/2} = \frac{1}{2}\left( \mathbb{I}_{2\times 2} + \vec{p}\cdot\vec{\sigma}\right),
\end{equation} 
where $\sigma_i$ are the Pauli spin matrices (Eq.~(\ref{eq:app:pauli-sigma})). The $\vec{p}$ represents the mean polarization of the ensemble, i.e,
\begin{equation}
\vec{p} = \mbox{Tr}\l[\rho_{1/2}\vec{\sigma}\r].
\end{equation}
For spin-$1$ particles, the SDM can be expressed, with $3\times 3$ identity matrix and linear combination along with a  spin-tensor with bilinear combinations of spin-$1$ matrices,  as~\cite{Bourrely:1980mr,Boudjema:2009fz},
\begin{equation}\label{eq:spin-production-desnity-matrix}
\rho_{1}(\lambda,\lambda^\prime)=\dfrac{1}{3}\l[\mathbb{I}_{3\times 3} +\dfrac{3}{2} \vec{p}.\vec{S}
+\sqrt{\dfrac{3}{2}} T_{ij}\big(S_iS_j+S_jS_i\big) \r],
\end{equation}
where $S_i$ are the spin basis for spin-$1$ given by,
\begin{eqnarray}
S_x=\frac{1}{\sqrt{2}}
\left(
\begin{array}{ccc}
0 & 1 & 0 \\
1 & 0 & 1 \\
0 & 1 & 0 \\
\end{array}
\right),~
S_y=\frac{i}{\sqrt{2}}
\left(
\begin{array}{ccc}
0 & -1 & 0 \\
1 & 0 & -1 \\
0 & 1 & 0 \\
\end{array}
\right)
,~
S_z=
\left(
\begin{array}{ccc}
1 & 0 & 0 \\
0 & 0 & 0 \\
0 & 0 & -1 \\
\end{array}
\right).
\end{eqnarray}
Here $\vec{p}=\{p_x,p_y,p_z\}$ is a $3$-component  vector  and the   
$T_{ij}$ are the elements of  a $2^{nd}$-rank symmetric traceless tensor, i.e.,
\begin{equation}
T_{ij}=T_{ji},~~\sum_{i} T_{ii} = 0
\end{equation}
leading to five independent elements.
The   parameters $p_i$ and 
$T_{ij}$ are all real and independent of each  other. 
We note that the tensor part here, in Eq.~(\ref{eq:spin-production-desnity-matrix}), is absent for spin-$1/2$ case since $\sigma_i\sigma_j+\sigma_j\sigma_i = 2\delta_{ij}\mathbb{I}_{2\times 2}$ and $\sum_i T_{ii}\sigma_i^2=0$ . Likewise,  any higher rank (e.g., rank-$3$ $T_{ijk}S_iS_jS_k$) tensor evaluate to zero due to spin-$1$ symmetry. 
The multi-pole parameters in the spherical tensor operator representation of the SDM can be related to the Cartesian vector and tensor polarizations.
The $\vec{p}$ measures 
 the mean spin vector as,
 \begin{equation}
 \vec{p}=\langle \hat{s} \rangle
 \end{equation}
 and $T_{ij}$ measures the mean rank-$2$ spin tensor as,
 \begin{equation}
 T_{ij}=\frac{1}{2}\sqrt{\frac{3}{2}} \left\{\left< \hat{s_i}\hat{s_j}+\hat{s_j}\hat{s_i} \right> -\frac{4}{3}\delta_{ij} \right\}.
 \end{equation}
 Thus, three $p_i$ are the vector polarizations and five $T_{ij}$ are the tensor polarizations of a 
 spin-$1$ particle. The degrees of vector polarization $\vec{p}$, and tensor polarization $T$, are
 \begin{equation}
 p=\sqrt{\vec{p}^2};~~ 0\leq p \leq 1
 \end{equation}
 and 
 \begin{equation}
 T=\sqrt{\sum_{ij}^{}(T_{ij})^2 };~~0\leq T\leq 1.
 \end{equation}
 The overall degree of polarization, which is proportional to it's distance to the unpolarized state, is given by,
 \begin{eqnarray}\label{eq:pol-degree-pol}
 d_{pol.}&=& \frac{1}{\sqrt{2s}}\sqrt{(2s+1)\mbox{Tr}\left[\rho_s^2\right]-1}, \nonumber\\
 &=&\sqrt{\frac{3}{4}p^2+T^2}~(\leq 1).
 \end{eqnarray}
 For an example, if the particle is such that it's spin is quantized along the $z$-axis and 
 $p_+,~p_0,~p_-$ be the probabilities of finding the particle with spin projections $1,~0,~-1$, 
 respectively along the quantization axis, then the measure of polarizations would be 
 \begin{equation}
 p_x=p_y=0,~~ p_z=(p_+ - p_-),~~ T_{ij}=0~ \text{if}~ i\ne j,~~T_{xx}=T_{yy}=-\frac{1}{2}T_{zz},~~ T_{zz}=\frac{1}{\sqrt{6}}(1-3p_0).
 \end{equation}
 The degrees of vector and tensor polarizations are then given by\footnote{Although $p$ and $T$ are independent, they have their upper limit according to Eq.~(\ref{eq:pol-degree-pol})},
 \begin{equation}
 p=|p_+ - p_-|,~~~T=\frac{1}{2}|1-3p_0|.
 \end{equation}
 
 For spin-$3/2$ case, the $4\times 4$ SDM can be expressed in the Cartesian basis extending the spin-$1$ SDM by a $3$-rank tensor with spin-$3/2$ matrices as~\cite{Song:2019pzr},
 \begin{equation}\label{eq:SDM-spin3by2}
 \rho_{3/2} = \frac{1}{4}\left[\mathbb{I}_{4\times 4} + 4\vec{p}.\vec{\Sigma} +\frac{4}{3}T_{ij}^{(2)}\Sigma_{ij} +\frac{4}{3}T_{ijk}^{(3)}\Sigma_{ijk}    \right]
 \end{equation}
 with 
 \begin{eqnarray}
 \Sigma_{ij}&=&\frac{1}{2}\left(\Sigma_i\Sigma_j+\Sigma_j\Sigma_i\right)-\frac{5}{4} \delta_{ij}, \\
 \Sigma_{ijk}&=&\frac{1}{6}\Big[\left(\Sigma_i\Sigma_j\Sigma_k+\Sigma_i\Sigma_k\Sigma_j
 +\Sigma_j\Sigma_i\Sigma_k+\Sigma_j\Sigma_k\Sigma_i+\Sigma_k\Sigma_i\Sigma_j
 +\Sigma_k\Sigma_j\Sigma_i\right)  \nonumber \\
 &&~~~~~~~~-\frac{5}{12}\left(\delta_{ij}\Sigma_k+\delta_{jk}\Sigma_i
 +\delta_{ki}\Sigma_j\right)\Big].
 \end{eqnarray}
 Here, $T^{(2)}$ and $T^{(3)}$ are the rank-$2$ and rank-$3$ spin tensors, respectively
 related to the tensor polarizations.
 The spin-$3/2$ matrices, $\Sigma_i$ are given by,
 \begin{eqnarray}
 \Sigma_x&=&\frac{1}{2}\left(
 \begin{array}{cccc}
 0 & \sqrt{3} & 0 & 0 \\
 \sqrt{3} & 0 & 2 & 0 \\
 0 & 2 & 0 & \sqrt{3} \\
 0 & 0 & \sqrt{3} & 0 \\
 \end{array}
 \right),~~
  \Sigma_y=\frac{i}{2}\left(
 \begin{array}{cccc}
 0 & -\sqrt{3} & 0 & 0 \\
 \sqrt{3} & 0 & -2 & 0 \\
 0 & 2 & 0 & -\sqrt{3} \\
 0 & 0 & \sqrt{3} & 0 \\
 \end{array}
 \right),\nonumber\\
 \Sigma_z&=&\frac{1}{2}\left(
 \begin{array}{cccc}
 3 & 0 & 0 & 0 \\
 0 & 1 & 0 & 0 \\
 0 & 0 & -1 & 0 \\
 0 & 0 & 0 & -3 \\
 \end{array}
 \right). 
 \end{eqnarray}
 The degree of polarization can be calculated similar to the case of spin-$1$ case, see Ref.~\cite{Song:2019pzr} for details. 
 
We now focus on the case of a spin-$1$ particle; let us call it $V$.    
The production density matrix of the spin-$1$ particle ($V$), when normalized, contains only the information of polarizations of the particle, thus can be equated to it's SDM given in Eq.~(\ref{eq:spin-production-desnity-matrix}). The normalized production density matrix is called the polarization density matrix which takes the form    
 \begin{eqnarray}
 \label{eq:polarization_matrix}
 \rho_{1}(\lambda,\lambda^\prime) = P_V(\lambda,\lambda^\prime)=
 \renewcommand{\arraystretch}{1.5}
 \left[
 \begin{tabular}{lll}
 $\frac{1}{3}+\frac{p_z}{2}+\frac{T_{zz}}{\sqrt{6}}$ &
 $\frac{p_x -ip_y}{2\sqrt{2}}+\frac{T_{xz}-iT_{yz}}{\sqrt{3}}$ &
 $\frac{T_{xx}-T_{yy}-2iT_{xy}}{\sqrt{6}}$ \\
 $\frac{p_x +ip_y}{2\sqrt{2}}+\frac{T_{xz}+iT_{yz}}{\sqrt{3}}$ &
 $\frac{1}{3}-\frac{2 T_{zz}}{\sqrt{6}}$ &
 $\frac{p_x -ip_y}{2\sqrt{2}}-\frac{T_{xz}-iT_{yz}}{\sqrt{3}}$ \\
 $\frac{T_{xx}-T_{yy}+2iT_{xy}}{\sqrt{6}}$ &
 $\frac{p_x +ip_y}{2\sqrt{2}}-\frac{T_{xz}+iT_{yz}}{\sqrt{3}}$ &
 $\frac{1}{3}-\frac{p_z}{2}+\frac{T_{zz}}{\sqrt{6}}$
 \end{tabular}\right]
 \end{eqnarray}
 after expansion of the 
 Eq.~(\ref{eq:spin-production-desnity-matrix}).
 The dynamic of a reaction decides the values of the polarizations $\vec{p}$ and $T_{ij}$. For a given reaction one has to relate Eq.~(\ref{eq:production_density_matrix}) to Eq.~(\ref{eq:spin-production-desnity-matrix})
 to measure the polarization parameters. 
 For the process  shown in Fig.~\ref{fig:V_production_decay}, one can calculate the polarization parameters $p_i$ and $T_{ij}$ in the following way. We first calculate the production density matrix in Eq.~(\ref{eq:production_density_matrix}) using helicity amplitudes of the production process and compare it after normalization to the 
 polarization density matrix $P_V(\lambda ,\lambda^\prime)$ in Eq.~(\ref{eq:polarization_matrix}) as,
 \begin{eqnarray}\label{eq:pol_prod_matrix}
 P_V(\lambda ,\lambda^\prime)&=&\dfrac{1}{\sigma_V} \rho(\lambda ,\lambda^\prime),\nonumber \\
 &=&\dfrac{1}{\sigma_V} \left[
 \begin{matrix}
 \rho(+,+)  & \rho(+,0) & \rho(+,-)  \\
 \rho(0,+)  & \rho(0,0) & \rho(0,-) \\
 \rho(-,+)  & \rho(-,0) & \rho(-,-) 
 \end{matrix}
 \right].
 \end{eqnarray}
Thus, the polarization parameters can 
be  extracted from the polarization matrix elements as,
\begin{eqnarray}\label{eq:pol_prod}
&&p_x= 
\frac{\l[\l(\rho(+,0)+\rho(0,+)\r)+\l(\rho(0,-)+\rho(-,0)\r) \r]}{\sqrt{2}\sigma_V},\nonumber\\
&&p_y =
\frac{i\l[\l(\rho(0,+)-\rho(+,0)\r)+\l(\rho(-,0)-\rho(0,-)\r) \r]}{\sqrt{2}\sigma_V},\nonumber\\
&&p_z =
\dfrac{\l[\rho(+,+)-\rho(-,-)\r]}{\sigma_V},\nonumber\\
&&T_{xy} =
\frac{i\sqrt{6} \l[\rho(-,+)-\rho(+,-) \r]}{4\sigma_V},\nonumber\\
&&T_{xz} =
\frac{\sqrt{3}\l[\l(\rho(+,0)+\rho(0,+)\r)-\l(\rho(0,-)+\rho(-,0)\r) \r]}{4\sigma_V},\nonumber\\
&&T_{yz} =
\frac{i\sqrt{3}\l[\l(\rho(0,+)-\rho(+,0)\r)-\l(\rho(-,0)-\rho(0,-)\r) \r]}{4\sigma_V},\nonumber\\
&&T_{xx}-T_{yy} =
\frac{\sqrt{6} \l[\rho(-,+)+\rho(+,-)\r]}{2\sigma_V},\nonumber\\
&&T_{zz}=
\dfrac{\sqrt{6}}{2}\l[\frac{\rho(+,+)+\rho(-,-)}{\sigma_V}-\frac{2}{3}\r],\nonumber\\
&&\hspace{0.5cm} = \dfrac{\sqrt{6}}{2}\l[\frac{1}{3}-\frac{\rho(0,0)}{\sigma_V}\r].
\end{eqnarray}
Using the traceless property of $T_{ij}$, i.e., $T_{xx}+T_{yy}+T_{zz}=0$, along with the
values of $T_{zz}$ and $T_{xx}-T_{yy}$ from above, one can calculate 
$T_{xx}$ and $T_{yy}$ separately, although they will not be independent parameters. Instead of
considering  $T_{xx}$ and $T_{yy}$ separately we consider $T_{xx}-T_{yy}$ as an independent
tensor polarization. These polarization parameters can also be obtained from the angular
information of the decayed products of the particle $V$, which is discussed in the next section.

The polarization parameters $p_i$ and $T_{ij}$ depend on the choice of reference frame; they posses
different values in different frame. The above formulation of polarizations is based on the helcity frame, which
is equivalent to the centre-of-mass frame (CM) of colliding beams in a given process. For an $e^+$-$e^-$ collider, the observables are calculated in the  CM frame, while for  hadronic collider  such as the LHC, the observables are calculated in the laboratory (Lab) frame as well as in CM frame if possible.
In the case of  hadronic collider, the CM frame and Lab frame are not the same due to the involvement of parton distribution functions (PDFs). The production density matrix
 (in Eq.~(\ref{eq:production_density_matrix})) and hence the polarization density matrix 
 (in Eq.~(\ref{eq:polarization_matrix})) receive an effective total rotation comprising boost
 and angular rotations leaving the trace invariant going from CM to Lab frame. Due to the rotation 
 of polarization density matrix, it's elements $p_i$ and $T_{ij}$ get transformed as~\cite{Bourrely:1980mr,V.:2016wba,Velusamy:2018ksp},
     \begin{eqnarray}\label{eq:cm-to-lab-pi}
     p_i^{Lab}&=&\sum_{j} R_{ij}^Y(\omega)p_j^{CM},\nonumber\\
     T_{ij}^{Lab}&=&\sum_{k,l} R_{ik}^Y(\omega)R_{jl}^Y(\omega)T_{kl}^{CM},
     \end{eqnarray} 
     where
     \begin{eqnarray}\label{eq:cm-to-lab-Tij}
     \cos \omega&=& \cos\theta_{CM} \cos\theta_{Lab} +\gamma_{CM} \sin\theta_{CM} \sin\theta_{Lab},\nonumber\\
     \sin \omega&=& \frac{m}{E}\left(\sin\theta_{CM} \cos\theta_{Lab} -\gamma_{CM} \cos\theta_{CM} \sin\theta_{Lab}\right).
     \end{eqnarray} 
    Here, $R_{ij}^Y$ is the usual rotational matrix w.r.t. $y$-direction and 
     $\gamma_{CM}=1/\sqrt{1-\beta_{CM}^2}$ with $\beta_{CM}$ being boost of the CM 
     frame.
The quantities $m$ and $E$ are the mass and energy of the particle $V$, respectively.

\section{Polarization asymmetries}\label{sect:pol-polarization_asymmetries}
\begin{figure}[h!]
    \begin{center}
        \includegraphics[scale=1]{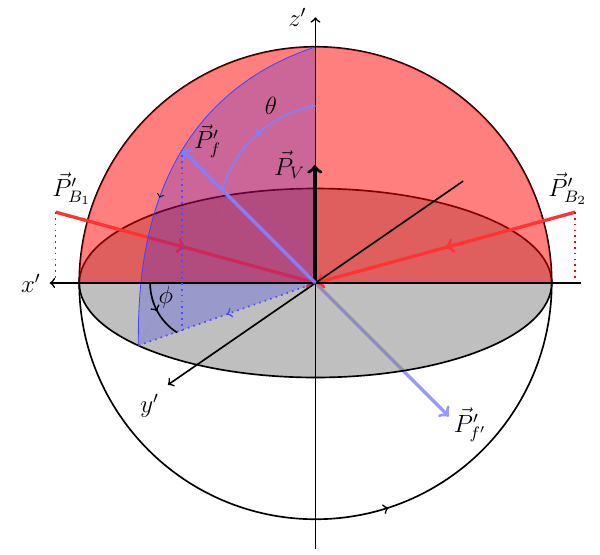}
    \end{center}
    \caption{\label{fig:helicity-frame-VPol} The reference frame showing the fermions decay
        angles in the  $V$ rest frame. The direction of $V$ in the Lab frame ($\vec{P}_V$) defines the $z^\prime$-axis (the prime is due to not being the colliding beam direction). The directions of decayed fermions are shown with $\vec{P}_f^\prime$ and $\vec{P}_{f^\prime}^\prime$ and they are in the decay plane shown by the upper transparent layer (light blue colour). The incoming particle $B_1$ 
        and $B_2$ are in the $xz$ plane shown by intermediated transparent layer (light red colour). The azimuthal angle $\phi$ of $f$ is measured w.r.t the $xz$ plane.  The co-ordinate system is right-handed, which defines the $y^\prime$-axis.} 
\end{figure}
For the spin-$1$ particle $V$ to be decayed to a pair of fermions $ff^\prime$
via the interaction vertex 
\begin{equation}
Vff^\prime: \gamma^\mu \left(L_f \ P_L + R_f \ P_R \right),~~ P_{L/R} = \frac{1}{2}\left(1\mp \gamma_5\right),
\end{equation}
the 
decay density matrix (normalized to one) is given by~\cite{Boudjema:2009fz}\footnote{Same choice of polarization vectors,  as in polarization density matrix in Eq.~(\ref{eq:polarization_matrix}), are used.},
\begin{equation}\label{eq:decay_density_matrix}
\renewcommand{\arraystretch}{1.5}
\Gamma_V(\lambda ,\lambda^\prime)=\left[
\begin{tabular}{lll}
$\frac{1+\delta+(1-3\delta)\cos^2\theta+2\alpha \cos\theta}{4}$ &
$\frac{\sin\theta(\alpha+(1-3\delta)\cos\theta)}{2\sqrt{2}} \ e^{i\phi}$&
$(1-3\delta)\frac{(1-\cos^2\theta)}{4} \ e^{i2\phi}$\\
$\frac{\sin\theta(\alpha+(1-3\delta)\cos\theta)}{2\sqrt{2}} \ e^{-i\phi}$&
$\delta+(1-3\delta)\frac{\sin^2\theta}{2}$ &
$\frac{\sin\theta(\alpha-(1-3\delta)\cos\theta)}{2\sqrt{2}} \ e^{i\phi}$\\
$(1-3\delta)\frac{(1-\cos^2\theta)}{4} \ e^{-i2\phi}$ &
$\frac{\sin\theta(\alpha-(1-3\delta)\cos\theta)}{2\sqrt{2}} \ e^{-i\phi}$ &
$\frac{1+\delta+(1-3\delta)\cos^2\theta-2\alpha\cos\theta}{4}$
\end{tabular} \right]. 
\end{equation}
Here $\theta$, $\phi$ are the polar and the azimuthal orientation of the  final 
state fermion $f$, in the rest frame of $V$ with it's would be momentum along  
$z$-direction, see Fig.~\ref{fig:helicity-frame-VPol}. The parameters $\alpha$, called analysing power,  and $\delta$ are given by,
\begin{equation}
\alpha=
\frac{2(R_f^2-L_f^2)\sqrt{1+(x_1^2-x_2^2)^2-2(x_1^2+x_2^2)}}
{12 L_fR_f x_1x_2+(R_f^2+L_f^2)[2-(x_1^2-x_2^2)^2+(x_1^2+x_2^2)]},
\end{equation}
\begin{equation}
\delta=\frac{4L_fR_f x_1x_2
    +(R_f^2+L_f^2)[(x_1^2+x_2^2)-(x_1^2-x_2^2)^2]} {12 L_fR_f x_1x_2+(R_f^2+L_f^2)[2-(x_1^2-x_2^2)^2+(x_1^2+x_2^2)]},
\end{equation}
with $x_1=m_f/m_V,~x_2=m_{f^\prime}/m_V$. For massless final state fermions,
$x_1\to0, \ x_2\to 0$; one obtains $\delta \to 0$ and 
$\alpha \to (R_f^2- L_f^2)/ (R_f^2+L_f^2)$. 
Using the expression of $P(\lambda,\lambda^\prime)$ from Eq.~(\ref{eq:polarization_matrix}) and 
$\Gamma(\lambda,\lambda^\prime)$ from Eq.~(\ref{eq:decay_density_matrix}), the  angular  
distribution in Eq.~(\ref{eq:norm_dist}) of the fermion $f$ becomes
\begin{eqnarray}\label{eq:angular_distribution}
\frac{1}{\sigma} \ \frac{d\sigma}{d\Omega} &=&\frac{3}{8\pi} \left[
\left(\frac{2}{3}-(1-3\delta) \ \frac{T_{zz}}{\sqrt{6}}\right) + \alpha \ p_z
\cos\theta 
+ \sqrt{\frac{3}{2}}(1-3\delta) \ T_{zz} \cos^2\theta
\right.\nonumber\\
&+&\left(\alpha \ p_x + 2\sqrt{\frac{2}{3}} (1-3\delta)
\ T_{xz} \cos\theta\right) \sin\theta \ \cos\phi \nonumber\\
&+&\left(\alpha \ p_y + 2\sqrt{\frac{2}{3}} (1-3\delta)
\ T_{yz} \cos\theta\right) \sin\theta \ \sin\phi \nonumber\\
&+&\left.(1-3\delta) \left(\frac{T_{xx}-T_{yy}}{\sqrt{6}} \right) \sin^2\theta
\cos(2\phi)
+\sqrt{\frac{2}{3}}(1-3\delta) \ T_{xy} \ \sin^2\theta \
\sin(2\phi) \right]. \nonumber\\
\end{eqnarray}
The above nice differential angular distribution is the {\em master equation} that is used to
probe all the polarization parameters of the particle $V$ from the data  in a real experiment 
or in a Monte-Carlo event simulator.
Using partial integration w.r.t $\theta$ and $\phi$ of the Eq.~(\ref{eq:angular_distribution})
one can construct several asymmetries which relate all the polarization parameters of $V$.

The asymmetries to probe the polarization parameters are given below.
We can obtain $p_x$ from the left-right asymmetry as,
\begin{eqnarray}\label{eq:pol_decay_Ax}
A_x&=&\frac{1}{\sigma}\l[
\int _{-\frac{\pi}{2}}^{\frac{\pi}{2}}  
\dfrac{d\sigma}{d\phi}d\phi
-\int _{\frac{\pi}{2}}^{\frac{3\pi}{2}}
\dfrac{d\sigma}{d\phi}
d\phi \r],\nonumber\\
&=&\frac{3 \alpha  p_x}{4}
\equiv  \dfrac{\sigma(\cos\phi>0)-\sigma(\cos\phi<0)}{\sigma(\cos\phi>0)+\sigma(\cos\phi<0)}.
\end{eqnarray}
The polarization parameters $p_y$ and $p_z$ are obtained from up-down and forward-backward
asymmetry, respectively as
\begin{eqnarray}\label{eq:pol_decay_Ay}
A_y&=&\frac{1}{\sigma}\l[
\int _{0}^{\pi}  
\dfrac{d\sigma}{d\phi}d\phi
-\int _{\pi}^{2\pi}
\dfrac{d\sigma}{d\phi}
d\phi \r],\nonumber\\
&=&\frac{3 \alpha  p_y}{4}
\equiv  \dfrac{\sigma(\sin\phi>0)-\sigma(\sin\phi<0)}{\sigma(\sin\phi>0)+\sigma(\sin\phi<0)},\\
%
A_z&=&\frac{1}{\sigma}\l[
\int _{0}^{\frac{\pi}{2}}  
\dfrac{d\sigma}{d\theta} d\theta
-\int _{\frac{\pi}{2}}^{\pi}
\dfrac{d\sigma}{d\theta}
 d\theta \r],\nonumber\\
&=&\frac{3 \alpha  p_z}{4}
\equiv  \dfrac{\sigma(\cos\theta>0)-\sigma(\cos\theta<0)}{\sigma(\cos\theta>0)+\sigma(\cos\theta<0)}.
\end{eqnarray}
All other polarization parameters are obtained from the following up-down-left-right mixed asymmetries:
\begin{eqnarray}\label{eq:pol_decay_Axy_to_Azz}
A_{xy}&=&\frac{1}{\sigma}\l[
\l(\int _{0}^{\frac{\pi}{2}} \dfrac{d\sigma}{d\phi}d\phi
+\int _{\pi}^{\frac{3\pi}{2}} \dfrac{d\sigma}{d\phi}d\phi\r)
-\l(\int _{\frac{\pi}{2}}^{\pi}\dfrac{d\sigma}{d\phi}d\phi 
+\int _{\frac{3\pi}{2}}^{2\pi} \dfrac{d\sigma}{d\phi}d\phi\r)
\r],\nonumber\\
&=& \frac{2}{\pi } \sqrt{\frac{2}{3}} (1-3 \delta ) T_{xy}
\equiv  \dfrac{\sigma(\sin 2\phi>0)-\sigma(\sin 2\phi<0)}{\sigma(\sin 2\phi>0)+\sigma(\sin 2\phi<0)},\\
A_{xz}&=&\frac{1}{\sigma}\l[
\l(\int _{\theta =0}^{\frac{\pi}{2} }\int _{\phi =-\frac{\pi}{2}}^{\frac{\pi}{2}} 
\dfrac{d\sigma}{d\Omega}d\Omega 
+\int _{\theta=\frac{\pi}{2}}^{\pi}\int _{\phi=\frac{\pi}{2}}^{\frac{3\pi}{2}}
\dfrac{d\sigma}{d\Omega}d\Omega
 \r) \r.\nonumber\\
&-&\l.\l(
\int _{\theta =0}^{\frac{\pi}{2} }\int _{\phi=\frac{\pi}{2}}^{\frac{3\pi}{2}}
\dfrac{d\sigma}{d\Omega}d\Omega
+\int _{\theta=\frac{\pi}{2}}^{\pi}  \int _{\phi=-\frac{\pi}{2}}^{\frac{\pi}{2}}
\dfrac{d\sigma}{d\Omega}d\Omega
\r)
\r]\nonumber\\
&=&\frac{2}{\pi } \sqrt{\frac{2}{3}} (1-3 \delta ) T_{xz}
\equiv  \dfrac{\sigma(\cos\theta\cos\phi>0)-\sigma(\cos\theta\cos\phi<0)}{\sigma(\cos\theta\cos\phi>0)+\sigma(\cos\theta\cos\phi<0)},\\
%
A_{yz}&=&\frac{1}{\sigma}\l[
\l(\int _{\theta =0}^{\frac{\pi}{2} }  \int _{\phi=0}^{\pi} 
\dfrac{d\sigma}{d\Omega}d\Omega 
+ \int _{\theta=\frac{\pi}{2}}^{\pi}\int _{\phi=\pi}^{2\pi}\dfrac{d\sigma}{d\Omega}d\Omega
 \r) \r.\nonumber\\
&-&\l.\l( 
\int _{\theta =0}^{\frac{\pi}{2} }\int _{\phi=\pi}^{2\pi}
\dfrac{d\sigma}{d\Omega}d\Omega
+\int _{\theta=\frac{\pi}{2}}^{\pi} \int _{\phi=0}^{\pi} 
\dfrac{d\sigma}{d\Omega}d\Omega
\r)
\r]\nonumber\\
&=& \frac{2 }{\pi }\sqrt{\frac{2}{3}} (1-3 \delta ) T_{\text{yz}}
\equiv \dfrac{\sigma(\cos\theta\sin\phi>0)-\sigma(\cos\theta\sin\phi<0)}{\sigma(\cos\theta\sin\phi>0)+\sigma(\cos\theta\sin\phi<0)},\\
%
\label{eq:pol_decay_Axxyy}
A_{x^2-y^2}&=&\dfrac{1}{\sigma}\l[ \l(\int_{-\frac{\pi}{4}}^{\frac{\pi}{4}}\dfrac{d\sigma}{d\phi}d\phi 
+ \int_{\frac{3\pi}{4}}^{\frac{5\pi}{4}}\dfrac{d\sigma}{d\phi}d\phi \r)
-\l( \int_{\frac{\pi}{4}}^{\frac{3\pi}{4}}\dfrac{d\sigma}{d\phi}d\phi 
+ \int_{\frac{5\pi}{4}}^{\frac{7\pi}{4}}\dfrac{d\sigma}{d\phi}d\phi \r)  \r] 
\nonumber\\
&=& \frac{1}{\pi }\sqrt{\frac{2}{3}} (1-3 \delta ) \left(T_{xx}-T_{yy}\right)
\equiv  \dfrac{\sigma(\cos 2\phi>0)
    -\sigma(\cos 2\phi<0)}{\sigma(\cos 2\phi>0)+\sigma(\cos 2\phi<0)},\\
%
\label{eq:pol_decay_Azz}
A_{zz}&=&\frac{1}{\sigma}\l[
\l(\int _{0}^{\frac{\pi}{3}}\dfrac{d\sigma}{d\theta} d\theta
+\int _{\frac{2\pi}{3}}^{\pi}\dfrac{d\sigma}{d\theta} d\theta \r)
-\int _{\frac{\pi}{3}}^{\frac{2\pi}{3}}\dfrac{d\sigma}{d\theta} d\theta
\r],\nonumber\\
&=& \frac{3}{8}\sqrt{\frac{3}{2}} (1-3 \delta ) T_{zz}
\equiv  \dfrac{\sigma(\sin 3\theta>0)-\sigma(\sin 3\theta<0)}{\sigma(\sin 3\theta>0)+\sigma(\sin 3\theta<0)}.
\end{eqnarray}
\begin{figure}
    \begin{center}
        \includegraphics[width=0.5\textwidth]{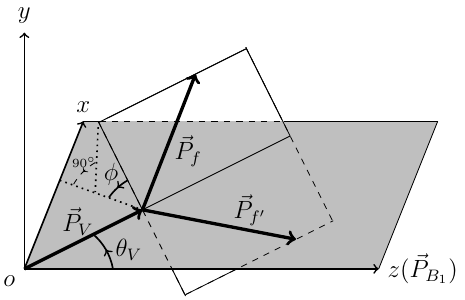}
    \end{center}
    \caption{\label{fig:lab-frame-VPol} The momentum configuration of the particles
        are shown in the Lab frame. The decay plane spanned by $\vec{P}_f$ and $\vec{P}_{f^\prime}$ makes an angle $\phi$ with the $xz$ plane. }  
\end{figure}

\begin{figure}
    \centering
    \includegraphics[width=0.6\textwidth]{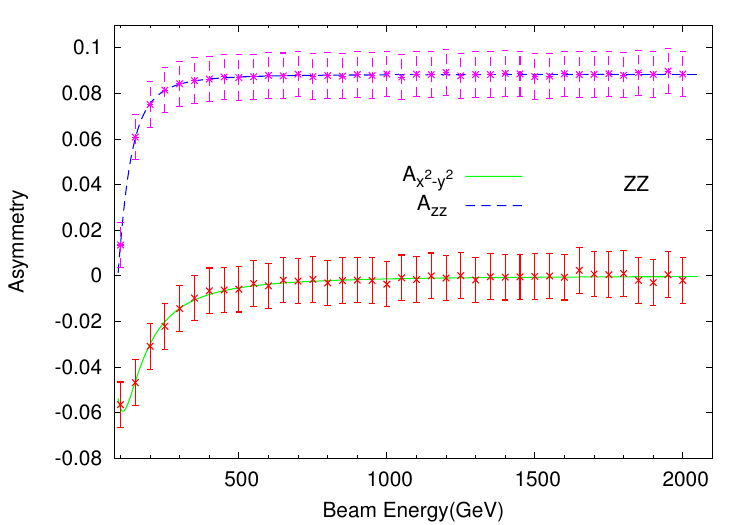}
    \includegraphics[width=0.6\textwidth]{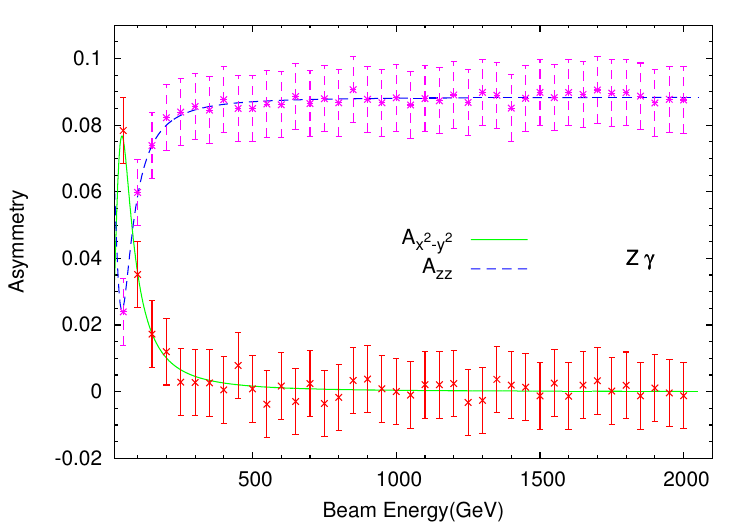}
    \includegraphics[width=0.6\textwidth]{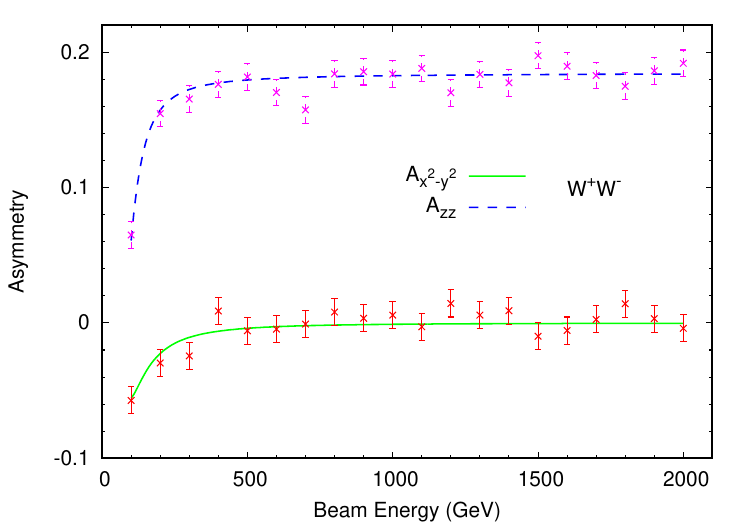}
    \caption{\label{fig:asymmetry_sanity_check} The SM values (analytic) of asymmetries
        $A_{x^2-y^2}$ ({\em solid/green line}) and $A_{zz}$ ({\em dashed/blue line}) as a function of beam
        energy in the $e^+e^-$ collider for $ZZ$ ({\em top-left-panel}), $Z\gamma$ ({\em top-right-panel})
        and  $W^+W^-$ ({\em bottom-panel})
        processes. The data points with error bar correspond to $10^4$ events
        generated by {\tt \MGvATNLO}.}
\end{figure}
While extracting the polarization asymmetries in a collider/event generator,  
we have to make sure that the analysis is done in the rest frame of $V$. The
initial beam defines the $z$-axis in the Lab, while the production plane of $V$
defines the $xz$ plane, i.e. $\phi=0$ plane, see Fig.~\ref{fig:lab-frame-VPol}. While boosting to the rest frame
of $V$, we keep the $xz$ plane invariant. The polar and the azimuthal angles of
the decay products of $V$ are measured with respect to the {\em would-be}
momentum of the particle $V$.

Thus, the polarization parameters of a spin-$1$ particle can be obtained at two levels:
At the production  level and the level of decay products.
As a demonstration of the two methods of obtaining polarization parameters, we look at three processes:
$e^+e^-\to~ZZ$, $e^+e^-\to~Z\gamma$ and  $e^+e^-\to W^+W^-$ in the SM. The polarization parameters are
constructed both at the production level using Eq.~(\ref{eq:pol_prod}) and
at the decay level using Eqs.~(\ref{eq:pol_decay_Ax})-(\ref{eq:pol_decay_Azz}).
The asymmetries $A_{x^2-y^2}$, $A_{zz}$   are calculated
analytically from the production part and shown as a function of beam energy in 
Fig.~\ref{fig:asymmetry_sanity_check}  with  lines.
For the same processes with $ZV\to f\bar fq\bar{q}$ and   $W^+W^- \to l^-\bar{\nu_l}q_u\bar{q_d}$, we generate events
using \MGvATNLO~\cite{Alwall:2014hca} with different values of beam 
energies. The
polarization asymmetries were constructed from these events and are shown in
Fig.~\ref{fig:asymmetry_sanity_check} with points. The statistical error 
bars shown correspond to $10^4$ events. We observe  agreement
between the asymmetries calculated at the production level (analytically) and
the decay level (using event generator). 
Any possible new physics in the production process of $Z$ and $W$ boson
is expected to change the cross section, kinematical distributions and the
values of the polarization parameters/asymmetries. We intend to use these
asymmetries to probe the standard and BSM physics. 

\section{Spin-spin correlations}\label{sect:pol-spin-spin-correlations}
The spin of two particles produced in a reaction could be correlated even if they do not
get produced polarized individually. In the SM the two top quarks are not produced polarized in top quark pair 
production at the LHC; but their spins are correlated~\cite{Mahlon:2010gw,Behring:2019iiv}. In a two going to two body reaction, if two particles with spin $s$ and $s^\prime$ are produced, there will be $4s(s+1)$ and $4s^\prime(s^\prime+1)$
individual polarizations and $4s(s+1)\times 4s^\prime(s^\prime+1)$ spin-spin correlator giving
a total of $(2s+1)^2(2s^\prime+1)^2-1$ number of spin observables.
For example, in a vector boson pair ($VV,V=W/Z$) production there are a total of $8+8+8\times 8=80$
spin observables. The spin-spin correlator can be obtained by constructing asymmetries
from the double angular distribution of the two particles' decay products.
These spin-spin correlators can be sensitive to new physics signals and can be used to probe them  at collider~\cite{Dixon:1999di,ATLAS:2018rgl,Linacre:2019evh}.

As an example of spin-spin correlation, one may consider top quark pair production and their leptonic decays. 
For spin-$1/2$ case, the polarization density matrix (expanding Eq.~(\ref{eq:spin-density-half})) and decay density matrix are given by~\cite{Boudjema:2009fz},
\begin{eqnarray}\label{eq:pol-spin-density-half-epjc2-expanded}
\rho_{1/2}=
\frac{1}{2}\left[
\begin{tabular}{cc}
$1+p_z$&$p_x - i p_y$\\
$p_x + i p_y$&$1-p_z$
\end{tabular}
\right],
\end{eqnarray}
\begin{equation}\label{eq:spin-half-decay_density_matrix}
\renewcommand{\arraystretch}{1.5}
\Gamma_{1/2}(\lambda ,\lambda^\prime)=\left[
\begin{tabular}{ll}
$\frac{1+\alpha \cos\theta}{2}$ & $\frac{\alpha\sin\theta}{2} \ e^{i\phi}$\\
$\frac{\alpha\sin\theta}{2} \ e^{-i\phi}$ & $\frac{1-\alpha \cos\theta}{2}$  \\
\end{tabular} \right]. 
\end{equation}
Thus, according to Eq.~\ref{eq:norm_dist}, the top-decayed leptons will have angular distribution as,
\begin{equation}\label{eq:chap1-spin1-decay-angular-dist}
\frac{1}{\sigma}\frac{d\sigma}{d\Omega}=\frac{1}{4\pi}\left[   
1+\alpha p_x \sin\theta\cos\phi +\alpha p_y\sin\theta\sin\phi +\alpha p_z\cos\theta
\r]
\end{equation}
in the rest frame of their respective mother top quark. 
The double $\cos\theta$ distribution of the two leptons will be given by,
\begin{equation}
\frac{1}{\sigma}\frac{d^2\sigma}{d\Omega_1d\Omega_2}
= C_N \left[1+\alpha_t p_z(t)\cos\theta_1 +\alpha_{\bar{t}} p_z(\bar{t})\cos\theta_2
+ \alpha_t \alpha_{\bar{t}} C_{t\bar{t}} \cos\theta_1 \cos\theta_2 + \dots
\r]
\end{equation}
with $C_N$ as an overall normalization constant and $C_{t\bar{t}}$ being one of the spin-spin correlator of $t$ and $\bar{t}$. 
The spin-spin correlator $C_{t\bar{t}}$ can be obtained from an asymmetry constructed as,
\begin{eqnarray}\label{eq:pol-ctt-corellator-asym}
A_{z_1 z_2} &=& \frac{1}{\sigma} \left[ \left( \int_{\theta_1=0}^{\frac{\pi}{2}} d\theta_1 \int_{\theta_2=0}^{\frac{\pi}{2}}\frac{d\sigma}{d\theta_1 d\theta_2} d\theta_2
+\int_{\theta_1=\frac{\pi}{2}}^{\pi} d\theta_1 \int_{\theta_2=\frac{\pi}{2}}^{\pi}\frac{d\sigma}{d\theta_1 d\theta_2} d\theta_2
\right)\right.\nonumber \\
&-& \left. \left( 
\int_{\theta_1=0}^{\frac{\pi}{2}} d\theta_1 \int_{\theta_2=\frac{\pi}{2}}^{\pi}\frac{d\sigma}{d\theta_1 d\theta_2} d\theta_2
+\int_{\theta_1=\frac{\pi}{2}}^{\pi} d\theta_1 \int_{\theta_2=0}^{\frac{\pi}{2}}\frac{d\sigma}{d\theta_1 d\theta_2} d\theta_2
\right) \right],\nonumber\\
&=& \left( C_N \alpha_t \alpha_{\bar{t}}\right) C_{t\bar{t}}
\equiv \frac{ \sigma \l(\cos\theta_1 \cos\theta_2 > 0  \r)-\sigma \l(\cos\theta_1 \cos\theta_2 < 0  \r) }{ \sigma \l(\cos\theta_1 \cos\theta_2 > 0  \r)+\sigma \l(\cos\theta_1 \cos\theta_2 < 0  \r) }.
\end{eqnarray}
All other spin-spin correlator can be obtained in a similar way.
The total number of spin observables for a general spin $s$ and $s^\prime$ system of particle( both of them being either $1$ or $1/2$) are listed in Table~\ref{tab:chap1-spin-correlation}.
\begin{table}\caption{\label{tab:chap1-spin-correlation} Total number of polarization parameters and correlator  in  spin-$s$ and spin-$s^\prime$ production  system.}
\centering
\renewcommand{\arraystretch}{1.5}
\begin{tabular}{|c|c|c|}\hline
$s$,$s^\prime$ & example   & total spin observables $=(2s+1)^2(2s^\prime+1)^2-1$ \\ \hline
$\frac{1}{2},\frac{1}{2}$& $t,\bar{t}$ & $3+3 + 3\times 3 = 15$ \\ \hline
$\frac{1}{2},1$ & $t,W$ & $3+8 + 3\times 8 = 35$ \\ \hline
$1,1$ & $Z,Z$ &  $8+8 + 8\times 8 = 80$ \\ \hline
\end{tabular}
\end{table}

\section{Summary}\label{sect:conclusion_chapter1}
A spin-$1$ particle offers eight polarization parameters providing eight more observables 
in addition to the total production cross section  to probe new physics. The polarization observables can
 discriminate among various new physics models. In a given new physics model, these polarization 
parameters can discriminate between different couplings, such as  vector-like, tensor-like, 
$CP$-even, $CP$-odd, etc. The polarization asymmetries $A_y$, $A_{xy}$ and $A_{yz}$
probe $CP$-odd couplings, while others probe $CP$-even couplings in a reaction, which 
will be discussed in detail in the next chapters in the context of anomalous triple gauge 
boson couplings. We  use these polarization asymmetries of $Z$ and $W$ along with the 
cross sections of their production to study anomalous triple gauge boson couplings in the neutral
sector as well as in the charge sector in the next chapters.
\chapter{The probe of aTGC in  $e^+e^-\to ZZ/Z\gamma$ and the role of $Z$ boson polarizations}\label{chap:epjc1}

\begingroup
\hypersetup{linkcolor=blue}
\minitoc
\endgroup
{\small\textit{\textbf{ The contents in this chapter are based on the published article in Ref.~\cite{Rahaman:2016pqj}.}}}
\vspace{1cm}

The possible trilinear gauge boson interactions in electroweak  theory 
are $WWZ$, $WW\gamma$, $ZZ\gamma$, $ZZZ$, $\gamma\gamma Z$, and 
$\gamma\gamma\gamma$, out of which the SM, after EWSB,  provides  only $WWZ$ 
and  $WW\gamma$ self-couplings. Other interactions among neutral gauge bosons
are not possible at the tree level in the SM, and hence they are anomalous. 
Thus any deviation from the SM prediction, either in strength or the tensorial
structure, would be a signal of  BSM physics.
In this chapter, we focus on the precise measurement of the neutral triple gauge boson 
couplings, in a model independent way, at the proposed International Linear 
Collider (ILC)~\cite{Djouadi:2007ik,Baer:2013cma,Behnke:2013xla}.
Different parametrization for the   neutral aTGC exists in the literature  for 
effective form factors~\cite{Gaemers:1978hg,Renard:1981es,Hagiwara:1986vm} as well as  effective operators~\cite{Larios:2000ni,Cata:2013sva,Degrande:2013kka}. 
We follow the effective form factor approach for the neutral aTGC  discussed in Ref.~\cite{Hagiwara:1986vm}.
The neutral aTGC have been widely studied in the literature 
~\cite{Czyz:1988yt,Baur:1992cd,Choudhury:1994nt,Choi:1994nv,
    Aihara:1995iq,Ellison:1998uy, Gounaris:1999kf,Gounaris:2000dn,Baur:2000ae,
    Rizzo:1999xj,Atag:2003wm,Ananthanarayan:2004eb,Ananthanarayan:2011fr,
    Ananthanarayan:2014sea,Poulose:1998sd,Senol:2013ym,Rahaman:2016pqj,Rahaman:2017qql,Ots:2006dv,
    Ananthanarayan:2003wi,Chiesa:2018lcs,Chiesa:2018chc,Boudjema:Desy1992,Ananthanarayan:2005ib,Rahaman:2018ujg} for various colliders:  in $e^+e^-$ 
collider~\cite{Czyz:1988yt,Choudhury:1994nt,
    Ananthanarayan:2004eb,Ananthanarayan:2011fr,Ananthanarayan:2014sea,Senol:2013ym,
    Rahaman:2016pqj,Rahaman:2017qql,Ots:2006dv,Ananthanarayan:2003wi,Boudjema:Desy1992,Ananthanarayan:2005ib},  $e\gamma$ 
collider~\cite{Choi:1994nv,Rizzo:1999xj,Atag:2003wm},  $\gamma\gamma$ 
collider~\cite{Poulose:1998sd}, hadron
collider~\cite{Baur:1992cd,Ellison:1998uy,Baur:2000ae,Chiesa:2018lcs,Chiesa:2018chc,Rahaman:2018ujg} and 
both $e^+e^-$ and hadron collider~\cite{Aihara:1995iq,Gounaris:1999kf,Gounaris:2000dn}.
For these effective anomalous vertices one can write an effective Lagrangian
and they have been given in \cite{Boudjema:Desy1992, Gounaris:1999kf,
Choi:1994nv, Choudhury:1994nt} up to differences in conventions and
parametrizations. The Lagrangian corresponding to the anomalous form factors in 
the neutral sector in \cite{Hagiwara:1986vm} is given by~\cite{Gounaris:1999kf},
\begin{eqnarray}\label{eq:LZVV-full}
{\cal L}_{ZVV} &=& \frac{g_e}{m_Z^2} \Bigg [
-\bigg[f_4^\gamma (\partial_\mu F^{\mu \beta})+
f_4^Z (\partial_\mu Z^{\mu \beta}) \bigg] Z_\alpha
( \partial^\alpha Z_\beta)+
\bigg[f_5^\gamma (\partial^\sigma F_{\sigma \mu})+
f_5^Z (\partial^\sigma Z_{\sigma \mu}) \bigg] \wtil{Z}^{\mu \beta} Z_\beta
\nonumber \\
&-&  \bigg[h_1^\gamma (\partial^\sigma F_{\sigma \mu})
+h_1^Z (\partial^\sigma Z_{\sigma \mu})\bigg] Z_\beta F^{\mu \beta}
-\bigg[h_3^\gamma  (\partial_\sigma F^{\sigma \rho})
+ h_3^Z  (\partial_\sigma Z^{\sigma \rho})\bigg] Z^\alpha
 \wtil{F}_{\rho \alpha}
\nonumber \\
&- & \left \{\frac{h_2^\gamma}{m_Z^2} \bigg[\partial_\alpha \partial_\beta
\partial^\rho F_{\rho \mu} \bigg]
+\frac{h_2^Z}{m_Z^2} \bigg[\partial_\alpha \partial_\beta
(\square +m_Z^2) Z_\mu\bigg] \right \} Z^\alpha F^{\mu \beta}\nonumber \\
&+& \left \{
\frac{h_4^\gamma}{2m_Z^2}\bigg[\square \partial^\sigma
F^{\rho \alpha}\bigg] +
\frac{h_4^Z}{2 m_Z^2} \bigg[(\square +m_Z^2) \partial^\sigma
Z^{\rho \alpha}\bigg] \right \} Z_\sigma \wtil{F}_{\rho \alpha }
 \Bigg ] ~ ,\nonumber\\ 
\end{eqnarray}
%
where $\wtil{Z}_{\mu \nu}=1/2 \epsilon_{\mu \nu \rho \sigma}Z^{\rho
\sigma}$ ($\epsilon^{0123}=+1$)  with
$Z_{\mu\nu}=\partial_\mu Z_\nu -\partial_\nu Z_\mu$ and similarly for
the photon tensor $F_{\mu\nu}$. The $g_e=e=\sqrt{4\pi\alpha_{EM}}$ is the electro magnetic coupling constant. 
The couplings $f_4^V, ~h_1^V,~ h_2^V$
correspond to the $CP$-odd tensorial structures, while $f_5^V, ~h_3^V ,~ h_4^V$
correspond to the $CP$-even ones. Further, the terms 
corresponding to $h_2^V$ and $h_4^V$ are of mass dimension-$8$,  while the others are dimension-$6$ in the Lagrangian.
In \cite{Ananthanarayan:2014sea} the authors have pointed out one more possible 
dimension-$8$ $CP$-even term  for $Z \gamma Z $ vertex, given by,
$${\cal L}_{aTGC} \supset \dfrac{g_e h_5^Z}{2m_Z^4} (\p^\tau F^{\alpha
\lambda})\wtil{Z}_{\alpha\beta}\p_\tau\p_\lambda Z^\beta.$$ 
In our present work, however,  we shall restrict ourselves to the
dimension-$6$ subset of the Lagrangian given in Eq.~(\ref{eq:LZVV-full}).
Besides the form factors, one can study the neutral aTGC with the  dimension-$8$ operators given in Eq.~(\ref{eq:intro-dim8-operator-ZVV}) as independent and translate the results  to the dimension-$6$ form factors using relations in Eqs.~(\ref{eq:intro-ntgc-operator-1}) \&~(\ref{eq:intro-ntgc-operator-2}).

In the theoretical side,
the tensorial structure for some of these anomalous couplings can be generated
at higher order loop within the framework of a renormalizable theory. For example,
a fermionic triangular diagram can generate $CP$-even couplings in the SM,  some simplified
fermionic model~\cite{Corbett:2017ecn}, the Minimal supersymmetric SM 
(MSSM)~\cite{Gounaris:2000tb,Choudhury:2000bw} and Little Higgs 
model~\cite{Dutta:2009nf}. On the other hand,   $CP$-odd
couplings can be generated  at $2$ loop in the MSSM~\cite{Gounaris:2000tb}.
A $CP$-violating $ZZZ$ vertex has been studied in 2HDM in Ref.~\cite{Corbett:2017ecn,Grzadkowski:2016lpv,Belusca-Maito:2017iob}.
Besides this, a non-commutative extension of the SM
(NCSM)~\cite{Deshpande:2001mu,Deshpande:2011uk} can also provide an anomalous
coupling structure in the neutral sector with a possibility of a trilinear 
$\gamma \gamma \gamma$ coupling as well~\cite{Deshpande:2001mu}.
We note that the dimension-$8$ operators which  contribute to the trilinear couplings 
also contribute to quartic gauge boson  couplings $WWVV$, $ZZZ\gamma$, $ZZ\gamma\gamma$ which appear in
triple gauge boson production~\cite{Senol:2016axw,Wen:2014mha} and vector boson scattering~\cite{Perez:2018kav}, for example. 
A  complete study of these operators will require one  to include
all these processes  involving triple gauge boson couplings as well as quartic gauge boson couplings. In the effective form factor
approach as we study in this paper, however, the triple and the quartic gauge boson couplings are independent of each other
and can be studied separately.

On the experimental side, the anomalous Lagrangian in Eq.~(\ref{eq:LZVV-full})
has been explored at the Large Electron-Positron collider (LEP)~\cite{Acciarri:2000yu,Abbiendi:2000cu,
Abbiendi:2003va,Achard:2004ds,Abdallah:2007ae}, the 
Tevatron~\cite{Abazov:2007ad,Aaltonen:2011zc,Abazov:2011qp}, and the
LHC~\cite{Chatrchyan:2012sga,Chatrchyan:2013nda,Aad:2013izg,
Khachatryan:2015kea,Khachatryan:2016yro,Sirunyan:2017zjc,Aaboud:2017rwm,Aaboud:2018jst}. The tightest bounds  on 
$f_i^V$($i=4,5$)~\cite{Sirunyan:2017zjc} and on 
$h_j^V$($j=3,4$)~\cite{Aaboud:2018jst} comes
from the CMS and ATLAS collaboration, respectively (see Table~\ref{tab:epjc1_aTGC_constrain_form_collider}). 
For the $ZZ$ process the total rate has been 
used~\cite{Chatrchyan:2012sga}, while for the  $Z\gamma$ process both
the cross section and the $p_T$ distribution of $\gamma$ has been 
used~\cite{Chatrchyan:2013nda,Aad:2013izg,Khachatryan:2015kea,
Khachatryan:2016yro} for obtaining the limits. All these analyses vary one parameter at
a time to find the $95$~\% confidence limits on the form factors. For the $Z\gamma$ process
the limits on the $CP$-odd form factors, $h_1^V, \ h_2^V$, are comparable to the 
limits on the $CP$-even form factors, $h_3^V, \ h_4^V$, respectively.
\begin{table}
\centering
\caption{\label{tab:epjc1_aTGC_constrain_form_collider} List of tightest limits on 
anomalous couplings of Eq.~(\ref{eq:LZVV-full}) available in literature.}
\renewcommand{\arraystretch}{1.5}
\begin{tabular*}{\columnwidth}{@{\extracolsep{\fill}}cl@{}}\hline
 Limits on couplings & Experiment\\ \hline
 $− 1.2\times 10^{-3} < f_4^\gamma < +1.3\times 10^{-3}$ &  $pp\to ZZ\to 4l$\\
  $− 1.2\times 10^{-3} < f_4^Z < +1.0\times 10^{-3} $ & $13$ TeV\\
 $− 1.2\times 10^{-3}  < f_5^\gamma < +1.3\times 10^{-3} $&  $35.9$ fb$^{-1}$\\
 $− 1.2\times 10^{-3}  < f_5^Z < 1.3\times 10^{-3} $ & CMS~\cite{Sirunyan:2017zjc}\\ \hline
 $-3.7\times 10^{-4}< h_3^\gamma< +3.7\times 10^{-4} $ & $pp\to Z\gamma\to\nu\bar{\nu}\gamma$ \\
 $-3.2\times 10^{-4}< h_3^Z< +3.3\times 10^{-4} $ & $13$ TeV\\
 $-4.4\times 10^{-7}< h_4^\gamma< +4.3\times 10^{-7} $ & $36.1$ fb$^{-1}$\\
 $-4.5\times 10^{-7}< h_4^Z< +4.4\times 10^{-7} $ & ATLAS~\cite{Aaboud:2018jst}\\
\hline
\end{tabular*}
\end{table}

To put simultaneous limits on all the form factors, one would need as many 
observables as possible, like differential rates, kinematic asymmetries, etc. 
Interestingly the $Z$ being a spin-$1$ particle it offers eight polarization
observables beyond the total cross section which are
discussed in chapter~\ref{chap:polarization} in details.
In this chapter, 
we  investigate all anomalous couplings (up to dimension-$6$ operators) 
of Eq.~(\ref{eq:LZVV-full}) in the processes  $e^+e^-\to ZZ/Z\gamma$ with
the help of the total cross section and the eight polarization asymmetries of the final state $Z$ boson. 


\section{Anomalous Lagrangian and their probe}
\label{sec:Lag}
The effective Lagrangian for the anomalous trilinear gauge boson interactions 
in the neutral sector is given in Eq.~(\ref{eq:LZVV-full}), which includes both 
dimension-$6$ and dimension-$8$ operators as found in the literature. 
For the present work 
we restrict our analysis to dimension-$6$ operators only. Thus, the anomalous 
Lagrangian of our interest is
\begin{eqnarray}\label{eq:LZZV-dim6}
{\cal L}_{ZVV}^{dim-6}&=&
 \frac{g_e}{m_Z^2} \Bigg [
-\bigg[f_4^\gamma (\partial_\mu F^{\mu \beta})+
f_4^Z (\partial_\mu Z^{\mu \beta}) \bigg] Z_\alpha 
( \partial^\alpha Z_\beta)
+\bigg[f_5^\gamma (\partial^\sigma F_{\sigma \mu})+
f_5^Z (\partial^\sigma Z_{\sigma \mu}) \bigg] \wtil{Z}^{\mu \beta} Z_\beta\nonumber \\
&-&  \bigg[h_1^\gamma (\partial^\sigma F_{\sigma \mu})
+h_1^Z (\partial^\sigma Z_{\sigma \mu})\bigg] Z_\beta F^{\mu \beta}
-\bigg[h_3^\gamma  (\partial_\sigma F^{\sigma \rho})
+ h_3^Z  (\partial_\sigma Z^{\sigma \rho})\bigg] Z^\alpha
 \wtil{F}_{\rho \alpha}
\Bigg ]. 
\end{eqnarray}
This yields anomalous vertices $ZZZ$ through $f^Z_{4,5}$ couplings, $\gamma ZZ$ 
through $f^\gamma_{4,5}$ and $h^Z_{1,3}$ couplings and $\gamma\gamma Z$ through
$h^\gamma_{1,3}$ couplings. There is no $\gamma\gamma\gamma$ vertex in the above
Lagrangian.
\begin{figure}[h!]
\begin{center}
\includegraphics[width=0.9\textwidth]{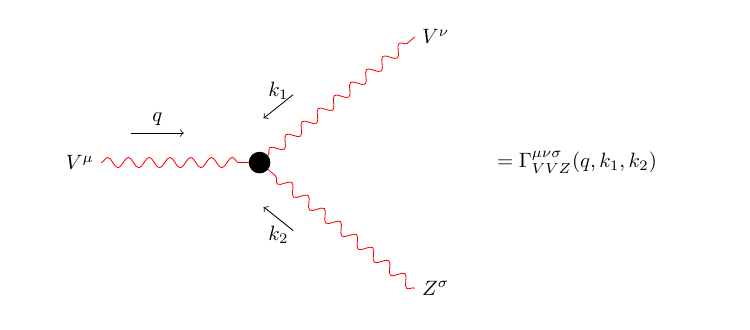}
\end{center}
\caption{\label{fig:aTGC_off-shell_Vertex} Feynman diagram for a general
anomalous triple gauge boson vertex with  $V=Z/\gamma$.}
\end{figure}
We use {\tt FeynRules}~\cite{Alloul:2013bka} to obtain the vertex tensors and 
they are given by,
\begin{eqnarray}\label{eq:aTGC_off-shel_vertex_azz}
\Gamma_{\gamma Z Z}^{\mu\nu\sigma}(q,k_1,k_2)&=&
\dfrac{g_e}{m_Z^2}\Bigg[f_4^\gamma \bigg(\left(k_2^{\nu}g^{\mu\sigma} +k_1^{\sigma} g^{\mu\nu}\right) q^2 -q^{\mu} \left(k_1^{\sigma} q^{\nu}+k_2^{\nu} q^{\sigma}\right)\bigg)\Bigg.\nonumber\\
&&+f_5^\gamma \left(q^{\mu}{q}_{\beta} \epsilon ^{\nu\sigma\alpha\beta} +q^2 \epsilon ^{\mu\nu\sigma\alpha}\right) (k_1-k_2)_{\alpha}\nonumber\\ 
&&+h_1^Z \bigg( k_2^{\mu}  q^{\nu} k_2^{\sigma} +  k_1^{\mu} k_1^{\nu} q^{\sigma} +\left(k_1^2-k_2^2\right) \left(q^{\nu}g^{\mu\sigma}-q^{\sigma}g^{\mu\nu}\right)\bigg.\nonumber\\
&&\bigg.-k_2^{\sigma} g^{\mu\nu} q.k_2-k_1^{\nu} g^{\mu\sigma}q.k_1  \bigg) \nonumber\\ 
&& \Bigg. - h_3^Z \left( k_1^{\nu} {k_1}_{\beta}\epsilon ^{\mu\sigma\alpha\beta} +{k_2}_{\beta} k_2^{\sigma} \epsilon ^{\mu\nu\alpha\beta}
+\left(k_2^2-k_1^2\right) \epsilon ^{\mu\nu\sigma\alpha}\right) {q}_{\alpha}\Bigg],
\end{eqnarray}
\begin{eqnarray}\label{eq:aTGC_off-shel_vertex_zzz}
\Gamma_{Z Z Z}^{\mu\nu\sigma}(q,k_1,k_2)&=&
\dfrac{g_e}{m_Z^2}\Bigg[f_4^Z\bigg(-q^{\mu} q^{\nu} k_1^{\sigma} -k_2^{\mu} q^{\nu} k_2^{\sigma} - k_2^{\mu}k_1^{\nu} k_1^{\sigma} -k_1^{\mu} k_2^{\nu} k_2^{\sigma}\bigg.\Bigg.\nonumber\\
&&-\left( q^{\mu} k_2^{\nu} +k_1^{\mu} k_1^{\nu} \right) q^{\sigma}
 + g^{\mu\nu}\left(q^2 k_1^{\sigma} +k_1^2 q^{\sigma} \right)\nonumber\\
&&\bigg. + g^{\mu\sigma}\left(q^2 k_2^{\nu} +k_2^2 q^{\nu} \right)+g^{\nu\sigma}\left(k_2^2 k_1^{\mu} +k_1^2 k_2^{\mu} \right)\bigg)\nonumber\\
&&- f_5^Z\bigg(\epsilon ^{\mu\nu\alpha\beta}(k_1-q)_{\alpha} {k_2}_{\beta} k_2^{\sigma}+\epsilon ^{\mu\nu\sigma\alpha}\left(\left(k_1^2-k_2^2\right){q}_{\alpha}\right.\bigg.\nonumber\\
&&\left.+\left(k_2^2-q^2\right){k_1}_{\alpha}
+\left(q^2-k_1^2\right){k_2}_{\alpha}\right)\nonumber\\
&&\Bigg.\bigg.+{k_1}_{\beta} k_1^{\nu}(k_2-q)_{\alpha}\epsilon ^{\mu\sigma\alpha\beta}+{q}_{\beta} q^{\mu}(k_2-k_1)_{\alpha}\epsilon ^{\nu\sigma\alpha\beta}\bigg)\Bigg],
\end{eqnarray}
\begin{eqnarray}\label{eq:aTGC_off-shel_vertex_aaz} 
 \Gamma_{\gamma \gamma Z}^{\mu\nu\sigma}(q,k_1,k_2)&=&
\dfrac{g_e}{m_Z^2}\Bigg[ h_1^\gamma \bigg(q^{\mu}q^{\nu} k_1^{\sigma}+q^{\sigma}k_1^{\mu} k_1^{\nu}-g^{\mu\nu}\left(q^2 k_1^{\sigma}+k_1^2 q^{\sigma} \right)\Bigg.\bigg.\nonumber\\
&&\bigg.+ g^{\mu\sigma}\left(k_1^2 q^{\nu}-q.k_1 k_1^{\nu} \right)  + g^{\nu\sigma}\left(q^2 k_1^{\mu}-q.k_1 q^{\mu} \right)\bigg)\nonumber\\
&& - h_3^\gamma \bigg({k_1}_{\beta} k_1^{\nu}{q}_{\alpha} \epsilon ^{\mu\sigma\alpha\beta} +q^{\mu} {k_1}_{\alpha}{q}_{\beta} \epsilon ^{\nu\sigma\alpha\beta} \bigg.\nonumber\\
&&+\Bigg.\bigg.\left(q^2 {k_1}_{\alpha} - k_1^2 {q}_{\alpha} \right) \epsilon ^{\mu\nu\sigma\alpha}\bigg)\Bigg].
\end{eqnarray}

The notations for momentum and Lorentz indices are shown in  
Fig.~\ref{fig:aTGC_off-shell_Vertex}. We are interested in possible trilinear
gauge boson vertices appearing in the processes $e^+e^-\to ZZ$ and 
$e^+e^-\to Z\gamma$ with final state gauge bosons being on-shell. 
For the process $e^+e^-\to ZZ$, the vertices $\gamma^\star Z Z$ and
$Z^\star Z Z$ appear with on-shell conditions $k_1^2=k_2^2=m_Z^2$. The terms 
proportional to $k_1^\nu$ and $k_2^\sigma$ in 
Eqs.~(\ref{eq:aTGC_off-shel_vertex_azz}) and 
(\ref{eq:aTGC_off-shel_vertex_zzz})
vanish due to the transversity of the polarization states. Thus, in the 
on-shell case, the vertices for $e^+e^-\to ZZ$ reduce to
\begin{eqnarray}\label{eq:aTGC_vertex_zz}
\Gamma_{V^\star Z Z}^{\mu\nu\sigma}(q,k_1,k_2)=
 -\dfrac{g_e}{m_Z^2}\left(q^2 - M_V^2 \right)\Bigg[f_4^V \left(q^{\sigma} g^{\mu\nu} +q^{\nu} g^{\mu\sigma} \right)
- f_5^V  \epsilon _{}^{\mu\nu\sigma\alpha}  (k_1 - k_2)_{\alpha}\Bigg].
\end{eqnarray}
For the process $e^+e^-\to Z\gamma$  the vertices $\gamma Z^\star Z$ and
$\gamma^\star \gamma Z$  appear with corresponding on-shell and polarization
transversity conditions. Putting these conditions in 
Eqs.~(\ref{eq:aTGC_off-shel_vertex_azz}) and 
(\ref{eq:aTGC_off-shel_vertex_aaz}) and some relabelling of momenta etc. in
Eq.~(\ref{eq:aTGC_off-shel_vertex_azz}) the relevant vertices  $Z^\star \gamma
Z$ and  $\gamma^\star \gamma Z$ can be represented together by,
\begin{eqnarray}\label{eq:aTGC_vertex_za}
\Gamma_{V^\star\gamma  Z}^{\mu\nu\sigma}(q,k_1,k_2)=
 \dfrac{g_e}{m_Z^2}\left(q^2 - M_V^2 \right)\Bigg[h_1^V  \left(k_1^{\mu} g^{\nu\sigma} - k_1^{\sigma} g^{\mu\nu} \right)
- h_3^V   \epsilon _{}^{\mu\nu\sigma\alpha}{k_1}_{\alpha}\Bigg].
\end{eqnarray}
The off-shell $V^\star$ is the propagator in our processes and 
couples to the massless electron current, as shown in 
Fig.~\ref{fig:Feynman-ZV}(c), (d).  After 
above-mentioned reduction of the vertices, there were some terms proportional to
$q^\mu$ that yield zero upon contraction with the electron current, hence they are 
dropped from the above expressions.
We note that although $h^Z$ and $f^\gamma$ appear together in the off-shell 
vertex of $\gamma ZZ$ in Eq.~(\ref{eq:aTGC_off-shel_vertex_azz}), they decouple 
after choosing separate processes; the $f^V$ appear only in $e^+e^-\to ZZ$,
while the $h^V$ appear only in $e^+e^-\to Z\gamma$. This decoupling simplifies our
analysis as we can study two processes independent of each other when 
we perform a global fit to the parameters in Section~\ref{sec:MCMC}.

\subsection{Helicity formalism and polarizations of $Z$}
\label{sec:asym_sen}
\begin{figure*}
\begin{center}
\includegraphics[width=1.0\textwidth]{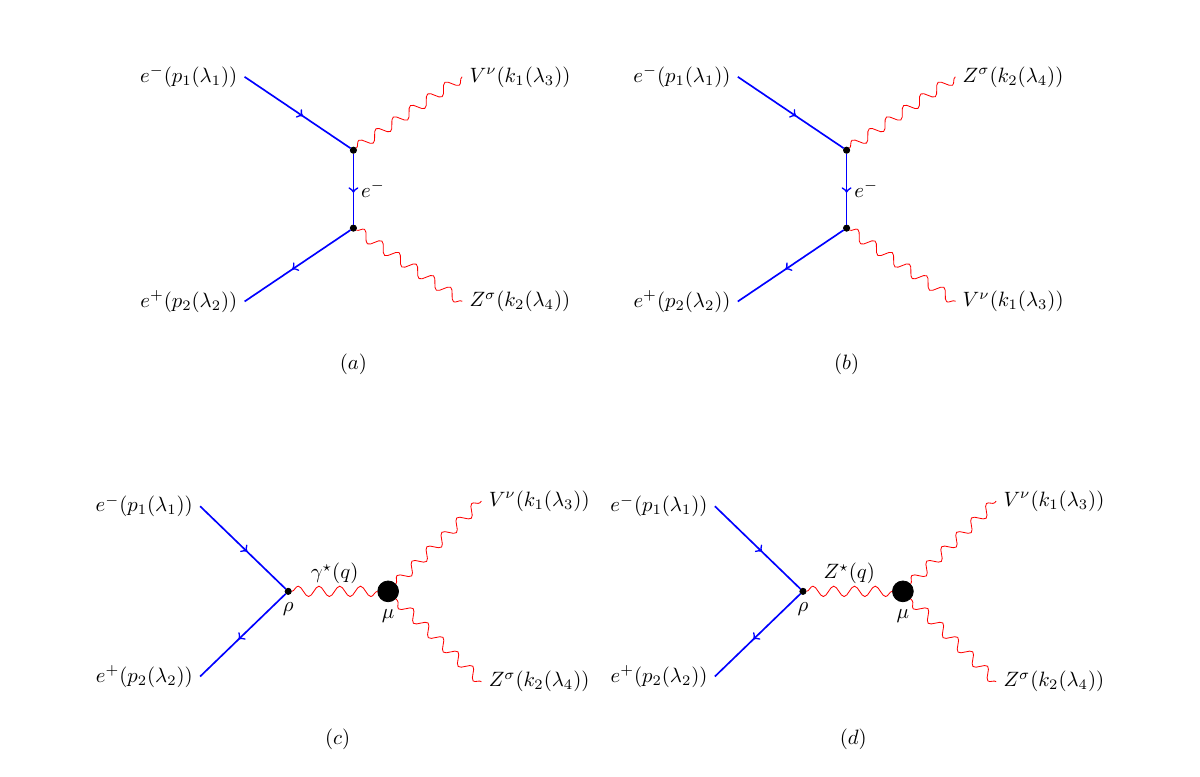}
\end{center}
\caption{\label{fig:Feynman-ZV} Feynman
diagrams for the production of $ZZ$ or $Z\gamma$ at 
$e^+e^-$ collider. }
\end{figure*}
%
We use tree level SM interactions along-with anomalous couplings 
shown in Eq.~(\ref{eq:LZZV-dim6}) for our analysis.
The Feynman diagrams for these processes are  given in the 
Fig.~\ref{fig:Feynman-ZV} where the anomalous 
vertices are 
shown with shaded blobs. 
We compute the  processes 
\begin{equation}\label{eq:eeZV-process}
e^-(p_1,\lambda_1) + e^+(p_2,\lambda_2) \to   V^\nu(k_1,\lambda_3) +  Z^\sigma(k_2,\lambda_4),
\end{equation}
in SM as well as in aTGC as is given in  Fig.~\ref{fig:Feynman-ZV}. The helicity amplitude for this process in tree level in the SM is given by,
\begin{eqnarray}\label{eq:matrix-element-sm}
\mathcal{M}_{SM}^{ZV} (\lambda_1,\lambda_2,\lambda_3,\lambda_4  )&=& \bar{v}(p_2,\lambda_2)  \Bigg[\left(\frac{-ig_Z}{2}\gamma^\sigma(C_LP_L+C_RP_R)\right)\frac{\slashed{l}_1}{t} \bigg(\Gamma_{e^+e^-V}^\nu\bigg)\nonumber\\
&+&\bigg(\Gamma_{e^+e^-V}^\nu\bigg)\frac{\slashed{l}_2}{u}
\bigg(\frac{-ig_Z}{2}\gamma^\sigma(C_LP_L+C_RP_R)\bigg)\Bigg] 
u(p_1,\lambda_1)\epsilon_\sigma^\star(k_2,\lambda_4)\epsilon_\nu^\star(k_1,\lambda_3), \nonumber\\
\end{eqnarray}
while  the aTGC  amplitudes with $Z$  and $\gamma$ mediator are given by,
\begin{eqnarray}\label{eq:matrix-element-aTGC}
\mathcal{M}_{Z(TGC)}^{ZV}(\lambda_1,\lambda_2,\lambda_3,\lambda_4  ) &=& \bar{v}(p_2,\lambda_2) \left(\frac{-ig_Z}{2}\gamma^\rho(C_LP_L+C_RP_R)\right) u(p_1,\lambda_1)\left(\dfrac{-g_{\rho\mu}+\frac{q_{\rho}q_{\mu}}{m_Z^2}}{q^2-m_Z^2}\right)\times  \nonumber\\
&&\bigg( \Gamma_{Z^\star ZV}^{\mu\sigma\nu}(q,k_2,k_1)  \bigg)\epsilon_\sigma^\star(k_2,\lambda_4)\epsilon_\nu^\star(k_1,\lambda_3)\hspace{1cm}\text{and} \nonumber \\
\mathcal{M}_{\gamma(TGC)}^{ZV} (\lambda_1,\lambda_2,\lambda_3,\lambda_4  )&=& \bar{v}(p_2,\lambda_2) \bigg(ig_e\gamma^\rho\bigg) u(p_1,\lambda_1)\left(\frac{-g_{\rho\mu}}{q^2}\right)\times \nonumber\\
&&\bigg( \Gamma_{\gamma^\star ZV}^{\mu\sigma\nu}(q,k_2,k_1)  \bigg)\epsilon_\sigma^\star(k_2,\lambda_4)\epsilon_\nu^\star(k_1,\lambda_3).
\end{eqnarray}
The momentum  $p_i,~k_i$ ($i=1,2$) and the helicities $\lambda_i$ in Eqs.~(\ref{eq:matrix-element-sm}) and (\ref{eq:matrix-element-aTGC}) are shown in the
Feynman diagrams in Fig.~\ref{fig:Feynman-ZV}. Various symbols used in the above equations
are given by,
\begin{eqnarray}
P_L=\frac{1-\gamma_5}{2},~
P_R\frac{1+\gamma_5}{2},~
l_1= p_1-k_1,~
l_2 = p_1-k_2,\nonumber\\
t = (p_1-k_1)^2,~
u = (p_1-k_2)^2~~\text{with}~
\slashed{a} = \gamma^\mu a_\mu.
\end{eqnarray}
The vertex for $e^+e^-V$ is 
\begin{equation}
\Gamma_{e^+e^-\gamma}^\nu=ig_e\gamma^\nu , ~
\Gamma_{e^+e^-Z}^\nu=\frac{-ig_Z}{2}\gamma^\nu(C_LP_L+C_RP_R)
\end{equation}
and the anomalous vertex $\Gamma_{\gamma^\star ZV}^{\mu\sigma\nu}(q,k_2,k_1)$, 
$\Gamma_{Z^\star ZV}^{\mu\sigma\nu}(q,k_2,k_1) $ are taken from the on-shell vertex
in Eqs.~(\ref{eq:aTGC_vertex_zz}) and (\ref{eq:aTGC_vertex_za}). The transverse and longitudinal polarization vector for $Z$ are chosen 
to be 
\begin{eqnarray}\label{eq:polvec-Z}
\epsilon^\mu(k,\pm)&=&\frac{1}{\sqrt{2}}\left\{0,\mp \cos\theta, -i, \pm  \cos\theta\right\},\nonumber\\
\epsilon^\mu(k,0)&=&\frac{1}{m_Z}\left\{|\vec{k}|, k_0\sin\theta, 0, k_0\cos\theta\right\}
\end{eqnarray}
with $\theta$ being the polar angle of $Z$ made with the $e^-$ direction which is taken along
the positive $z$-direction. For the photon, the transverse polarizations are same as for the $Z$ with no longitudinal polarization. The kinematics for both processes are given in appendix~\ref{appendix:helicity_amplitude}.  

The total helicity amplitude including SM and aTGC will be 
\begin{eqnarray}\label{eq:total-matrix-element}
\mathcal{M}_{tot}^{ZV}(\lambda_{e^-},\lambda_{e^+},\lambda_\gamma,\lambda_Z) =\mathcal{M}_{SM}^{ZV} (\lambda_{e^-},\lambda_{e^+},\lambda_\gamma,\lambda_Z)+ \mathcal{M}_{Z(TGC)}^{ZV}(\lambda_{e^-},\lambda_{e^+},\lambda_\gamma,\lambda_Z)\nonumber\\ +\mathcal{M}_{\gamma(TGC)}^{ZV}(\lambda_{e^-},\lambda_{e^+},\lambda_\gamma,\lambda_Z),
\end{eqnarray} 
denoting 
\begin{equation}
\lambda_1=\lambda_{e^-},~\lambda_2=\lambda_{e^+},~\lambda_3=\lambda_{\gamma},~
\lambda_4=\lambda_{Z}.
\end{equation}
The helicity amplitudes for the anomalous part together with SM
contributions for both $ZZ$ and $Z\gamma$  processes are given in appendix~\ref{appendix:helicity_amplitude}.

To calculate the polarization observables we calculate the production density matrix
in Eq.~(\ref{eq:production_density_matrix}) as,
\begin{equation}\label{eq:ZV-rho-modM}
\rho(\lambda_Z,\lambda_Z^\prime) =\frac{1}{S}\frac{\beta}{64\pi^2 \hat{s}} \int_{d\Omega}^{}d\Omega \frac{1}{2\times2}{\cal\:\sum}_{\lambda_{e^-},\lambda_{e^+},\lambda_{\gamma}} 
\bigg( \mathcal{M}_{tot}^{ZV}(\lambda_{e^-},\lambda_{e^+},\lambda_\gamma,\lambda_Z^\prime) \bigg)^\dagger
\mathcal{M}_{tot}^{ZV}(\lambda_{e^-},\lambda_{e^+},\lambda_\gamma,\lambda_Z) .
\end{equation}
The $1/S$ factor is the final state symmetry factor which is $1/2$ for $ZZ$ process and 
$1$ for $Z\gamma$ process.
The helicities of $e$, $\gamma$ and $Z$ can take values $\lambda_e,\lambda_\gamma\in\{-1,1\}$ and   $\lambda_Z\in\{-1,0,1\}$.
The density matrix given above is used to calculate all the polarization observables and
the total cross section in both processes using the technique discussed in 
chapter~\ref{chap:polarization}  which are given 
in~ appendix\ref{appendix:Expresson_observables}.

\subsection{Parametric dependence of observables}
The dependences of the observables on the anomalous couplings for the $ZZ$ and 
$Z\gamma$ processes are given in Tables~\ref{tab:parameter_dependent_pol_zz} and~\ref{tab:parameter_dependent_pol_za}, respectively.
\begin{table}[h]
\centering
\caption{\label{tab:parameter_dependent_pol_zz} Dependence of the polarization 
observables on the anomalous coupling in $ZZ$ final state.}
\renewcommand{\arraystretch}{1.5}
\begin{tabular*}{\columnwidth}{@{\extracolsep{\fill}}lll@{}}
\hline
Observables & Linear terms & Quadratic terms\\ \hline
$ \sigma $& $f_5^Z,f_5^\gamma$ & $(f_4^\gamma)^2, (f_5^\gamma)^2, (f_4^Z)^2, 
(f_5^Z)^2, f_4^\gamma f_4^Z, f_5^\gamma f_5^Z $  \\ 
$\sigma \times A_x$& $ f_5^\gamma ,f_5^Z $ & $-$\\ 
$\sigma \times A_y $& $f_4^\gamma ,f_4^Z $ & $-$ \\ 
$\sigma \times A_{xy} $& $f_4^Z,f_4^\gamma$ & $f_4^Zf_5^\gamma ,f_4^\gamma
f_5^Z,f_4^\gamma f_5^\gamma,f_4^Zf_5^Z $ \\ 
$\sigma\times A_{x^2-y^2}$& $f_5^Z,f_5^\gamma$ & 
$(f_4^\gamma)^2, (f_5^\gamma)^2, (f_4^Z)^2, 
(f_5^Z)^2, f_4^\gamma f_4^Z, f_5^\gamma f_5^Z $  \\
$\sigma\times A_{zz} $& $f_5^Z,f_5^\gamma$ & 
$(f_4^\gamma)^2, (f_5^\gamma)^2, (f_4^Z)^2, 
(f_5^Z)^2, f_4^\gamma f_4^Z, f_5^\gamma f_5^Z $  \\  \hline
\end{tabular*}
\end{table}
\begin{table}[h]
\centering
\caption{\label{tab:parameter_dependent_pol_za} Dependence of the polarization 
observables on the anomalous coupling in $Z\gamma$ final state.}
\renewcommand{\arraystretch}{1.5}
\begin{tabular*}{\columnwidth}{@{\extracolsep{\fill}}lll@{}}\hline
Observables & Linear terms & Quadratic terms\\ \hline
$ \sigma $& $h_3^Z,h_3^\gamma$ & $(h_1^\gamma)^2,(h_3^\gamma)^2,(h_1^Z)^2,(h_3^Z)^2,h_1^\gamma h_1^Z,h_3^\gamma h_3^Z $  \\ 
$\sigma \times A_x$&  $h_3^Z,h_3^\gamma$ & $(h_1^\gamma)^2,(h_3^\gamma)^2,(h_1^Z)^2,(h_3^Z)^2,h_1^\gamma h_1^Z,h_3^\gamma h_3^Z $ \\ 
$\sigma \times A_y $& $h_1^\gamma ,h_1^Z $ & $-$ \\ 
$\sigma \times A_{xy} $& $h_1^\gamma ,h_1^Z $ & $-$ \\ 
$\sigma\times A_{x^2-y^2}$& $h_3^\gamma ,h_3^Z $ & 
$- $  \\ 
$\sigma\times A_{zz} $&$h_3^Z,h_3^\gamma$ & $(h_1^\gamma)^2,(h_3^\gamma)^2,(h_1^Z)^2,(h_3^Z)^2,h_1^\gamma h_1^Z,h_3^\gamma h_3^Z $   \\ \hline
\end{tabular*}
\end{table}
In the SM, the helicity amplitudes are real, 
thus the production density matrix elements in 
Eq.~(\ref{eq:production_density_matrix}) are all  real. This implies  
$A_y$, $A_{xy}$ and $A_{yz}$ are all zero in the SM: see 
Eq.~(\ref{eq:pol_prod}). The asymmetries $A_z$ and $A_{xz}$ are also zero 
for the SM couplings due to the forward-backward symmetry of the $Z$ boson in the 
c.m. frame, owing to the presence of both $t$- and $u$-channel
diagrams and unpolarized initial beams.
After including anomalous couplings, $A_{y}$ and $A_{xy}$ receive a 
non-zero contribution, while $A_z$, $A_{xz}$ and $A_{yz}$ 
remain zero for the unpolarized initial beams.

From the list of non-vanishing asymmetries, only $A_y$ and $A_{xy}$ are $CP$-odd,
while the others
are $CP$-even. All the $CP$-odd observables are linearly dependent upon the $CP$-odd
couplings, like $f_4^V$ and $h_1^V$, while all the $CP$-even observables have
only quadratic dependence on the $CP$-odd couplings. In the SM, the $Z$ boson's
couplings respect CP symmetry; thus $A_y$ and $A_{xy}$ vanish. 
Hence, any significant deviation of $A_{y}$ and $A_{xy}$ 
from zero at the collider will indicate a clear sign of $CP$-violating new 
physics interactions. Observables that have only a linear dependence on the
anomalous couplings yield a {\it single interval limits} on these couplings.
On the other hand, any quadratic appearance (like $(f_5^V)^2$ in $\sigma$) may
yield more than one interval of the couplings, while putting limits. For the 
case of $ZZ$ process, $A_x$ and $A_y$ do not have any quadratic dependence; hence they yield the cleanest limits on the $CP$-even and -odd parameters, respectively.
Similarly, for the $Z \gamma $ process, we have $A_y$, $A_{xy}$, and $A_{x^2-y^2}$, 
which have only a linear dependence and provide clean limits. 
These clean limits, however, may not be the strongest limits as we will see 
in the following sections. 

\subsection{Sensitivity of observables to anomalous couplings}
Sensitivity of an observable $\mathcal{O}$ dependent on parameter $\vec{f}$ is
defined as
\begin{equation}\label{eq:sensitivity}
{\cal S}(\mathcal{O}(\vec{f}))=\dfrac{|\mathcal{O}(\vec{f})-\mathcal{O}(\vec{f}=0)|}{\delta \mathcal{O}},
\end{equation}
where $\delta \mathcal{O}=\sqrt{(\delta \mathcal{O}_{stat.})^2 + (\delta
\mathcal{O}_{sys.})^2}$ is the estimated error in $\mathcal{O}$. 
If the observable is an asymmetry, $A=(N^+ - N^-)/(N^+ + N^-)$, the 
error is given by, 
\begin{equation}\label{eq:error_in_asymmetry}
\delta A= \sqrt{\frac{1-A^2}{{\cal L}\sigma}+\epsilon_A^2},
\end{equation}
where $N^+ + N^-=N_T=L\sigma$, ${\cal L}$ being the integrated luminosity of the 
collider.  The error in the cross section $\sigma$ will be given by,
\begin{equation}\label{eq:error_in_sigma}
\delta\sigma= \sqrt{\frac{\sigma}{{\cal L}} + (\epsilon \sigma)^2}. 
\end{equation}
Here $\epsilon_A$ and $\epsilon$ are the systematic fractional errors in $A$ and 
$\sigma$,  respectively, while remaining one are statistical errors.
\subsubsection{One-parameter sensitivity}
\begin{figure*}[h]
\centering
\includegraphics[width=0.49\textwidth]{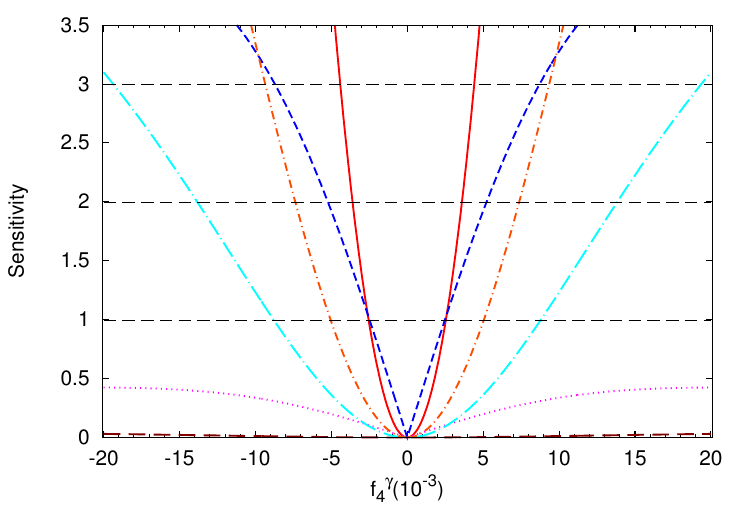}
\includegraphics[width=0.49\textwidth]{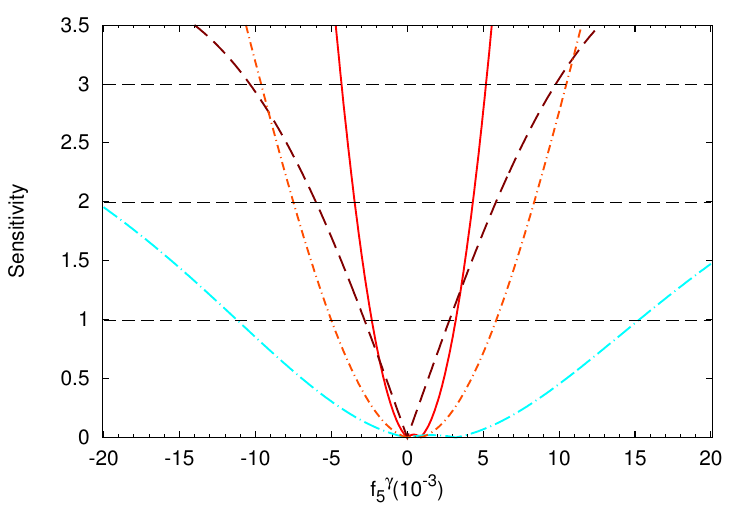}
\includegraphics[width=0.49\textwidth]{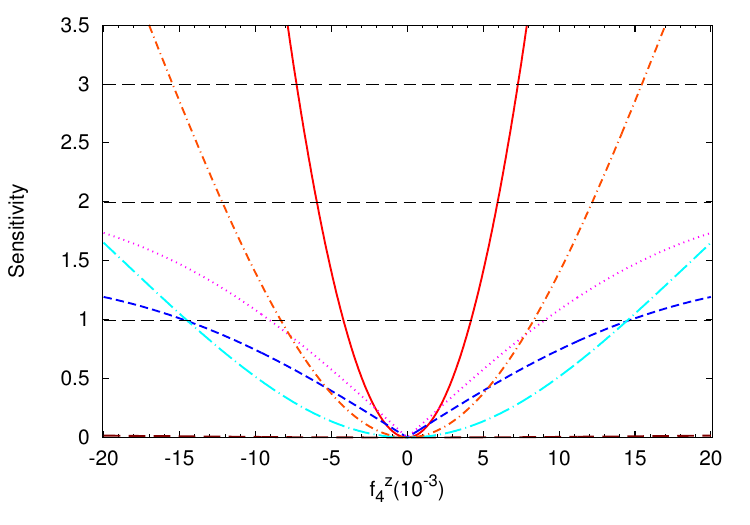}
\includegraphics[width=0.49\textwidth]{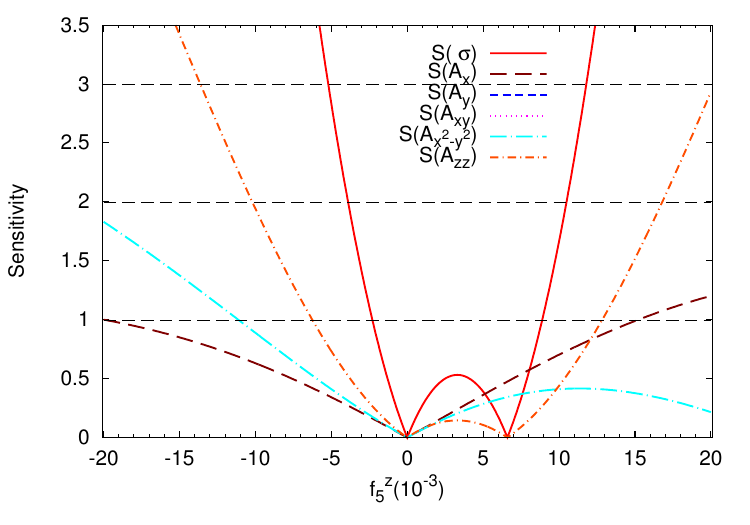}
\caption{\label{fig:one_parameter_sensitivity_zz} Sensitivity of the cross section
and asymmetries to anomalous couplings for the process $e^+e^-\rightarrow ZZ$
with $\sqrt{s}=500$ GeV and ${\cal L}=100$ fb$^{-1}$. }
\end{figure*}
\begin{figure*}[h]
\centering
\includegraphics[width=0.49\textwidth]{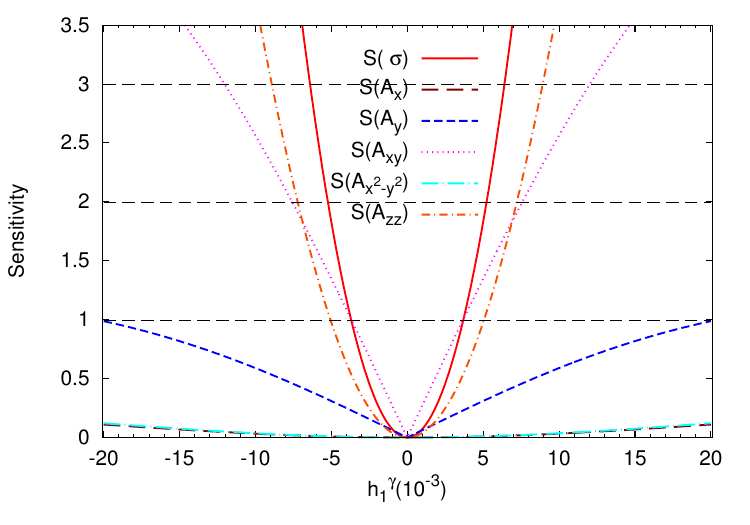}
\includegraphics[width=0.49\textwidth]{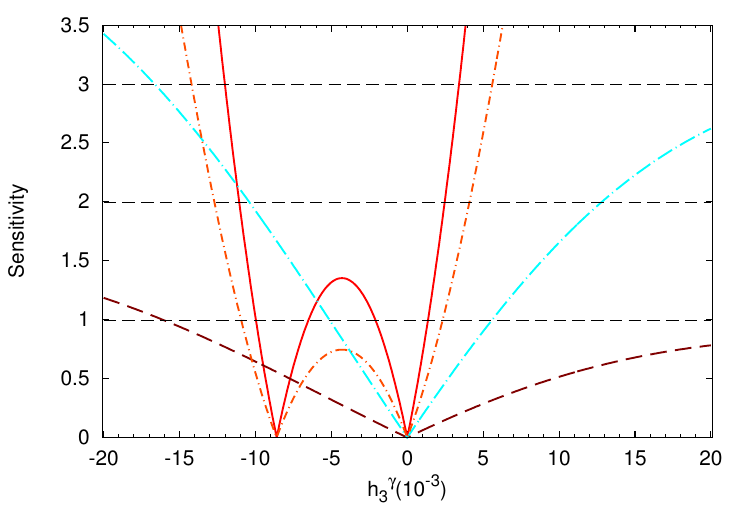}
\includegraphics[width=0.49\textwidth]{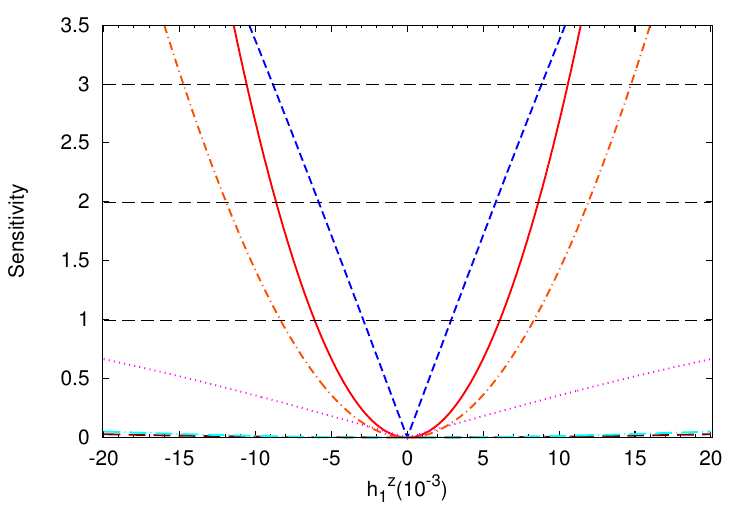}
\includegraphics[width=0.49\textwidth]{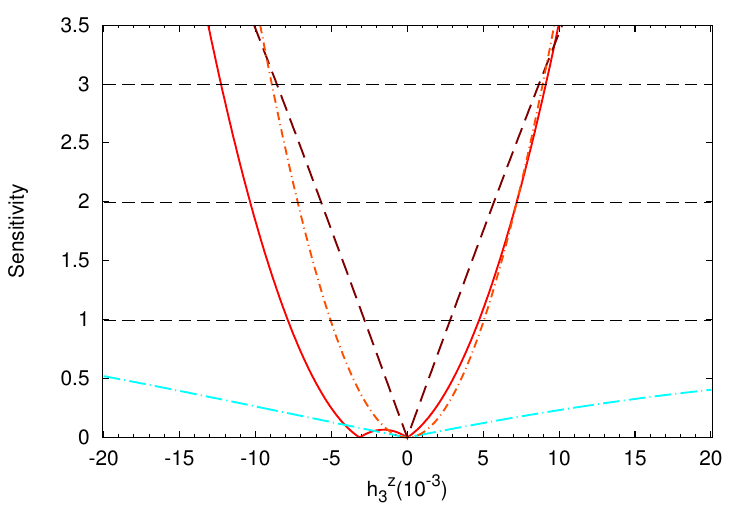}
\caption{\label{fig:one_parameter_sensitivity_za} Sensitivity of the cross section
and asymmetries to anomalous couplings for the process 
$e^+e^-\rightarrow Z\gamma$ with $\sqrt{s}=500$ 
GeV, ${\cal L}=100$ fb$^{-1}$, and $10^\circ \le\theta_\gamma\le170^\circ$.}
\end{figure*}

For numerical calculations, we choose ILC running at c.m. energy 
$\sqrt{s}=500$ GeV and integrated luminosity ${\cal L}=100$ fb$^{-1}$.
We use $\epsilon_A=\epsilon=0$ for most of our analysis. However,
 we do discuss the impact of systematic errors on our results.
 With this
choice the sensitivity of all the polarization asymmetries of $Z$ boson
discussed in chapter~\ref{chap:polarization},  and the cross section have been
calculated varying one parameter at a time. These sensitivities are shown
in Figs.~\ref{fig:one_parameter_sensitivity_zz} and~\ref{fig:one_parameter_sensitivity_za} 
for the $ZZ$ and $Z\gamma$ processes, respectively, for each observable.
In the $e^+e^-\to Z\gamma$ process we have taken a cut-off on the polar angle,
$10^\circ \le\theta_\gamma\le170^\circ$ to keep away from the beam pipe.  
For these limits, the analytical expressions  shown in~\ref{appendix:Expresson_observables}  are used.

We see that in the $ZZ$ process, the tightest constraint on 
$f_4^\gamma$ at $1\sigma$ level comes from the asymmetry $A_{y}$ owing to 
its linear and strong dependence on the coupling. For $f_5^\gamma$, 
both $A_x$ and the cross section $\sigma_{ZZ}$, give comparable limits at 
$1\sigma$ but $\sigma_{ZZ}$ 
gives a tighter limit at higher values of sensitivity. This is because the
quadratic term in  $\sigma_{ZZ}$ comes with a higher power of energy/momenta and
hence a larger sensitivity. Similarly, the strongest limit on 
$f_4^Z$ and $f_5^Z$ as well comes from $\sigma_{ZZ}$. Though the cross section
gives the tightest constrain on most of the coupling in $ZZ$ process, our 
polarization asymmetries also provide comparable limits. 
Another noticeable fact is that $\sigma_{ZZ}$ has a linear as well as quadratic 
dependence on $f_5^Z$ and the sensitivity curve is symmetric about a point 
larger than zero. 
Thus, when we do a parameter estimation exercise, we will always have 
a bias toward a positive value of $f_5^Z$. The same is the case with the coupling 
$f_5^\gamma$, but the strength of the linear term is small and 
the sensitivity plot with $\sigma_{ZZ}$ looks almost symmetric about 
$f_5^\gamma=0$.

In the $Z\gamma$ process, the tightest constraint on $h_1^\gamma$ comes from 
$A_{xy}$, on $h_3^\gamma$ it comes from $\sigma_{Z \gamma }$, on it $h_1^Z$ comes 
from $A_y$, and on $h_3^Z$ it comes from $A_x$. The cross section $\sigma_{Z \gamma}$
and $A_{zz}$ has a linear as well as quadratic dependence on $h_3^\gamma$,
and $\sigma_{Z \gamma}$ and they give two intervals at $1\sigma$ level. Other 
observables can help resolve the 
degeneracy when we use more than one observables at a time. Still, the
cross section prefers a negative value of $h_3^\gamma$, and it will be seen again
in the multi-variate analysis.
The coupling $h_3^Z$ also has a quadratic appearance in the cross section,  and it
yields a bias toward negative values of $h_3^Z$.

\begin{table}[h]
\centering
\caption{\label{tab:aTGC_constrain_form_1sigma_sensitivity} List of tightest 
limits on anomalous couplings at $1~\sigma$ level and the corresponding
observable obtained for $\sqrt{s}=500$ GeV and ${\cal L}=100$ fb$^{-1}$.}
\renewcommand{\arraystretch}{1.5}
\begin{footnotesize}
\begin{tabular*}{\textwidth}{@{\extracolsep{\fill}}llllll@{}}\hline
\multicolumn{3}{c}{$ZZ$ process} &
\multicolumn{3}{c}{$Z \gamma$ process}\\ \hline
Coupling & Limits & Comes from &
Coupling & Limits & Comes from\\ \hline
$|f_4^\gamma|$ & $ \le 2.4 \times 10^{-3}$ & $A_y$ & 
$|h_1^\gamma|$  & $\le 3.6\times 10^{-3}$ & $A_{xy}$, $\sigma$ \\ 
$|f_4^Z|$ &  $\le 4.2\times 10^{-3}$ &  $\sigma$ & 
$|h_1^Z|$ & $\le2.9\times 10^{-3}$ & $A_y$   \\ 
$f_5^\gamma$ & $\in[-2.3, 2.7] \times 10^{-3}$& $A_x$,
$\sigma$ & 
$h_3^\gamma$  & $\in [-2.1, 1.3] \times 10^{-3} $ &  $\sigma$ \\ 
 & & & &or $\in[-9.9, -6.5] \times 10^{-3} $& \\ 

$f_5^Z$ & $\in [-2.3, 8.8] \times 10^{-3}$ & $\sigma$ & 
$|h_3^Z|$ & $\le 2.8\times 10^{-3}$ & $A_x$ \\ \hline
\end{tabular*}
\end{footnotesize}
\end{table}
The tightest limits on the anomalous couplings (at $1\sigma$ level), obtained
using one observable at a time and varying one coupling at a time, are listed in 
Table~\ref{tab:aTGC_constrain_form_1sigma_sensitivity}  
along with the corresponding observables. 
A comparison between 
Tables~\ref{tab:epjc1_aTGC_constrain_form_collider} and~\ref{tab:aTGC_constrain_form_1sigma_sensitivity} shows that 
an $e^+e^-$ collider running at $500$ GeV and $100$ fb$^{-1}$ provides
better limits on the anomalous coupling ($f_i^V$) in the $ZZ$ process than the $7$ 
TeV LHC at $5$ fb$^{-1}$. For the $Z\gamma$ process the experimental limits are
available from $8$ TeV LHC with $19.6$ fb$^{-1}$ luminosity 
(Table~\ref{tab:epjc1_aTGC_constrain_form_collider}) and they are comparable
to the single observable limits shown in
Table~\ref{tab:aTGC_constrain_form_1sigma_sensitivity}. 
These limits can be further improved if we use all the
observables in a $\chi^2$ kind of analysis.

We can further see that the sensitivity curves for $CP$-odd observables, $A_y$ 
and $A_{xy}$, has no or a very mild dependence on the $CP$-even couplings. The
mild dependence comes through the cross section $\sigma$,  sitting in 
the denominator of the asymmetries. We see that $CP$-even observables provide 
a tight constraint on $CP$-even couplings and
$CP$-odd observables provide a tight constraint on the $CP$-odd couplings. Thus, not
only  can we study the two processes independently, it is possible to study the
$CP$-even and $CP$-odd couplings almost independent of each other.
To this end, we shall perform a two-parameter sensitivity analysis in the next
subsection.

A note on the systematic error is in order. The sensitivity of an observable is
inversely  proportional to the size of its estimated error, 
Eq.~(\ref{eq:sensitivity}). Including the systematic error will
increase the size of the estimated error and hence decrease the
sensitivity. For example, including $\epsilon_A=1~\%$ to ${\cal L}=100$ fb$^{-1}$ 
increases $\delta A$ by a factor of $1.3$ and dilutes the sensitivity by the same factor. This modifies the best limit on $|f_4^\gamma|$, coming from
$A_y$, to $2.97\times 10^{-3}$ (dilution by a factor of $1.3$); see
Table~\ref{tab:aTGC_constrain_form_1sigma_sensitivity}. 
For the cross section, adding $\epsilon=2~\%$ systematic error increases
$\delta\sigma$ by a factor of $1.5$. The best limit on $|f_4^Z|$, coming 
from the cross section, changes to $5.35\times 10^{-3}$, dilution by a factor
of $1.2$. Since the inclusion of the above systematic errors modifies the limits
on the couplings only by $20~\%$ to $30~\%$, we shall restrict ourselves to the statistical error for simplicity for rest of the analysis.

\subsubsection{Two-parameter sensitivity}
\begin{figure*}
\centering
\includegraphics[width=0.325\textwidth]{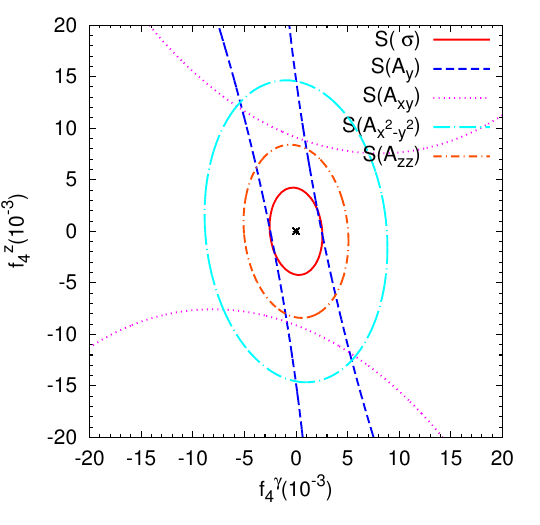}
\includegraphics[width=0.325\textwidth]{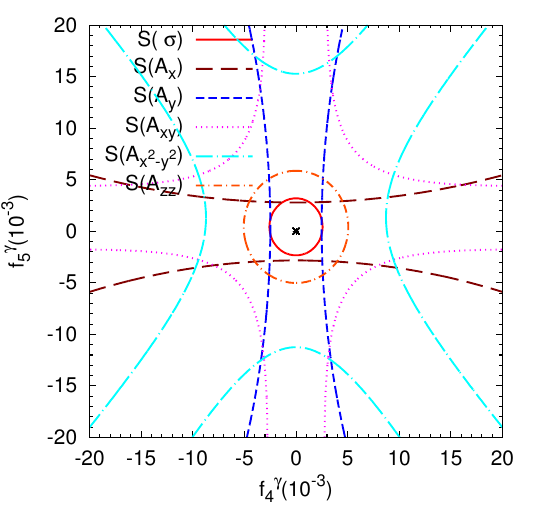}
\includegraphics[width=0.325\textwidth]{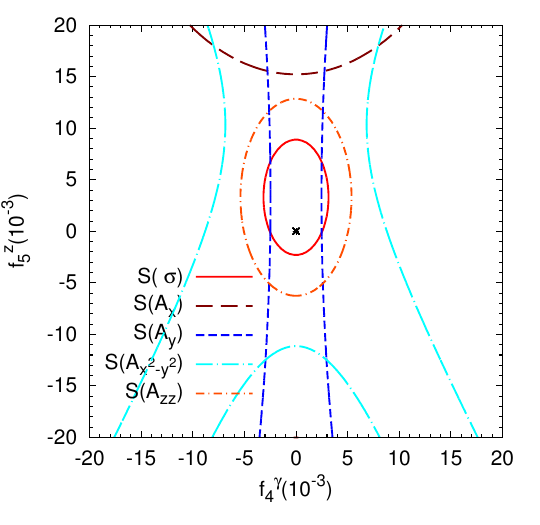}
\includegraphics[width=0.325\textwidth]{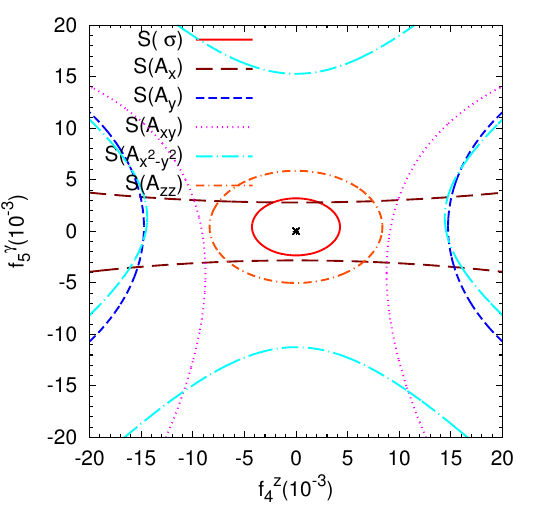}
\includegraphics[width=0.325\textwidth]{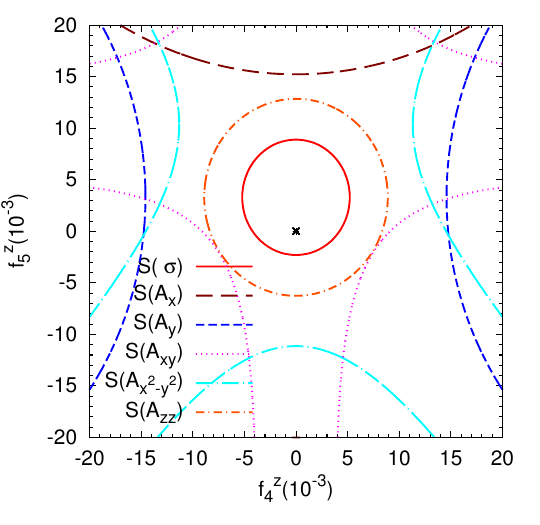}
\includegraphics[width=0.325\textwidth]{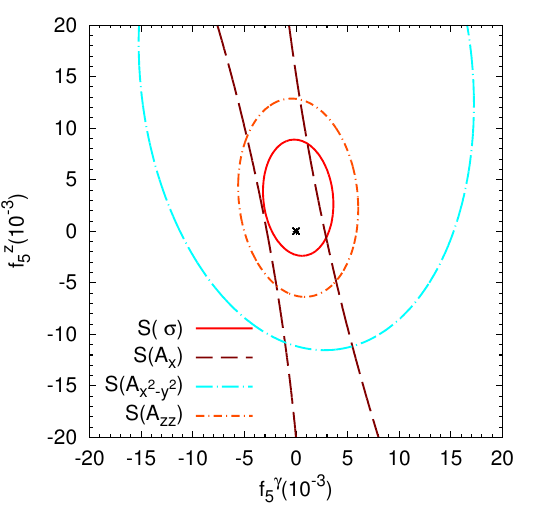}
\caption{\label{fig:Two_parameter_sensitivity_zz} $1\sigma$ sensitivity
contours ($\Delta\chi^2=1$) for  cross section and asymmetries obtained
by varying two parameters at a time and keeping the others at zero for the 
$ZZ$ process at $\sqrt{s}=500$ GeV and ${\cal L}=100$ 
fb$^{-1}$.}
\end{figure*}
\begin{figure*}
\centering
\includegraphics[width=0.325\textwidth]{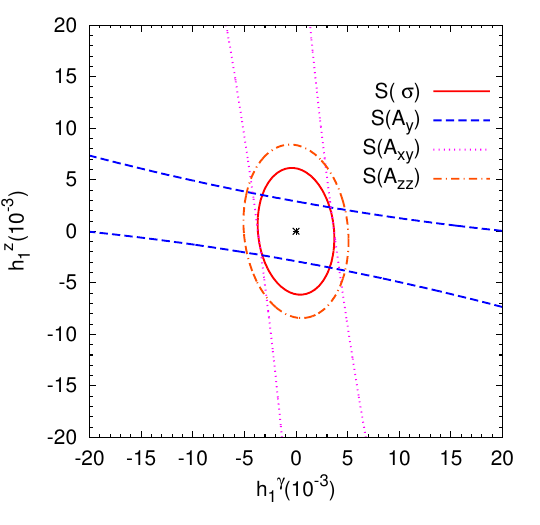}
\includegraphics[width=0.325\textwidth]{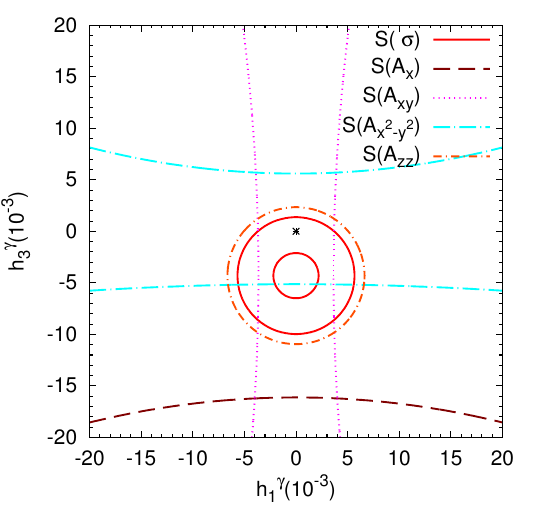}
\includegraphics[width=0.325\textwidth]{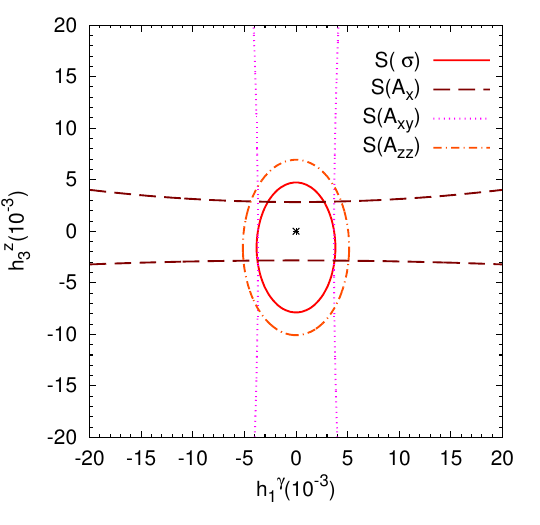}
\includegraphics[width=0.325\textwidth]{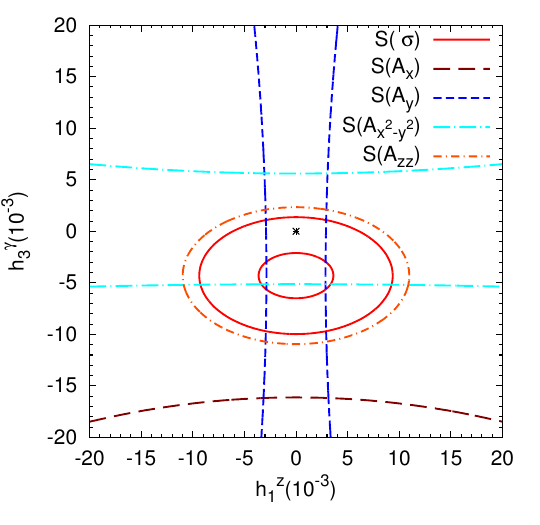}
\includegraphics[width=0.325\textwidth]{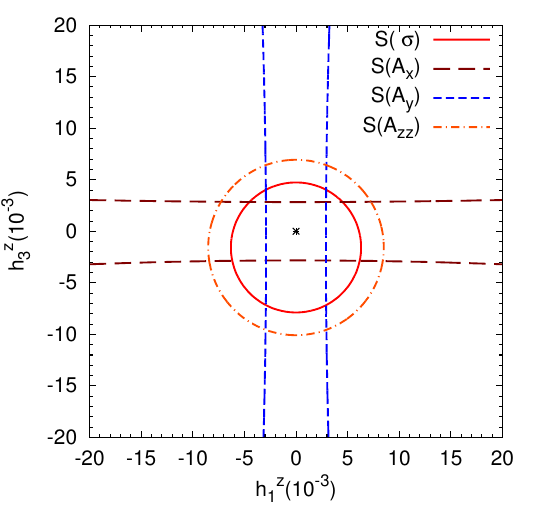}
\includegraphics[width=0.325\textwidth]{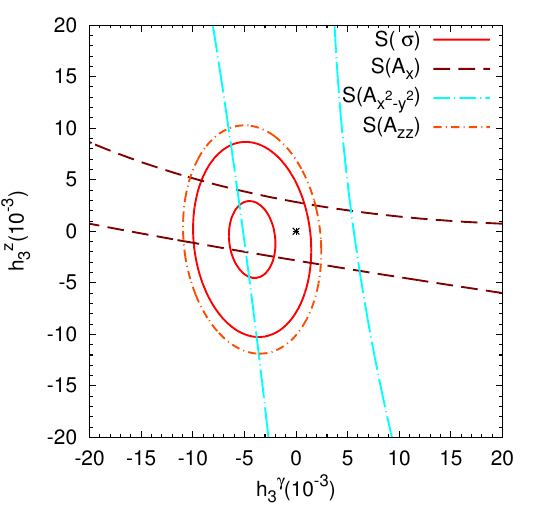}
\caption{\label{fig:Two_parameter_sensitivity_za}  $1\sigma$ sensitivity 
contours ($\Delta\chi^2=1$) for cross section and asymmetries obtained
by varying two parameters at a time and keeping the others at zero for the 
$Z\gamma$ process at $\sqrt{s}=500$ GeV, ${\cal L}=100$ fb$^{-1}$,
and $10^\circ \le\theta_\gamma\le170^\circ$.}
\end{figure*}
We vary two couplings at a time, for each observable, and plot the 
${\cal S}=1$ (or $\Delta\chi^2=1$) contours in the corresponding parameter 
plane. These contours are shown in Fig.~\ref{fig:Two_parameter_sensitivity_zz}
and Fig.~\ref{fig:Two_parameter_sensitivity_za} for $ZZ$ and $Z\gamma$
processes, respectively. Asterisk ($\star$) marks in the middle of these plots
denote the SM value, i.e., the $(0,0)$ point. Each panel corresponds to two 
couplings
that are varied and all others are kept at zero. The shapes of the contours, for a
given observable, are a reflection of its dependence on the couplings as shown
in Tables~\ref{tab:parameter_dependent_pol_zz} and~\ref{tab:parameter_dependent_pol_za}. For example, let us look at the 
middle-top panel of Fig.~\ref{fig:Two_parameter_sensitivity_zz}, i.e. the
$(f_4^\gamma-f_5^\gamma)$ plane. The contours corresponding to the 
cross section (solid/red) and 
$A_{zz}$ (short-dash-dotted/orange) are circular in shape due to their quadratic
dependence on these two couplings with the same sign. The small linear dependence on
$f_5^\gamma$ makes these circles move toward a small positive value, as already
observed in the one-parameter analysis above. The $A_y$ contour 
(short-dash/blue) depends only on $f_4^\gamma$ in the numerator and a mild
dependence on $f_5^\gamma$ enters through the cross section, sitting in the
denominator of the asymmetries. The role of two couplings are exchanged for 
the $A_x$ contour (big-dash/black). The $A_{xy}$ contour (dotted/magenta) is 
hyperbolic in shape, indicating a dependence on the product 
$f_4^\gamma f_5^\gamma$, while a small shift toward positive $f_5^\gamma$ value
indicates a linear dependence on it. Similarly the symmetry about $f_4^\gamma=0$
indicates no linear dependence on it for $A_{xy}$. All these observations can
be confirmed by looking at  Table~\ref{tab:parameter_dependent_pol_zz} and
the expressions in \ref{appendix:Expresson_observables}. Finally, the shape of the $A_{x^2-y^2}$ contour 
(big-dash-dotted/cyan)
indicates a quadratic dependence on two couplings with opposite sign. Similarly,
all other panels can be read. Note that taking any one of the coupling
to zero in these panels gives us the $1\sigma$ limit on the other couplings as
found in the one-parameter analysis above.

In the contours for the $Z \gamma $ process,
Fig.~\ref{fig:Two_parameter_sensitivity_za}, one new kind of shape appears: the
annular ring corresponding to $\sigma_{Z\gamma}$ in middle-top, left-bottom, and right-bottom panels. This shape
corresponds to a largely linear dependence of the cross section on $h_3^\gamma$
along with the quadratic dependence. By putting the other couplings to zero in
above-mentioned panels, one obtains two disjoint internals for $h_3^\gamma$ at 
$1\sigma$ level as found before in the one-parameter analysis. The plane
containing two $CP$-odd couplings, i.e. the left-top panel, has two sets of
slanted contours corresponding to $A_{y}$ (short-dash/blue) and $A_{xy}$ 
(dotted/magenta), the $CP$-odd observables. These observables depend upon both
the couplings linearly and hence the slanted (almost) parallel lines. The rest of the panels can be read in the same way.

Till here, we have used only one observable at a time for finding the limits.
A combination of all the observables would provide a much tighter limit on
the couplings than provided by any one of them alone. Also, the shape, the position, 
and the orientation of the allowed region would  change if the other two 
parameters were set to some value other than zero. A more comprehensive analysis
requires varying all the parameters and using all the observables to find the
parameter region of  low $\chi^2$ or high likelihood. The likelihood mapping of
the parameter space is performed using the MCMC method in
the next section.

\section{Likelihood mapping of parameter space}
\label{sec:MCMC}
In this section we perform a mock analysis of parameter estimation of anomalous coupling using {\em pseudo data} generated by {\tt MadGraph5}. We choose 
two benchmark points for coupling parameters as follows:
\begin{eqnarray}
\mbox{{\tt SM}} &:& f_{4,5}^V = 0.000, \ \ h_{1,3}^V = 0.000 \ \ 
\mbox{and}\nonumber\\
\mbox{{\tt aTGC}} &:& f_{4,5}^V = 0.005, \ \ h_{1,3}^V = 0.005 \ .
\end{eqnarray}
For each of these benchmark points we generate events in {\tt MadGraph5} 
 for {\em pseudo data} corresponding to ILC running at $500$ GeV and integrated 
 luminosity of ${\cal L}=100$ fb$^{-1}$. The
likelihood of a given point $\vec{f}$ in the parameter space is defined by,
\begin{equation}
{\cal L}(\vec{f}) = \prod_i \exp\left[- \ 
\frac{\left({\cal O}_i(\vec{f})-{\cal O}_i(\vec{f}_0)\right)^2}
{2\left(\delta {\cal O}_i(\vec{f}_0)\right)^2}\right]  ,
\end{equation}
where $\vec{f}_0$ defines the benchmark point. The product runs over the list
of observables we have: the cross section and five non-zero asymmetries.
We use the MCMC method to map the likelihood of the parameter space for each of the
benchmark point and for both processes. 
The one-dimensional marginalized 
distributions and the two-dimensional contours on the anomalous couplings are 
drawn from the Markov chains using the {\tt GetDist} package~\cite{Antony:GetDist,Lewis:2019xzd}.

\subsection{MCMC analysis for $e^+e^-\to ZZ$}
Here we look at the process $e^+e^-\to ZZ$ followed by the decays $Z\to l^+l^-$
and $Z\to q\bar{q}$, with $l^-=e^-,~\mu^-$ in the \\ {\tt MadGraph5} simulations. 
The total cross section for this whole process would be
\begin{equation}
\sigma= \sigma(e^+e^-\to ZZ)\times 2 \ Br(Z\to l^+l^-) \ Br(Z\to q\bar{q}).
\end{equation}
The theoretical values of $\sigma(e^+e^-\to ZZ)$ and all the asymmetries are
obtained using expressions given in Appendix B and shown in the second column 
of Tables \ref{tab:marginalised_stat_observables_zz_sm} and 
\ref{tab:marginalised_observables_zz_bsm} for benchmark points {\tt SM} and
{\tt aTGC}, respectively. The {\tt MadGraph5} simulated values for these
observables are given in the third column of the two tables mentioned for
two benchmark points. Using these simulated values as {\em pseudo data} we
perform the likelihood mapping of the parameter space and obtain the posterior
distributions for parameters and the observables.
The last two columns of Tables \ref{tab:marginalised_stat_observables_zz_sm} and
\ref{tab:marginalised_observables_zz_bsm} show the 68~\% and 95~\% Bayesian
confidence interval (BCI) of the observables used. One naively expects  68~\%
BCI to roughly has the same size as the $1\sigma$ error in the {\em pseudo data}. 
However, we note that the  68~\% BCI for all the asymmetries is much narrower than expected, for both benchmark points. This can be understood from the
fact that the cross section provides the strongest limit on any parameter, as
noticed in the earlier section, thus limiting the range of values for the
asymmetries. However, this must allow 68~\% BCI of the cross section to match
the expectation. This indeed happens for the {\tt aTGC} case 
(Table~\ref{tab:marginalised_observables_zz_bsm}), but for the {\tt SM} case,
even the cross section is narrowly constrained compared to a naive expectation.
The reason for this can be found in the dependence of the cross section on the
parameters. For most of the parameter space, the cross section is larger than
the {\tt SM} prediction, and only for a small range of parameter space, it can be
smaller.
\begin{table*}[h!]
    \centering
    \caption{\label{tab:marginalised_stat_observables_zz_sm} 
        List of observables shown for the process $e^+e^-\to ZZ$ for the  benchmark 
        point {\tt SM} with $\sqrt{s}=500$ GeV: theoretical values (column 2), 
        {\tt MadGraph5} simulated value for ${\cal L}=100$ fb$^{-1}$ (column 3), 68~\% (column
        4) and 95~\% (column 5) Bayesian confidence intervals (BCI). }
    \renewcommand{\arraystretch}{1.5}
    \begin{footnotesize}
        \begin{tabular*}{\textwidth}{@{\extracolsep{\fill}}lllll@{}}\hline
            Observables    &Theoretical ({\tt SM}) & MadGraph ({\tt SM}, prior)  
            &68~\% BCI (posterior)& 95~\% BCI (posterior)\\ 
            \hline
            $\sigma$      & $ 38.096$ fb  & $ 38.16\pm 0.62 $ fb    &
            $38.61^{+0.31}_{-0.53}$fb        & $38.61^{0.83}_{-0.74}$fb     \\ 
            $A_x$         & $ 0.00099$ & $0.0023\pm 0.0161$   & $-0.0021\pm 0.0087
            $     & $-0.0021^{+0.016}_{-0.017}  $    \\
            $A_y$         & $0 $           & $-0.0016\pm 0.0161 $ & $-0.0005\pm 0.0090
            $     & $-0.0005^{+0.017}_{-0.017}  $    \\ 
            $A_{xy}$      & $0 $           & $0.0004\pm 0.0161$  & $0.0001\pm 0.0036
            $     & $0.0001^{+0.0071}_{-0.0071}$      \\ 
            $A_{x^2-y^2}$ & $-0.02005 $  & $-0.0189\pm 0.0161$   &
            $-0.0166^{+0.0032}_{-0.0018}$    & $-0.0166^{+0.0043}_{-0.0052}$     \\ 
            $A_{zz}$      & $0.17262 $    & $0.1745\pm 0.0159 $    &
            $0.1691^{+0.0035}_{-0.0022}$     & $0.1691^{+0.0051}_{-0.0056}$      \\
            \hline
        \end{tabular*}
    \end{footnotesize}
\end{table*}
\begin{table*}[h!]
    \centering
    \caption{\label{tab:marginalised_observables_zz_bsm} 
        List of observables shown for the process $e^+e^-\to ZZ$ for the  benchmark 
        point {\tt aTGC} with $\sqrt{s}=500$ GeV. The rest of the details are the same as in 
        Table~\ref{tab:marginalised_stat_observables_zz_sm}. }
    \renewcommand{\arraystretch}{1.5}
    \begin{footnotesize}
        \begin{tabular*}{\textwidth}{@{\extracolsep{\fill}}lllll@{}}\hline
            Observables    &Theoretical ({\tt aTGC})& MadGraph ({\tt aTGC}, prior) & 
            68~\% BCI (posterior) 
            & 95~\% BCI (posterior) \\ \hline
            $\sigma$      & $43.307 $ fb & $43.33\pm 0.6582 $ fb      & $43.40\pm 0.66
            $ fb        & $43.40 \pm 1.3$ fb         \\  
            $A_x$         & $-0.02954 $ & $-0.0308\pm 0.0151$    &
            $-0.0240^{+0.0087}_{-0.013}$ & $-0.0240^{+0.021}_{-0.020}  $   \\ 
            $A_y$         & $0.03424 $  & $0.0308\pm 0.0151$     &
            $0.0230^{+0.013}_{-0.0085} $ & $0.0230^{+0.020}_{-0.022}   $    \\  
            $A_{xy}$      & $ 0.00574$ & $0.0056\pm 0.0152$    &
            $0.0041^{+0.0076}_{-0.0063}$ & $0.0041^{+0.015}_{-0.015}   $    \\
            $A_{x^2-y^2}$ & $-0.00941 $ & $-0.0119 \pm 0.0152$   &
            $-0.0116^{+0.0071}_{-0.0032}$& $-0.0116^{+0.0093}_{-0.012}$     \\
            $A_{zz}$      & $ 0.14057$   & $0.1382 \pm 0.0150$     & $0.1401\pm 0.0035
            $ & $0.1401^{+0.0069}_{-0.0067}$     \\\hline
        \end{tabular*}
    \end{footnotesize}
\end{table*}
\begin{figure*}[h!]
    \centering
    \includegraphics[width=0.49\textwidth]{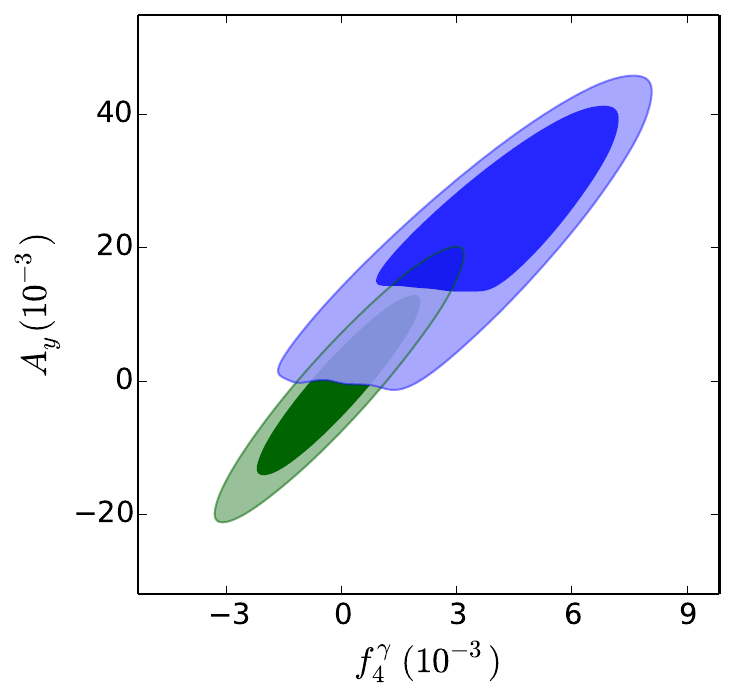}
    \includegraphics[width=0.49\textwidth]{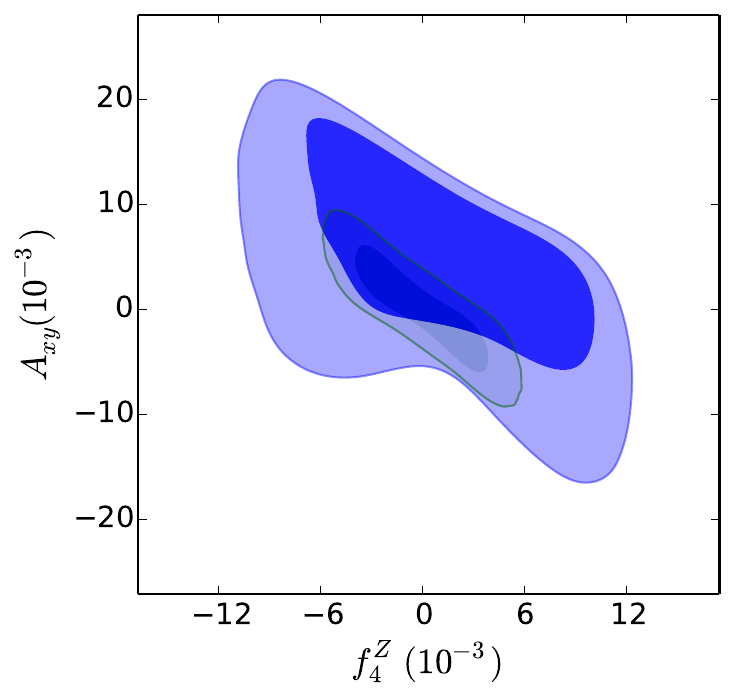}
    \includegraphics[width=0.49\textwidth]{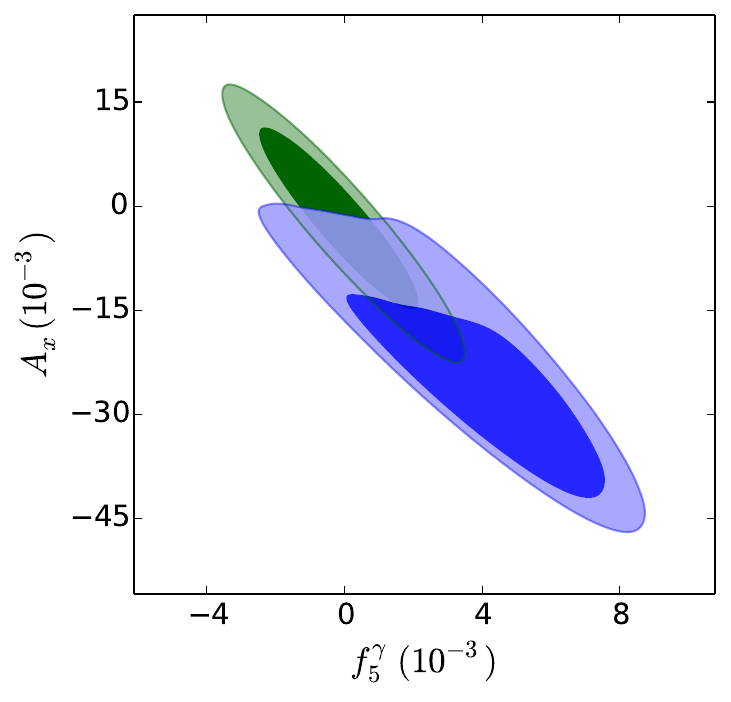}
    \includegraphics[width=0.49\textwidth]{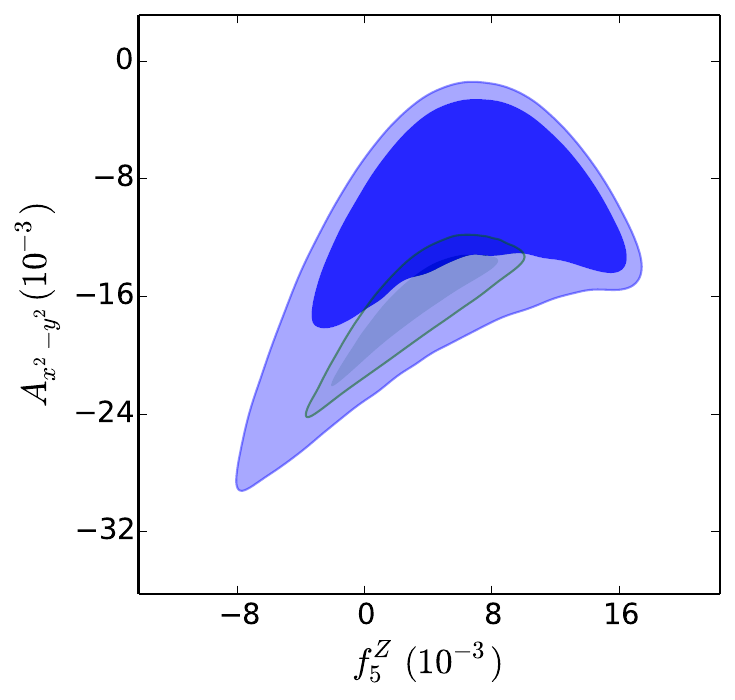}
    \caption{\label{fig:correlations_observables_in_mcmc_zz} 
        Two-dimensional marginalized contours showing most correlated observable for
        each parameter of the  process $e^+e^-\rightarrow ZZ$. The upper transparent layer ({\em blue})
        contours correspond to {\tt aTGC}, while the lower layer ({\em green}) contours
        correspond to {\tt SM}. The darker shade shows $68~\%$ contours, while the lighter
        shade is for $95~\%$ contours.}
\end{figure*}
\begin{figure}[h]
    \centering
    \includegraphics[width=0.5\textwidth]{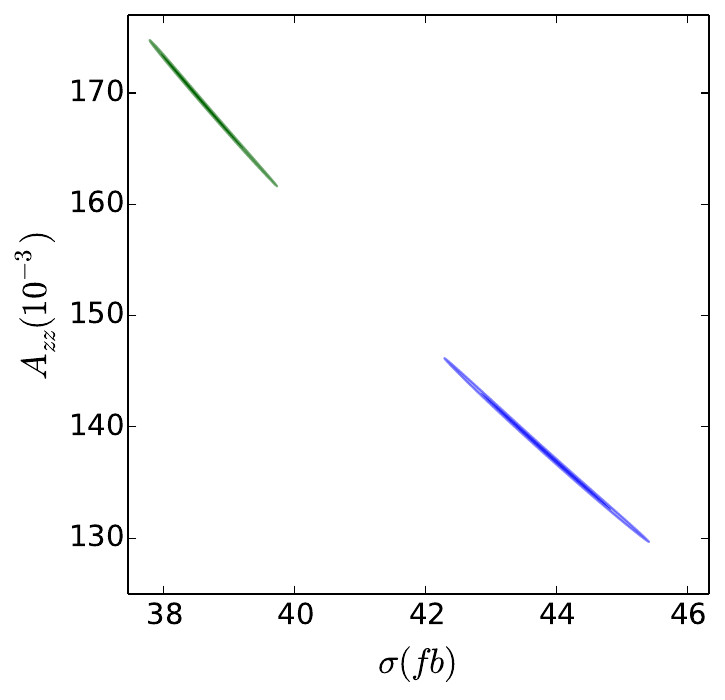}
    \caption{\label{fig:sigmatzz_zz_withc} Two-dimensional marginalized contours showing correlation between $A_{zz}$ and $\sigma$ in the $ZZ$ process. The rest of the details are the same as in Fig.~\ref{fig:correlations_observables_in_mcmc_zz}.}
\end{figure}
 This was already pointed out while discussing multi-valued 
sensitivity in Fig.~\ref{fig:one_parameter_sensitivity_zz}. We found the
lowest possible value of the cross section to be $37.77$ fb, obtained for 
$f_4^{\gamma,Z}\approx 0$, $f_5^\gamma \sim 2\times10^{-4}$, and $f_5^Z \sim
3.2\times10^{-3}$.
Thus, for  most of the parameter space the anomalous couplings cannot 
emulate the negative statistical fluctuations in the cross section making the
likelihood function, effectively, a one-sided Gaussian function. This 
forces the mean of posterior distribution to a higher value. We also note that 
the upper bound of the 68~\% BCI for cross section ($38.92$ fb) is comparable to 
the expected $1\sigma$ upper bound ($38.78$ fb). Thus we have an overall 
narrowing of the range of the posterior
distribution of the cross section values. This, in turns, leads to a narrow range
of parameters allowed and hence narrow ranges for the asymmetries in the case 
of {\tt SM} benchmark point. For the {\tt aTGC} benchmark point, it is possible
to emulate the negative fluctuations in cross section by varying the 
parameters, thus the corresponding posterior distributions compare with the 
expected $1\sigma$ fluctuations. The narrow ranges for the posterior distribution 
for all the asymmetries are due to the tighter constraints on the parameters coming
from the cross section and correlation between the observables. 

\begin{figure}[h!]
\centering
\includegraphics[width=0.35\textwidth]{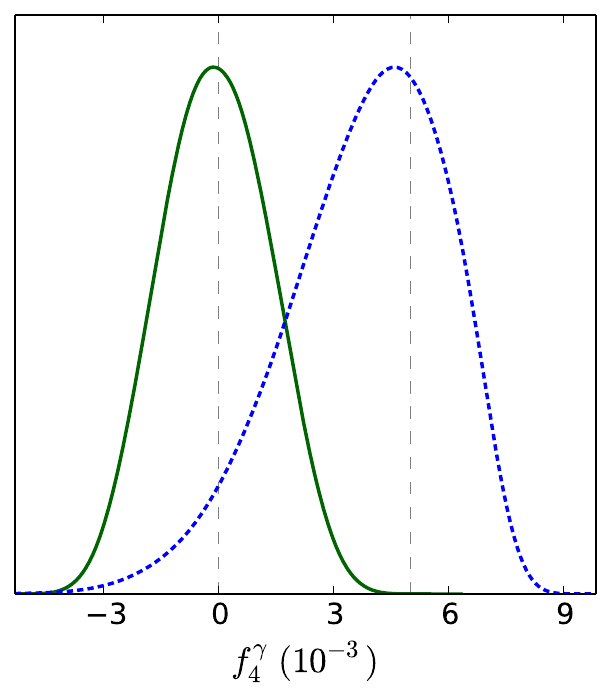}
\includegraphics[width=0.35\textwidth]{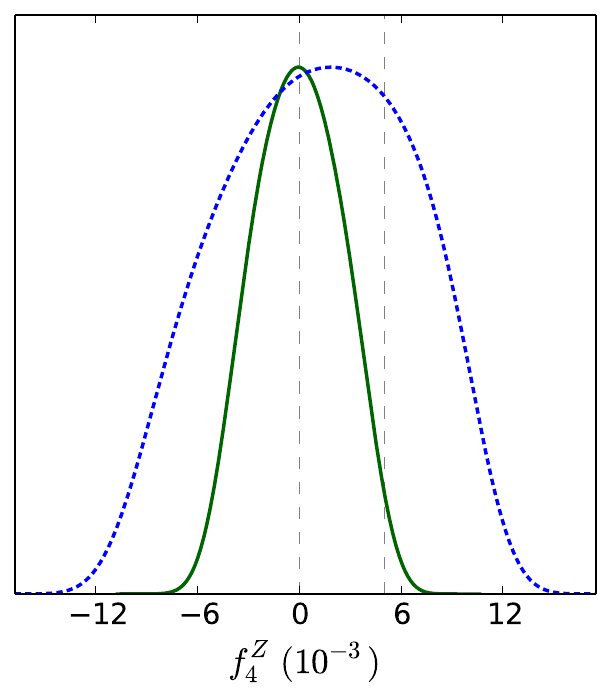}
\includegraphics[width=0.35\textwidth]{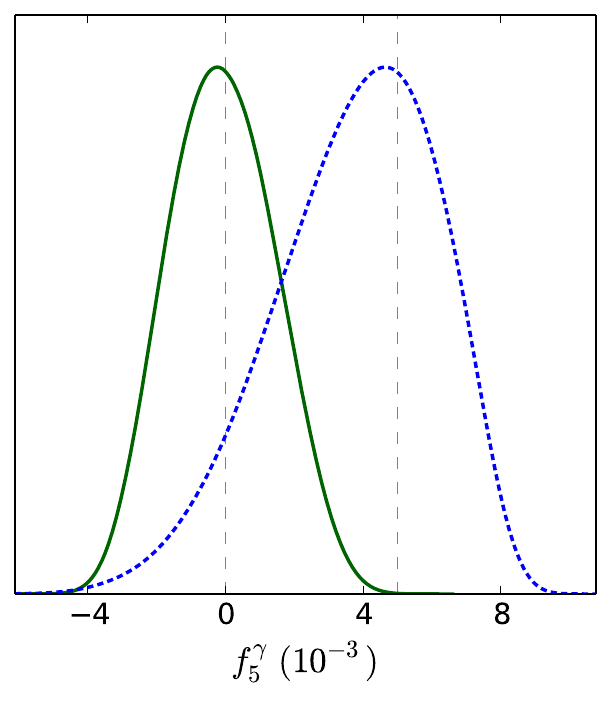}
\includegraphics[width=0.35\textwidth]{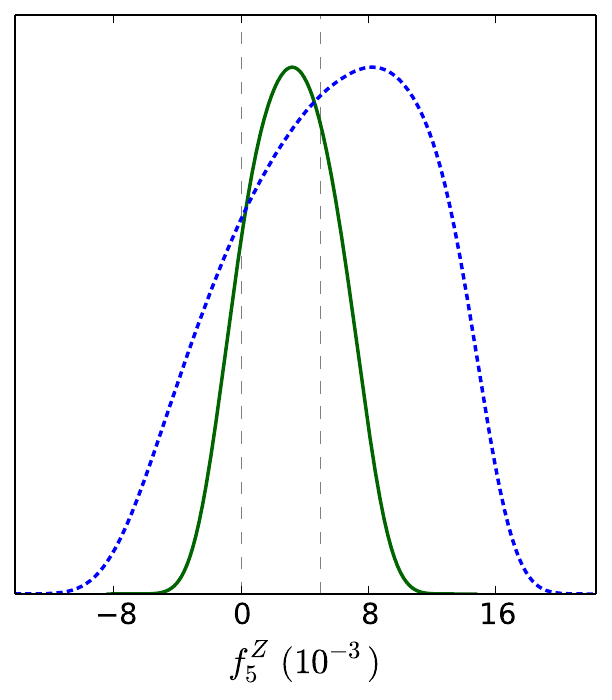}
\caption{\label{fig:one_parameter_mcmc_parameter_zz} 
One-dimensional marginalized posterior distribution for the parameters of the 
process $e^+e^-\rightarrow ZZ$. {\em Solid (green)} lines are for {\tt SM} and 
{\em dashed (blue)} lines are for {\tt aTGC} hypothesis. The values of the parameters
for the benchmark points are shown by vertical lines for reference.}
\end{figure}
\begin{figure*}[h!]
\centering
\includegraphics[width=0.32\textwidth]{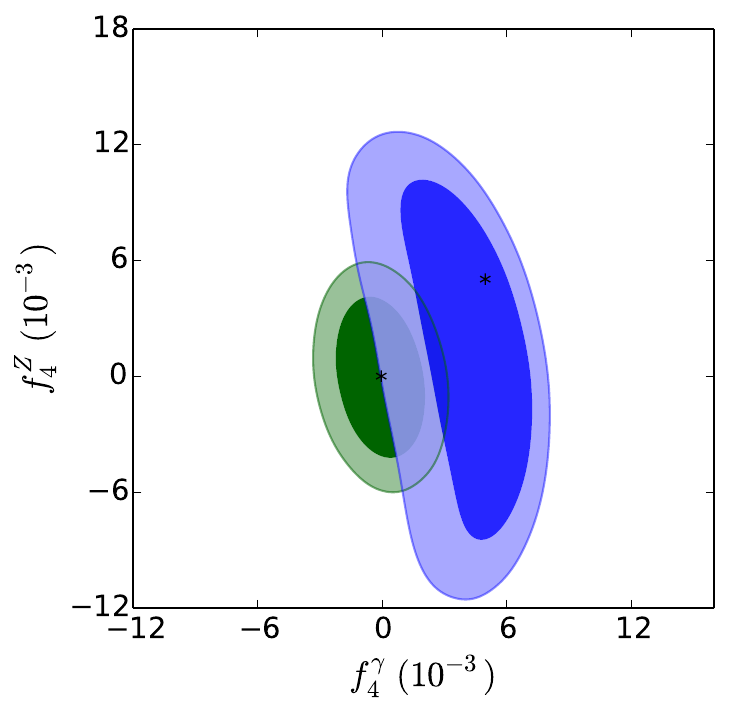}
\includegraphics[width=0.32\textwidth]{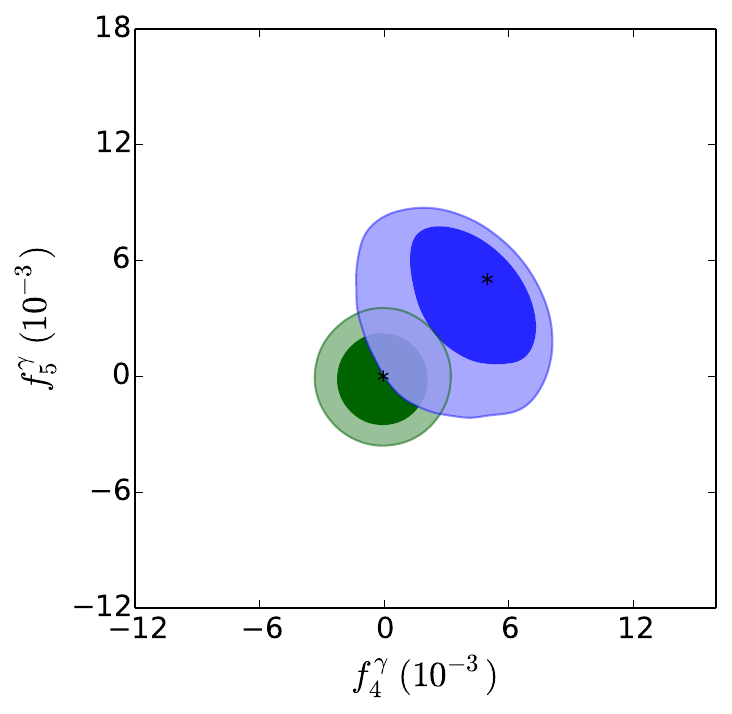}
\includegraphics[width=0.32\textwidth]{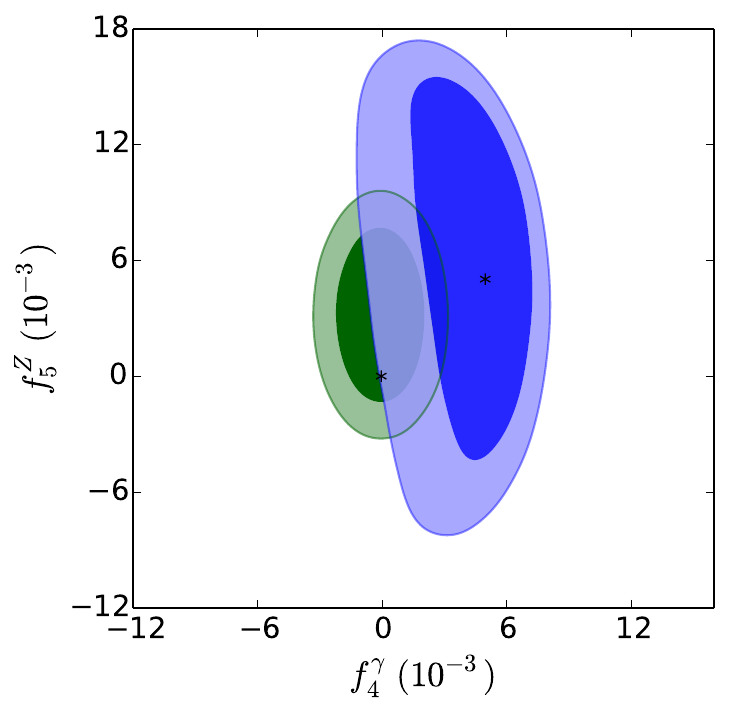}
\includegraphics[width=0.32\textwidth]{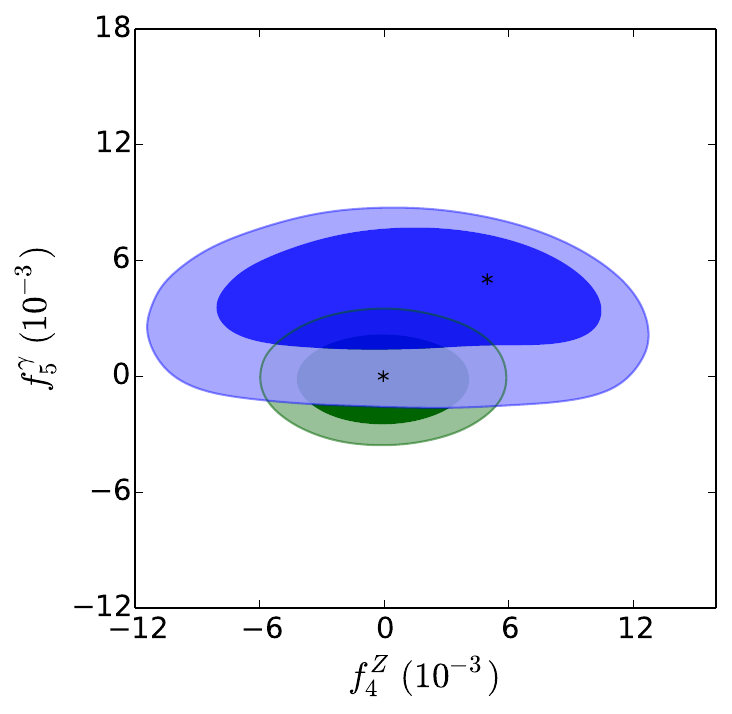}
\includegraphics[width=0.32\textwidth]{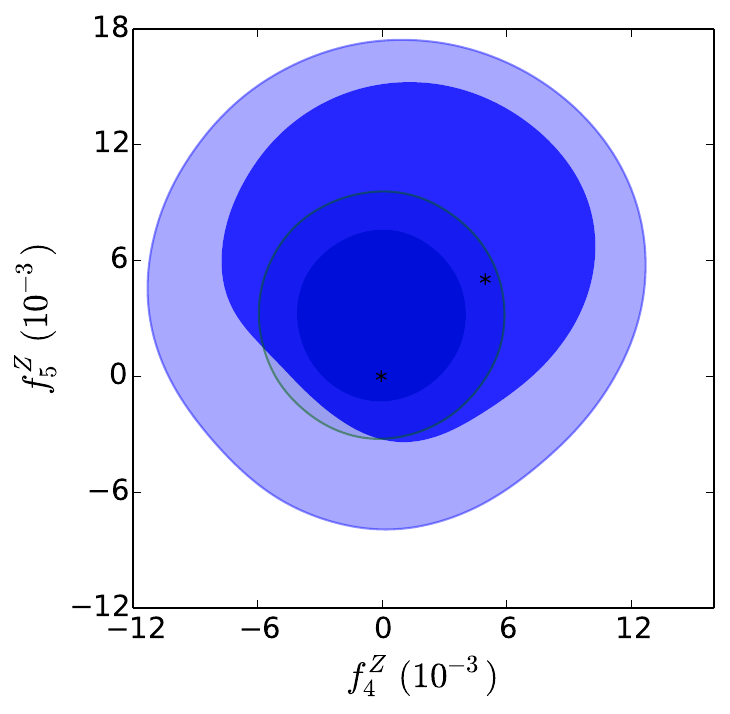}
\includegraphics[width=0.32\textwidth]{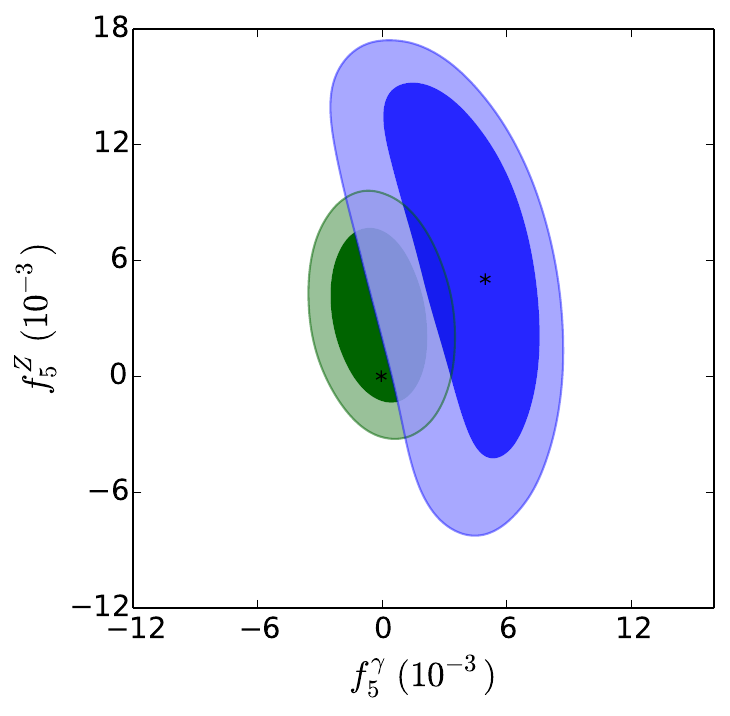}
\caption{\label{fig:correlations_parameters_in_mcmc_zz} 
Two-dimensional marginalized contours showing correlations between parameters
of the  process $e^+e^-\rightarrow ZZ$. The other details are the same as in 
Fig.~\ref{fig:correlations_observables_in_mcmc_zz}.
}
\end{figure*}
\begin{table*}[t]
    \centering
    \caption{\label{tab:marginalised_parameters_zz} 
        The list of best-fit points, posterior 68~\% and 95~\% BCI  for the parameters for 
        the process $e^+e^-\to ZZ$ for both {\tt SM} and {\tt aTGC} benchmark points.
    }
    \renewcommand{\arraystretch}{1.5}
    \begin{footnotesize}
        \begin{tabular*}{\textwidth}{@{\extracolsep{\fill}}lllllll@{}}\hline
            \multicolumn{1}{c}{}&\multicolumn{3}{c}{{\tt SM} Benchmark}
            & \multicolumn{3}{c}{{\tt aTGC} Benchmark}\\ \hline
            $f^V_i$ &   68~\% BCI &  95~\% BCI & Best-fit & 68~\% BCI  &  95~\% BCI & Best-fit
            \\\hline
            $f_4^\gamma$   & $-0.0001\pm 0.0014  $  & $-0.0001^{+0.0027}_{-0.0027}$
            &$-0.0002$ & $0.0038^{+0.0026}_{-0.0016}$& $0.0038^{+0.0037}_{-0.0042}$&$0.0044$
            \\ 
            $f_4^Z$        & $0.0000\pm 0.0026   $  & $0.0000^{+0.0049}_{-0.0049}$
            &$-0.0002$  & $0.0010^{+0.0065}_{-0.0055}$& $0.0010^{+0.0098}_{-0.011} $
            &$0.0050$  \\ 
            $f_5^\gamma$   & $-0.0001\pm 0.0015  $  & $-0.0001^{+0.0030}_{-0.0029}$ &
            $-0.0002$ & $0.0038^{+0.0029}_{-0.0019}$& $0.0038^{+0.0042}_{-0.0047}$ &$0.0057$
            \\ 
            $f_5^Z$        & $0.0032\pm 0.0028   $  & $0.0032^{+0.0053}_{-0.0053}$
            &$0.0000$  & $0.0057^{+0.0074}_{-0.0051}$& $0.0057^{+0.010}_{-0.011}
            $&$0.0037$  \\ \hline
        \end{tabular*}
    \end{footnotesize}
\end{table*}
We are using a total of six observables, five asymmetries and one cross section
for our analysis of two benchmark points; however, we have only four free 
parameters. This invariably leads to some correlations between the observables
apart from the expected correlations between parameters and observables.
Figure~\ref{fig:correlations_observables_in_mcmc_zz} shows the
most prominently correlated observable for each of the parameters. The $CP$
nature of  observables is reflected in the parameter it is strongly correlated
with. We see that $A_y$ and $A_{xy}$ are linearly dependent upon both $f_4^\gamma$ and
$f_4^Z$; however, $A_y$ is more sensitive to $f_4^\gamma$ as shown in 
Fig.~\ref{fig:one_parameter_sensitivity_zz} as well. Similarly, for the other
asymmetries and parameters, one can see a correlation which is consistent with the
sensitivity plots in Fig.~\ref{fig:one_parameter_sensitivity_zz}.
The strong (and negative) correlation between $A_{zz}$ and $\sigma$ shown in Fig.~\ref{fig:sigmatzz_zz_withc} 
indicates that any one of them is sufficient for the analysis, in principle.
However, in practice, the cross section puts a much stronger limit than $A_{zz}$, which
explains the much narrower BCI for it as compared to the $1\sigma$ expectation.

Finally, we come to the discussion of the parameter  estimation. The marginalized
one-dimensional posterior distributions for the parameters of $ZZ$ production
process are shown in Fig.~\ref{fig:one_parameter_mcmc_parameter_zz}, while
the corresponding BCI along with best-fit points are listed in 
Table~\ref{tab:marginalised_parameters_zz} for both  benchmark points. 
The vertical lines near zero correspond to the true value of parameters for 
{\tt SM} and the other vertical lines correspond to {\tt aTGC}. The best-fit
points are very close to the true values except for $f_5^Z$ in the  {\tt aTGC} 
benchmark point due
to the multi-valuedness of the cross section. The 95~\% BCI of the parameters for two 
benchmark points overlap, and it appears as if they cannot be resolved. To see 
the resolution better, we plot two-dimensional posteriors in 
Fig.~\ref{fig:correlations_parameters_in_mcmc_zz}, with the benchmark points
shown with an asterisk. Again we see that the 95~\% contours do overlap as these
contours are obtained after marginalizing over non-shown parameters in each 
panel. Any higher-dimensional representation is not possible on paper, but we
have checked three-dimensional scatter plot of points on the Markov chains and 
conclude that the shape of the {\em good likelihood} region is ellipsoidal for
the {\tt SM} point with the true value at its centre. The corresponding
three-dimensional
shape for the {\tt aTGC} point is like a part of an ellipsoidal shell. Thus
in full four-dimension, there will not be any overlap 
(see Section \ref{ssec:separation})
and we can distinguish the
two chosen benchmark points as it is quite obvious from the corresponding 
cross sections. However, left to only the cross section we would have the 
entire ellipsoidal shell as possible range of parameters for the {\tt aTGC} 
case. The presence of asymmetries in our analysis helps narrow down to a part
of the ellipsoid and hence aids the parameter estimation for the $ZZ$ production
process.

\subsection{MCMC analysis for $e^+e^-\to Z\gamma$ }
Next we look at the process $e^+e^-\to Z\gamma$ and $Z\to l^+l^-$ with 
$l^-=e^-, \mu^-$ in the {\tt MadGraph5} simulations. The total cross section 
for this whole process is given by,
\begin{equation}
\sigma= \sigma(e^-e^+\to Z\gamma)\times Br(Z\to l^+l^-).
\end{equation}
The theoretical values of the cross section and asymmetries (using expression 
in~\ref{appendix:Expresson_observables}) are given in the second column of
Tables \ref{tab:marginalised_observables_za_sm} and
\ref{tab:marginalised_observables_za_bsm} for {\tt SM} and {\tt aTGC} points,
respectively. The tables contain the {\tt MadGraph5} simulated data for 
${\cal L}=100$ fb$^{-1}$ along with 68~\% and 95~\% BCI for the observables obtained from
the MCMC analysis. For the {\tt SM} point, 
Table~\ref{tab:marginalised_observables_za_sm}, we notice that the 68~\% BCI for
all the observables are narrower than the $1\sigma$ range of the {\em psuedo
    data} from {\tt MadGraph5}. This is again related to the correlations between
observables and the fact that the cross section has a lower bound of about 111
fb obtained for $h_3^\gamma\sim -4.2\times 10^{-3}$ with other parameters close
to zero. This lower bound of the cross section leads to narrowing of 68~\% BCI for
$\sigma$ and hence for other asymmetries too, as observed in the $ZZ$ 
production process. The 68~\% BCI for 
$A_{x^2-y^2}$ and $A_{zz}$ are particularly narrow. For $A_{zz}$, this is 
related to the strong correlation between $(\sigma - A_{zz})$, while for 
$A_{x^2-y^2}$ the slower dependence on $h_3^\gamma$ along with strong 
dependence of $\sigma$ on $h_3^\gamma$ is the cause of a narrow 68~\% BCI.
\begin{table*}[h!]
\centering
\caption{\label{tab:marginalised_observables_za_sm}
List of observables shown for the process $e^+e^-\to Z \gamma$ for the {\tt SM} 
point with $\sqrt{s}=500$ GeV and ${\cal L}=100$ fb$^{-1}$. The rest of the details are the 
same as in Table~\ref{tab:marginalised_stat_observables_zz_sm}.
}
\renewcommand{\arraystretch}{1.5}
\begin{footnotesize}
\begin{tabular*}{\textwidth}{@{\extracolsep{\fill}}lllll@{}}\hline
 Observables    & Theoretical ({\tt SM})& MadGraph ({\tt SM}, prior) & 
68~\% BCI (posterior) & 95~\% BCI (posterior)  \\ \hline
$\sigma$      & $ 112.40$ fb & $112.6\pm 1.06 $  fb   &
$112.64^{+0.64}_{-0.91}$ fb      & $112.6^{+1.5}_{-1.4}$ pb\\
$A_x$         & $0.00480 $ & $0.0043\pm 0.0094 $ & $0.0041\pm 0.0088
$     & $0.0041^{+0.017}_{-0.018}   $ \\
$A_y$         & $0 $          & $-0.0011\pm 0.0094$ & $-0.0009\pm 0.0088
$     & $-0.0009 \pm 0.017  $ \\
$A_{xy}$      & $ 0$          & $0.0003\pm 0.0094$ & $0.0001\pm 0.0065
$     & $0.0001 \pm 0.012  $ \\
$A_{x^2-y^2}$ & $0.00527 $ & $0.0056\pm 0.0094$ &
$-0.0001^{+0.0064}_{-0.0034}$    & $-0.0001^{+0.0079}_{-0.0096}$\\
$A_{zz}$      & $ 0.17819$   & $0.1781\pm 0.0092$   &
$0.1771^{+0.0043}_{-0.0031}$     & $0.1771^{+0.0066}_{-0.0070}$ \\\hline
\end{tabular*}
\end{footnotesize}
\end{table*}
\begin{table*}[h!]
\centering
\caption{\label{tab:marginalised_observables_za_bsm} 
List of observables shown for the process $e^+e^-\to Z\gamma$ for the 
{\tt aTGC} point with $\sqrt{s}=500$ GeV and ${\cal L}=100$ fb$^{-1}$. The rest of the 
details are the same as in Table~\ref{tab:marginalised_stat_observables_zz_sm}.
}
\renewcommand{\arraystretch}{1.5}
\begin{footnotesize}
\begin{tabular*}{\textwidth}{@{\extracolsep{\fill}}lllll@{}}\hline
Observables & Theoretical ({\tt aTGC})&MadGraph ({\tt aTGC}, prior) & 
68~\% BCI (posterior) & 95~\% BCI  (posterior)\\\hline
$\sigma$      & $122.0 $ fb  & $122.4 \pm 1.11$ fb   &  $122.3\pm 1.0
$ fb    & $122.3 \pm 2.0$ fb          \\
$A_x$         & $0.02404 $   & $0.0252\pm 0.0090$  &  $0.0263\pm 0.0093
$   & $0.0263 \pm 0.018   $        \\
$A_y$         & $-0.01775 $  & $-0.0165\pm 0.0090$ &  $-0.0172\pm 0.0092
$   & $-0.0172 \pm 0.018  $        \\
$A_{xy}$      & $ -0.01350$  & $-0.0104\pm 0.0090$ &
$-0.0109^{+0.0069}_{-0.011}$   & $-0.0109^{+0.017}_{-0.015}  $\\
$A_{x^2-y^2}$ & $0.01440 $   & $0.0133\pm 0.0090$  &
$0.0121^{+0.0055}_{-0.0010}$   & $0.0121^{+0.0068}_{-0.012} $\\
$A_{zz}$      & $0.13612 $    & $0.1361\pm 0.0089$   &  $0.1351\pm 0.0041
$   & $0.1351^{+0.0080}_{-0.0079}$\\\hline
\end{tabular*}
\end{footnotesize}
\end{table*}
\begin{figure*}[h!]
\centering
\includegraphics[width=0.49\textwidth]{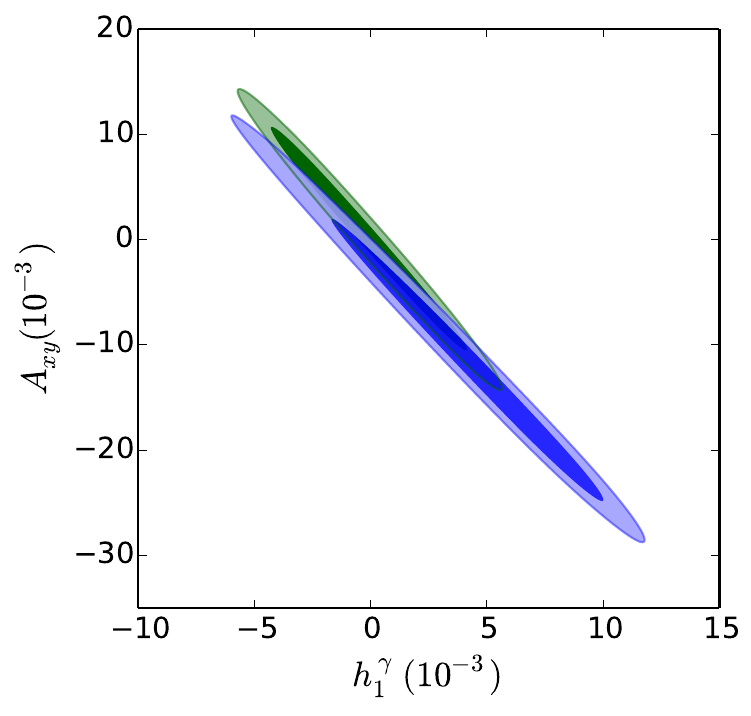}
\includegraphics[width=0.49\textwidth]{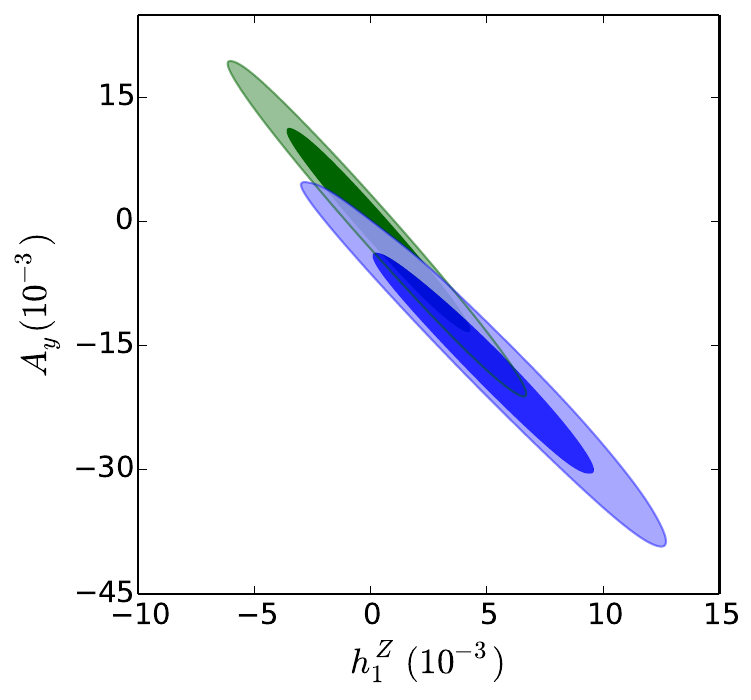}
\includegraphics[width=0.49\textwidth]{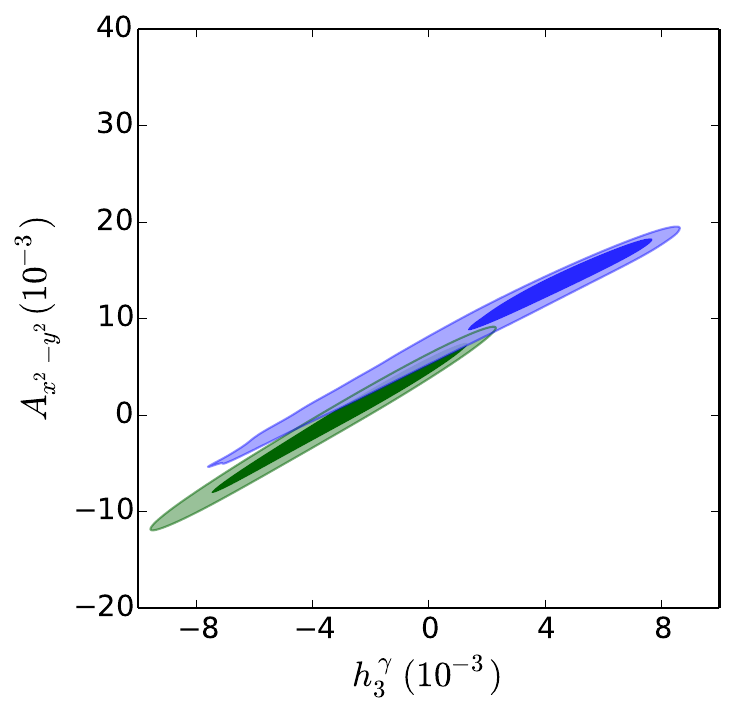}
\includegraphics[width=0.49\textwidth]{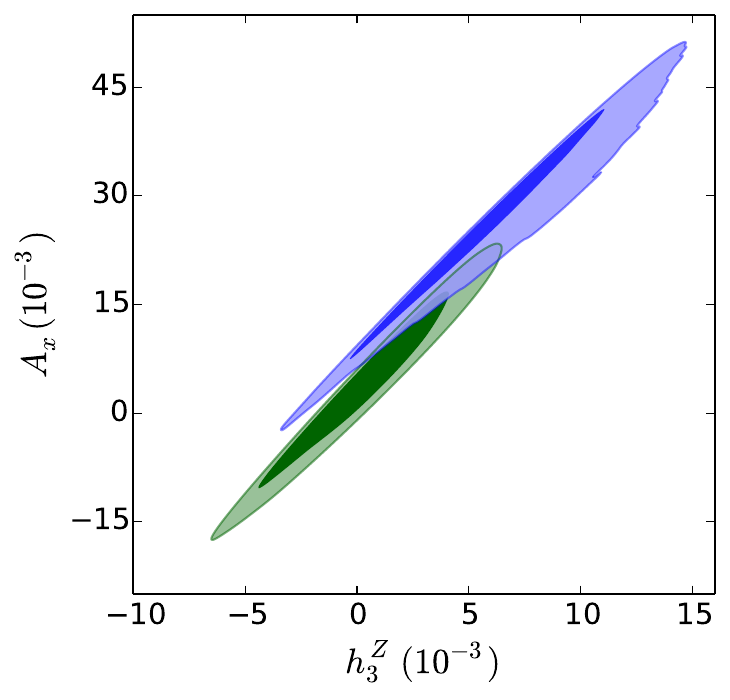}
\caption{\label{fig:correlations_observables_in_mcmc_za} 
Two-dimensional marginalized contours showing most correlated observables for each
parameter of the process $e^+e^-\rightarrow Z\gamma$ for two benchmark points. 
The rest of the details are the same as in
Fig.~\ref{fig:correlations_observables_in_mcmc_zz}.
}
\end{figure*}
\begin{table*}[h!]
\centering
\caption{\label{tab:marginalised_parameters_za_sm} 
The list of best-fit points, posterior $68~\%$ and $95~\%$ BCI  for the parameters for the process
$e^+e^-\to Z\gamma$ for both  benchmark points.
}
\renewcommand{\arraystretch}{1.5}
\begin{footnotesize}
\begin{tabular*}{\textwidth}{@{\extracolsep{\fill}}lllllll@{}}\hline
\multicolumn{1}{c}{}&\multicolumn{3}{c}{{\tt SM} Benchmark}
& \multicolumn{3}{c}{{\tt aTGC} Benchmark}\\ \hline
 $h^V_i$    &  68~\% BCI                     &  95~\% BCI &Best-fit
&  68~\% BCI &  95~\% BCI & Best-fit  \\ \hline
$h_1^\gamma$  & $-0.0001\pm 0.0026         $     & $-0.0001^{+0.0048}_{-0.0047}$
&$ -0.0002$ & $0.0039^{+0.0047}_{-0.0031}$   & $0.0039^{+0.0068}_{-0.0075}$
&$0.0040 $\\
$h_1^Z$       & $0.0003\pm 0.0028          $     & $0.0003^{+0.0054}_{-0.0054}$
& $0.0001 $& $0.0050\pm 0.0033          $   &
$0.0050^{+0.0064}_{-0.0063}$&$0.0047 $\\
$h_3^\gamma$  & $-0.0030^{+0.0036}_{-0.0020}$    & $-0.0030^{+0.0045}_{-0.0054}$
& $0.0002 $& $0.00348^{+0.0036}_{-0.00086}$ &
$0.00348^{+0.0047}_{-0.0076}$&$0.0056 $\\
$h_3^Z$       & $0.0004\pm 0.0028          $     & $0.0004^{+0.0053}_{-0.0055}$
& $-0.0002 $& $0.0062^{+0.0030}_{-0.0035}$   &
$0.0062^{+0.0070}_{-0.0062}$&$0.0052 $\\\hline
\end{tabular*}
\end{footnotesize}
\end{table*}
\begin{figure*}[h!]
\centering
\includegraphics[width=0.35\textwidth]{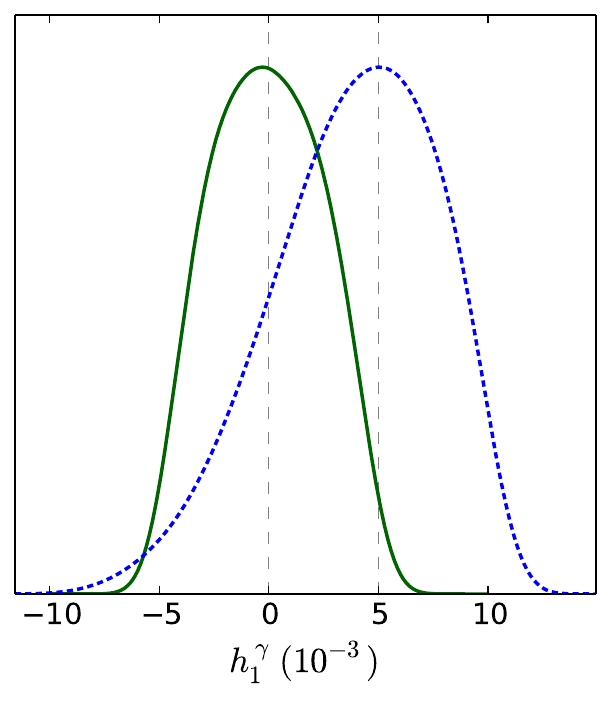}
\includegraphics[width=0.35\textwidth]{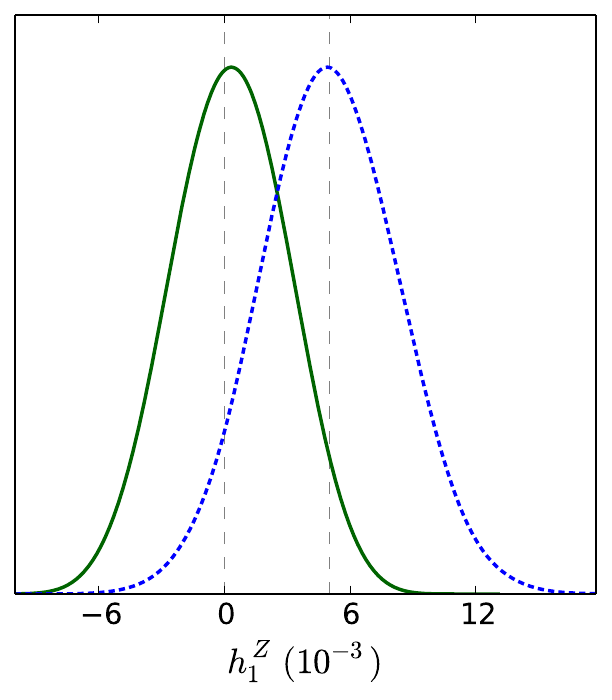}
\includegraphics[width=0.35\textwidth]{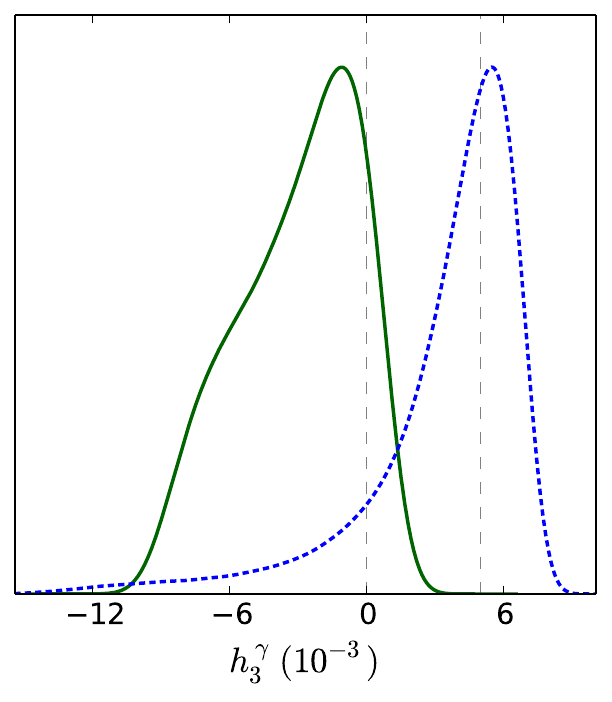}
\includegraphics[width=0.35\textwidth]{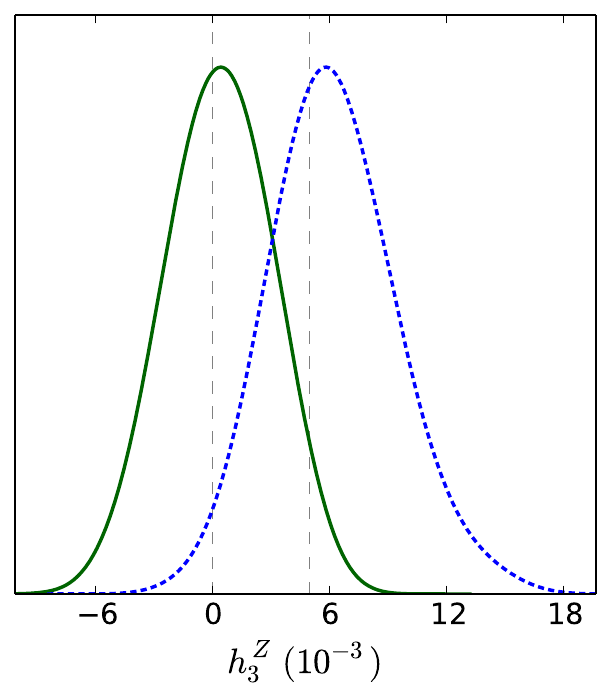}
\caption{\label{fig:one_parameter_mcmc_parameter_za} 
Posterior one-dimensional marginalized distributions for parameters of the
process $e^+e^-\rightarrow Z\gamma$ for {\tt SM} ({\em green/solid}) and
{\tt aTGC} ({\em blue/dashed}) points. Vertical lines denote the values of the
benchmark points. }
\end{figure*}
\begin{figure*}[h!]
\centering
\includegraphics[width=0.32\textwidth]{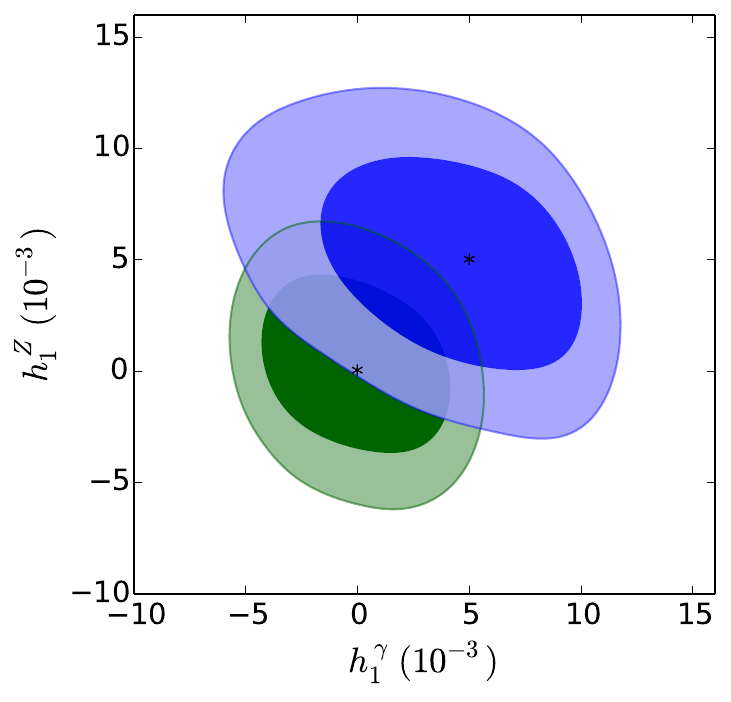}
\includegraphics[width=0.32\textwidth]{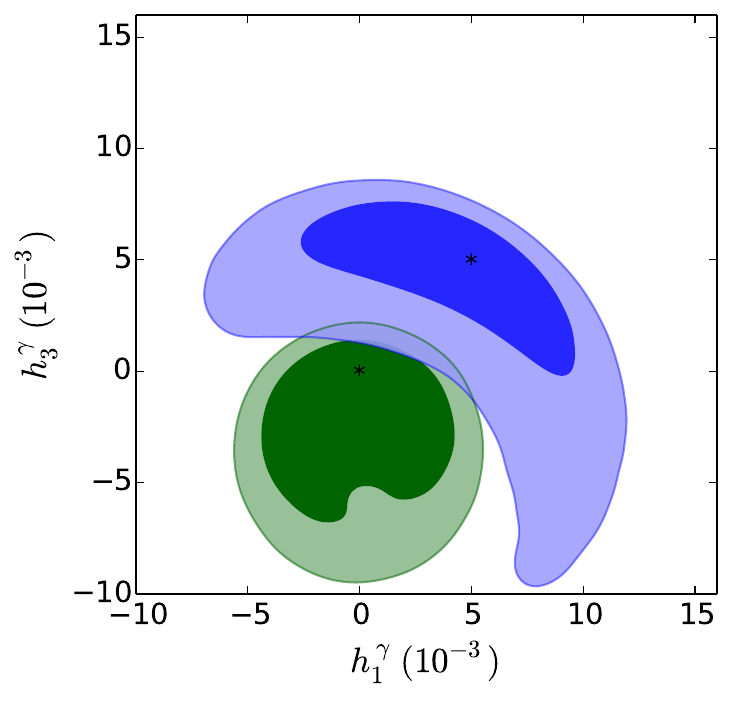}
\includegraphics[width=0.32\textwidth]{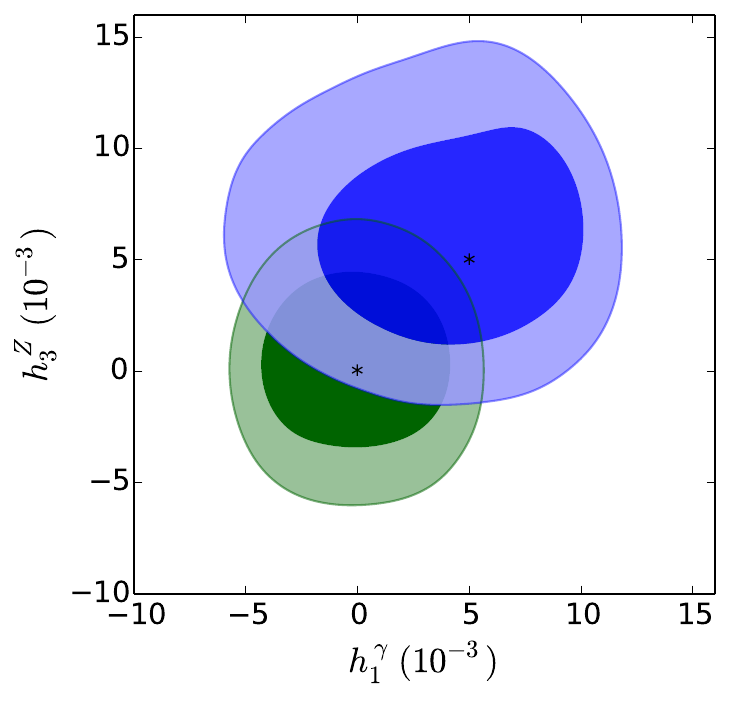}
\includegraphics[width=0.32\textwidth]{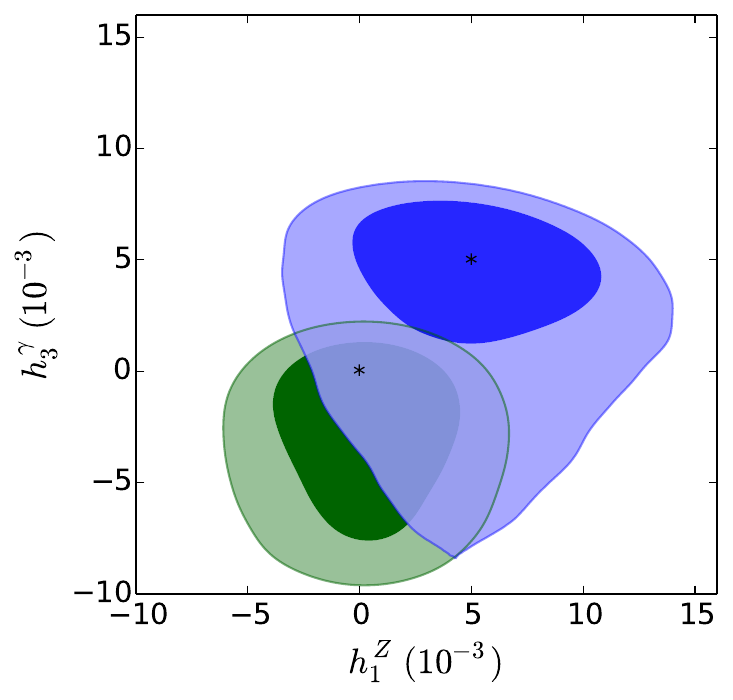}
\includegraphics[width=0.32\textwidth]{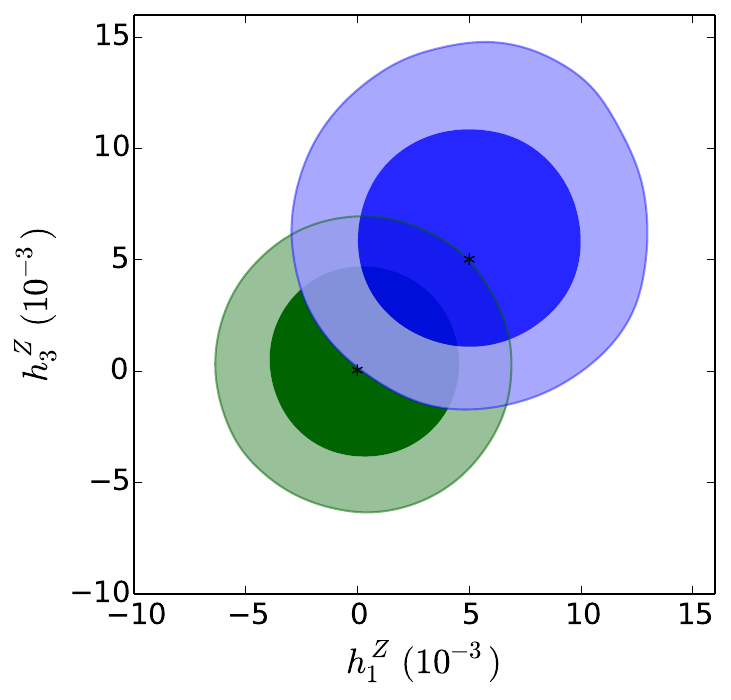}
\includegraphics[width=0.32\textwidth]{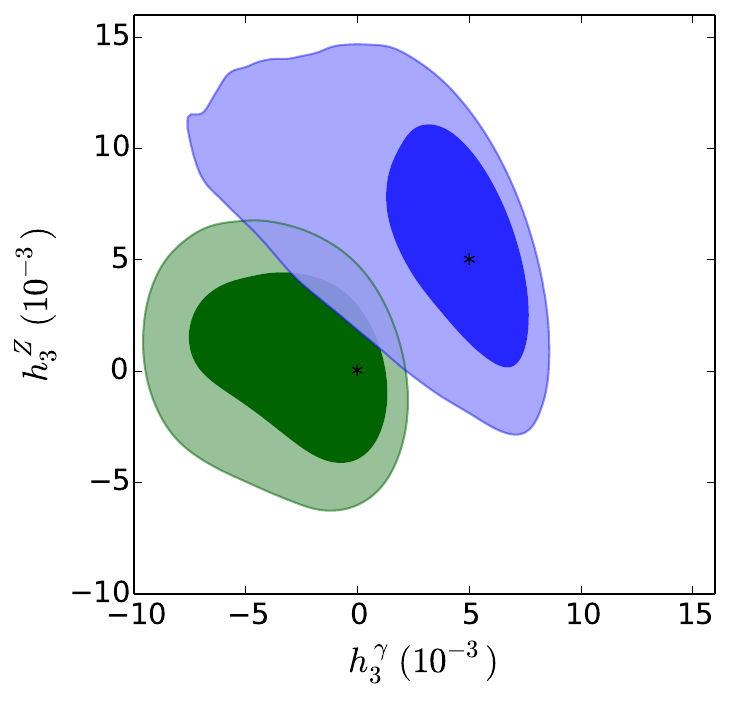}
\caption{\label{fig:correlations_parameters_in_mcmc_za} 
Two-dimensional contours for all pairs of the parameters in the  process 
$e^+e^-\rightarrow Z\gamma$. The Upper transparent layers ({\em blue}) are for
{\tt aTGC} and the lower layers ({\em green}) for the {\tt SM} showing the 68~\% BC 
({\em dark shades}) and  95~\% BC ({\em light shades}) contours.}
\end{figure*}

For the {\tt aTGC} point, there is enough room for the negative fluctuation in
the cross section, and hence  no narrowing of the 68~\% BCI is observed
for it; see Table~\ref{tab:marginalised_observables_za_bsm}. The 68~\% BCIs for
$A_x$ and $A_y$ are comparable to the corresponding $1\sigma$ intervals, while
the  68~\% BCIs for other three asymmetries are certainly narrower than 
$1\sigma$ intervals. This narrowing, as discussed earlier, is due to the 
parametric dependence of the observables and their correlations.  
Each of the parameters has a strong correlation with one of the asymmetries, as
shown in Fig.~\ref{fig:correlations_observables_in_mcmc_za}. The narrow 
contours indicate that if one can improve the errors on the asymmetries, it will 
improve the parameter extraction. The steeper is the slope of the narrow contour the larger will be its improvement. We note that $A_x$ and $A_y$
have a steep dependence on the corresponding parameters. Thus even small variations in the parameters lead to large variations in the asymmetries. 
For $A_{xy}$ and $A_{x^2-y^2}$ the parametric dependence is weaker, leading to
their smaller variation with the parameters and hence narrower 68~\% BCI.

For the parameter extraction, we look at their one-dimensional marginalized 
posterior distribution function, shown in 
Fig.~\ref{fig:one_parameter_mcmc_parameter_za} for the 
two benchmark points. The best-fit points along with 68~\% and 95~\% BCI are listed
in Table~\ref{tab:marginalised_parameters_za_sm}. The best-fit points are very
close to the true values of the parameters, and so are the means of the BCI for
all parameters except $h_3^\gamma$. For it, there is a downward movement in the
value owing to the multi-valuedness of the cross section. Also, we note that
the 95~\% BCI for the two benchmark points largely overlap, making them seemingly
un-distinguishable at the level of one-dimensional BCIs. To highlight the 
difference between two benchmark points, we look at two-dimensional BC contours
as shown in Fig.~\ref{fig:correlations_parameters_in_mcmc_za}. The 68~\% BC
contours (dark shades) can be roughly compared with the contours of 
Fig.~\ref{fig:Two_parameter_sensitivity_za}. The difference is that 
Fig.~\ref{fig:correlations_parameters_in_mcmc_za} has all four parameters 
varying and all six observables are used simultaneously. The 95~\% BC contours
for the two benchmark points overlap despite the fact that the cross section can
distinguish them very clearly. In full four-dimensional parameter space,
the two contours do not overlap, and in the next section, we try to establish this.

\subsection{Separability of  benchmark aTGCs}
\label{ssec:separation}
\begin{figure*}[h!]
\centering
\includegraphics[width=0.495\textwidth]{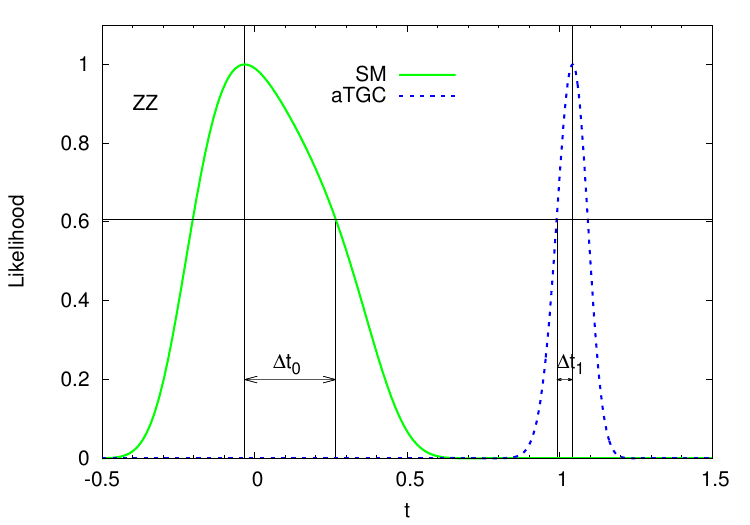}
\includegraphics[width=0.495\textwidth]{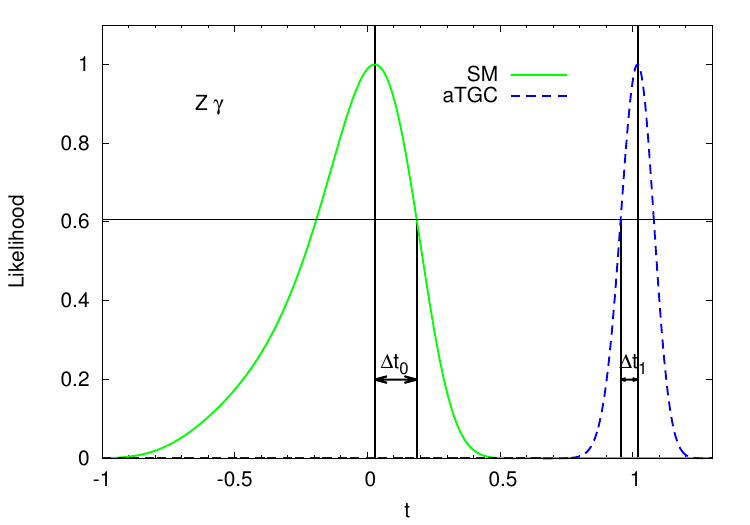}
\caption{\label{fig:likelyhood_ratio} Likelihood ratio for the separability of 
benchmark points for $ZZ$ (left) and  $Z\gamma$ (right) final state: {\tt SM} 
{\em pseudo data} are in solid ({\em green}) and {\tt aTGC} {\em pseudo data} are in 
dotted ({\em blue}) lines.}
\end{figure*}
To depict the separability of the two benchmark points pictorially, we vary
all four parameters for a chosen process as a linear function of one parameter,
$t$, as
\begin{equation}
\vec{f}(t) = (1-t) \ \vec{f}_{\tt SM} + t \ \vec{f}_{\tt aTGC},
\end{equation}
such that $\vec{f}(0) = \vec{f}_{\tt SM}$ is the coupling for the
{\tt SM} benchmark point and $\vec{f}(1) = \vec{f}_{\tt aTGC}$ is the
coupling for the {\tt aTGC} point. In Fig.~\ref{fig:likelyhood_ratio} we show
the normalized likelihood for the point $\vec{f}(t)$ assuming the {\tt SM}
{\em pseudo data}, ${\cal L}(\vec{f}(t)|\mbox{\tt SM})$, in  solid/green line
and assuming the {\tt aTGC} {\em pseudo data},
${\cal L}(\vec{f}(t)|\mbox{\tt aTGC})$, in dashed/blue line.
The left panel is for the  $ZZ$ production process and  the right panel
is for the $Z\gamma$ process. The horizontal lines correspond to the normalized
likelihood being $e^{-\frac{1}{2}}$, while the full vertical lines correspond to
the maximum value, which is normalized to $1$. It is clearly visible that the
two benchmark points are quite well separated in terms of the likelihood
ratios. We have ${\cal L}(\vec{f}_{\tt aTGC}|\mbox{\tt SM})\sim 8.8 \times
10^{-19}$
for the $ZZ$ process, and it means that the relative likelihood for the {\tt SM}
{\em pseudo data} being generated by the {\tt aTGC} parameter value is
$8.8 \times 10^{-19}$, i.e. negligibly small. Comparing the likelihood ratio to
$e^{-n^2/2}$ we can say that the data is $n\sigma$ away from the model point.
In this case, {\tt SM} {\em pseudo data} is $9.1\sigma$ away from the
{\tt aTGC} point for the $ZZ$ process. Similarly we have
${\cal L}(\vec{f}_{\tt SM}|\mbox{\tt aTGC})\sim 1.7 \times 10^{-17}$, i.e. the
{\tt aTGC} {\em pseudo data} is $8.8\sigma$ away from the {\tt SM} point for the 
$ZZ$ process.
For the $Z\gamma$ process we have
 ${\cal L}(\vec{f}_{\tt aTGC}|\mbox{\tt SM})\sim 1.7 \times 10^{-24}
(10.5\sigma)$ and
 ${\cal L}(\vec{f}_{\tt SM}|\mbox{\tt aTGC})\sim 1.8 \times 10^{-25}
(10.7\sigma)$.
In all cases, the two benchmark points are well separable as clearly seen in 
Fig.~\ref{fig:likelyhood_ratio}.

\section{Summary}
\label{sec:conclusion-epjc1}
Among all the polarization asymmetries, three of them,
$A_y$, $A_{xy}$, and $A_{yz}$, are $CP$-odd and can be used to measure $CP$-Odd couplings
in the production process. On the other hand, $A_z$, $A_{xz}$, and $A_{yz}$ are 
$P$-odd observables, while $A_x$, $A_{x^2-y^2}$, and $A_{zz}$ are $CP$- and $P$-even.
The anomalous trilinear gauge couplings in the neutral sector,
Eq.~(\ref{eq:LZZV-dim6}), are studied here using these asymmetries along with the cross section. The one
and two parameter sensitivity of these asymmetries, together with cross section,
are explored, and the one-parameter limits using one observable are listed in 
Table~\ref{tab:aTGC_constrain_form_1sigma_sensitivity} for an unpolarized
$e^+e^-$ collider. For finding the best and simultaneous limits on the anomalous
couplings, we  performed a likelihood mapping using the MCMC method and the
obtained limits are listed in Tables~\ref{tab:marginalised_parameters_zz} and
\ref{tab:marginalised_parameters_za_sm} for $ZZ$ and $Z\gamma$ processes,
respectively. 
The observables are calculated up to the quadratic
in dimension-$6$ form factors. In practice, one should consider the effect of dimension-$8$ contribution at linear
order. However, we choose to work with only dimension-$6$ in couplings with a partial contribution
up to quadratic so as to compare the results with the current LHC constraints on dimension-$6$ parameters. 
With appropriate polarized initial beams, the anomalous couplings can be further constrained,
 which will be discussed in the next chapter.

\chapter{The role of beam polarizations along with $Z$ boson polarizations to probe aTGC in  $e^+e^-\to ZZ/Z\gamma$}\label{chap:epjc2}
\begingroup
\hypersetup{linkcolor=blue}
\minitoc
\endgroup
{\small\textit{\textbf{ The contents in this chapter are based on the published article in Ref.~\cite{Rahaman:2017qql}. }}}
\vspace{1cm}

The future ILC~\cite{Djouadi:2007ik,Baer:2013cma,
Behnke:2013xla} will be a precision testing machine~\cite{MoortgatPick:2005cw} 
which will have the possibility of  polarized initial beams. Two types of 
polarization,  namely longitudinal and transverse, for both initial beams 
($e^-$ and $e^+$) will play an important role in precise measurement of various 
parameters, like the coupling among gauge bosons, Higgs coupling to the top quark, and Higgs 
coupling to the gauge boson. Beam polarization has the ability to enhance the 
relevant signal to background ratio and hence the sensitivity of 
observables~\cite{MoortgatPick:2005cw,Andreev:2012cj,Ananthanarayan:2010bt,
Osland:2009dp,Pankov:2005kd}.  It can also be used to separate 
$CP$-violating couplings from a $CP$-conversing one~\cite{MoortgatPick:2005cw,
Kittel:2011rk,Dreiner:2010ib,Bartl:2007qy,Rao:2006hn,Bartl:2005uh,Czyz:1988yt,
Choudhury:1994nt,Ananthanarayan:2004eb,Ananthanarayan:2011fr,Ananthanarayan:2003wi} if $CP$-violation 
is present in  Nature. These potentials of the beam polarizations have been 
explored, for example, to study $\tau$ polarization~\cite{Dreiner:2010ib}, top 
quark polarization~\cite{Groote:2010zf} and its anomalous
 couplings~\cite{Amjad:2015mma}, littlest 
Higgs model~\cite{Ananthanarayan:2009dw}, $WWV$ 
couplings~\cite{Ananthanarayan:2011ga,Andreev:2012cj,Ananthanarayan:2010bt}, 
Higgs couplings to gauge bosons~\cite{Kumar:2015eea,Rindani:2010pi,
Biswal:2009ar,Rindani:2009pb}. 

In this chapter, we study the effect of beam polarizations (longitudinal only) on
neutral aTGC  using the polarization observables of $Z$ in $e^+e^-\to ZZ/Z\gamma$ processes (as studied in the previous chapter, chapter~\ref{chap:epjc1}).
The neutral aTGC has been studied earlier with unpolarized beam in Refs.~\cite{Boudjema:Desy1992,Baur:1992cd,
Ellison:1998uy,Baur:2000ae,Ananthanarayan:2005ib,Aihara:1995iq,Gounaris:2000dn,Poulose:1998sd,Senol:2013ym} as 
well as with polarized beams in Refs.~\cite{Ots:2006dv,Czyz:1988yt,Choudhury:1994nt,
Ananthanarayan:2004eb,Ananthanarayan:2014sea,
Gounaris:1999kf,Choi:1994nv,Rizzo:1999xj,Atag:2003wm,Ananthanarayan:2003wi}. 
Some of these studies have used  a fixed beam polarizations to enhance the 
sensitivity of observables, while others have used two different sets of beam 
polarizations (opposite choices) to construct the observables. We see the implication in
both the approaches.

\section{Beam polarizations and polarization observables}
\label{sec:epjc2-beampol}
The spin density matrix of a spin-$1/2$ particle, as introduced in Eq.~(\ref{eq:spin-density-half}), is given by,   
\begin{equation}\label{eq:spin-density-half-epjc2}
\rho_{1/2} = \frac{1}{2}\left( \mathbb{I}_{2\times 2} + \vec{p}\cdot\vec{\sigma}\right),
\end{equation} 
with  $\sigma_i$ being  the Pauli spin matrices. After expansion, the above equation
takes the form
\begin{eqnarray}\label{eq:spin-density-half-epjc2-expanded}
\rho_{1/2}=
\frac{1}{2}\left[
\begin{tabular}{cc}
$1+p_z$&$p_x - i p_y$\\
$p_x + i p_y$&$1-p_z$
\end{tabular}
\right].
\end{eqnarray}
Thus the polarization density matrices for $e^-$ and $e^+$ beams, in terms of longitudinal and transverse polarizations, are given by,
\begin{eqnarray}\label{eq:electron_pol_matrix}
P_{e^-}(\lambda_{e^-},\lambda_{e^-}^\prime)=
\frac{1}{2}\left[
\begin{tabular}{cc}
$(1+\eta_3)$&$\eta_T$\\
$\eta_T$&$(1-\eta_3)$
\end{tabular}
\right] \hspace{0.5cm}\textrm{and}
\end{eqnarray}
\begin{eqnarray}\label{eq:positron_pol_matrix}
P_{e^+}(\lambda_{e^+},\lambda_{e^+}^\prime)=
\frac{1}{2}\left[
\begin{tabular}{cc}
$(1+\xi_3)$&$\xi_Te^{-i\delta}$\\
$\xi_Te^{i\delta}$&$(1-\xi_3)$
\end{tabular}
\right],
\end{eqnarray}
where $\eta_3$ and $\eta_T$ ($\xi_3$ and $\xi_T$) are longitudinal and 
transverse polarization of $e^-$ ($e^+$)  with $\delta$ being the azimuthal 
angle between two transverse polarizations. The positive $x$-axis is taken 
along the transverse polarization of $e^-$ and positive $z$-axis along its
momentum.

\begin{figure}
\centering
\includegraphics[width=0.5\textwidth]{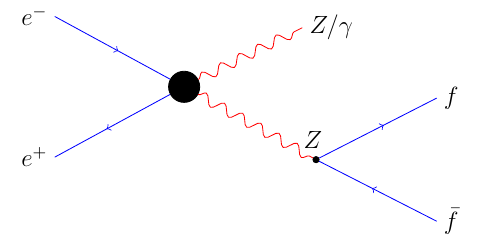}
\caption{\label{fig:Z_production_decay} Feynman diagram  
for production of $Z$ boson and its decay to a pair of 
fermions.} 
\end{figure}
The density matrix for the production of $Z$ boson in the above process 
(Fig.~\ref{fig:Z_production_decay} ) would be
\begin{eqnarray}\label{eq:density_matrix_beam_pol}
\rho(\lambda_Z,\lambda_Z^\prime)=
{\cal\:\sum}_{\lambda_{e^-},\lambda_{e^-}^\prime,\lambda_{e^+},\lambda_{e^+}^\prime} 
{\cal M}^\dagger(\lambda_{e^-}^\prime,\lambda_{e^+}^\prime,\lambda_Z^\prime)\times
{\cal M}(\lambda_{e^-},\lambda_{e^+},\lambda_Z) \times \nonumber\\
P_{e^-}(\lambda_{e^-},\lambda_{e^-}^\prime)\times
P_{e^+}(\lambda_{e^+},\lambda_{e^+}^\prime).
\end{eqnarray}
We note that the different helicities can take the following values:
\begin{eqnarray}
\lambda_Z,\lambda_Z^\prime\in\{-1,0,1\} \ \text{and} \ 
\lambda_{e^\pm},\lambda_{e^\pm}^\prime\in\{-1,1\}. 
\end{eqnarray}
For the present work, we  restrict ourselves  only to the longitudinal beam 
polarizations, i.e. $\eta_T=0=\xi_T$. With the chosen beam polarizations, we construct the complete set of eight polarization observables for the $Z$ boson 
along with the total cross section in the processes $e^+e^-\to ZZ/Z\gamma$. 
Among the $8$ polarization asymmetries of $Z$ boson in the given processes, the 
asymmetries $A_z$, $A_{xz}$, $A_{yz}$ are zero
in the SM (as has been seen in chapter~\ref{chap:epjc1})  even with polarized beam  owing to the 
forward-backward symmetry of produced $Z$. To make these asymmetries
non-zero we redefine the polarization observables ${\cal O}\in\{p_z,T_{xz},T_{yz}\}$ (corresponding to
$A_z$, $A_{xz}$, $A_{yz}$) as
\begin{eqnarray}
{\cal O}\to \wtil{{\cal O}}= \frac{1}{\sigma_{Z}} \left[\int^{c_{\theta_0}}_{0} 
{\rm Comb}({\cal O},\rho(\lambda,\lambda')) dc_{\theta_Z}
-\int^{0}_{-c_{\theta_0}}
{\rm Comb}({\cal O},\rho(\lambda,\lambda')) dc_{\theta_Z}\right],
\end{eqnarray}
where $c_{\theta_0}$ is the beam pipe cut and
${\rm Comb}({\cal O},\sigma(\lambda,\lambda'))$ is the combination of production
density matrix corresponding the polarization observable ${\cal O}$ (given in Eq.~(\ref{eq:pol_prod}) ). For example, 
with ${\cal O}=p_z$ one has 
$${\rm Comb}(P_z,\rho(\lambda,\lambda'))=\rho(+1,+1) -\rho(-1,-1)$$
and the corresponding modified polarization is given by,
\begin{eqnarray}
\wtil{p}_z=\frac{1}{\sigma_{Z}} \l[\int^{c_{\theta_0}}_{0}    
\l[\rho(+1,+1) -\rho(-1,-1) \r] dc_{\theta_Z}
 -\int^{0}_{-c_{\theta_0}}\l[\rho(+1,+1) -\rho(-1,-1) \r]
dc_{\theta_Z}\r]. 
\end{eqnarray} 
The asymmetries $\wtil{A}_z$ corresponding to the modified polarization 
$\wtil{P}_z$ is given by,
\begin{eqnarray}
\wtil{A}_z\equiv \frac{1}{\sigma}\l(\sigma(c_{\theta_Z}\times c_{\theta_f}> 0)
-\sigma(c_{\theta_Z} \times c_{\theta_f}< 0)\r).
\end{eqnarray}
Similarly $A_{xz}$ and $A_{yz}$ related to $T_{xz}$ and $T_{yz}$ are modified as,
\begin{eqnarray}\label{eq:epjc2-tilde-AxzAyz}
\wtil{A}_{xz}\equiv \frac{1}{\sigma}\l(\sigma(c_{\theta_Z}\times c_{\theta_f}c_{\phi_f}> 0)
-\sigma(c_{\theta_Z} \times c_{\theta_f}c_{\phi_f}< 0)\r),\nonumber\\
\wtil{A}_{yz}\equiv \frac{1}{\sigma}\l(\sigma(c_{\theta_Z}\times c_{\theta_f}s_{\phi_f}> 0)
-\sigma(c_{\theta_Z} \times c_{\theta_f}s_{\phi_f}< 0)\r).
\end{eqnarray}
Redefining these asymmetries increases the total number of the non-vanishing 
observables to put simultaneous limit on the anomalous coupling and we expect 
limits  tighter than reported earlier in chapter~\ref{chap:epjc1}. 

\begin{figure}
\centering
\includegraphics[width=0.495\textwidth]{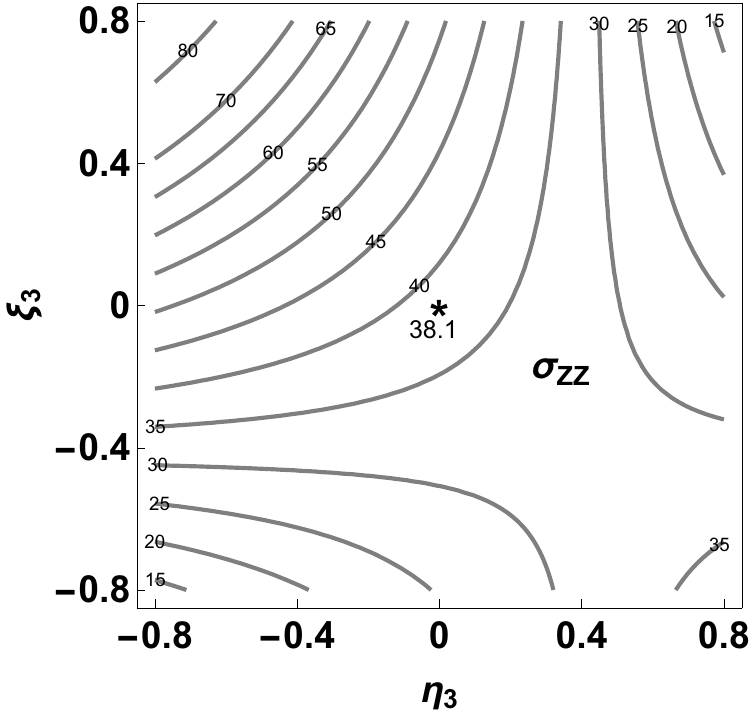}
\includegraphics[width=0.495\textwidth]{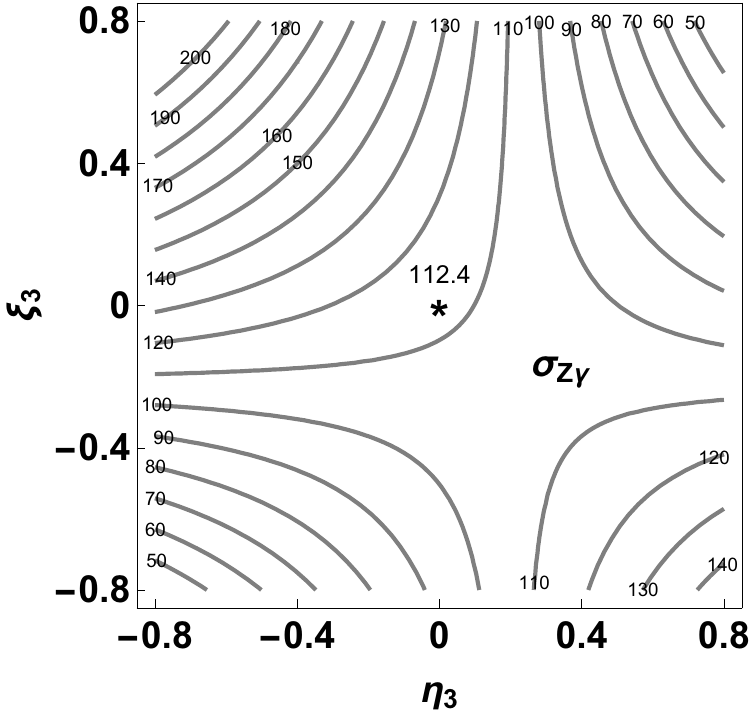}
\caption{\label{fig:beampol_cross-sections} 
The SM cross section (in fb) for the process $e^+e^-\to ZZ/Z\gamma$ as a function of
longitudinal beam polarizations $\eta_3$ (for $e^-$) and $\xi_3$ (for $e^+$)
at $\sqrt{s}=500$ GeV. }
\end{figure}
The total cross section (or the total number of events) of a process plays an
important role in determining the sensitivity and the limits on the anomalous 
couplings. A tighter limit on the anomalous couplings can be obtained if the 
cross section can be enhanced. Beam polarization can enhance the 
cross section, and hence  it is important to see how it depends on beam 
polarization. Fig.~\ref{fig:beampol_cross-sections} shows the dependence of the 
cross sections $\sigma_{ZZ}$  and  $\sigma_{Z\gamma}$  on the longitudinal beam 
polarizations $\eta_3$ and $\xi_3$ at $\sqrt{s}=500$ GeV. The asterisk mark 
on the middle of the plots represents the unpolarized case. We notice that
the cross section in  the two processes are larger for a negative value of $\eta_3$ 
and a positive value of $\xi_3$. The sensitivity on the cross section is expected to be 
high in the left-top corner of the $\eta_3-\xi_3$ plane. This would convince 
us to set beam polarizations at the left-top corner for analysis. But the
cross section is not the only observable; the asymmetries have different 
behaviour on beam polarizations. For example, $A_x$ peaks at the right-bottom
corner, i.e. we have an opposite behaviour compared to cross section, while $A_z$ has
a similar dependence as the cross section on the beam polarizations in both the 
processes. Processes involving $W^\pm$ are also expected to have a higher 
cross section  at the left-top corner of $\eta_3-\xi_3$ plane as $W$ couple 
to the left chiral electron. Anomalous couplings are expected to change the 
dependence of all the observables, including the cross section, on the beam 
polarizations. To explore this, we study the effect of beam polarizations on 
the sensitivity of cross section and other observables to anomalous couplings 
in the next section.

\section{Effect of beam polarization on the sensitivity}
\label{sec:sensitivity_likelihood}
\begin{figure}
    \centering
    \includegraphics[width=0.325\textwidth]{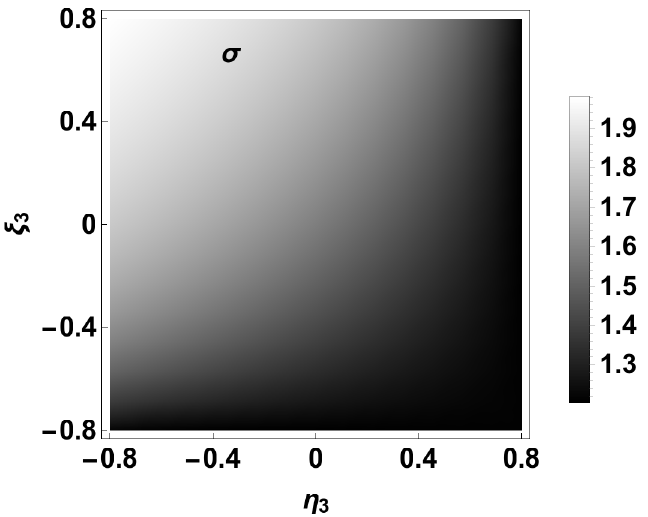}
    \includegraphics[width=0.325\textwidth]{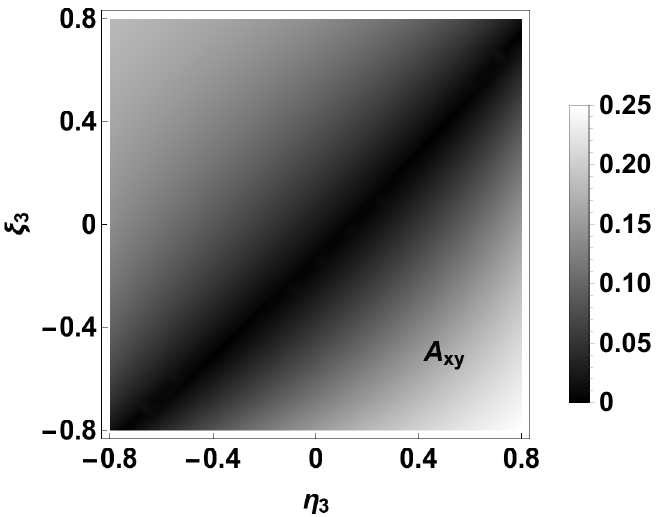}
    \includegraphics[width=0.325\textwidth]{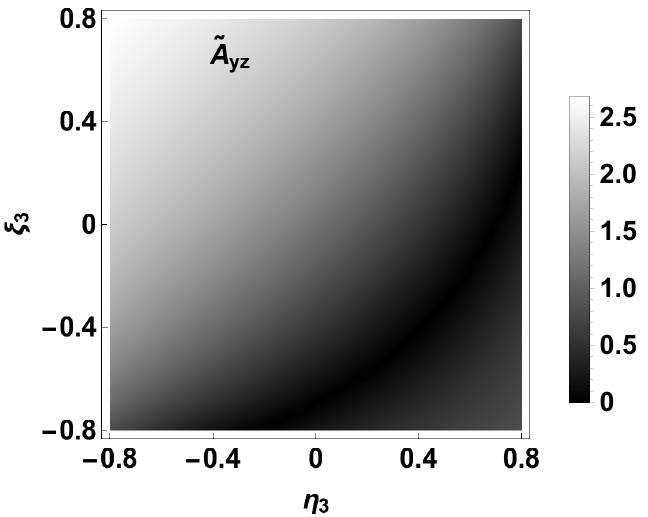}
    \caption{\label{fig:sensitivity_zz}Effect of beam polarizations on sensitivity
        of cross section $\sigma$, $A_{xy}$ and $\wtil{A}_{yz}$ in the process 
        $e^+e^-\to ZZ$ for anomalous couplings $\vec{f}=\{+3,+3,+3,+3\}\times 10^{-3}$
        at $\sqrt{s}=500$ GeV and ${\cal L}=100$ fb$^{-1}$.}
\end{figure}

The sensitivity of an observables ${\cal O}$ depending on anomalous couplings 
$\vec{f}$ given in Eq.~(\ref{eq:sensitivity}) with a given beam polarizations $\eta_3$ and $\xi_3$ will now be given by,
\begin{equation}\label{eq:epjc2-sensitivity-beam-pol}
{\cal S}({\cal O}(\vec{f},\eta_3,\xi_3))=\dfrac{|{\cal O}(\vec{f},\eta_3,\xi_3)
    -{\cal O}(\vec{0},\eta_3, \xi_3)|}{|\delta{\cal O}(\eta_3, \xi_3)|} ,
\end{equation}
where $\delta{\cal O}=\sqrt{(\delta{\cal O}_{stat.})^2+
(\delta{\cal O}_{sys.})^2}$ is the estimated error in ${\cal O}$. 
The estimated error to cross section would be
\begin{equation}
\delta\sigma(\eta_3, \xi_3)=\sqrt{\frac{\sigma(\eta_3, \xi_3)}{{\cal L}} +
     \epsilon_\sigma^2 \sigma(\eta_3, \xi_3)^2 } ,
\end{equation}
whereas the estimated error to the asymmetries would be
\begin{equation}
\delta A(\eta_3, \xi_3)=\sqrt{\frac{1-A(\eta_3, \xi_3)^2}
    {{\cal L}\sigma(\eta_3, \xi_3)} + \epsilon_A^2 } .
\end{equation}
Here ${\cal L}$ is the integrated luminosity, $\epsilon_\sigma$ and 
$\epsilon_A$ are the systematic fractional error in cross section and 
asymmetries, respectively. In these analyses we take ${\cal L}=100$ fb$^{-1}$,
$\epsilon_\sigma=0.02$ and $\epsilon_A=0.01$ as a benchmark. 
We study the sensitivity of all the observables to the aTGC for some benchmark
values and see the effect of beam polarization on them. 
Choosing a benchmark value for the anomalous
couplings to be
$$\vec{f}=\{f_4^\gamma ,f_4^Z, f_5^\gamma, f_5^Z \} =\{+3,+3,+3,+3\}\times 
10^{-3} ,$$
we show the sensitivities for $\sigma$, $A_{xy}$ and $\wtil{A}_{yz}$ in 
Fig.~\ref{fig:sensitivity_zz} as a function of beam polarizations. The 
sensitivities for the cross section and $\wtil{A}_{yz}$ peak at the left-top corner
of the plots. For $A_{xy}$ sensitivity peak at the right-bottom corner, it is 
not much smaller in the left-top corner either. The sensitivities of all other 
asymmetries (not shown here) except $\wtil{A}_{z}$ peaks at the left-top corner
although the exact dependence on the beam polarization may differ. Thus, the 
combined sensitivity of all the observables is high on the left-top corner
of the polarization plane making $(\eta_3,\xi_3)=(-0.8,+0.8)$ the best choice
for the chosen benchmark coupling. This best choice, however, strongly depends
upon the values of the anomalous couplings. 
We note that the best choice of the beam polarization is mainly decided by the
behaviour of the cross section because most of the asymmetries also have 
similar dependences on the beam polarizations. This, however, does not mean that
the cross section provides the best sensitivity or limits. For example, in
Fig.~\ref{fig:sensitivity_zz} we can see that $\wtil{A}_{yz}$ has a better
sensitivity than the cross section.
For the $Z\gamma$ process with  the benchmark point 
$$\vec{h}=\{h_1^\gamma,h_1^Z,h_3^\gamma,h_3^Z\}=\{+3,+3,+3,+3\}\times10^{-3}$$
one obtains similar conclusions: the sensitivities of all observables peak at left-top corner of $\eta_3-\xi_3$ plane (not shown) except for $\wtil{A}_{z}$.

\begin{figure}
    \centering
    \includegraphics[width=0.325\textwidth]{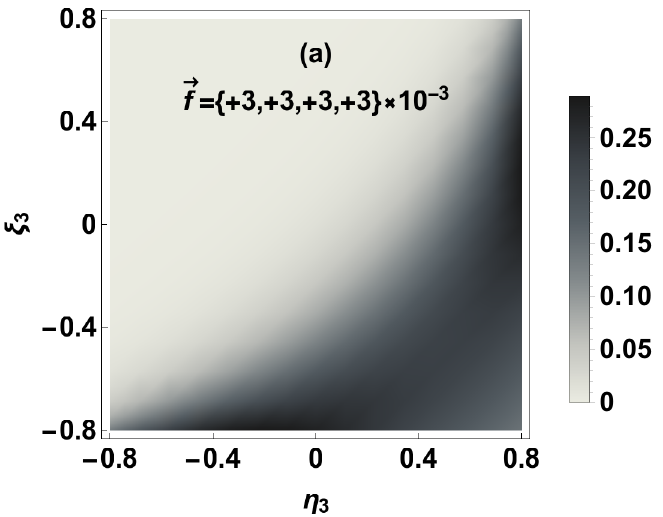}
    \includegraphics[width=0.325\textwidth]{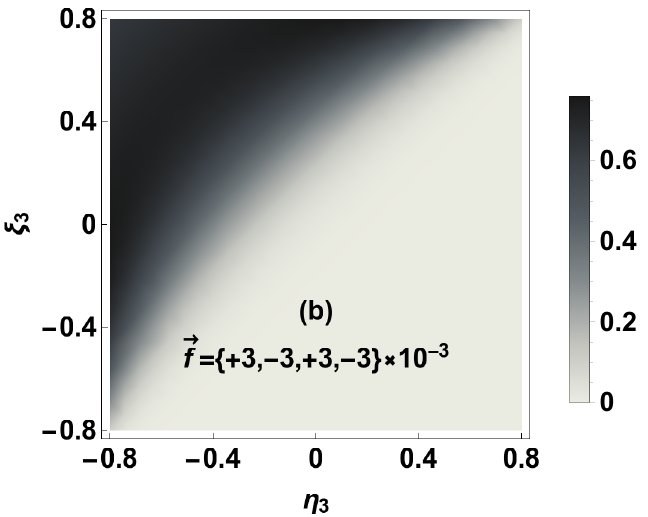}
    \includegraphics[width=0.325\textwidth]{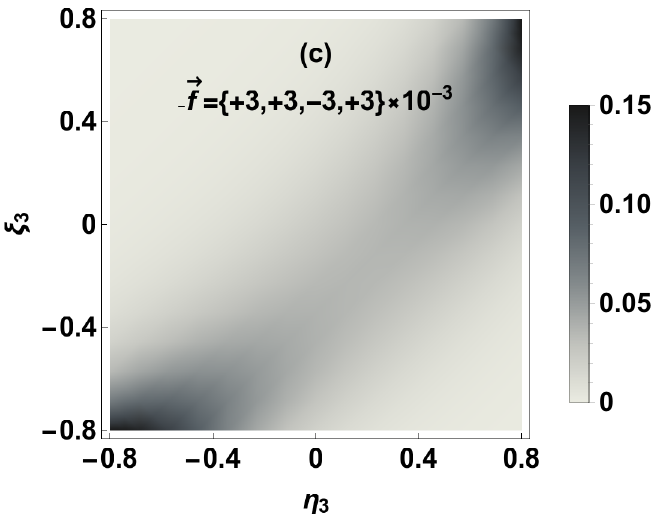}
    \caption{\label{fig:x2_all_Obszz} Likelihood 
        $\text{L}({\{\cal O}\},\vec{f};\eta_3,\xi_3)$ for three different benchmark 
        anomalous couplings at $\sqrt{s}=500$ GeV and ${\cal L}=100$ fb$^{-1}$ in 
        $ZZ$ process.}
\end{figure}
For a complete analysis we need to use all the observables simultaneously.
To this end we define a likelihood function considering the set of all the 
observables depending on the anomalous coupling $\vec{f}$ as, 
\begin{eqnarray}\label{eq:likelihood_zz}
\text{L}({\{\cal O}\},\vec{f};\eta_3,\xi_3)=
\exp{\bigg[-\frac{1}{2}\sum_{i} {\cal S}({\cal O}_i(\vec{f},\eta_3, 
\xi_3))^2 \bigg]} ,
\end{eqnarray}
$i$ runs over the set of observables in a process. Maximum sensitivity of 
observables requires the likelihood to be minimum. The likelihood defined here
is proportional to the $p$-value and hence the best choice of beam polarizations 
comes from the {\em minimum} likelihood or maximum distinguishability.

The beam polarization dependence of the likelihood for the $ZZ$ process at the
above chosen anomalous couplings is given in Fig.~\ref{fig:x2_all_Obszz}(a).
The minimum of the likelihood falls in the left-top corner of the 
$\eta_3-\xi_3$ plane as expected as most of the observables has higher 
sensitivity at this corner. For different anomalous couplings, the minimum 
likelihood changes its position in the $\eta_3-\xi_3$ plane.
We have checked the likelihood for $16$ different corners of 
$$\vec{f}_{\pm\pm\pm\pm}=\{\pm3,\pm3,\pm3,\pm3\}\times 10^{-3}$$ 
and they have different dependences on  $\eta_3 , \xi_3$. Here we present the
likelihood for three different choices of the anomalous couplings in 
Fig.~\ref{fig:x2_all_Obszz}. In Fig.~\ref{fig:x2_all_Obszz}(b), the minimum of
the likelihood falls in the right-bottom corner where most of the observables have
higher sensitivity. In Fig.~\ref{fig:x2_all_Obszz}(c) low likelihood falls in 
both  diagonal corners in the $\eta_3-\xi_3$ plane. This is because some of the 
observables prefer the left-top corner, while others prefer the right-bottom corner 
of the polarization plane for higher sensitivity. We have a similar behaviour
for the likelihood in the $Z\gamma$ process.

\section{Average likelihood and best choice of beam polarization}\label{sec:epjc2-average-like-best-choich}
In the previous section, we observed that, as the anomalous couplings change, the minimum likelihood region changes 
accordingly and hence the best choice of beam polarizations. So the best choice
for the beam polarizations depends on the new physics in the process. If one
knows the new physics one could tune the beam polarizations to have the best 
sensitivity for the analysis. But in order to have a suitable choice of beam 
polarizations irrespective of the possible new physics one needs to minimize 
the likelihood averaged over all the anomalous couplings. 
The likelihood function averaged  over a volume in parameter space $V_{\vec{f}}$ would be 
defined as,
\begin{equation}\label{eq:average_likelihood_zz}
L(V_{\vec{f}},{\{\cal O}\};\eta_3,\xi_3)=\int_{V_{\vec{f}}}
\text{L}({\{\cal O}\},\vec{f};\eta_3,\xi_3) d\vec{f}.
\end{equation}
This quantity is nothing but the {\em weighted volume} of the parameter space that is
statistically consistent with the SM. The size of this weighted volume determines
the limits on the parameters. The beam polarizations with the minimum averaged 
likelihood (or minimum weighted volume) is expected to be the average best 
choice for any new physics in the process. For numerical analysis, we
choose the volume to be a hypercube in the $4$ dimensional parameter space 
with sides equal to $2\times 0.05$ (much larger than the 
available limits on them) in both the processes. The contribution to the average
likelihood from the region outside this volume is negligible. 

\begin{figure}
\centering
\includegraphics[width=0.6\textwidth]{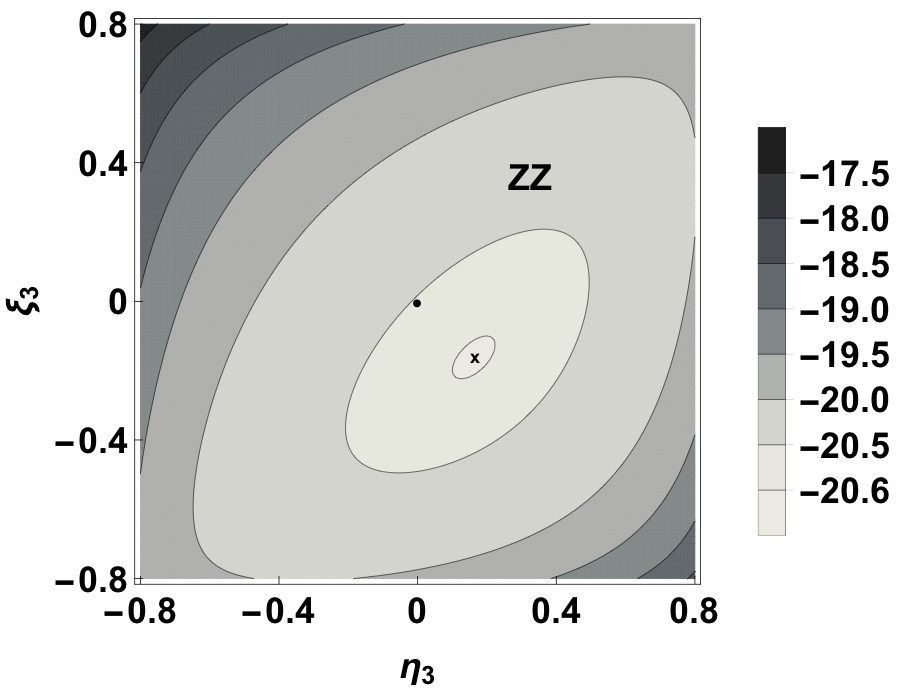}
\caption{\label{fig:x2_all_ObsZZ_Int}The log of average likelihood, 
$\log[ L(V_{\vec{f}},{\{\cal O}\};\eta_3,\xi_3)]$ as a function of beam
polarization is shown for the $ZZ$ process at $\sqrt{s}=500$ GeV and
${\cal L}=100$ fb$^{-1}$. The dot at the centre is the ($0,0$) point, while 
cross mark at $P_{ZZ}=(+0.16,-0.16)$ is the minimum likelihood point and 
hence the best choice of beam polarizations for $ZZ$ process.}
\end{figure}
The average likelihood $L(V_{\vec{f}},{\{\cal O}\};\eta_3,\xi_3)$  in the $ZZ$
process as a function of  beam polarization is shown in 
Fig.~\ref{fig:x2_all_ObsZZ_Int} on $\log$-scale. The dot on the middle of the 
the plot represents the unpolarized case and the cross mark at 
$P_{ZZ}=( +0.16,-0.16)$ represents the minimum averaged likelihood point, i.e.,
the best choice of beam polarizations. 
The unpolarized point, the best point and the points within 
two central contour 
in Fig.~\ref{fig:x2_all_ObsZZ_Int} have the same order of average likelihood 
and expected to give similar limits on anomalous couplings. The polarization
point from darker contours corresponds to larger values of average likelihood, 
and it is expected to give relatively looser limits on anomalous couplings.
To explore this, we estimate simultaneous limits using 
Markov-Chain--Monte-Carlo (MCMC) method at $P_{ZZ}$, unpolarized beam, and few 
other benchmark choices of beam polarizations. The limits thus obtained on the 
anomalous couplings for the  $ZZ$ process are listed in 
Table~\ref{tab:coupling_limit_mcmc_ZZ}. We note that the limits for the 
best choice of polarizations ($P_{ZZ}$) are best but comparable to other 
nearby benchmark beam polarization including the unpolarized beams.
This is due to the fact that the average likelihood is comparable for these 
cases. Further, the limits for $(+0.4,-0.4)$ and $(-0.4,+0.4)$
are increasingly bad, as these points correspond to the third and fourth contours,
i.e., we have an increasingly larger average likelihood. The point $(-0.8,+0.8)$ has the
largest average likelihood and the corresponding limits are the worst in 
Table~\ref{tab:coupling_limit_mcmc_ZZ}. 
We also note that the limits for the unpolarized case in
Table~\ref{tab:coupling_limit_mcmc_ZZ} are better than the ones reported in
Ref.~\cite{Rahaman:2016pqj}, when adjusted for the systematic errors. This
improvement  here is due to the inclusion of three new non-vanishing asymmetries
$\wtil{A}_z$, $\wtil{A}_{xz}$ and $\wtil{A}_{yz}$. Of these, $\wtil{A}_{xz}$ has 
a linear dependence on $f_5^{\gamma,Z}$ with larger sensitivity to $f_5^Z$ leading
to about $30~\%$ improvement in the limit. Similarly, the $CP$-odd asymmetry
$\wtil{A}_{yz}$ has a linear dependence on $f_4^{\gamma,Z}$ with larger
sensitivity to $f_4^Z$ and this again leads to about  $30~\%$ improvement in the
corresponding limit. The asymmetry $\wtil{A}_z$ has a quadratic dependence on all
four parameters and has too poor sensitivity for all of them to be useful.

\begin{table}[t!]
    \centering
    \caption{\label{tab:coupling_limit_mcmc_ZZ} List of simultaneous limits on the
        anomalous couplings obtained for $\sqrt{s}=500$ GeV and ${\cal L}=100$ fb$^{-1}$
        for different $\eta_3$ and $\xi_3$ from MCMC in $ZZ$ process.}
    \renewcommand{\arraystretch}{1.5}
    \begin{footnotesize}
        \begin{tabular*}{\textwidth}{@{\extracolsep{\fill}}lllllllllllllll@{}}\hline
            \multicolumn{1}{c}{\begin{tabular}{l} Beam\\polarizations \end{tabular}} &
            \multicolumn{8}{c}{   Limits on couplings ($ 10^{-3}$)}\\ \hline
            \multicolumn{1}{c}{}  &     \multicolumn{2}{c}{$f_4^\gamma $} &     \multicolumn{2}{c}{$f_4^Z$} &     \multicolumn{2}{c}{$f_5^\gamma$} &     \multicolumn{2}{c}{$f_5^Z$}\\\hline
            ~~~~~$(\eta_3,\xi_3)$ & $68~\%$ & $95~\%$ & $68~\%$ & $95~\%$ & $68~\%$ & $95~\%$ & $68~\%$ & $95~\%$ & Comments \\ \hline
            $-0.80,+0.80$ & $_{-9.3}^{+7.3} $& $_{-12.0}^{+13.0} $& $_{-14.0}^{+15.0} $& $_{-19.0}^{+18.0}  $& $\pm 7.3 $& $\pm 13.0  $ & $\pm 11.0 $& $_{-18.0}^{+19.0}  $ &\\
            $-0.40,+0.40$ & $\pm3.1 $& $_{-5.7}^{+5.8} $& $\pm 4.4 $& $_{-8.4}^{+8.2}  $& $\pm3.3 $& $_{-6.2}^{+6.3}  $& $_{-5.2}^{+4.5}  $& $_{-8.5}^{+9.3}  $&\\ \hline
            ~~~$0.00,~~~0.00$ & $\pm1.7 $& $\pm3.3 $& $\pm2.5 $& $\pm4.8 $ & $\pm1.9 $& $_{-3.6}^{+3.7}$ & $_{-2.7}^{+2.3}$& $_{-4.6}^{+5.1} $& Unpolarized point\\ 
            $+0.09,-0.10$ & $\pm 1.7$& $\pm 3.2 $& $\pm2.4 $& $_{-4.6}^{+4.7}  $& $\pm1.8 $&
            $_{-3.4}^{+3.5} $& $_{-2.6}^{+2.2} $& $_{-4.5}^{+4.9} $ & $P_{Z\gamma}$, 
            best for $Z\gamma$ \\
            $+0.12,-0.12$ & $\pm 1.6$& $\pm 3.1$& $\pm2.4 $& $\pm4.7 $& $\pm1.8 $&
            $_{-3.4}^{+3.5} $& $_{-2.6}^{+2.2} $& $_{-4.5}^{+5.0} $ & $P_{best}$, combined best \\ 
            $+0.16,-0.16$ & $\pm 1.6$& $\pm3.1 $& $\pm2.4 $& $\pm4.7 $& $\pm1.8 $&
            $_{-3.4}^{+3.5} $& $_{-2.7}^{+2.3} $& $_{-4.5}^{+5.1} $ & $P_{ZZ}$, best for $ZZ$\\ \hline
            $+0.40,-0.40$ & $\pm1.9 $& $\pm3.7$& $\pm3.2 $& $_{-6.2}^{+6.1} $& $\pm2.1 $& $_{-4.1}^{+4.0} $& $_{-3.7}^{+3.1} $& $_{-6.0}^{+6.7} $&\\
            $+0.80,-0.80$ & $_{-6.2}^{+5.3} $& $_{-9.3}^{+9.8} $& $_{-12.0}^{+9.7} $& $_{-17.0}^{+18.0} $& $\pm5.4 $& $_{-9.9}^{+9.5} $& $\pm9.9 $& $_{-18.0}^{+17.0} $&\\\hline
        \end{tabular*}
    \end{footnotesize}
\end{table}
\begin{figure}[t!]
    \centering
    \includegraphics[width=0.6\textwidth]{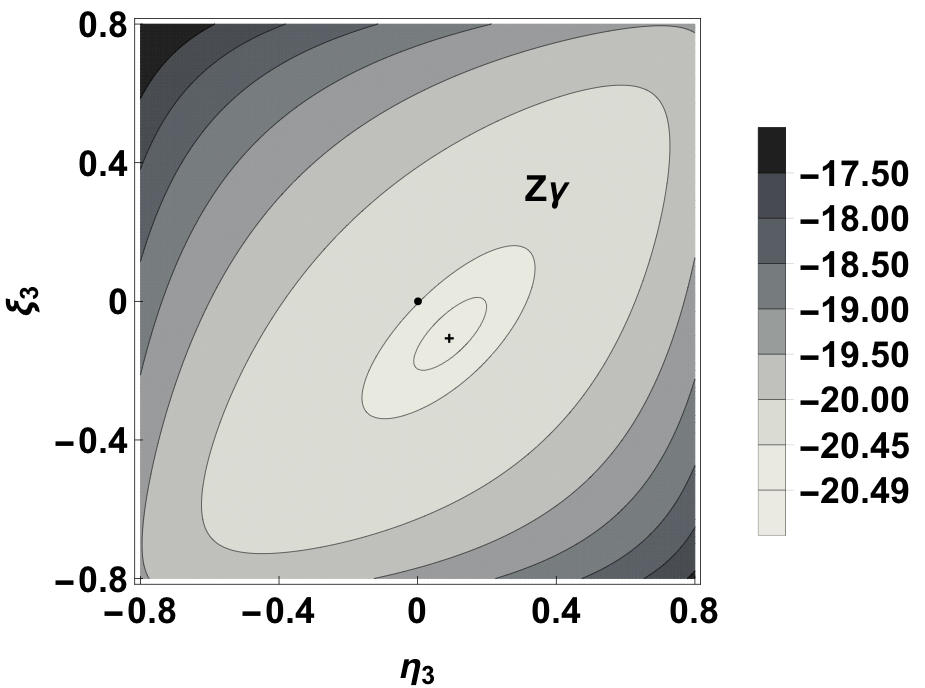}
    \caption{\label{fig:x2_all_ObsZA_Int} 
        Same as Fig.~\ref{fig:x2_all_ObsZZ_Int} but for the $Z\gamma$ process.
        The plus mark at $P_{Z\gamma}=(+0.09,-0.10)$ is the lowest likelihood point 
        and hence the best choice of beam polarizations for $Z\gamma$ process.}
\end{figure}
\begin{table}[h!]
    \centering
    \caption{\label{tab:coupling_limit_mcmc_ZA} List of simultaneous limits on the 
        anomalous couplings obtained for $\sqrt{s}=500$ GeV and ${\cal L}=100$ fb$^{-1}$
        for different $\eta_3$ and $\xi_3$ from MCMC in $Z\gamma$ process.}
    \renewcommand{\arraystretch}{1.5}
    \begin{footnotesize}
        \begin{tabular*}{\textwidth}{@{\extracolsep{\fill}}lllllllllllllll@{}}\hline
            \multicolumn{1}{c}{\begin{tabular}{l} Beam\\polarizations \end{tabular}} &
            \multicolumn{8}{c}{   Limits on couplings ($ 10^{-3}$)}\\ \hline
            \multicolumn{1}{c}{}  &     \multicolumn{2}{c}{$h_1^\gamma $} &     \multicolumn{2}{c}{$h_1^Z$} &     \multicolumn{2}{c}{$h_3^\gamma$} &     \multicolumn{2}{c}{$h_3^Z$}\\\hline
            ~~~~~$(\eta_3,\xi_3)$ & $68~\%$ & $95~\%$ & $68~\%$ & $95~\%$ & $68~\%$ & $95~\%$ & $68~\%$ & $95~\%$ & Comments \\ \hline
            $-0.80,+0.80$ & $_{-9.3}^{+7.7} $& $\pm 13.0$& $\pm 11.0$& $_{-19.0}^{+18.0}  $& $\pm 7.5 $& $\pm 13.0  $ & $\pm 11.0 $& $\pm 19.0$ &\\
            $-0.40,+0.40$ & $\pm3.9 $& $_{-7.5}^{+7.4} $& $\pm 6.5 $& $\pm 12.0$& $_{-3.7}^{+4.4}$& $_{-8.0}^{+7.1}  $& $\pm 6.6$& $_{-12.0}^{+13.0}  $ &\\ \hline
            ~~~$0.00,~~~0.00$ & $\pm1.6 $& $\pm3.1 $ & $\pm3.7 $& $_{-7.0}^{+7.1}$& $_{-1.4}^{+1.6} $& $_{-3.0}^{+2.8} $ & $\pm 3.6$& $\pm 7.1 $ & Unpolarized point\\ 
            $+0.09,-0.10$ & $\pm 1.5$& $\pm2.9 $& $\pm3.6 $& $\pm7.0 $& $_{-1.3}^{+1.4}$&
            $_{-2.8}^{+2.6} $& $\pm 3.6$& $_{-7.1}^{+7.0}$& $P_{Z\gamma}$, best for $Z\gamma$\\
            $+0.12,-0.12$ & $\pm 1.5$& $\pm2.9 $& $\pm3.7 $& $\pm7.1 $& $\pm 1.4$&
            $_{-2.8}^{+2.6} $& $\pm 3.6$& $\pm 7.1$ & $P_{best}$, combined best\\
            $+0.16,-0.16$ & $\pm 1.5$& $\pm3.0 $& $\pm3.7 $& $_{-7.3}^{+7.2} $& $_{-1.3}^{+1.5}
            $& $_{-2.8}^{+2.6}$ & $\pm 3.7$& $_{-7.3}^{+7.1}$ & $P_{ZZ}$, best for $ZZ$\\\hline
            $+0.40,-0.40$ & $\pm2.4 $& $\pm4.6$& $\pm5.2 $& $\pm 10.0$& $_{-2.2}^{+2.5} $& $_{-4.7}^{+4.3} $& $\pm 5.2 $& $\pm 10.0 $&\\
            $+0.80,-0.80$ & $\pm 5.8$& $_{-9.9}^{+10.0} $& $_{-13.0}^{+11.0} $& $_{-18.0}^{+19.0} $& $_{-7.2}^{+5.8} $& $_{-9.7}^{+10.0} $& $_{-15.0}^{+13.0}$& $_{-18.0}^{+19.0} $ &\\\hline
        \end{tabular*}
    \end{footnotesize}
\end{table}
We do a similar analysis for the $Z\gamma$ process. The average likelihood  
$L(V_{\vec{h}},{\{\cal O}\};\eta_3,\xi_3)$ is shown in 
Fig.~\ref{fig:x2_all_ObsZA_Int} on log-scale.
Here also the dot on the middle of the plot is for unpolarized case while the
plus mark at $P_{Z\gamma}=( +0.09,-0.10)$ is for the minimum averaged likelihood 
and hence the best choice of beam polarizations. The corresponding simultaneous 
limits on the anomalous couplings $h_i$ are presented in 
Table~\ref{tab:coupling_limit_mcmc_ZA}.Again we notice that the limits obtained
for the best choice of the beam polarizations $P_{Z\gamma}$ are tighter than any 
other point on the polarization plane,  yet comparable to the nearby polarization
points within the two central contours in Fig.~\ref{fig:x2_all_ObsZA_Int}, including the unpolarized point. This again is due to the comparable
values of the averaged likelihood of the two
central contours containing $P_{Z\gamma}$ and the unpolarized point. The limits at the points $(+0.4,-0.4)$ 
and $(-0.4,+0.4)$ are  worse as they fall in the fourth and fifth contour containing much 
larger likelihood values. Like the $ZZ$ case the point $(-0.8,+0.8)$ has the
largest average likelihood and the corresponding limits are the worst. 
The simultaneous limits for the unpolarized case (also the $P_{Z\gamma}$) turns out
to be much better than the ones reported in Ref.~\cite{Rahaman:2016pqj} for 
$h_{1,3}^\gamma$ due to the inclusions of new asymmetries in the present
analysis. The $CP$-odd asymmetry $\wtil{A}_{yz}$ has linear dependence on
$h_1^{\gamma,Z}$ with a large sensitivity towards $h_1^\gamma$ leading to an
improvement in the corresponding limit by a factor of two compare to earlier 
report when adjusted for systematic errors. The limit on $h_3^\gamma$ improves 
by a factor of $3$ owing to the asymmetry $\wtil{A}_{xz}$. The limits on 
$h_{1,3}^Z$ remain comparable.

\begin{figure}[h!]
    \centering
    \includegraphics[width=0.6\textwidth]{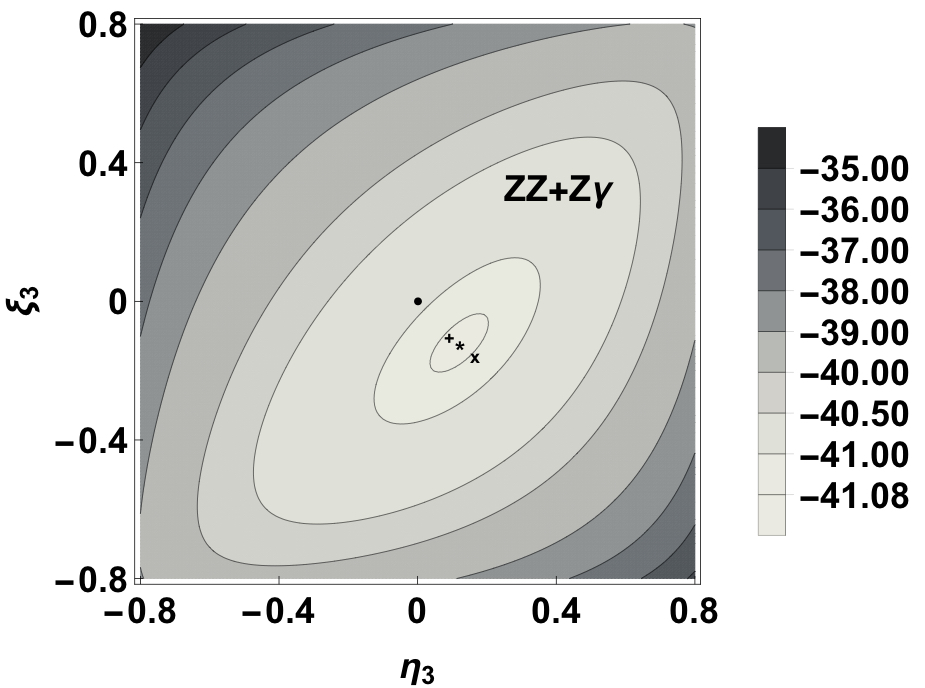}
    \caption{\label{fig:x2_all_ObsZZZA_Int}The log of average likelihood, 
        $\log[L(V_{\{\vec{f},\vec{h}\}},{\{\cal O}\};\eta_3,\xi_3)]$, is shown 
        considering both the processes $ZZ$ and $Z\gamma$ at $\sqrt{s}=500$ GeV, 
        ${\cal L}=100$ fb$^{-1}$. The {\em asterisk mark} at $P_{best}=(+0.12,-0.12)$ is the
        combined best choice for beam polarizations while the other points are for 
        $ZZ$ ({\em cross mark}) and $Z\gamma$ ({\em plus mark}). }
\end{figure}
The combined analysis of the processes $ZZ$ and $Z\gamma$ is expected to change
the best choice of beam polarizations and limits accordingly. For the average 
likelihood for these two processes the volume, in which one should  average, 
will change to $V_{\vec{f}/\vec{h}} \to V_{\vec{F}}$, where 
$\vec{F}=\{\vec{f},\vec{h}\}$ and observables from both  processes should be 
added to the likelihood defined in Eq.~\ref{eq:likelihood_zz}. The combined averaged
likelihood showing dependence on the beam polarizations for the two processes 
considered here is shown in Fig.~\ref{fig:x2_all_ObsZZZA_Int}. The dot on the 
middle of the plot is for the unpolarized case and asterisk mark at 
$P_{best}=(+0.12,-0.12)$ is the combined best choice of beam polarizations. 
Other points are due to $P_{ZZ}$ and $P_{Z\gamma}$. The combined best choice 
point sits in between $P_{ZZ}$ and $P_{Z\gamma}$. The limits, presented in 
Table~\ref{tab:coupling_limit_mcmc_ZZ} and~\ref{tab:coupling_limit_mcmc_ZA}, 
at the combined best choice of the beam polarizations are slightly weaker than 
the limit at the best choice points but comparable in both  processes as 
expected. Thus the combined best choice can be a good benchmark beam 
polarizations for the process $ZZ$ and $Z\gamma$ to study at ILC.

\begin{figure}[t]
    \centering
    \includegraphics[width=0.49\textwidth]{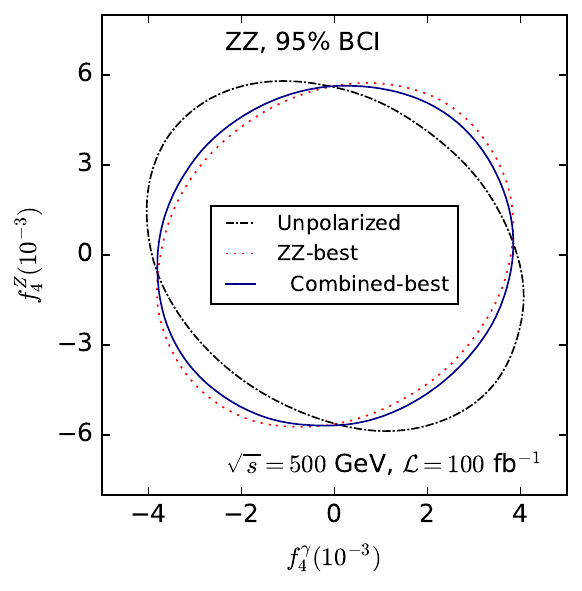}
    \includegraphics[width=0.49\textwidth]{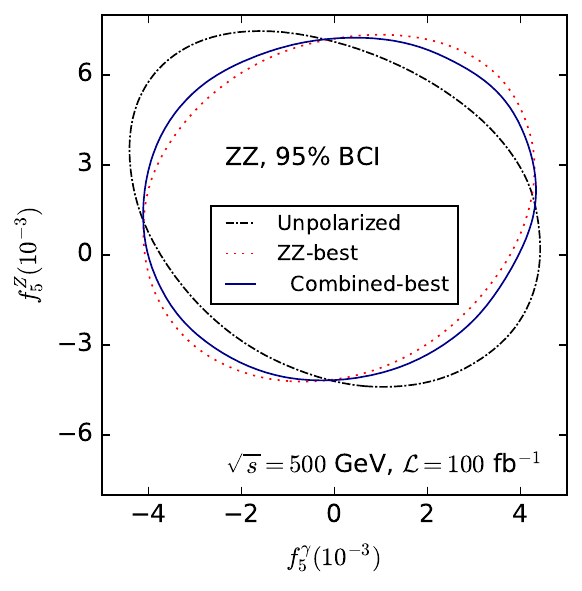}
    \caption{\label{fig:mcmc-best-choice-zz} Two dimensional marginalised contours at $95~\%$
    confidence level (C.L.) from MCMC in $ZZ$ production in $f_4^\gamma$-$f_4^Z$ and $f_5^\gamma$-$f_5^Z$ planes for unpolarized case, best choice for $ZZ$ process
        and combined best choice  of beam polarization including both processes. }
\end{figure}
\begin{figure}[h!]
    \centering
    \includegraphics[width=0.49\textwidth]{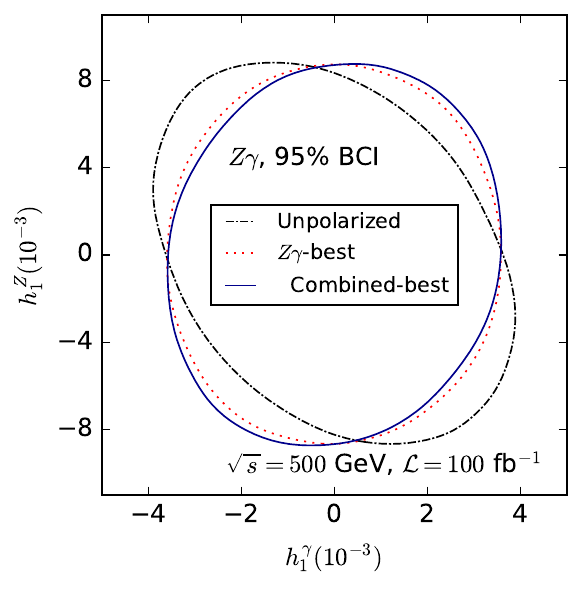}
    \includegraphics[width=0.49\textwidth]{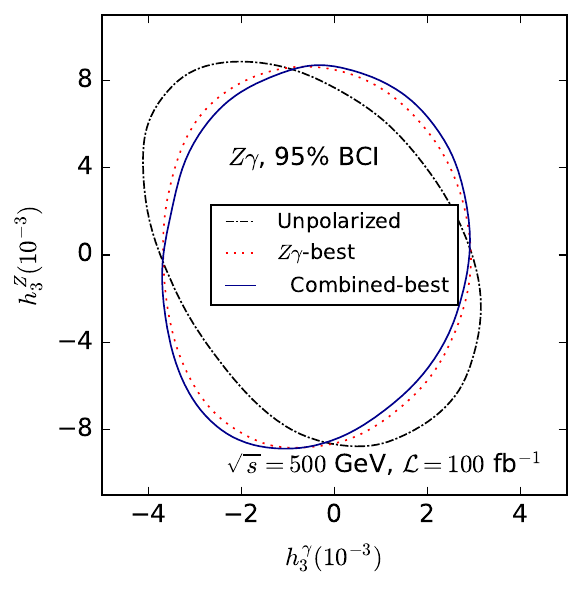}
    \caption{\label{fig:mcmc-best-choice-za} Two dimensional marginalised contours at $95~\%$
        C.L. from MCMC in $Z\gamma$ production in $h_1^\gamma$-$h_1^Z$ and $h_3^\gamma$-$h_3^Z$ planes for unpolarized case, best choice for $Z\gamma$ process
        and combined best choice  of beam polarization including both processes. }
\end{figure}
The best choice of beam polarizations, obtained here, depends 
on the size of the estimated error of the observables and hence on the 
systematics $\epsilon_\sigma$ and $\epsilon_A$. Numerical analysis shows that 
the best choice points, for both  processes separately and combined, move away 
from the unpolarized point along the cross diagonal axis towards the right-bottom 
corner on the $\eta_3 - \xi_3$ plane when $\epsilon_\sigma$ or
$\epsilon_A$ or both are increased. For example, if we double $\epsilon_\sigma$
and $\epsilon_A$ both, i.e. we take $\epsilon_\sigma=0.04$ and $\epsilon_A=0.02$, 
the best choice points $P_{ZZ}$, $P_{Z\gamma}$ and $P_{best}$ become 
($+0.20,-0.20$),  ($+0.13,-0.12) $ and ($+0.17,-0.16 $), respectively.
On the other hand, the best choice points move towards the unpolarized point as
the systematics are decreased. For example, when the systematics are reduced by
$1/2$, i.e. for $\epsilon_\sigma=0.01$ and $\epsilon_A=0.005$, the best choice 
points for $ZZ$, $Z\gamma$ and for combined process  move to ($+0.15,-0.15 $),
($+0.08,-0.08 $) and ($+0.11,-0.11 $), respectively. However, the best choice 
points do not move further closer to the unpolarized point when the size of 
systematics becomes smaller than the statistical one.

Similar analysis as presented in Fig.~\ref{fig:x2_all_ObsZZZA_Int}
can be done by combining many processes, as one should 
do, to choose a suitable beam polarizations at ILC. For many processes
with different couplings, the volume in which one should do the 
average will change to $V_{\vec{f}/\vec{h}}= V_{\vec{F}}$, where $\vec{F}$ would be the
set of all couplings for all the processes considered. The set of observables
$\{\mathcal{O}\}$ would include all the relevant observables from all the
processes combined in the expression for the likelihood.

The best choice of beam polarization in both processes not only gives tighter constraints
on the anomalous couplings but also changes the correlation among the couplings. 
In Figs.~\ref{fig:mcmc-best-choice-zz} and~\ref{fig:mcmc-best-choice-za}, we show
correlations among the anomalous couplings in both processes in marginalised contours
at $95~\%$ BCI from MCMC for the unpolarized case as well as three best choices of beam 
polarizations. The correlations got reduced in the best choices of beam polarization
apart from tightening the limits on them.
\section{Results with  beam polarizations combined with their opposite values}
\begin{figure}[!ht]
    \centering
    \includegraphics[width=0.496\textwidth]{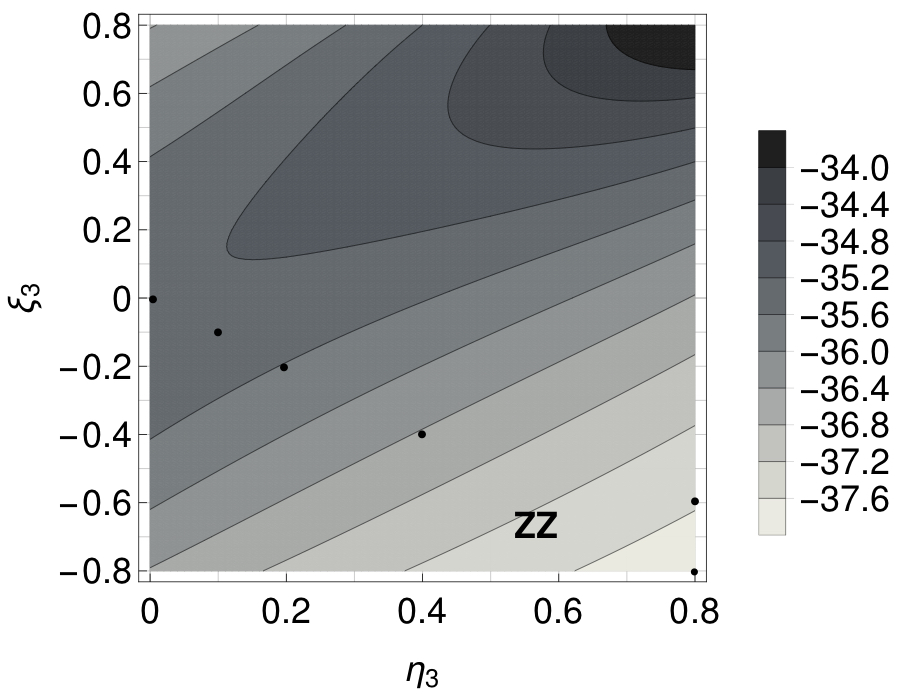}
    \includegraphics[width=0.496\textwidth]{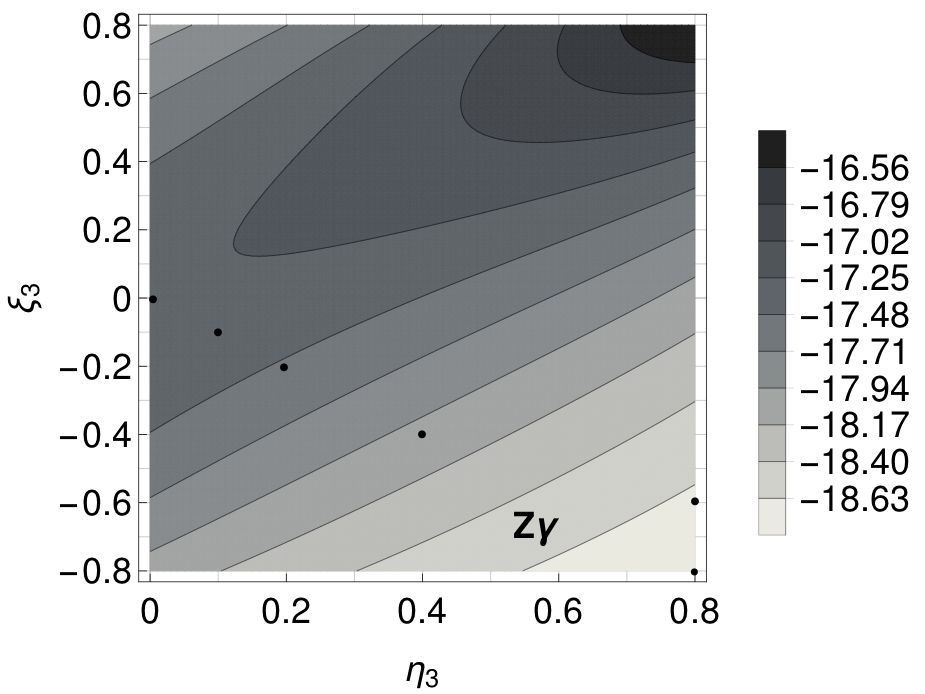}
    \caption{\label{fig:averageLikZZZA}The log of average likelihood, 
        $\log[ L(V_{\vec{f}},{\{\cal O}\};\eta_3,\xi_3)]$ as a function of beam
        polarization $(\pm\eta_3,\pm\xi_3)$ is shown for the $ZZ$ ({\em left-panel}) and $Z\gamma$ ({\em right-panel}) process  for $\sqrt{s}=500$ GeV and
        ${\cal L}=100$ fb$^{-1}$. The dots ({\tiny $\bullet$}) on the plots are the choice of polarizations for obtaining simultaneous limits given in Table~\ref{tab:flippedBeamPol} and Figs.~\ref{fig:triangle-ZZ} \&~\ref{fig:triangle-ZA}.}
\end{figure}
The above analyses of obtaining best choice of beam polarization and the limits on the couplings 
is done using a  fixed choice of beam polarizations. However, an 
$e^+e^-$ machine will run with longitudinal beam polarizations switching between $(\eta_3,\xi_3)$ and $(-\eta_3,-\xi_3)$~\cite{MoortgatPick:2005cw}.
For integrated luminosity of  $100$ fb$^{-1}$, one will have half the luminosity ($50$ fb$^{-1}$) available for each polarization
configuration. We combine the beam polarization $(+\eta_3,+\xi_3)$ and its opposite $(-\eta_3,-\xi_3)$
 at the level of $\chi^2$ 
 as, 
\begin{equation}\label{eq:beampol-totChi2}
\chi^2_{tot}(\pm\eta_3,\pm\xi_3)= \sum_{i}^{N} \left(\chi^2\left[obs_i(+\eta_3,+\xi_3)\right] +\chi^2\left[obs_i(-\eta_3,-\xi_3)\right] \right) ,
\end{equation}
where $N=9$ is the total number of observables. 
\begin{table}[!h]\caption{\label{tab:flippedBeamPol}List of simultaneous limits at $95~\%$ C.L. on the 
        anomalous couplings ($10^{-3}$) obtained for $\sqrt{s}=500$ GeV and ${\cal L}=100$ fb$^{-1}$
        for different beam polarization $(\pm\eta_3,\pm\xi_3)$ from MCMC in $ZZ$ and $Z\gamma$ processes.}
    \renewcommand{\arraystretch}{1.50}
    \centering
    \begin{tabular}{ccccccc}\hline
        $f^V,h^V$      &$    (   0.0,0.0)     $&$  (\pm 0.1,\mp 0.1)$&$ (\pm 0.2,\mp 0.2)$&$(\pm 0.4,\mp 0.4)$&$(\pm 0.8,\mp 0.6)  $&$(\pm 0.8,\mp 0.8) $\\ \hline
        $f_4^\gamma    $&$ _{ -3.3  }^{+ 3.3 }$&$ _{ -3.0  }^{ +3.0}$&$ _{ -2.9 }^{ +2.9}$&$ _{ -2.6 }^{ +2.6}$&$ _{ -2.1 }^{+ 2.1}$&$ _{ -2.0 }^{+ 2.0}$\\  \hline
        $f_4^Z         $&$ _{ -4.8  }^{+ 4.8 }$&$ _{ -4.4  }^{ +4.4}$&$ _{ -4.3 }^{ +4.3}$&$ _{ -3.9 }^{ +4.0}$&$ _{ -3.6 }^{+ 3.6}$&$ _{ -3.4 }^{+ 3.4}$\\  \hline
        $f_5^\gamma    $&$ _{ -3.6  }^{+ 3.7 }$&$ _{ -3.3  }^{ +3.3}$&$ _{ -3.2 }^{ +3.1}$&$ _{ -2.8 }^{ +2.6}$&$ _{ -2.3 }^{+ 2.1}$&$ _{ -2.1 }^{+ 2.0}$\\  \hline
        $f_5^Z         $&$ _{ -4.6  }^{+ 5.1 }$&$ _{ -2.8  }^{ +6.0}$&$ _{ -2.8 }^{ +5.8}$&$ _{ -2.6 }^{ +5.3}$&$ _{ -2.5 }^{+ 4.7}$&$ _{ -2.4 }^{+ 4.4}$\\  \hline\hline
        $ h_1^\gamma   $&$ _{ -3.1  }^{+ 3.1 }$&$ _{ -2.7  }^{ +2.7}$&$ _{ -2.5 }^{ +2.6}$&$ _{ -2.3 }^{ +2.3}$&$ _{ -2.0 }^{+ 2.0}$&$ _{ -1.9 }^{+ 1.9}$\\  \hline
        $ h_1^Z        $&$ _{ -7.0  }^{+ 7.0 }$&$ _{ -6.0  }^{ +6.1}$&$ _{ -5.6 }^{ +5.5}$&$ _{ -4.4 }^{ +4.4}$&$ _{ -3.5 }^{+ 3.4}$&$ _{ -3.2 }^{+ 3.3}$\\  \hline
        $ h_3^\gamma   $&$ _{ -2.8  }^{+ 2.6 }$&$ _{ -2.9  }^{ +2.0}$&$ _{ -2.7 }^{ +2.0}$&$ _{ -2.4 }^{ +1.8}$&$ _{ -2.0 }^{+ 1.7}$&$ _{ -1.9 }^{+ 1.6}$\\  \hline
        $ h_3^Z        $&$ _{ -7.1  }^{+ 7.0 }$&$ _{ -5.8  }^{ +6.0}$&$ _{ -5.2 }^{ +5.4}$&$ _{ -4.1 }^{ +4.2}$&$ _{ -3.1 }^{+ 3.2}$&$ _{ -2.9 }^{+ 3.0}$\\  \hline
        
    \end{tabular}
\end{table}
\begin{figure}[ht]
    \centering
    \includegraphics[width=1.0\textwidth]{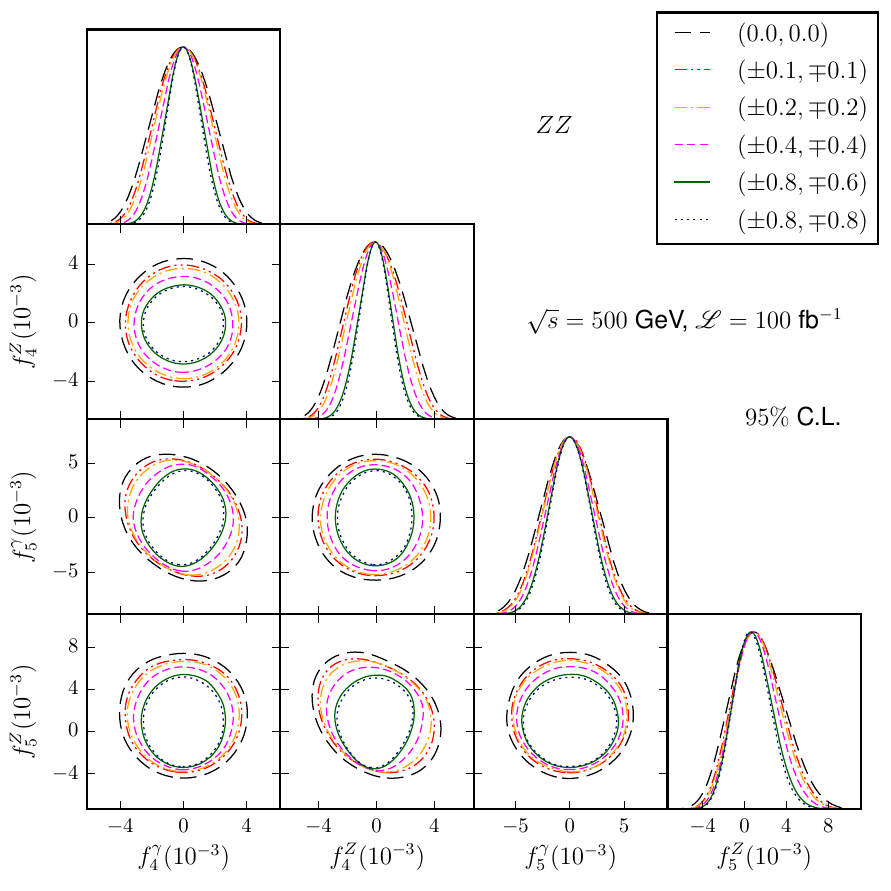}
    \caption{\label{fig:triangle-ZZ}All the one dimensional marginalised projections and  two dimensional marginalised contours at $95~\%$
        C.L. in triangular array from MCMC in $ZZ$ production for $\sqrt{s}=500$ GeV and ${\cal L}=100$ fb$^{-1}$
        for different beam polarizations $(\pm\eta_3,\pm\xi_3)$. }
\end{figure}
\begin{figure}[ht]
    \centering
    \includegraphics[width=1.0\textwidth]{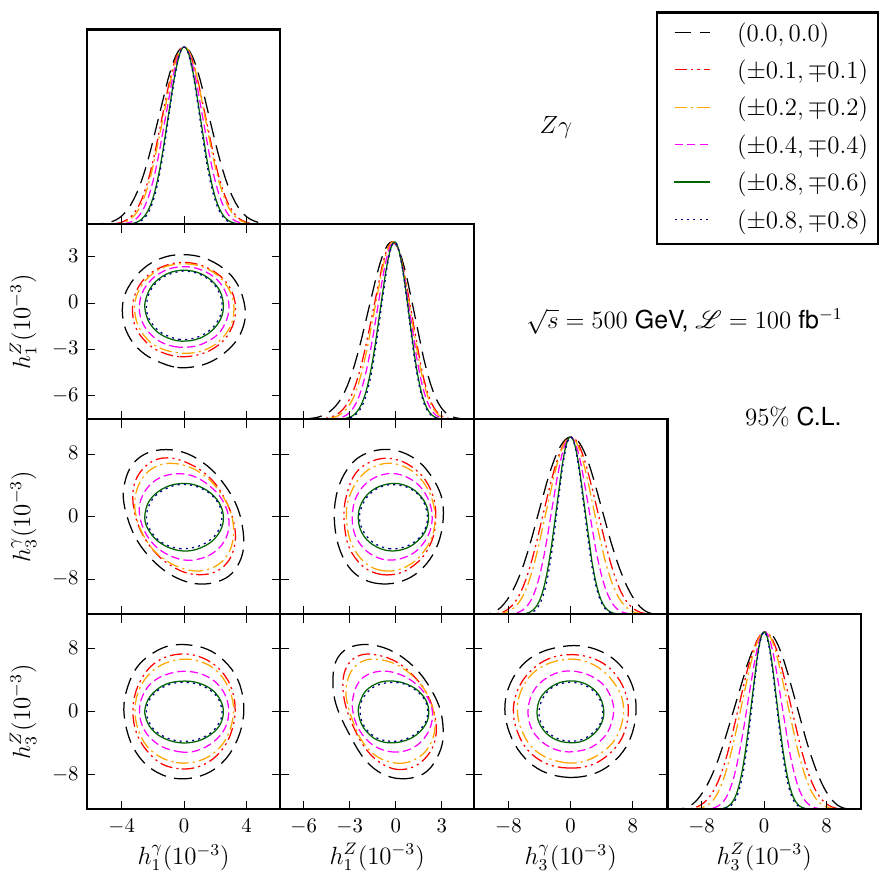}
    \caption{\label{fig:triangle-ZA} All the one dimensional marginalised projections and  two dimensional marginalised contours at $95~\%$
        C.L. in triangular array from MCMC in $Z\gamma$ production for $\sqrt{s}=500$ GeV and ${\cal L}=100$ fb$^{-1}$
        for different beam polarizations $(\pm\eta_3,\pm\xi_3)$. }
\end{figure}

We calculate the {\em weighted-volume} in Eq.~(\ref{eq:average_likelihood_zz}) using the total $\chi^2$
given in Eq.~(\ref{eq:beampol-totChi2}) for both $ZZ$ and $Z\gamma$ production processes and they are 
shown in Fig.~\ref{fig:averageLikZZZA} as a function of beam polarization
 $(\pm\eta_3,\pm\xi_3)$. The {\em weighted-volume} or the 
averaged likelihood decreases along the $\pm\eta_3=\mp\xi_3$ line and the beam polarization $(\pm 
0.8,\mp 0.8)$ poses  the minimum values for both $ZZ$ and $Z\gamma$ processes and their combined one. 
There are constant lines for a constant values of the {\em weighted-volume} implying that each beam 
polarization points on a given line will provide similar limit on the couplings.
Though, the point $(\pm 0.8, \mp 0.8)$ is the best choice for beam polarization, the point 
$(\pm 0.8, \mp 0.6)$ is the best within the limitation for positron polarization, i.e., $|\xi_3|<0.6$.

We estimate simultaneous limits on the couplings in both processes using MCMC with the combined
$\chi^2$ given  in Eq.~(\ref{eq:beampol-totChi2}) for a set of beam polarizations $(0,0)$, $(\pm 0.1, \mp 0.1)$,
 $(\pm 0.2, \mp 0.2)$, $(\pm 0.4, \mp 0.4)$,  $(\pm 0.8, \mp 0.6)$, and $(\pm 0.8, \mp 0.8)$. The simultaneous limits at $95~\%$ BCI
on the anomalous couplings are shown in Table~\ref{tab:flippedBeamPol} for both processes.
It can be seen that the limits with beam polarization combined with the opposite values given in 
Table~\ref{tab:flippedBeamPol} are better than the limits with fixed beam polarization given in 
Tables~\ref{tab:coupling_limit_mcmc_ZZ} \&~\ref{tab:coupling_limit_mcmc_ZA} with the same luminosity
of $100$ fb$^{-1}$.
The one dimensional marginalised projections and  two dimensional marginalised contours at $95~\%$
BC in triangular array from MCMC obtained for the same set of beam polarizations as in Table~\ref{tab:flippedBeamPol} 
are shown in Fig.~\ref{fig:triangle-ZZ} and Fig.~\ref{fig:triangle-ZA} for $ZZ$ and $Z\gamma$ processes, respectively. We observe that as the amplitude of beam polarizations are increased, 
the correlations reduce ( $f_4^\gamma$-$f_5^\gamma$, $f_4^Z$-$f_5^Z$ in Fig.~\ref{fig:triangle-ZZ}
and $h_1^\gamma$-$h_3^\gamma$, $h_1^Z$-$h_3^Z$ in Fig.~\ref{fig:triangle-ZA}) along with the limits
getting tighter.  

\section{Summary}\label{sec:conclusion-epjc2}
To summarize,  
we studied the effects of beam polarization on polarization asymmetries and
corresponding sensitivities towards anomalous couplings in this chapter. Using the {\em minimum
averaged} likelihood, we found the best choice of the beam polarization for the
two processes for fixed beam polarization as well as when opposite beam polarization are combined together. Here, the list of observables 
includes the cross section along with eight polarization asymmetries for 
the $Z$ boson. Simultaneous limits on
anomalous couplings were obtained using the MCMC method for a set of benchmark 
beam polarizations for both fixed choices and  combined with flipped choices.
The simultaneous limits for a fixed choice of beam polarizations are presented in 
Tables~\ref{tab:coupling_limit_mcmc_ZZ} and \ref{tab:coupling_limit_mcmc_ZA}, 
while for choice for  polarizations combined with opposite values are presented
in Table~\ref{tab:flippedBeamPol}. 
The limits obtained for the unpolarized case are better than the ones
reported in chapter~\ref{chap:epjc1}. This is because the present analysis
includes three new observables $\wtil{A}_z$, $\wtil{A}_{xz}$ and
$\wtil{A}_{yz}$. These new asymmetries yield better limits on $f^Z_{4,5}$
and $h^\gamma_{1,3}$, while we have comparable (yet better) limits on $f^\gamma_{4,5}$
and $h^Z_{1,3}$.
In the fixed beam polarization case, the best choices
of beam polarizations are somewhere closer to the unpolarized point. In the combined case, however,
the best choices of beam polarization appear to be as  maximum as can be, and that is same
for two processes separately as well as combinedly.


\chapter{The probe of aTGC in $ZZ$ production at the LHC and the role of $Z$ boson polarizations}\label{chap:ZZatLHC}
\begingroup
\hypersetup{linkcolor=blue}
\minitoc
\endgroup
{\small\textit{\textbf{ The contents in this chapter are based on the published article in Ref.~\cite{Rahaman:2018ujg}. }}}
\vspace{1cm}

In the previous two chapters, we studied the neutral aTGC at a future linear collider, the ILC. It is natural to see the implication of the aTGC using the polarization observables at the current collider LHC, which already have collected enough data to put stringent limits on the aTGC.
In this chapter, we see the prospects of aTGC in $ZZ$ production in $4$-lepton final state at the LHC.
The neutral aTGC appearing in the $ZZ$ production at dimension-$6$ are given by the subset
\begin{eqnarray}\label{eq:ZZ-LZZV}
{\cal L}_{ZZV}=
\frac{e}{m_Z^2} \l[-
\left[f_4^\gamma \left(\partial_\mu F^{\mu \beta}\right)+f_4^Z \left(\partial_\mu Z^{\mu \beta}\right) 
\right] Z_\alpha 
\left( \partial^\alpha Z_\beta\right)
+\left[f_5^\gamma \left(\partial^\sigma F_{\sigma \mu}\right)+
f_5^Z \left(\partial^\sigma Z_{\sigma \mu}\right) \right] \wtil{Z}^{\mu \beta} Z_\beta
\r]\nonumber\\ 
\end{eqnarray}
of the full Lagrangian given in Eq.~(\ref{eq:LZZV-dim6})
containing only four parameters $f_4^V$ and $f_5^V$.
There has been a lot of study of these neutral aTGC for a hadron collider ~\cite{Baur:1992cd,Ellison:1998uy,Baur:2000ae,Chiesa:2018lcs,Chiesa:2018chc,Aihara:1995iq,Gounaris:1999kf,Gounaris:2000dn} with different techniques. These neutral aTGCs have also been searched at the LHC
in different processes~\cite{Chatrchyan:2012sga,Chatrchyan:2013nda,Aad:2013izg,
    Khachatryan:2015kea,Khachatryan:2016yro,Aaboud:2017rwm,Sirunyan:2017zjc} including the $ZZ$ production~\cite{Chatrchyan:2012sga,Sirunyan:2017zjc} using cross section in suitable kinematical cuts. 
The stringent limits  on these aTGC   has been  obtained in  $ZZ$ production itself at the LHC~\cite{Sirunyan:2017zjc}.   The tightest limits at  $95~\%$ C.L. for
$\sqrt{s}=13$ TeV and ${\cal L}=35.9$ fb$^{-1}$ are 
\begin{align}\label{eq:CMS-limit}
− 0.0012 < f_4^Z < 0.0010,~~ − 0.0010 < f_5^Z < 0.0013,\nonumber\\
− 0.0012 < f_4^\gamma < 0.0013,~~ − 0.0012 < f_5^\gamma < 0.0013,
\end{align}
obtained by varying one parameter at a time and using only the cross section as observable.
We note that these ranges of couplings do not violate unitarity bound up to an energy scale of $10$ TeV.  
Whereas a size as large as  ${\cal O}(\pm 0.1)$ of the couplings can be  allowed if the unitarity violation 
is assumed to take place at   the energy scale of $3$ TeV, a typical energy range explored by the current $13$ TeV LHC.
Our strategy, here, is to see the significance of the polarization
observables on top of the cross section in probing the aTGC. 

The leading order (LO) result of the $ZZ$ pair production cross section  is way below
the result measured at the LHC~\cite{Aaboud:2017rwm,Sirunyan:2017zjc}. However, the existing 
next-to-next-to-leading  order (NNLO)~\cite{Heinrich:2017bvg,Cascioli:2014yka} results are  
comparable with the measured values at CMS~\cite{Sirunyan:2017zjc} and ATLAS~\cite{Aaboud:2017rwm}.
We, however, obtain the cross section at next-to-leading order (NLO) in the SM and in aTGC using \MGvATNLO~\cite{Alwall:2014hca}  
and have used the SM $k$-factor to match to the NNLO value.
The details of these calculations are described in section~\ref{sect:signal-background}.


The LHC being a symmetric collider, most of the polarization of $Z$ in $ZZ$ pair production
are either zero or  close to zero  except   the polarization $T_{xz}$, $T_{xx}-T_{yy}$,  and $T_{zz}$. 
For better significance, we used the tilde asymmetry $\wtil{A}_{xz}$ 
corresponding to $T_{xz}$ as given in Eq.~(\ref{eq:epjc2-tilde-AxzAyz})
with  $c_{\theta_Z}$ being measured in the Lab frame.
To get the momentum direction of $Z$ boson,
one needs a reference axis ($z$-axis), but we can not assign a direction at the LHC because it is a
symmetric collider. So we consider the direction of the boost   of the  $4l$ final state to be the proxy
for reference $z$-axis. In $q\bar{q}$ fusion, the quark is supposed to have larger momentum
then the anti-quark at the LHC, thus above proxy statistically stands for the direction of the quark
and $c_{\theta_Z}$ is measured w.r.t. the boost.

\section{The signal and  background}\label{sect:signal-background}
\begin{figure}[h]
    \centering
    \includegraphics[width=1\textwidth]{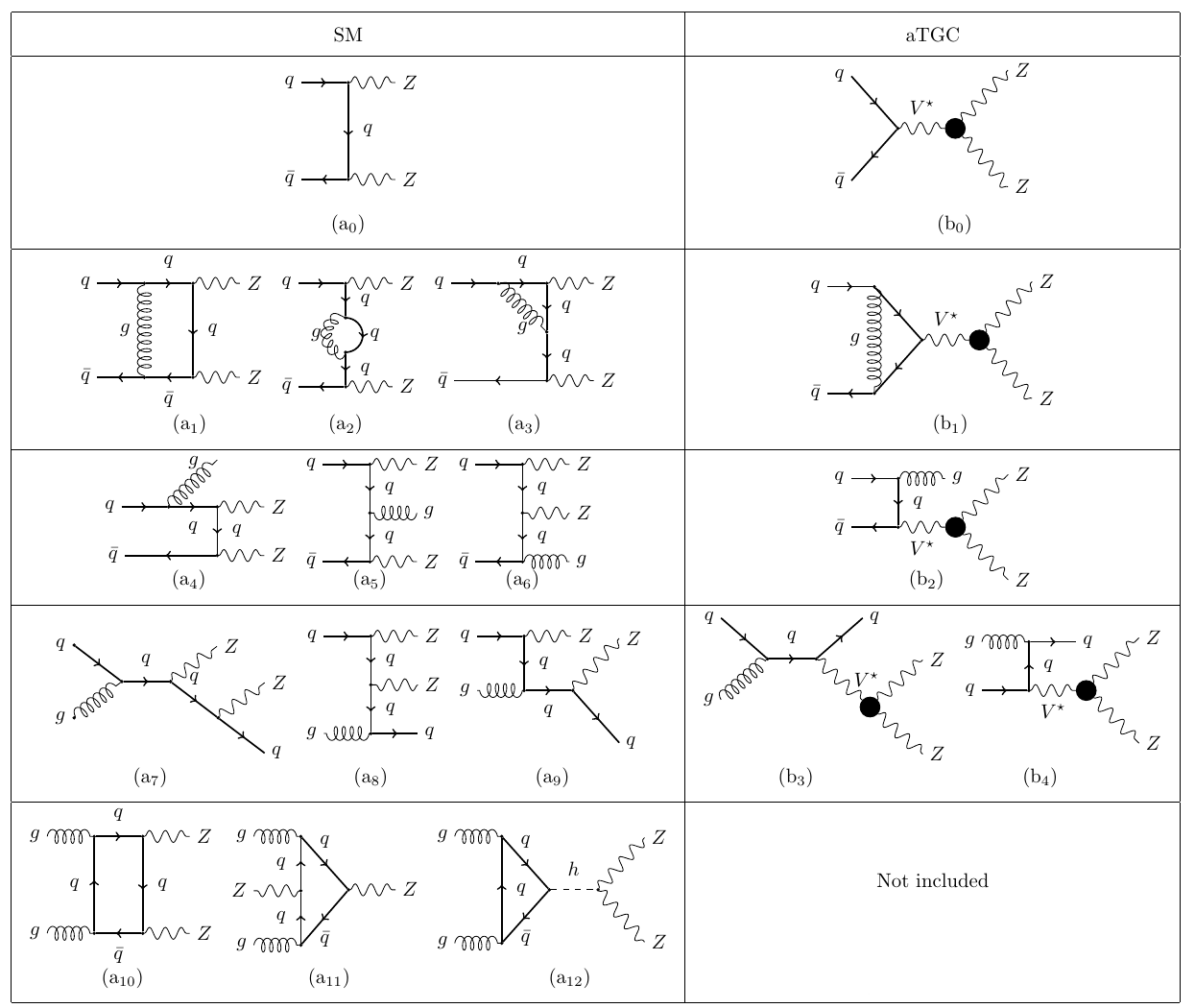}
    \caption{\label{fig:Feynman-diagram-all}
        Representative Feynman diagrams for $ZZ$ pair production at the LHC in the SM 
        ($q\bar{q}$  and $gg$ initiated) as well as in aTGC ($q\bar{q}$ initiated) 
        at tree level together with NLO in QCD.
    }
\end{figure}
\begin{figure}
    \centering
    \includegraphics[width=0.496\textwidth]{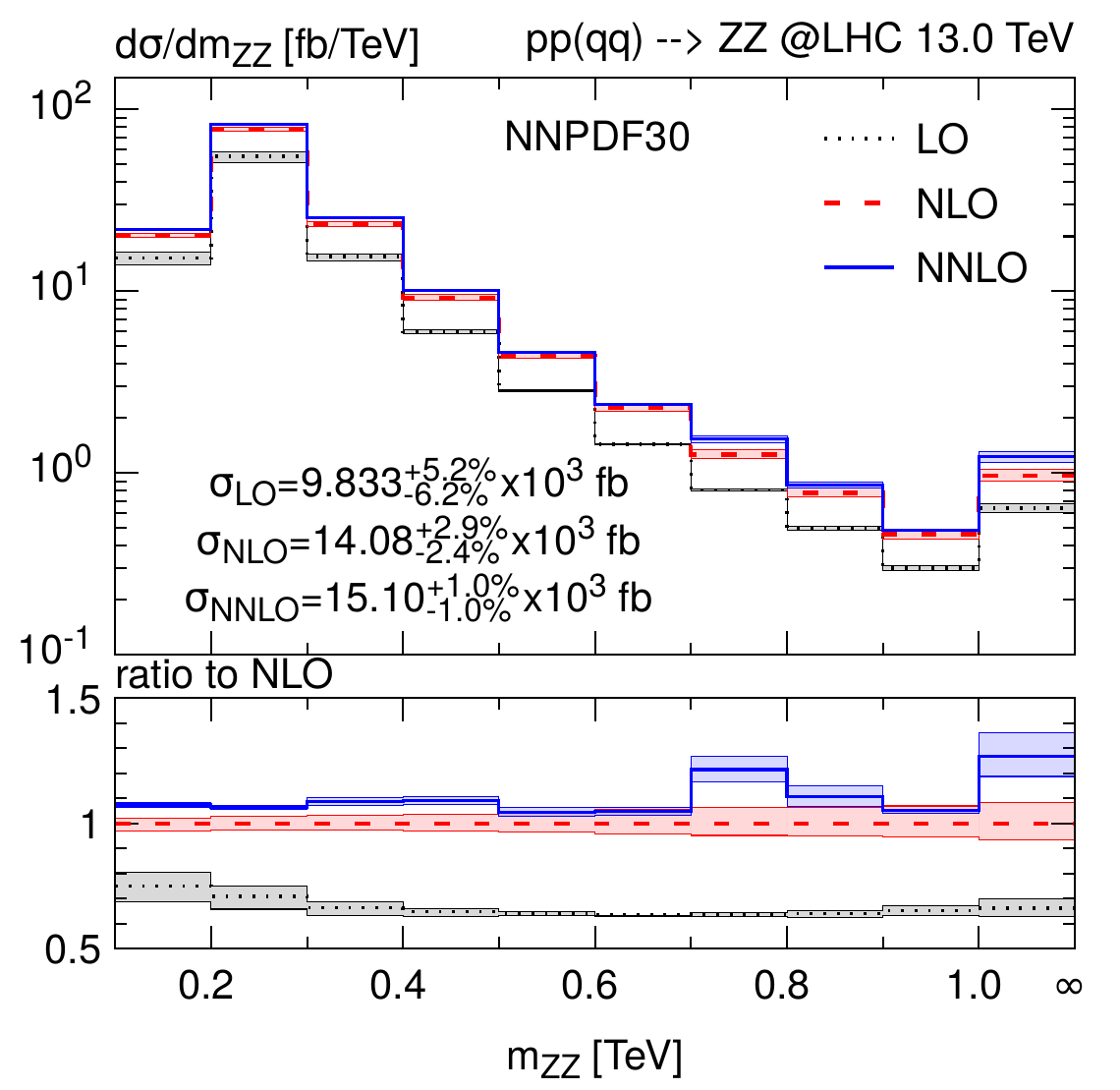}    
    \includegraphics[width=0.496\textwidth]{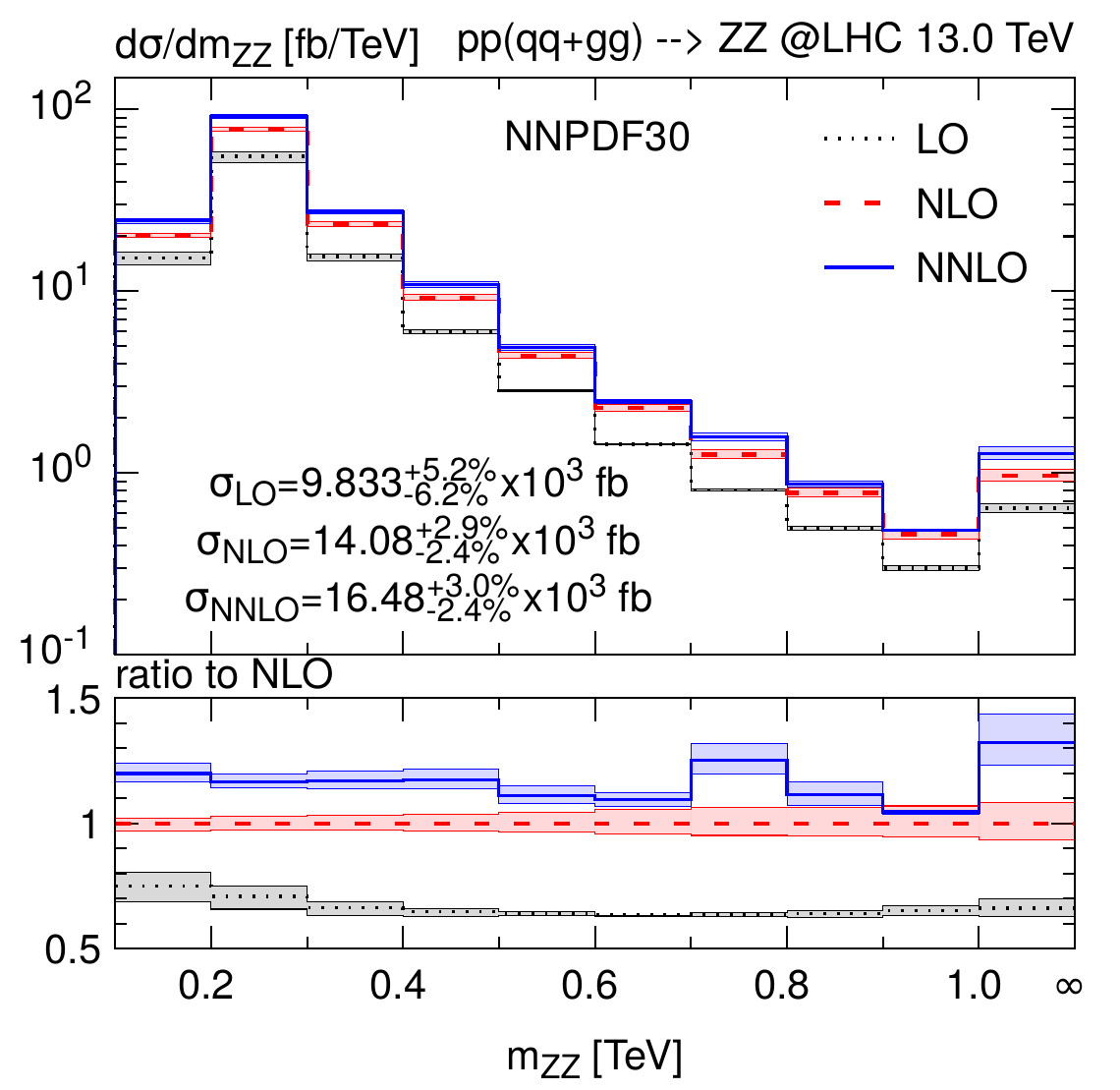}
    \caption{\label{fig:mZZ-matrx}
        The differential distributions of $m_{ZZ}$ in the $ZZ$ production at the LHC at $\sqrt{\hat{s}}=13$ TeV in LO, NLO and NNLO  obtained using {\tt MATRIX}. In the {\em left-panel} $q\bar{q}$
        initiated results are shown, while in the {\em right-panel} $q\bar{q}+gg$ initiated results are shown. }
\end{figure}

We are interested in studying anomalous triple gauge boson couplings in $ZZ$ pair
production at the LHC. The tree level standard model contribution to this process comes from the representative
diagram (a$_0$) in Fig.~\ref{fig:Feynman-diagram-all}, while the tree level aTGC contribution is shown in the diagram (b$_0$). 
Needless to say, the tree level cross section in the SM is way below the
measured cross section at the LHC, because  QCD corrections are very high in this process.
In the SM, at NLO (${\cal O}(\alpha_s)$), virtual  contributions come from the representative diagrams (a$_{1}$--a$_{3}$) 
and real contributions come  from (a$_{4}$--a$_{9}$) in the    $q\bar{q}$ initiated sub-process.
The  $gg$ initiated sub-process  appears
at $1$-loop level,   the diagrams (a$_{10}$--a$_{12}$), and contributes at ${\cal O}(\alpha_s^2)$. 
The LO, NLO and NNLO results from theoretical calculation  available in   literature~\cite{Heinrich:2017bvg,Cascioli:2014yka} and our estimate in {\tt MATRIX}~\cite{Grazzini:2015hta,Kallweit:2018nyv,Cascioli:2014yka,Grazzini:2017mhc,Cascioli:2011va,Denner:2016kdg,Gehrmann:2015ora,Catani:2012qa,Catani:2007vq}
for $ZZ$ production cross section  at $\sqrt{s}=13$ TeV for a  $pp$ collider are listed
in Table~\ref{tab:ZZ-signal}. The recent experimental
measurement from CMS~\cite{Sirunyan:2017zjc}
and ATLAS~\cite{Aaboud:2017rwm} are also shown for comparison.
\begin{table}[h]\caption{\label{tab:ZZ-signal} The theoretical estimates  and experimental measurements of the
        $ZZ$ production  cross section at $\sqrt{s}=13$ TeV at the LHC. The uncertainties in the theoretical 
        estimates come from scale variation.}
    \renewcommand{\arraystretch}{1.50}
    \begin{tabular*}{\columnwidth}{@{\extracolsep{\fill}}llll@{}} \hline
    Obtained at & $\sigma_{\text{LO}}$ [pb]& $\sigma_{\text{NLO}}$ [pb] & $\sigma_{\text{NNLO}}$ [pb] \\ \hline
{\tt MATRIX} & $9.833_{-6.2\%}^{+5.2\%}$ & $14.08_{-2.4\%}^{+2.9\%}$ & $16.48_{-2.4\%}^{+3.0\%}$ \\ \hline
        Heinrich {\em et} al.~\cite{Heinrich:2017bvg} & $9.890_{-6.1\%}^{+4.9\%}$ & $14.51_{-2.4\%}^{+3.0\%}$ & $16.92_{-2.6\%}^{+3.2\%}$ \\ \hline
        Cascioli {\em et} al.~\cite{Cascioli:2014yka}& $9.887_{-6.1\%}^{+4.9\%}$ & $14.51_{-2.4\%}^{+3.0\%}$ & $16.91_{-2.4\%}^{+3.2\%}$\\ \hline
    \end{tabular*}
    \begin{tabular*}{\columnwidth}{@{\extracolsep{\fill}}lll@{}}
        CMS~\cite{Sirunyan:2017zjc} &~~~~~~~~~~~~  $17.2\pm 0.5 (stat.) \pm 0.7 (syst.) 
        \pm 0.4 (lumi.) $  \\ \hline
        ATLAS~\cite{Aaboud:2017rwm} &~~~~~~~~~~~~ $17.3 \pm 0.6(stat.) \pm 0.5(syst.) \pm 0.6(lumi.)$ \\ \hline
    \end{tabular*}
\end{table}
The cross section at
NLO  receives as much as $\sim 46~\%$ correction over LO and further the NNLO cross section
receives  $\sim 16~\%$  correction over the NLO result. At NNLO the $q\bar{q}$ sub-process receives
$10~\%$ correction~\cite{Cascioli:2014yka} over NLO and the $gg$ initiated ${\cal O}(\alpha_s^3)$ sub-process receives $70~\%$
correction~\cite{Caola:2015psa} over it's ${\cal O}(\alpha_s^2)$ result. 
The higher order corrections to the cross section vary w.r.t. $\sqrt{\hat{s}}$ or $m_{ZZ}$  as shown
in Fig.~\ref{fig:mZZ-matrx} with only $q\bar{q}$ initiated processes in the {\em left-panel} and $q\bar{q}+gg$ initiated processes in the {\em right-panel} obtained at  {\tt MATRIX}~\cite{Grazzini:2015hta,Kallweit:2018nyv,Cascioli:2014yka,Grazzini:2017mhc,Cascioli:2011va,Denner:2016kdg,Gehrmann:2015ora,Catani:2012qa,Catani:2007vq}. 
The lower panels display the respective bin-by-bin ratios to the NLO central predictions. 
The NLO to LO ratio does not appear to be constant over the range of 
$m_{ZZ}$. Thus a simple  $k$-factor with LO events can not be used 
as proxy for NLO events. 
We use results obtained at \MGvATNLO including   NLO  QCD  corrections for our analysis.
The LO and NLO results obtained in \MGvATNLO~v2.6.2 with PDF (parton-distribution-function) sets NNPDF23 are 
\begin{eqnarray}\label{eq:result-mg5-zz}
\sigma_{{\cal O}(\alpha_{s}^0)}^{q\bar{q}\to ZZ} &=& 9.341_{-5.3\%}^{+4.3\%}  ~~\text{pb},\nonumber\\  
\sigma_{{\cal O}(\alpha_s)}^{q\bar{q}\to ZZ} &=& 13.65_{-3.6\%}^{+3.2\%}  ~~\text{pb}, \nonumber\\ 
\sigma_{{\cal O}(\alpha_s^2)}^{gg\to ZZ} &=& 1.142_{-18.7\%}^{+24.5\% } ~~\text{pb},\nonumber\\   
\sigma_{mixed_1}^{q\bar{q}+gg\to ZZ}&=&
\sigma_{{\cal O}(\alpha_s)}^{q\bar{q}\to ZZ}+ \sigma_{{\cal O}(\alpha_s^2)}^{gg\to ZZ}\nonumber\\
 &=&14.79_{-4.7\%}^{+4.8\%} ~~\text{pb}. 
\end{eqnarray}
The  errors in the subscript and superscript on the  cross section are the uncertainty 
from scale variation. 
The total cross section combining the $q\bar{q}$  sub-process at ${\cal O}(\alpha_s^2)$
 with $gg$ at ${\cal O}(\alpha_s^3)$  is given by,
\begin{eqnarray}
\sigma_{mixed_2}^{q\bar{q}+gg\to ZZ}&=&
\underbrace{\sigma_{{\cal O}(\alpha_s)}^{q\bar{q}\to ZZ}\times 1.1}_{{\cal O}(\alpha_s^2)} \ \ + \ \ \underbrace{ \sigma_{{\cal O}(\alpha_s^2)}^{gg\to ZZ}\times 1.7}_{{\cal O}(\alpha_s^3)} \nonumber\\
&=&16.96_{-5.3\%}^{+5.6\%} ~~\text{pb}. 
\end{eqnarray}
The {\tt MadGraph5} results are below the {\tt MATRIX} results due the difference in PDF sets.
The  aTGC  has also a substantial NLO QCD correction and they come from 
the diagram (b$_{2}$) at 1 loop level and from   (b$_{2}$--b$_{4}$) as the real radiative process. 
The aTGC effect is not included in the $gg$ process where the aTGC may come from a similar 
diagram with   $h\to ZZ$ in Fig.~\ref{fig:Feynman-diagram-all}(a$_{12}$) but $h$ replaced
with a $Z$. As an example of NLO QCD correction of aTGC in this process, we obtain cross section
at $\sqrt{s}=13$ TeV  with all couplings $f_i^V=0.001$. The  cross section for only aTGC part, $(\sigma^{\text{aTGC}}-\sigma^{\text{SM}})$
at LO and NLO are  $71.82$~fb $(0.77~\%)$ and $99.94$~fb $(0.73~\%)$, respectively. Thus  NLO 
result comes with a substantial amount ($\sim 39~\%$) of QCD correction over LO at this given aTGC point.

The signal  consists of $4l$ ($2e2\mu/4e/4\mu$) final state which includes  
 $ZZ$,  $Z\gamma^\star$, and $\gamma^\star \gamma^\star$ processes. The signal events are generated  
 in \MGvATNLO\\with PDF sets NNPDF23 in the SM as well as in the aTGC as $pp\to VV \to 2e2\mu$  
 ($V=Z/\gamma^\star$) at NLO in QCD in $q\bar{q}$, $qg$ as well as in $1$-loop $gg$ initiated process
with the following basic cuts (in accordance with Ref.~\cite{Sirunyan:2017zjc}),
 \begin{itemize}
     \item $p_T^l>10$  GeV,
     hardest  $p_T^l>20$  GeV,
     and second hardest  $p_T^l>12$  GeV,
     \item $|\eta_e|<2.5$, $|\eta_{\mu}|<2.4$,
     \item $\Delta R (e,\mu)>0.05$, $\Delta R (l^+,l^-)>0.02$.
 \end{itemize}
To select the $ZZ$ final state from the above generated signal we further put a constraint on invariant
mass of same flavoured oppositely charged leptons pair with
 \begin{itemize}
\item $60$ GeV $< m_{l^+l^-}< 120$ GeV. 
\end{itemize}
The  $2e2\mu$ cross section up to a factor of two is used as the proxy for the $4l$ cross section
for the ease of event generation and related handling.

The background event consisting $t\bar{t}Z$ and $WWZ$ with leptonic decay 
are generated at LO in \MGvATNLO~   with NNPDF23 with the same sets of 
cuts as applied to the signal, and their cross section is matched to NLO in QCD with a $k$-factor of $1.4$\footnote{This $k$-factor for the backgrounds along with the NLO to NNLO $k$-factor for the signal is of-course an approximation as they really depend on kinematic and angular distributions.}. 
This  $k$-factor estimation was done at the production level.
 We have estimated the total cross section of the signal in the SM to be
\begin{eqnarray}
\sigma(pp\to ZZ\to 4l)_{{\cal O}(\alpha_s)}^{q\bar{q}} &=& 28.39~~\text{fb}, \nonumber \\
\sigma(pp\to ZZ\to 4l)_{{\cal O}(\alpha_s^2)}^{gg} &=&  1.452~~\text{fb}, \nonumber \\
\sigma(pp\to ZZ\to 4l)_{mixed_1}^{q\bar{q}+gg} &=&  29.85~~\text{fb}, \nonumber \\
\sigma(pp\to ZZ\to 4l)_{mixed_2}^{q\bar{q}+gg} &=&  33.70~~\text{fb}.
\end{eqnarray}
  The background cross section  at NLO  is estimated to be 
 \begin{align}
 \sigma(pp\to t\bar{t}Z+WWZ\to 4l+ \cancel{\it{E}}_{T})_{NLO}= 0.020~~\text{fb}. 
 \end{align}
 The values of various parameters  used   for the generation of signal and background are
 \begin{itemize}
     \item $m_Z=91.1876$ GeV, $M_H=125.0$ GeV,
     \item $G_F=1.16639\times 10^{-5}$ GeV$^{-2}$, $\alpha_{EM}=1/132.507$,\\ $\alpha_s=0.118$,
     \item $\Gamma_Z=2.441404$ GeV, $\Gamma_H=6.382339$ MeV.
 \end{itemize}
The renormalization and factorization scale is set to $\sum M_i^T/2$, $M_i^T$ are the transverse mass  of all final state particles.

In our  analysis, the total cross section in the SM including the aTGC is taken 
as\footnote{$mixed_1 \approx q\bar{q}({\cal O}(\alpha_s)) + gg({\cal O}(\alpha_s^2))$, 
$mixed_2 \approx q\bar{q}({\cal O}(\alpha_s^2)) + gg({\cal O}(\alpha_s^3))$}
\begin{align}\label{eq:sigma-setup}
\sigma_{\text{Tot}}=\sigma^{\text{SM}}_{mixed_2} + (\sigma_{\text{NLO}}^{\text{aTGC}} 
- \sigma_{\text{NLO}}^{\text{SM}}),
\end{align}
 the SM is considered at order $mixed_2$, whereas the aTGC contribution
along with it's interference with the SM are considered at NLO in QCD (as the NNLO contribution is 
not known with aTGC).

We will use polarization asymmetries as described in the previous section in our
analysis. Assuming that the NNLO effect cancels away because of the ratio
of two cross section, we will use the asymmetries  as
\begin{align}\label{eq:asymmetry-setup}
A_i=\frac{\Delta\sigma_i^{mixed_1}}{\sigma^{mixed_1}}.
\end{align}
We use total cross section at $mixed_2$ order   and asymmetries 
at $mixed_1$ order to put constrain  on the anomalous couplings. We note that the
$Z$ boson momentum is required to be reconstructed to obtain it's polarization asymmetries, which require the
right pairing of two oppositely charged leptons coming from a same $Z$ boson in $4e/4\mu$  channel. The right paring of
leptons for the $Z$ boson in the same flavoured channel is possible with $\sim 95.5~\%$ for $m_{4l}>300$ GeV and
$\sim 99~\%$ for $m_{4l}>700$ GeV for both SM and aTGC by requiring a smaller value of $|m_Z-m_{l^+l^-}|$. This small miss pairing is neglected as it allows to use  the $2e2\mu$ channel as a proxy for the full $4l$ final state with good enough accuracy.

\section{Probe of the anomalous couplings}\label{sect:expressions-limits}
The observables in this process, we have, are the cross section and the polarization asymmetries.
We use these observables in a suitable kinematical cut region for  $m_{4l}$ and $\Delta R$ (signal region) to study the
sensitivity on aTGC and obtain limits on them.

\subsection{Effect of aTGC in kinematic distributions}
\begin{figure}[h]
    \centering
    \includegraphics[width=0.48\textwidth]{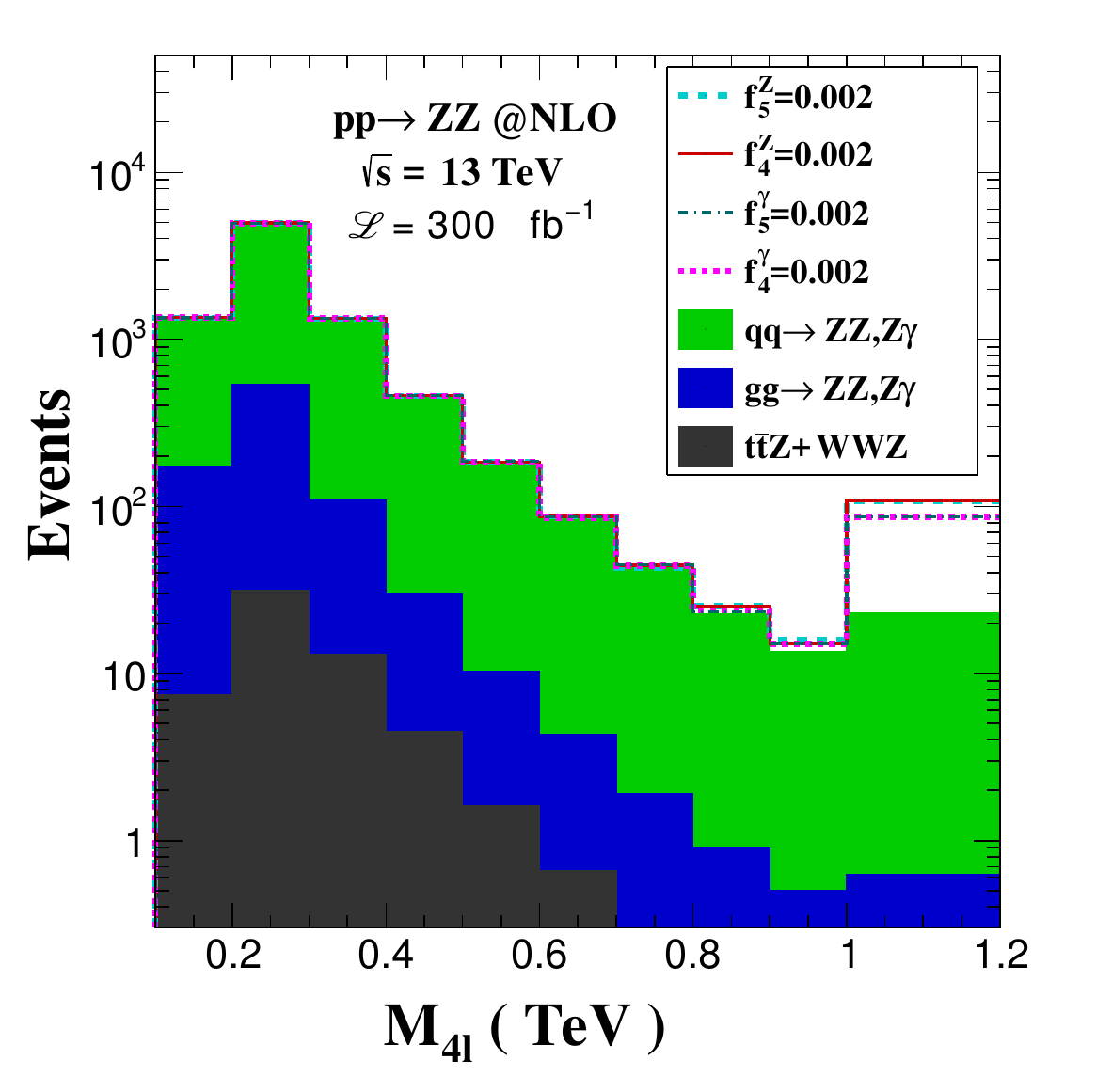}    
    \includegraphics[width=0.48\textwidth]{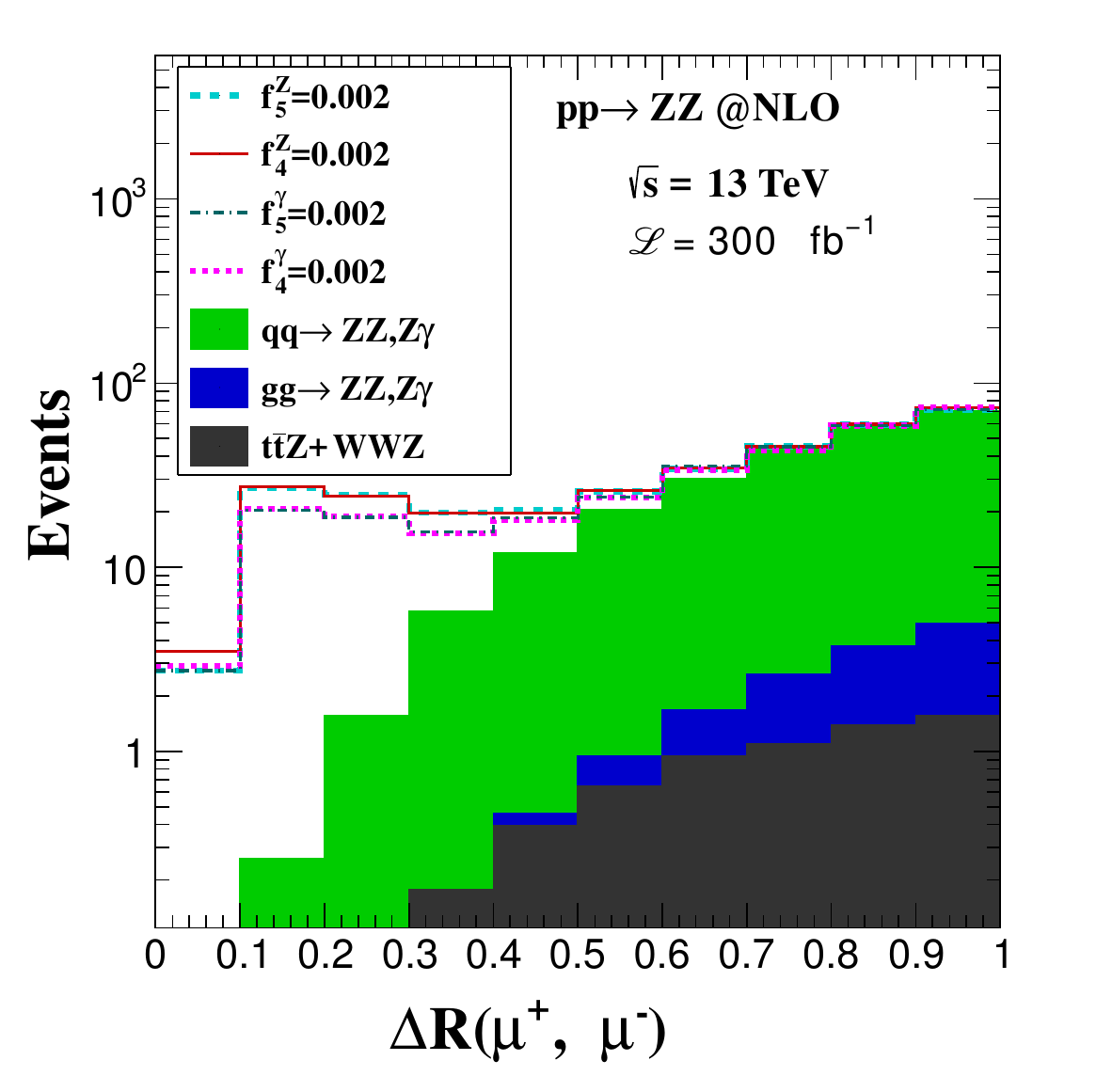}
    \caption{\label{fig:m4lDrll}
    The differential distributions of  $m_{4l}$ ({\em left-panel}) and $\Delta R(\mu^+,\mu^-)$ ({\em right-panel})  in
    the $ZZ$ production at the LHC at $\sqrt{s}=13$ TeV and  
        ${\cal L}=300$ fb$^{-1}$ at NLO in QCD. 
The {\em light-shaded} region with maximum heights (green shaded), the {\em dark-shaded} region (blue shaded)
and the {\em light-shaded} region with smallest heights (grey shaded) correspond to 
$q\bar{q}$ SM contribution,   $gg$ SM contributions and the background, respectively.
The aTGC contributions are shown with different line types (colours). }
\end{figure}
\begin{figure}[h]
    \centering
    \includegraphics[width=0.495\textwidth]{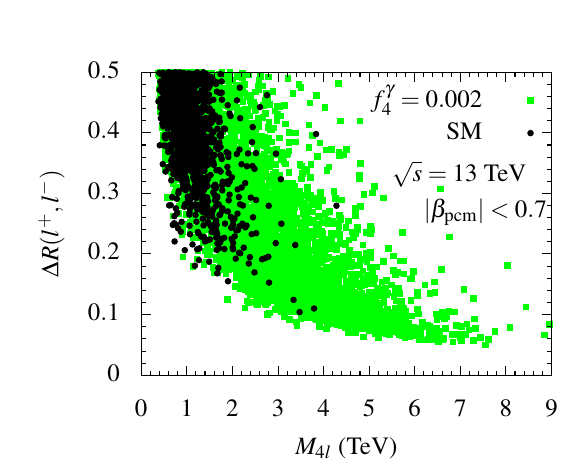}
    \includegraphics[width=0.49\textwidth]{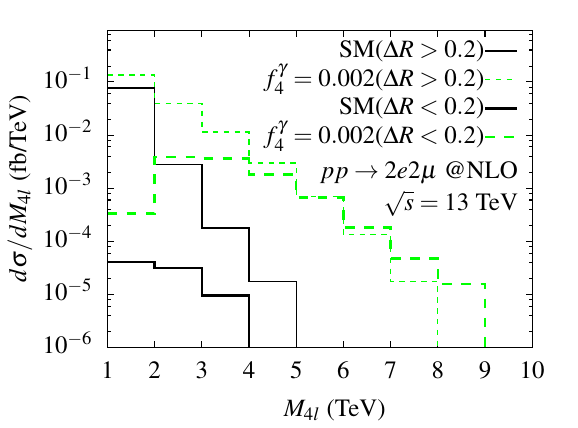}
    \caption{\label{fig:m4lvsDrr}
        $m_{4l}$ vs $\Delta R$  scattered plot ({\em left}) and $m_{4l}$ distribution for $\Delta R(l^+,l^-) \gtrless 0.2$ ({\em right}) in
        $ZZ$ production at the LHC at $\sqrt{s}=13$ TeV for the SM and for aTGC with $f_4^\gamma=0.002$.}
\end{figure}
The effect of aTGC on the observables varies with energy scale. 
We study the effect of aTGC on various observables in their distribution and determine the signal region.
In Fig.~\ref{fig:m4lDrll}  
we show four lepton invariant mass ($m_{4l}$) or  centre-of-mass energy ($\sqrt{\hat{s}}$)
distribution ({\em left-panel})  and $\Delta R$ distribution of $\mu^+\mu^-$ pair ({\em right-panel})   
at $\sqrt{s}=13$ TeV  for the  SM along with background 
$t\bar{t}Z+WWZ$ and some benchmark aTGC points for events 
normalized to luminosity  $300$ fb$^{-1}$ using {\tt \sc{MadAnalysis5}}~\cite{Conte:2012fm}.  
The   $gg$ contribution  is at it's LO (${\cal O}(\alpha_s^2)$), while all other contributions are 
shown at NLO (${\cal O}(\alpha_s)$).
The $q\bar{q} \to ZZ,~Z\gamma$ contribution is shown in {\em green band}, $gg \to ZZ,Z\gamma$ is 
in {\em blue band} and the background $t\bar{t}Z+WWZ$ contribution is shown in {\em grey band}. 
 The aTGC contribution for various choices are   shown in {\em dashed}/cyan ($f_5^Z=0.002$),   {\em solid}/red ($f_4^Z=0.002$),
  {\em dashed-dotted}/dark-green ($f_5^\gamma=0.002$)  and 
 {\em small-dashed}/magenta ($f_4^\gamma=0.002$).   
 For the $m_{4l}$ distribution in left, all events above $1$ TeV are added  in the last bin.
 All the aTGC benchmark i.e., $f_i^V=0.002$ are not visibly different 
than the SM $q\bar{q}$ contribution upto $\sqrt{\hat{s}}=0.8$ TeV and there are significant 
excess of events in the last bin, i.e., above $\sqrt{\hat{s}}=1$ TeV. 
This is due to momentum dependence~\cite{Rahaman:2016pqj} of the interaction vertex
that leads to increasing contribution at higher momentum transfer.
 In the distribution of $\Delta R (\mu^+,\mu^-)$ in the {\em right-panel},
 the effect of aTGC is higher for lower $\Delta R$  (below $0.5$). 
 In the $ZZ$ process,  the $Z$ bosons are highly boosted for larger $\sqrt{\hat{s}}$
 and their decay products are collimated leading to a smaller $\Delta R$ separation between
 the decay leptons. To see this kinematic effect, we plot events in $m_{4l}$ - $\Delta R$
 plane in Fig.~\ref{fig:m4lvsDrr} ({\em left-panel}). Here,  we choose a minimum $\Delta R$ between $e$ pair and $\mu$ pair
 event by event. We note that additional events coming from aTGC contributions have higher 
 $m_{4l}$ and lower $\Delta R$ between leptons. For $\Delta R<0.2$ most of the events
 contribute to the $m_{4l}>1$ TeV bin and they are dominantly coming from aTGC,  Fig.~\ref{fig:m4lvsDrr} ({\em right-panel}). Thus
 we can choose $m_{4l}>1$ TeV to be the signal region.

In this analysis, the set of observables consist of the cross section and polarization 
asymmetries $\wtil{A_{xz}}$, $A_{x^2-y^2}$, and $A_{zz}$.
The signal region for the cross section $\sigma$ is chosen to be $m_{4l} >1$ TeV as 
we have discussed in the previous 
section. In case of asymmetries, we choose the signal region as $m_{4l} >0.3$ TeV for $\wtil{A_{xz}}$ and 
$m_{4l} >0.7$ TeV for $A_{x^2-y^2}$ and $A_{zz}$ as the effect of aTGC is found to be best
in these region corresponding to these asymmetries. 
The expression for the cross section and the polarization asymmetries as a function of couplings are obtained by
numerical fitting the data generated by \MGvATNLO. The events are generated   for
different set of values of  the couplings $f_i^V=(f_4^\gamma,~f_4^Z,~f_5^\gamma,~f_5^Z)$ and then
various cross sections, i.e., the total cross section and the numerator of the asymmetries, ${\cal O} $, are fitted as
\begin{align}\label{eq:gen-fit}
{\cal O} = {\cal O}_0 + f_i^V \times {\cal O}_i + f_i^V\times f_j^V \times {\cal O}_{ij},     
\end{align}
in general, where ${\cal O}_0$ is the value of corresponding cross sections in the SM.
 The observables, considered here, are all $CP$-even in nature which leads to the modification of Eq.~(\ref{eq:gen-fit})
 as 
 \begin{align}\label{eq:cp-even-fit}
 {\cal O} = {\cal O}_0 + f_5^V \times {\cal O}_5^V + f_4^\gamma  f_4^Z\times {\cal O}_4^{\gamma,Z}
 + f_5^\gamma  f_5^Z\times {\cal O}_5^{\gamma,Z} + (f_i^V)^2 \times {\cal O}_i^{VV},
 \end{align}
as the $f_4^V$ are $CP$-odd, while $f_5^V$ are $CP$-even couplings reducing the unknown from $15$ 
to $9$ to be solved.
The numerical expressions of the cross section and the asymmetries as a function of the couplings
are  given in appendix~\ref{app:zzlhc-a}.
The observables are obtained up to   ${\cal O}(\Lambda^{-4})$, i.e., quadratic in dimension-$6$. 
In practice, one should consider the effect of dimension-$8$ contribution at linear order. However, we choose to work
with only dimension-$6$  in couplings with a contribution up to quadratic so as to compare the results with the  current LHC constraints on
dimension-$6$ parameters~\cite{Sirunyan:2017zjc}. 
A note on keeping terms up to quadratic in  couplings, and not terminating at linear order, is presented in appendix~\ref{app:zzlhc-b}.

\subsection{Sensitivity of observables to the  couplings}
\begin{figure}[ht]
    \centering
    \includegraphics[width=0.45\textwidth]{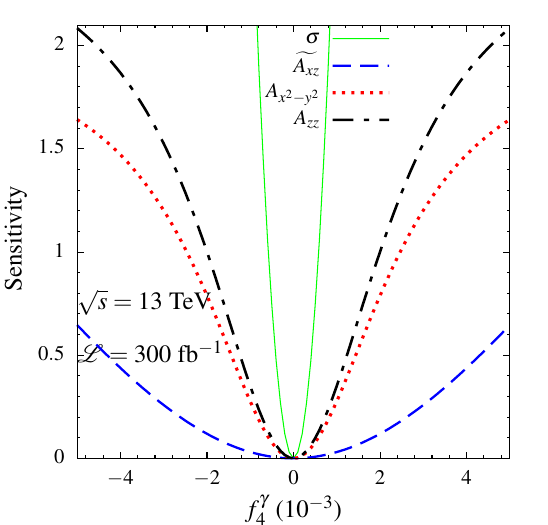}
    \includegraphics[width=0.45\textwidth]{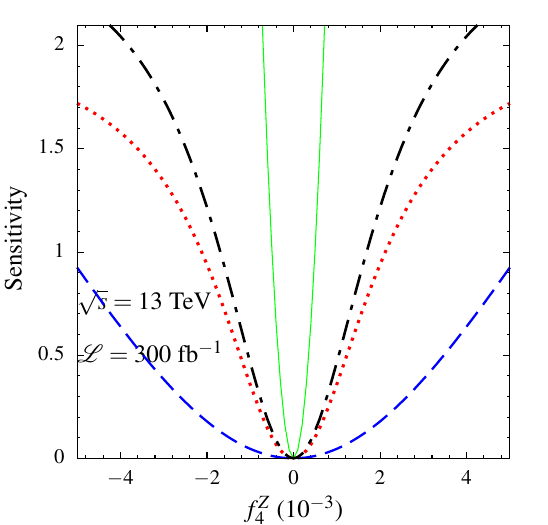}
    \includegraphics[width=0.45\textwidth]{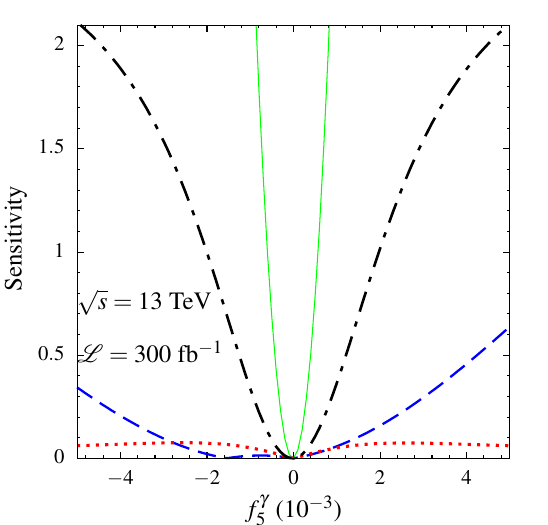}
    \includegraphics[width=0.45\textwidth]{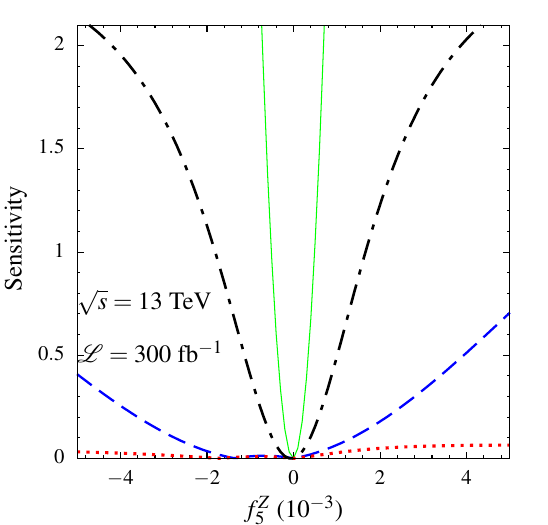}
    \caption{\label{fig:sensitivity-L300}The  sensitivity  of the cross section and the polarization observables
        to the anomalous  couplings at $\sqrt{s}=13$ TeV and  ${\cal L}=300$ fb$^{-1}$ in $ZZ$ production at the LHC. }
\end{figure}
We studied the  sensitivities (see Eq.~(\ref{eq:sensitivity}) for definition) of all the observables to the couplings and show them
in Fig.~\ref{fig:sensitivity-L300} for ${\cal L}=300$ fb$^{-1}$. 
We consider systematic uncertainties of  $\epsilon_\sigma=5~\%$~\cite{Sirunyan:2017zjc}  and 
$\epsilon_A=2~\%$  as a benchmark.
We find asymmetries to be less sensitive than the cross section to the couplings
 and thus cross section wins in putting limits on the couplings. 
 The sensitivity curves of all the couplings in each observable are symmetric about zero as $f_4^V$ (being $CP$-odd) does not appear in linear in any observables, and also the linear contribution from $f_5^V$ are negligibly small compared to their quadratic contribution (see appendix~\ref{app:zzlhc-a}). 
For example, the coefficient of $f_5^V$ are $\sim 1$ in $\sigma (m_{4l} > 1~\text{TeV})$ (Eq.~(\ref{eq:sigma-1TeV})), while the
coefficient of $(f_5^V)^2$ are $\sim 5\times 10^{4}$. Thus even at 
 $f_5^V=10^{-3}$ the quadratic contribution is $50$ times stronger than the linear one.
 Although the asymmetries are not strongly sensitive to the couplings as the cross section, they are useful in 
the measurement of the anomalous couplings, which will be discussed in the next section.
 
  It is noteworthy to mention that the sensitivity
 of $A_{x^2-y^2}$ are flat and negligible for $CP$-even couplings $f_5^V$, while they  vary   significantly
  for $CP$-odd couplings $f_4^V$. Thus the asymmetry 
 $A_{x^2-y^2}$, although a $CP$-even observables, is able to distinguish between $CP$-odd and $CP$-even interactions
 in the $ZZ$ production at the LHC.
  \begin{table}\caption{\label{tab:single-limits} One parameter limits ($10^{-3}$) at $95~\%$ C.L. on anomalous couplings
          in $ZZ$ production at the LHC at $\sqrt{s}=13$ TeV for various luminosities.}
      \renewcommand{\arraystretch}{1.50}
      \begin{tabular*}{\columnwidth}{@{\extracolsep{\fill}}lllll@{}}\hline
          param / ${\cal L}$ & $35.9$ fb$^{-1}$ & $150$ fb$^{-1}$ & $300$ fb$^{-1}$ & $1000$ fb$^{-1}$\\\hline
          $f_4^\gamma$ &$ _{-1.20}^{+1.22}$&$ _{-0.85}^{+0.85}$&$ _{-0.72 }^{+ 0.72}$&$ _{-0.55 }^{+ 0.55}$     \\\hline
          $f_5^\gamma$ &$ _{-1.23}^{+1.21}$&$ _{-0.87}^{+0.84}$&$ _{-0.74 }^{+ 0.71}$&$ _{-0.57 }^{+ 0.54}$     \\\hline
          $f_4^Z$      &$ _{-1.03}^{+1.04}$&$ _{-0.72}^{+0.73}$&$ _{-0.61 }^{+ 0.62}$&$ _{-0.47 }^{+ 0.47}$     \\\hline
          $f_5^Z$      &$ _{-1.05}^{+1.03}$&$ _{-0.74}^{+0.72}$&$ _{-0.63 }^{+ 0.61}$&$ _{-0.49 }^{+ 0.46}$     \\\hline
      \end{tabular*}
  \end{table}

We use the total $\chi^2$ as (Eq.~(\ref{eq:sensitivity}))
\begin{equation}
\chi^2(f_i)=\sum_{j} \left[ {\cal S}{\cal O}_j(f_i) \right]^2
\end{equation}
to obtain the single parameter limits on the couplings
  by varying one parameter at a time and keeping all other  to their SM values.
 The single parameter limits thus  obtained on all the anomalous couplings at $95~\%$ C.L.  
  for four benchmark luminosities ${\cal L}=35.9$ fb$^{-1}$,  
 $150$ fb$^{-1}$, $300$ fb$^{-1}$ and $1000$ fb$^{-1}$ are presented in Table~\ref{tab:single-limits}. 
 The  limit at ${\cal L}=35.9$ fb$^{-1}$ given in the first column of Table~\ref{tab:single-limits}
 are comparable to the tightest limit available at the LHC  by CMS~\cite{Heinrich:2017bvg} given in
 Eq.~(\ref{eq:CMS-limit}).  
\subsection{Simultaneous limits on the aTGC}
\begin{figure}[ht]
    \centering
    \includegraphics[width=0.48\textwidth]{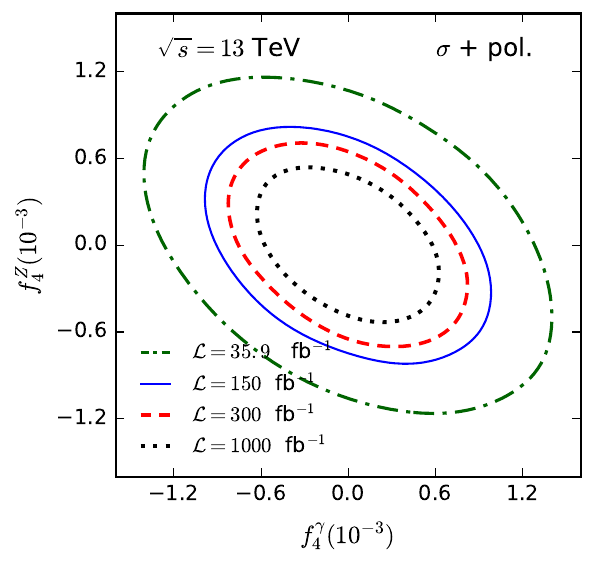}    
    \includegraphics[width=0.48\textwidth]{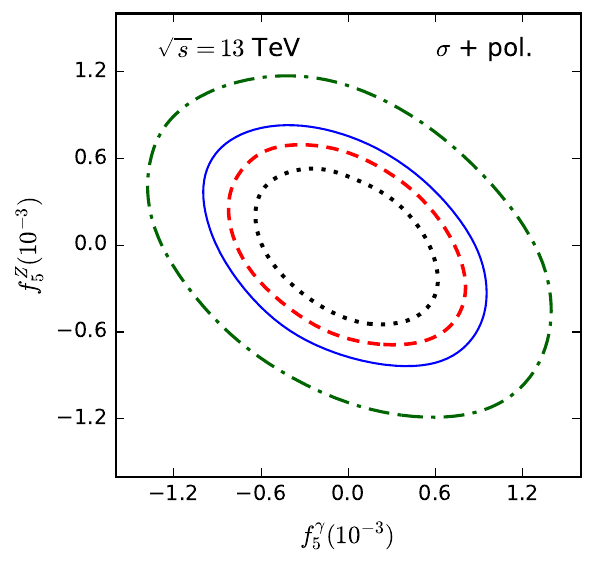}
    \caption{\label{fig:2dmcmcplot-sigwithpol} Two dimensional marginalised contours at
        $95~\%$ BCI from MCMC using the cross section $\sigma$ along with polarization asymmetries (pol.)  
        at $\sqrt{s}=13$ TeV for various luminosities in $ZZ$ production at the LHC. }
\end{figure}
\begin{table}[h]\caption{\label{tab:simul-limits} Simultaneous limits ($10^{-3}$) at $95~\%$ C.L. on anomalous couplings
in $ZZ$ production at the LHC at $\sqrt{s}=13$ TeV for various luminosities from MCMC. }
\renewcommand{\arraystretch}{1.50}
\begin{tabular*}{\columnwidth}{@{\extracolsep{\fill}}lllll@{}}\hline
param / ${\cal L}$ & $35.9$ fb$^{-1}$   & $150$ fb$^{-1}$       & $300$ fb$^{-1}$       & $1000$ fb$^{-1}$\\\hline
$f_4^\gamma$      &$ _{-1.15}^{+1.17}$  &$ _{-0.81}^{+0.81}$ &$ _{-0.68 }^{+ 0.67}$&$ _{-0.52 }^{+ 0.52}$     \\\hline
$f_5^\gamma$      &$ _{-1.13}^{+1.50}$  & $ _{-0.83}^{+0.78}$&$ _{-0.68 }^{+ 0.66}$&$ _{-0.53 }^{+ 0.51}$     \\\hline
$f_4^Z$                 &$ _{-0.96}^{+0.95}$&$ _{-0.67}^{+0.67}$ &$ _{-0.58 }^{+ 0.58}$&$ _{-0.44 }^{+ 0.45}$     \\\hline
$f_5^Z$                 &$ _{-0.98}^{+0.95}$&$ _{-0.69}^{+0.68}$ &$ _{-0.57 }^{+ 0.57}$&$ _{-0.45 }^{+ 0.43}$     \\\hline
\end{tabular*}
\end{table}
 A likelihood-based analysis using the total $\chi^2$  with the MCMC method is done by 
varying all the parameters simultaneously to extract 
simultaneous limits on all the anomalous couplings for the four benchmark luminosity chosen. 
 The two dimensional marginalised contours
 at $95~\%$ C.L.   in the  $f_4^\gamma$ -$f_4^Z$ and $f_5^\gamma$ -$f_5^Z$ planes
 are shown  in  Fig.~\ref{fig:2dmcmcplot-sigwithpol} 
  for the four benchmark luminosities chosen, using 
the  cross section together with the polarization asymmetries, i.e, using ($\sigma$ + pol.).
 The  outer most contours are for ${\cal L}=35.9$ fb$^{-1}$ and 
 the innermost contours are for ${\cal L}=1000$ fb$^{-1}$.
 The corresponding simultaneous limits on the aTGC couplings for four benchmark 
luminosities are presented in Table~\ref{tab:simul-limits}. The simultaneous limits are usually
less tight than the one-dimensional limits, but find the opposite in some case, which can
be seen comparing Table~\ref{tab:simul-limits}
with  Table~\ref{tab:single-limits}. 
The reason for this is the following. The cross section, the dominant observable, has a very little
linear dependence, while it has a large quadratic dependence on the couplings (see
 Eq.~(\ref{eq:sigma-1TeV})). As a result, when one obtains the limit on one parameter
 in the multi-parameter analysis, a slight deviation on any other parameter from zero 
 (SM point)  tightens the limit on the former  coupling.
 
\subsection{Role of the polarization asymmetries in parameter extraction}
\begin{figure}[h!]
    \centering
    \includegraphics[width=0.24\textwidth]{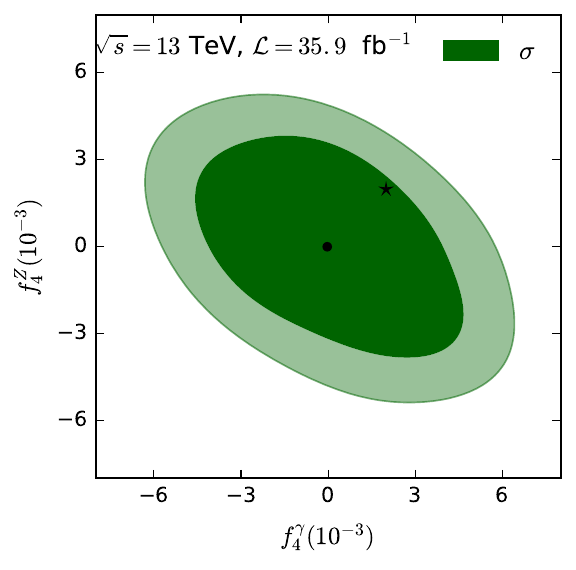}    
    \includegraphics[width=0.24\textwidth]{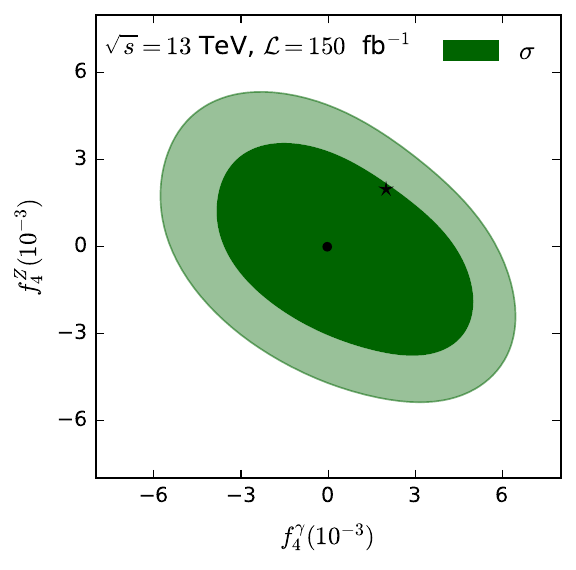}    
    \includegraphics[width=0.24\textwidth]{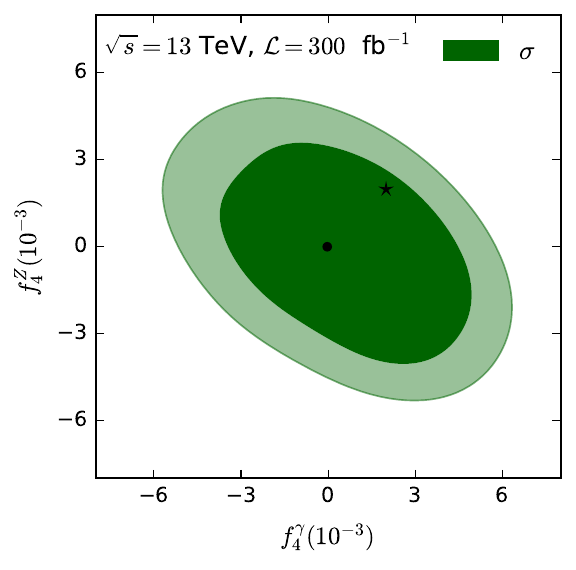}
    \includegraphics[width=0.24\textwidth]{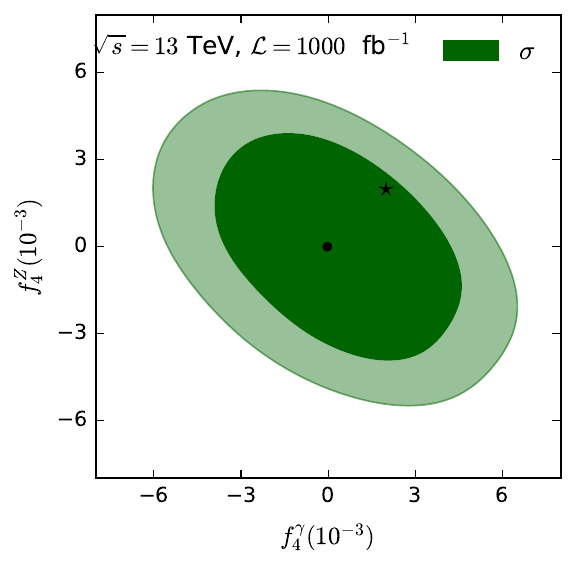}
    \includegraphics[width=0.24\textwidth]{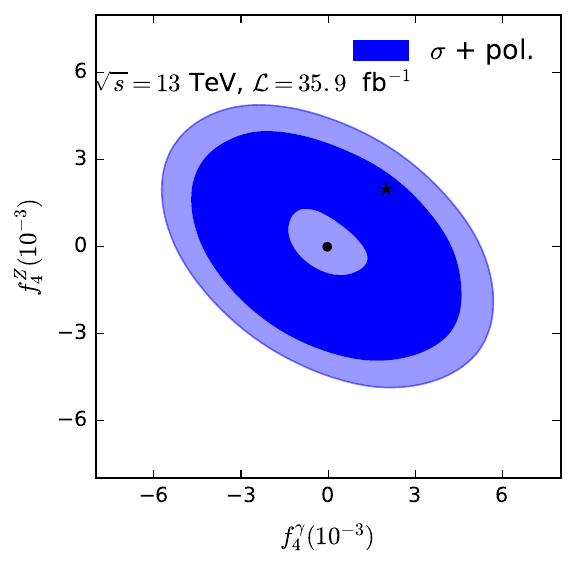}    
    \includegraphics[width=0.24\textwidth]{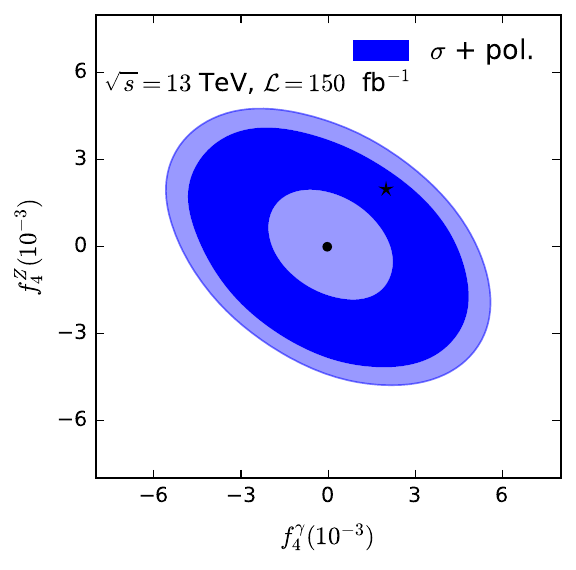}    
    \includegraphics[width=0.24\textwidth]{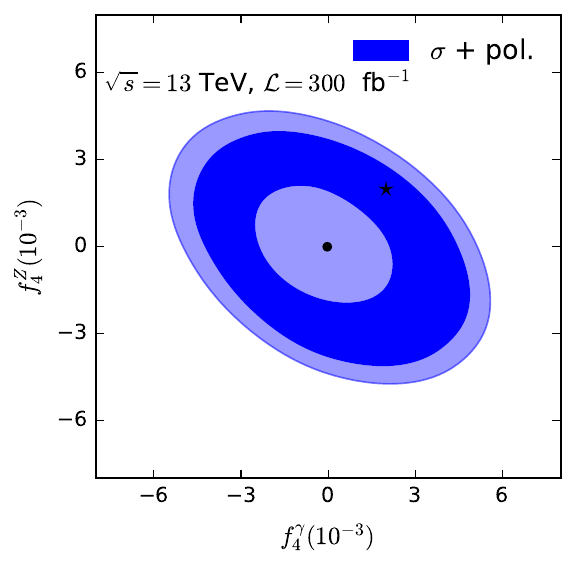}
    \includegraphics[width=0.24\textwidth]{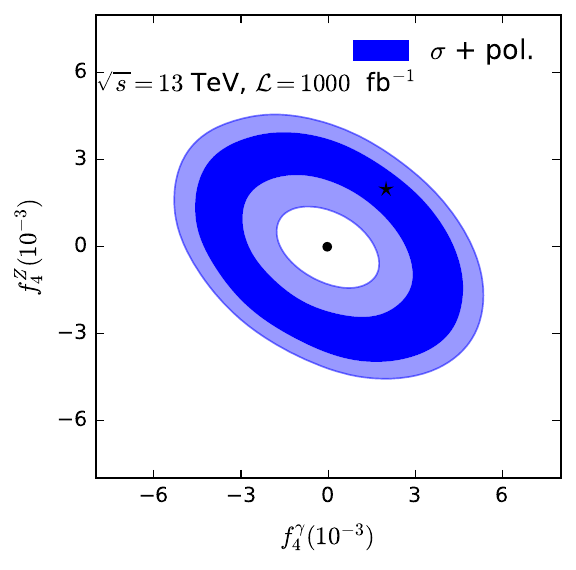}
    \includegraphics[width=0.24\textwidth]{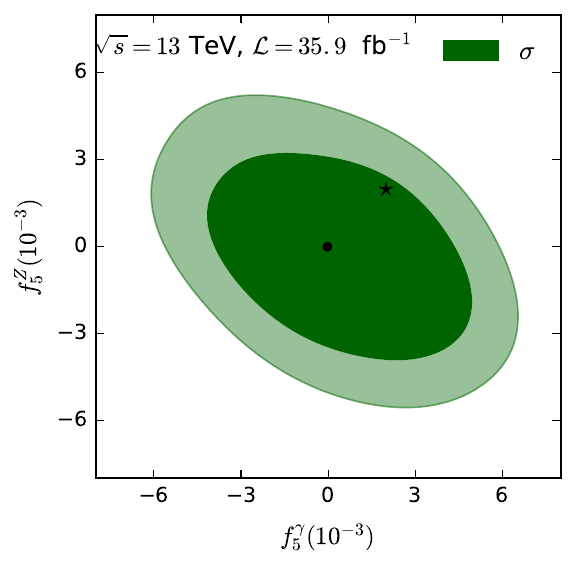}    
    \includegraphics[width=0.24\textwidth]{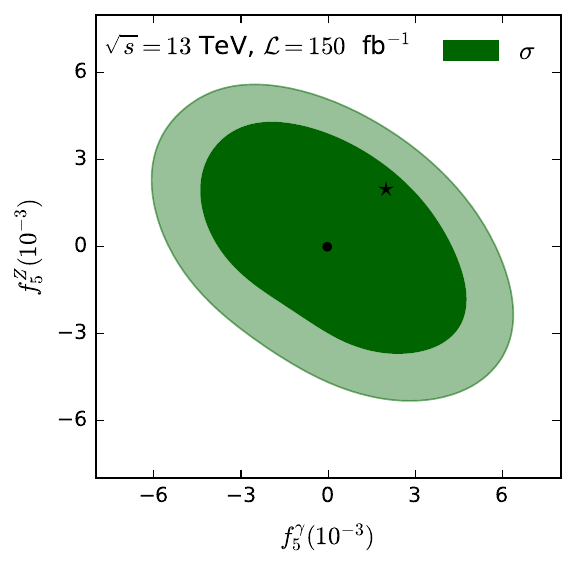}    
    \includegraphics[width=0.24\textwidth]{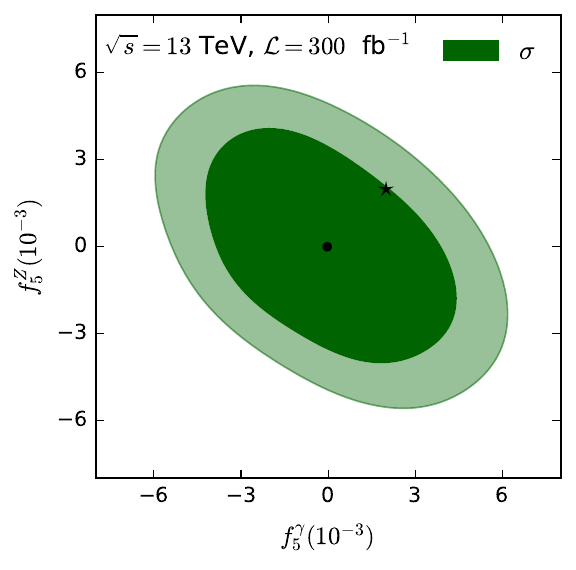}
    \includegraphics[width=0.24\textwidth]{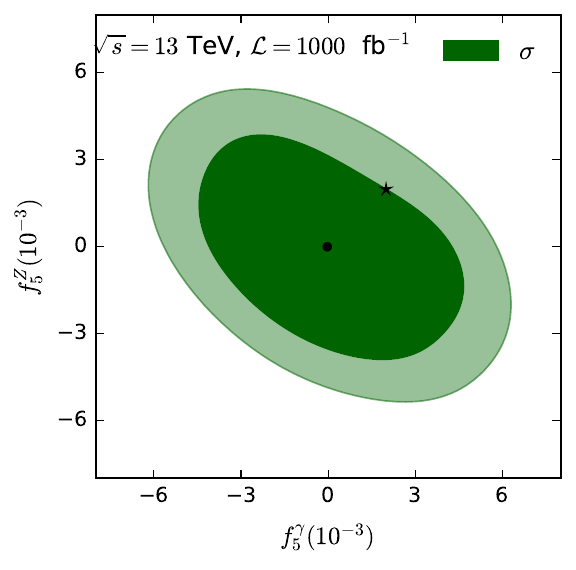}
    \includegraphics[width=0.24\textwidth]{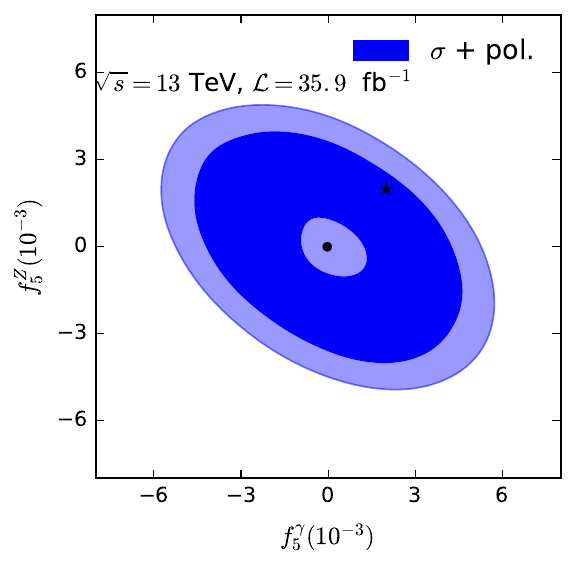}    
    \includegraphics[width=0.24\textwidth]{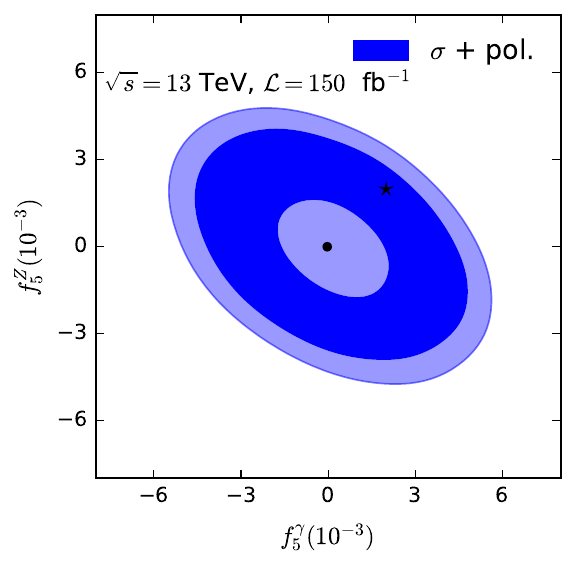}    
    \includegraphics[width=0.24\textwidth]{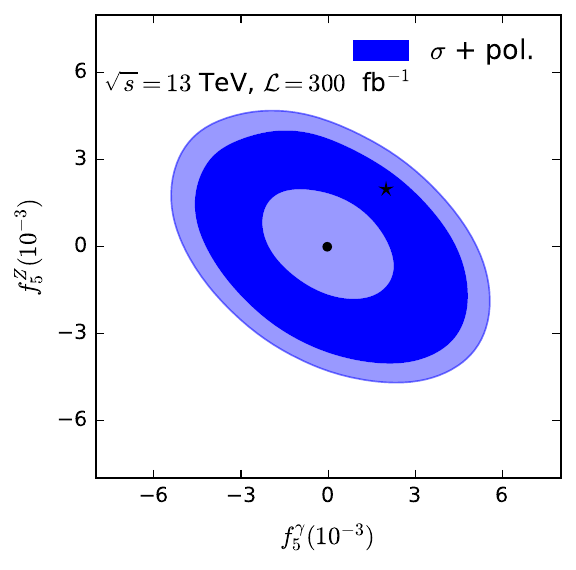}
    \includegraphics[width=0.24\textwidth]{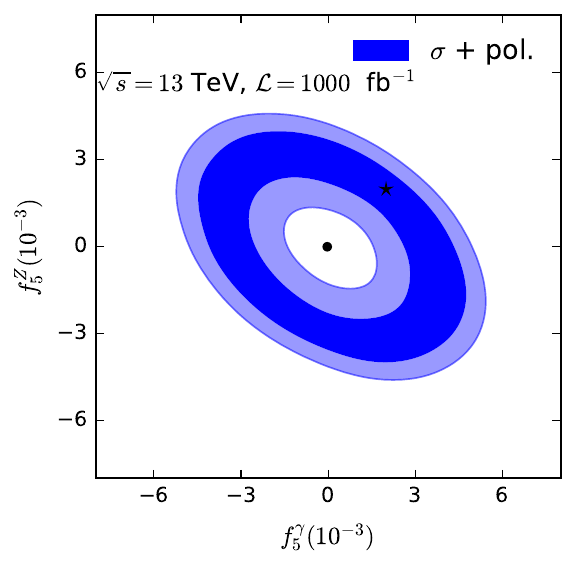}
    \caption{\label{fig:2dmcmcplot-bench2Lall} Comparison of $\sigma$ vs  
        ($\sigma$ + pol.)  in two dimensional marginalised contours from MCMC for aTGC 
        benchmark $f_i^V=0.002$ in $f_4^\gamma$-$f_4^Z$ panel and $f_5^\gamma$-$f_5^Z$ panel   
        at $\sqrt{s}=13$ TeV for various luminosities in $ZZ$ production at the LHC. }
\end{figure}
The inclusion of polarization asymmetries with  the
cross section has no significant effect in constraining the anomalous couplings. 
The asymmetries  may still be useful  in extracting 
parameters if excess events were found at the LHC. 
To explore 
this, we do a toy analysis of parameter extraction using 
the data for all aTGC couplings $f_i^V=0.002$ (well above current limit)
and  use the MCMC method to extract back these parameters. 
We deliberately choose the benchmark couplings with large values so as to emulate 
a situation where a deviation from the SM is observed. 
In Fig.~\ref{fig:2dmcmcplot-bench2Lall}, we show  two-dimensional marginalized contours
for the four benchmark luminosities for the benchmark aTGC couplings 
in $f_4^\gamma$-$f_4^Z$ and $f_5^\gamma$-$f_Z^Z$ planes for the set of observables
$\sigma$ and ($\sigma$ + pol.) for comparison.
The {\em darker-shaded} regions are for $68~\%$ C.L., while {\em lighter-shaded} regions are for
$95~\%$ C.L. The dot ({\tiny $\bullet$}) and the star ($\star$) mark  in the plot are for the SM ($0,0$) and 
aTGC benchmark ($0.002,0.002$) points, respectively.
We note that the SM point is inside the $68~\%$ C.L. contours even at a high luminosity of 
${\cal L}=1000$ fb$^{-1}$  if we use only cross section as observable, see  row-$1$ and $3$
of Fig.~\ref{fig:2dmcmcplot-bench2Lall}. The distinction between the SM and the aTGC get improved when polarization asymmetries are included, 
i.e., the SM point is outside the $95~\%$ C.L. contour for luminosity of much less than  ${\cal L}=1000$ fb$^{-1}$, see row-$2$ and $4$
of the figure.
As the luminosity increases, from the left column to the right, the contours for ($\sigma$ + pol.) shrink around
the star ($\star$) mark maintaining the shape of a ring giving better exclusion of the SM  from aTGC benchmark. 
Polarization asymmetries are thus useful in the measurement of the anomalous couplings 
if excess events are found at the LHC.

\section{Summary}\label{sect:conclusion}
In summary, we studied anomalous triple gauge boson couplings
in the neutral sector in $ZZ$ pair production at the LHC and investigated
the role of $Z$ boson polarizations in this chapter. The QCD corrections in this process
are very high and can  not be ignored. We obtained the cross section and the asymmetries
at higher order in QCD. The aTGC contributes more in the higher $\sqrt{\hat{s}}$
region as they are momentum dependent. The major background $t\bar{t}Z+WWZ$  are
negligibly small, and they vanish in the signal regions. Although the asymmetries
are not as sensitive as the cross section to the couplings, the asymmetry
$A_{x^2-y^2}$ is able to distinguish between $CP$-even and $CP$-odd couplings.
We estimated the one parameter as well as simultaneous limits on the couplings using all 
the observables   based on the total $\chi^2$ for luminosities $35.9$ fb$^{-1}$, $150$ 
fb$^{-1}$, $300$ fb$^{-1}$ and $1000$ fb$^{-1}$. Our one parameter limits are comparable to the best 
available limits obtained at the LHC~\cite{Sirunyan:2017zjc}.
The asymmetries are instrumental in extracting the parameters
should a deviation from the SM is observed at the LHC. We 
performed a toy analysis of parameter extraction
with a benchmark aTGC coupling point with $f_i^V=0.002$ and found that the polarizations observables along with the cross
section can exclude the SM from the aTGC point better than the cross section can do alone.
In this work, the observables for the aTGC are obtained at ${\cal O }(\alpha_s)$, 
while they are obtained in the next order in the SM. 
The NNLO result in aTGC, when available, is expected to improve the limits on 
the couplings.

\chapter{The probe of aTGC in  $e^+e^-\to W^+W^-$ and the role of $W$ boson  polarizations along with beam polarizations}\label{chap:eeWW}
\begingroup
\hypersetup{linkcolor=blue}
\minitoc
\endgroup
{\small\textit{\textbf{ The contents in this chapter are based on the published article and preprints in Refs.~\cite{Rahaman:2017aab,Rahaman:2019mnz}. }}}
\vspace{1cm}

The non-abelian gauge symmetry $SU(2)\times U(1)$ of the Standard Model 
allows the $WWV$ ($V=\gamma, Z$) couplings after the electroweak symmetry breaking by the Higgs field. 
To test the SM $WWV$ couplings, 
one has to hypothesize BSM couplings  and make sure they do not
appear at all, or they are severely constrained. There are two approaches to 
study BSM $WWV$ couplings; one is effective operator, approach another is 
effective form factor approach, as discussed in section~\ref{sec:intro-EFT}.
 In the EFT approach, 
the dimension-$6$ operators  contributing to  $WWV$ couplings are~\cite{Hagiwara:1993ck,Degrande:2012wf} 
\begin{eqnarray}\label{eq:opertaors-dim6}
    {\cal O}_{WWW}&=&\mbox{Tr}[W_{\mu\nu}W^{\nu\rho}W_{\rho}^{\mu}] ,\nonumber\\
    {\cal O}_W&=&(\D_\mu\Phi)^\dagger W^{\mu\nu}(\D_\nu\Phi) ,\nonumber\\
    {\cal O}_B&=&(\D_\mu\Phi)^\dagger B^{\mu\nu}(\D_\nu\Phi) ,\nonumber\\
    {\cal O}_{\wtil{WWW}}&=&\mbox{Tr}[\wtil{W}_{\mu\nu}W^{\nu\rho}W_{\rho}^{\mu}] ,\nonumber\\
    {\cal O}_{\wtil W}&=&(\D_\mu\Phi)^\dagger \wtil{ W}^{\mu\nu}(\D_\nu\Phi) ,
\end{eqnarray}
which respect the SM gauge symmetry. 
Among these operators, ${\cal O}_{WWW}$,  ${\cal O}_W$ and  ${\cal O}_B$ are
$CP$-even, while ${\cal O}_{\wtil{WWW}}$  and ${\cal O}_{\wtil W}$ are
$CP$-odd. 
These effective operators, 
after EWSB, also provides $ZZV$, $HZV$ couplings which can be examined in various 
processes, e.g. $ZV$ production, $WZ$ production, $HV$ production processes. The couplings
in these processes may contain some other effective operator  as well.

In the form factor approach, the 
most general  Lagrangian for the $WWV$ couplings is given by~\cite{Hagiwara:1986vm},
\begin{eqnarray}
    {\cal L}_{WWV} &=&ig_{WWV}\left(g_1^V(W_{\mu\nu}^+W^{-\mu}-
    W^{+\mu}W_{\mu\nu}^-)V^\nu
    +ig_4^VW_\mu^+W^-_\nu(\partial^\mu V^\nu+\partial^\nu V^\mu)\right.\nonumber\\
    &-&ig_5^V\epsilon^{\mu\nu\rho\sigma}(W_\mu^+\partial_\rho W^-_
    \nu-\partial_\rho W_\mu^+W^-_\nu)V_\sigma
    +\frac{\lambda^V}{m_W^2}W_\mu^{+\nu}W_\nu^{-\rho}V_\rho^{\mu}\nonumber\\
    &+&\left.\frac{\wtil{\lambda^V}}{m_W^2}W_\mu^{+\nu}W_\nu^{-\rho}\wtil{V}_\rho^{\mu}
    +\kappa^V W_\mu^+W_\nu^-V^{\mu\nu}+\wtil{\kappa^V}W_\mu^+W_\nu^-\wtil{V}^{\mu\nu}
    \right) .
    \label{eq:WW-LagWWV}
\end{eqnarray}
Here $W_{\mu\nu}^\pm = \partial_\mu W_\nu^\pm - 
\partial_\nu W_\mu^\pm$, $V_{\mu\nu} = \partial_\mu V_\nu - 
\partial_\nu V_\mu$, 
$\wtil{V}^{\mu\nu}=1/2\epsilon^{\mu\nu\rho\sigma}V_{\rho\sigma}$,
and the overall coupling constants are defined as
$g_{WW\gamma}=-g\sin\theta_W$ and $g_{WWZ}=-g\cos\theta_W$, $\theta_W$ being the weak
mixing angle. In the SM
$g_1^V=1$, $\kappa^V=1$ and other couplings are zero. The anomalous part in $g_1^V$,
$\kappa^V$ would be $\Delta g_1^V=g_1^V-1$, $\Delta\kappa^V=\kappa^V-1$, respectively. 
The couplings $g_1^V$, $\kappa^V$
and $\lambda^V$  are $CP$-even (both $C$ and $P$-even), 
while $g_4^V$ (odd in $C$, even in $P$), $\wtil{\kappa^V}$
and $\wtil{\lambda^V}$  (even in $C$, odd in $P$) are $CP$-odd. On the other hand 
$g_5^V$ is both $C$ and $P$-odd making it  $CP$-even. 
We note that the coupling ($c_i^{\cal L}$) of the Lagrangian
in Eq.~(\ref{eq:WW-LagWWV}) are related to  the couplings of the operators
in Eq.~(\ref{eq:opertaors-dim6}) through the relations given in Eq.~(\ref{eq:intro-Operator-to-Lagrangian})  
when $SU(2)\times U(1)$ gauge invariance is assumed.

For convenience, we label the anomalous couplings of the three scenarios  as follows:
The couplings of the operators in Eq.~(\ref{eq:opertaors-dim6}), the couplings of effective vertices
in ${\cal L}_{WWV}$ in Eq.~(\ref{eq:WW-LagWWV}) and the vertex couplings translated from the 
operators in Eq.~(\ref{eq:intro-Operator-to-Lagrangian}) are labelled as $c_i^{\cal O}$, $c_i^{\cal L}$,
and $c_i^{{\cal L}_g}$, respectively.   The couplings in the three scenarios are thus,
\begin{eqnarray}
c_i^{\cal O}&=&\{ c_{WWW}, c_{W}, c_B, c_{\wtil{WWW}}, c_{\wtil{W}} \}\label{eq:ciO} ,\\
c_i^{\cal L}&=&\{ \Delta g_1^V,g_4^V,g_5^V,\lambda^V,\wtil{\lambda^V},\Delta\kappa^V, \wtil{\kappa^V} \},~~~ V=\gamma,Z \label{eq:ciL},\\
c_i^{{\cal L}_g}&=& \{ \lambda^V, \wtil{\lambda^V}, 
\Delta\kappa^\gamma, \wtil{\kappa^\gamma}, \Delta g_1^Z,  \Delta\kappa^Z, \wtil{\kappa^Z} \}\label{eq:ciLg} .
\end{eqnarray}

In the theoretical side, 
these anomalous gauge boson self couplings may be obtained from some high scale new 
physics such as MSSM~\cite{Lahanas:1994dv,Arhrib:1996rj,Argyres:1995ib}, extra dimension~\cite{FloresTlalpa:2010rm,Lopez-Osorio:2013xka}, Georgi-Machacek model~\cite{Arroyo-Urena:2016gjt}, etc.
by integrating out the heavy degrees of freedom.
Some of these couplings can also be obtained at loop level within the SM~\cite{Argyres:1992vv,Papavassiliou:1993ex}. 

There has been a lot of studies to probe the 
anomalous $WWZ/\gamma$ couplings  in the effective operators  method as well
as in the effective vertex factor approach  subjected to  $SU(2)\times U(1)$ invariance for various colliders:  for $e^+$-$e^-$ linear 
collider~\cite{Gaemers:1978hg,Hagiwara:1986vm,Bilchak:1984ur,Hagiwara:1992eh,
    Wells:2015eba,Buchalla:2013wpa,Zhang:2016zsp,Berthier:2016tkq,Bian:2015zha,Bian:2016umx,
    Choudhury:1996ni,Choudhury:1999fz,Rahaman:2019mnz}, for Large Hadron electron collider (LHeC)
~\cite{Biswal:2014oaa,Cakir:2014swa,Li:2017kfk}, $e$-$\gamma$ collider~\cite{Kumar:2015lna} and for
LHC~\cite{Baur:1987mt,Dixon:1999di,Falkowski:2016cxu,Azatov:2017kzw,Azatov:2019xxn,Bian:2015zha,Campanario:2016jbu,Bian:2016umx,Butter:2016cvz,Baglio:2017bfe,Li:2017esm,Bhatia:2018ndx,Chiesa:2018lcs}. 
Some $CP$-odd $WWV$ couplings have been studied in Refs.~\cite{Choudhury:1999fz,Li:2017esm}. 
Direct measurement of these charged aTGC have been performed  at the LEP~\cite{Abbiendi:2000ei,Abbiendi:2003mk,
    Abdallah:2008sf,Schael:2013ita}, Tevatron~\cite{Aaltonen:2007sd,Abazov:2012ze},  LHC~\cite{Aaboud:2017cgf,
    Sirunyan:2017bey,Aaboud:2017fye,Khachatryan:2016poo,
    Aad:2016ett,Aad:2016wpd,Chatrchyan:2013yaa,
    RebelloTeles:2013kdy,ATLAS:2012mec,Chatrchyan:2012bd,Aad:2013izg,
    Chatrchyan:2013fya,Sirunyan:2017jej,Sirunyan:2019gkh,Sirunyan:2019dyi,Sirunyan:2019bez} and Tevatron-LHC~\cite{Corbett:2013pja}.
The  tightest one parameter limit
obtained on the  anomalous couplings from   experiments are given in 
Table~\ref{tab:aTGC_constrain_form_collider}. 
The tightest limits on operator couplings ($c_i^{\cal O}$) are obtained in 
Ref.~\cite{Sirunyan:2019gkh} for $CP$-even
ones and in Ref.~\cite{Aaboud:2017fye} for $CP$-odd ones. 
These limits translated to  $c_i^{{\cal L}_g}$ using Eq.~(\ref{eq:intro-Operator-to-Lagrangian}) are also given in
Table~\ref{tab:aTGC_constrain_form_collider}. The tightest limits on the  couplings $g_4^Z$ and 
$g_5^Z$ are obtained  in Ref.~\cite{Abdallah:2008sf,Abbiendi:2003mk} considering the 
Lagrangian in Eq.~(\ref{eq:WW-LagWWV}).
\begin{table}[!ht]
    \centering
    \caption{\label{tab:aTGC_constrain_form_collider} The list of tightest limits obtained on the
        anomalous couplings of dimension-$6$ operators in Eq.~(\ref{eq:opertaors-dim6}) and
   effective vertices in Eq.~(\ref{eq:WW-LagWWV}) in $SU(2)\times U(1)$ gauge (except $g_4^Z$ and $g_5^Z$) at $95\%$ C.L.  from experiments.}
    \renewcommand{\arraystretch}{1.5}
    \begin{tabular*}{\textwidth}{@{\extracolsep{\fill}}lll@{}}\hline
        $c_i^{\cal O}$            & Limits (TeV$^{-2}$)   & Remark\\\hline 
        $\frac{c_{WWW}}{\Lambda^2}$                    & $[-1.58,+1.59]$ &CMS $\sqrt{s}=13$ TeV, ${\cal L}=35.9$ fb$^{-1}$, $SU(2)\times U(1)$~\cite{Sirunyan:2019gkh} \\
        $\frac{c_{W}}{\Lambda^2}$                     & $[-2.00,+2.65]$ &CMS~\cite{Sirunyan:2019gkh} \\
        $\frac{c_{B}}{\Lambda^2}$                    & $[-8.78,+8.54]$ &CMS~\cite{Sirunyan:2019gkh} \\    
        $ \frac{c_{\widetilde{WWW}}}{\Lambda^2}$    &$[-11,+11]$  &ATLAS $\sqrt{s}=7(8)$ TeV, ${\cal L}=4.7(20.2)$ fb$^{-1}$ ~\cite{Aaboud:2017fye}\\
        $ \frac{c_{\widetilde{W}}}{\Lambda^2}$      &$[-580,580]$  &ATLAS~\cite{Aaboud:2017fye} \\
        \hline
        $c_i^{{\cal L}_g}$ & Limits ($\times 10^{-2}$) & Remark\\ \hline
        $\lambda^V$ &  $[-0.65,+0.66]$ &CMS~\cite{Sirunyan:2019gkh}\\
        $\Delta\kappa^\gamma$ &$[-4.4,+6.3]$&CMS $\sqrt{s}=8$ TeV, ${\cal L}=19$ fb$^{-1}$, $SU(2)\times U(1)$~\cite{Sirunyan:2017bey}\\
        $\Delta g_1^Z$ & $[-0.61,+0.74]$ &  CMS~\cite{Sirunyan:2019gkh}\\
        $\Delta\kappa^Z$ & $[-0.79,+0.82]$ &CMS~\cite{Sirunyan:2019gkh}\\
        $\wtil{\lambda^V}$ & $[-4.7,+4.6]$ &ATLAS~\cite{Aaboud:2017fye}\\
        $\wtil{\kappa^Z}$  & $[-14,-1]$ & DELPHI (LEP2), $\sqrt{s}=189$-$209$ GeV, ${\cal L}=520$ pb$^{-1}$~\cite{Abdallah:2008sf}\\
        \hline
        $c_i^{{\cal L}}$  & Limits ($\times 10^{-2}$) & Remark\\ \hline
        $g_4^Z$ & $[-59,-20]$ &DELPHI~\cite{Abdallah:2008sf}\\ 
        $g_5^Z$  & $[-16,+9.0]$ &OPAL (LEP), $\sqrt{s}=183$-$209$ GeV, ${\cal L}=680$ pb$^{-1}$~\cite{Abbiendi:2003mk} \\
        \hline
    \end{tabular*}
\end{table}

The  $W^+W^-$ production is one of the important processes to be studied
at the ILC~\cite{Djouadi:2007ik,Baer:2013cma,
    Behnke:2013xla} for precision test~\cite{MoortgatPick:2005cw} as well as 
for BSM physics. This process has been studied earlier for SM phenomenology
as well as for various BSM physics with and without beam 
polarization~\cite{Hagiwara:1986vm,Gounaris:1992kp,Ananthanarayan:2009dw,Ananthanarayan:2010bt,Ananthanarayan:2011ga,Andreev:2012cj}.
Here we intend to study $WWV$ anomalous couplings in $e^+e^-\to W^+W^-$ at $\sqrt{s}=500$ GeV  and integrated luminosity of 
${\cal L}=100$ fb$^{-1}$ using the cross section, forward-backward asymmetry and $8$ 
polarizations asymmetries  of $W^-$ for a set of choices of longitudinally  polarized $e^+$ and $e^-$ beams
in the channel $W^- \to l^- \bar{\nu_l}$ ($l=e,\mu$)\footnote{For simplicity we do not include tau decay mode as the tau decays to neutrino within the beam pipe 
    giving extra missing momenta affecting the reconstruction of the events.} and $W^+\to hadrons$.
The polarization of $Z$ and $W$ are being used widely recently for various BSM studies~\cite{Renard:2018tae,Renard:2018bsp,Renard:2018lqv,Renard:2018jxe,Renard:2018blr,Aguilar-Saavedra:2017zkn,Behera:2018ryv} along with studies with anomalous gauge boson couplings~\cite{Rahaman:2016pqj,Rahaman:2017qql,Abbiendi:2000ei,Rahaman:2018ujg}. Recently the polarizations of $W/Z$ has been measured in $WZ$ production at the LHC~\cite{Aaboud:2019gxl}.
Besides the final state polarizations, the initial state beam polarizations at the ILC can be used to enhance the relevant signal to background ratio~\cite{MoortgatPick:2005cw,Andreev:2012cj,Ananthanarayan:2010bt,Osland:2009dp,Pankov:2005kd}. It also has the ability to distinguish between $CP$-even and  $CP$-odd couplings~\cite{MoortgatPick:2005cw,
    Kittel:2011rk,Dreiner:2010ib,Bartl:2007qy,Rao:2006hn,Bartl:2005uh,Czyz:1988yt,
    Choudhury:1994nt,Ananthanarayan:2004eb,Ananthanarayan:2011fr,Ananthanarayan:2003wi}.
 We note that  an 
$e^+e^-$ machine will run with longitudinal beam polarizations switching between $(\eta_3,\xi_3)$ and $(-\eta_3,-\xi_3)$~\cite{MoortgatPick:2005cw},
where $\eta_3(\xi_3$) is the longitudinal polarization of $e^-$ ( $e^+$).
For integrated luminosity of  $100$ fb$^{-1}$, one will have half the luminosity available for each polarization
configurations. The most common observables, the cross section for example, 
studied in literature with beam polarizations are the total cross section
\begin{equation}\label{eq:sigma_T}
\sigma_T(\eta_3,\xi_3)=\sigma(+\eta_3,+\xi_3)+\sigma(-\eta_3,-\xi_3)
\end{equation}
and the difference
\begin{equation}\label{eq:sigma_A}
\sigma_A(\eta_3,\xi_3)=\sigma(+\eta_3,+\xi_3)-\sigma(-\eta_3,-\xi_3) .
\end{equation}
We find that  combining the two opposite beam polarizations at the level of $\chi^2$ rather than 
combining them as  in Eq.~(\ref{eq:sigma_T}) \&~(\ref{eq:sigma_A}), we can constrain the anomalous 
couplings better  in this analysis, see Sect.~\ref{sec:WW-beampolEffect} for explanation. 

We note that there exist $64$ polarization correlations~\cite{Hagiwara:1986vm} 
apart from $8+8$ polarizations for $W^+$ and $W^-$. The measurement of these 
correlations require identification of light quark flavours in the
above channel, which is not possible, hence  we are not including polarization correlations in our analysis.
In the case of both the $W$s decaying leptonicaly, there are two  missing neutrinos and  
reconstruction of polarization observables suffers combinatorial ambiguity. 
Here we aim to work with a set of observables that can be reconstructed 
uniquely and test their ability to probe the anomalous couplings including partial contribution
up to ${\cal O}(\Lambda^{-4})$\footnote[2]{ We calculate cross section
    up to ${\cal O}( 
    \Lambda^{-4})$, i.e., quadratic in dimension-$6$ (as linear approximation is not valid, e.g., see  appendix~\ref{app:zzlhc-b})  and linear in dimension-$8$ couplings choosing dimension-$8$ couplings to be zero to
    compare our result with current LHC constraints on
    dimension-$6$ parameters~\cite{Sirunyan:2019gkh,Aaboud:2017fye}.}.

\section{Observables and effect of beam polarizations}\label{sec:2}
\begin{figure}[!h]
    \centering
    \includegraphics[width=0.8\textwidth]{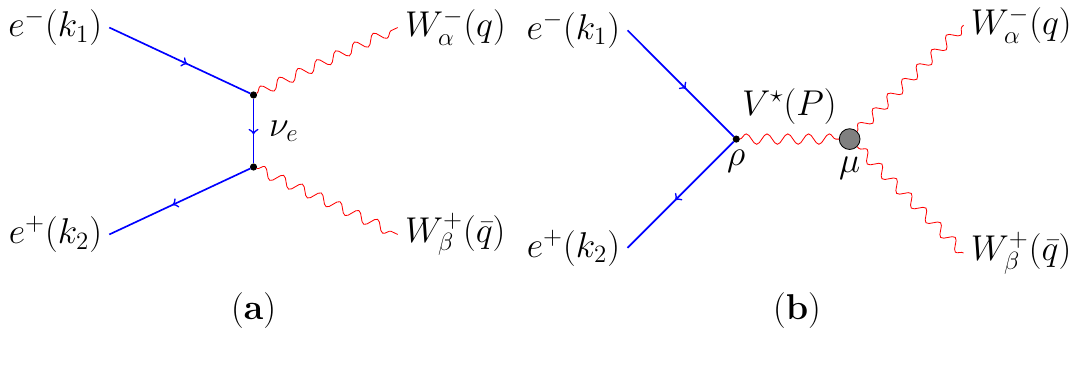}
    \caption{\label{fig:Feynman-ee-ww}Feynman diagrams of $e^+e^-\to W^+W^-$,
        (a) $t$-channel and (b) $s$-channel with anomalous $W^+W^-V$ 
        ($V=\gamma,Z$) vertex contribution shown by the shaded blob.} 
\end{figure}
We study $W^+W^-$ production at ILC running at $\sqrt{s}=500$ GeV and 
integrated luminosity ${\cal L}=100$ fb$^{-1}$ using longitudinal polarization  
of $e^-$ and $e^+$ beams giving $50$ fb$^{-1}$ to each choice of beam polarization.
The Feynman diagrams for the process are  shown in 
Fig.~\ref{fig:Feynman-ee-ww} where Fig.~\ref{fig:Feynman-ee-ww}(\textbf{a}) corresponds
to the $\nu_e$ mediated  $t$-channel diagram and the 
Fig.~\ref{fig:Feynman-ee-ww}(\textbf{b}) corresponds to the $V~(Z/\gamma)$ mediated $s$-channel 
diagram containing the aTGC contributions represented by the shaded blob. The decay mode is chosen to be
\begin{equation}
W^+\to q_u~\bar{q}_d \ \   ,~~~~~  W^-\to l^-~\bar{\nu}_l , ~~~l=e,\mu ,
\end{equation}
where $q_u$ and $q_d$ are up-type and down-type quarks, respectively.
We use complete set of eight spin-$1$  observables of $W^-$ boson (see chapter~\ref{chap:polarization})
along with the production rate.
Owing to the $t$-channel process (Fig.~\ref{fig:Feynman-ee-ww}\textbf{a}) and absence of a $u$-channel 
process,  like in $ZV$ productions in chapters~\ref{chap:epjc1} \&~\ref{chap:epjc2},
the $W^\pm$ produced are not 
forward-backward symmetric. We include the forward-backward asymmetry of the $W^-$, defined as 
\begin{equation}
A_{fb}=\frac{1}{\sigma_{W^+W^-}}\Bigg[\int_0^1 \frac{d\sigma_{W^+W^-}}{d\cos\theta_{W^-}} 
-\int_{-1}^0 \frac{d\sigma_{W^+W^-}}{d\cos\theta_{W^-}}    \Bigg] ,
\end{equation}
 to the set of observables making a total of ten observables including
the cross section as well. Here $\theta_{W^-}$ is the production angle of 
the $W^-$ w.r.t. the $e^-$ beam direction and $\sigma_{W^+W^-}$ is the production cross 
section. 
The  asymmetries of the $W^-$  can be measured in a real collider
from the final state lepton $l^-$. One has to calculate the asymmetries 
in the rest frame of $W^-$ which require the missing $\bar{\nu_l}$ momenta to be
reconstructed. At an $e^+$ $e^-$ collider, as studied here, reconstructing the missing 
$\bar{\nu_l}$ is possible because only one missing particle is involved and
no PDFs
are involved, i.e., initial momentums are known. But 
for a collider where PDFs are involved, reconstructing the actual missing momenta
may not be  possible.

\begin{figure}[ht!]
    \centering
    \includegraphics[width=0.485\textwidth]{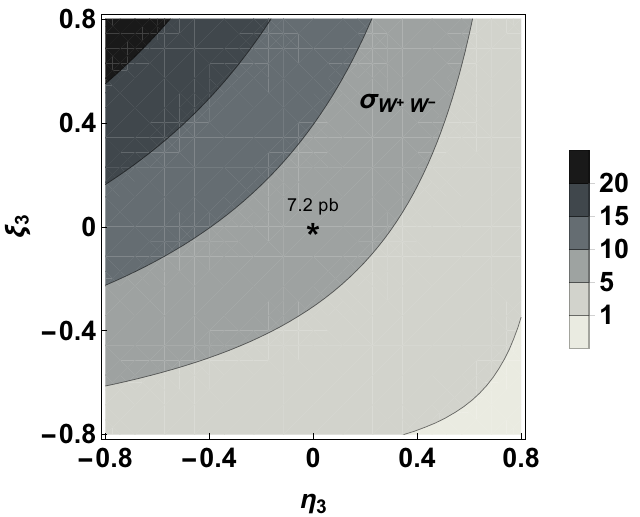}
    \includegraphics[width=0.506\textwidth]{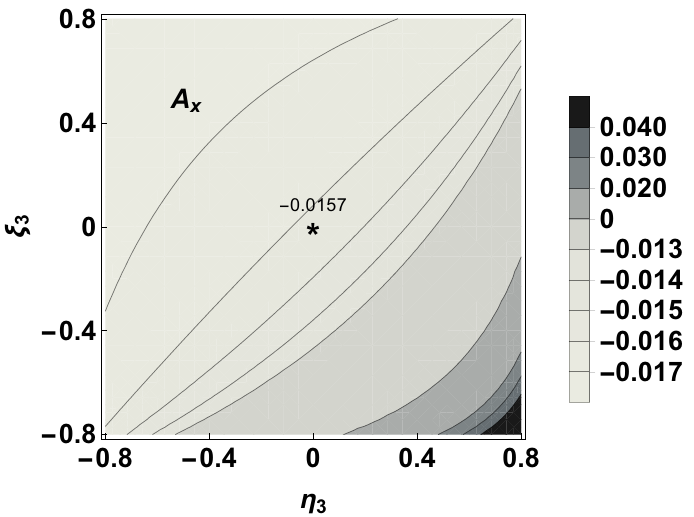}
    \caption{\label{fig:Sigma_and_Afb_eta3xi3} The production cross section $\sigma_{W^+W^-}$ in pb ({\em left-panel})
        and the polarization asymmetry $A_x$ ({\em right-panel}) in the SM as a function of longitudinal beam polarizations
        $\eta_3$ (for $e^-$) and $\xi_3$ (for $e^+$)   at $\sqrt{s}=500$ GeV. 
        The asterisks mark represent the unpolarized points and the number near it correspond to the SM
        values for corresponding observables with unpolarized beams.} 
\end{figure}
We explore the dependence of the cross section and asymmetries on the 
longitudinal polarization $\eta_3$ of $e^-$ and $\xi_3$ of $e^+$.   
In Fig.~\ref{fig:Sigma_and_Afb_eta3xi3} we show the production cross section 
$\sigma_{W^+W^-}$  and $A_x$ as a function of beam polarization as an example. 
The cross section decreases along $\eta_3=-\xi_3$ path from $20$ pb on the 
left-top corner to $7.2$ pb at the unpolarized point and further to $1$ pb in the 
right-bottom corner. This is due to the fact that  the $W^\pm$ 
couples to left chiral $e^-$ i.e., it requires $e^-$ to be negatively 
polarized and $e^+$ to be positively polarized for the higher cross section.
The variation 
of $A_{fb}$ (not shown) with beam polarization is the same as the cross section but 
 very slow above the line $\eta_3=\xi_3$. From this, we can expect 
that a positive $\eta_3$ and a negative $\xi_3$ will reduce the SM contributions to  
observables increasing the $S/\sqrt{B}$ ratio ($S=$ signal, $B=$ background).
Some other asymmetries, like $A_x$, have opposite dependence on the beam
polarizations compared to the cross section, their modulus get reduce for negative $\eta_3$ and positive
$\xi_3$.
\section{Probe of the anomalous couplings}\label{sec:3}
\begin{figure}[!h]
    \centering
    \includegraphics[width=0.7\textwidth]{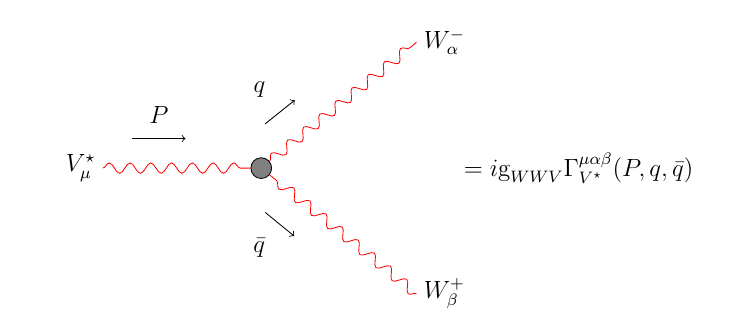}
    \caption{\label{fig:wwv_vertex}The $WWV$ vertex showing anomalous contribution 
        represented the shaded blob on top of SM. The momentum $P$ is incoming to the vertex, 
        while $q$ and $\bar{q}$ are outgoing from the vertex.} 
\end{figure}
The $W^+W^-V$ vertex (Fig.~\ref{fig:wwv_vertex})
for the Lagrangian in Eq.~(\ref{eq:WW-LagWWV}) for on-shell $W$s would be 
$ig_{WWV}\Gamma_V^{\mu\alpha\beta}$ \cite{Gaemers:1978hg,Hagiwara:1986vm}
and it is given by,
\begin{eqnarray}
\Gamma_V^{\mu\alpha\beta}&=&f_1^V(q-\bar q)^\mu g^{\alpha\beta}-
\frac{f_2^V}{m_W^2}(q-\bar q)^\mu P^\alpha P^\beta
+f_3^V(P^\alpha g^{\mu\beta}-P^\beta g^{\mu\alpha})
+if_4^V(P^\alpha g^{\mu\beta}+P^\beta g^{\mu\alpha})\nonumber\\&&
+if_5^V\epsilon^{\mu\alpha\beta\rho}(q-\bar{q})_\rho
-f_6^V\epsilon^{\mu\alpha\beta\rho}P_\rho 
+\frac{\wtil{f_7^V}}{m_W^2}
\left(\bar{q}^\alpha\epsilon^{\mu\beta\rho\sigma} + 
q^\beta\epsilon^{\mu\alpha\rho\sigma}\right)q_\rho\bar{q}_\sigma ,
\label{eq:wwv_vertex}
\end{eqnarray}
where $P,q,\bar q$ are the four-momenta of $V,W^-,W^+$, respectively. The 
momentum conventions are shown in  Fig.~\ref{fig:wwv_vertex}. 
The form factors $f_i$s have been  obtained from the Lagrangian in Eq.~(\ref{eq:WW-LagWWV}) using  
{\tt FeynRules}~\cite{Alloul:2013bka} to be
\begin{eqnarray}\label{eq:reltn_f_Lagrn}
&&f_1^V=g_1^V + \frac{s}{2m_W^2}\lambda^V, \hspace{0.2cm}
f_2^V=\lambda^V,\hspace{0.2cm}
f_3^V=g_1^V + \kappa^V + \lambda^V , \nonumber\\
&&f_4^V=g_4^V,\hspace{0.2cm}
f_5^V=g_5^V,\hspace{0.2cm}
f_6^V=\widetilde{\kappa^V} +
\left(1-\frac{s}{2m_W^2} \right)\widetilde{\lambda^V},\hspace{0.2cm}
\wtil{f_7^V}=\widetilde{\lambda^V} .
\end{eqnarray}

\begin{figure}[!htb]
    \centering
    \includegraphics[width=0.496\textwidth]{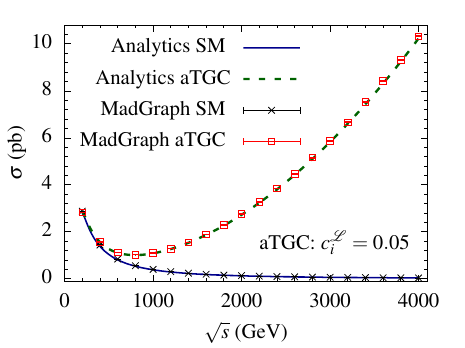}
    \includegraphics[width=0.496\textwidth]{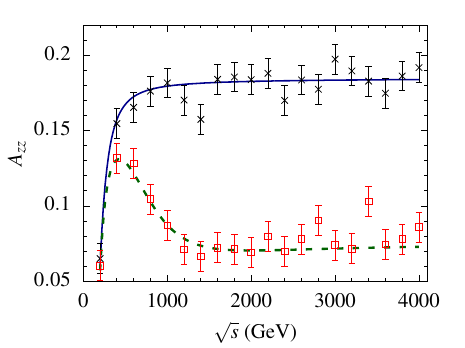}
    \caption{\label{fig:SanityCheck}  The cross section $\sigma$ including the decays in pb ({\em left-panel})
        and the  asymmetry $A_{zz}$ ({\em right-panel}) in the SM   and {\tt aTGC} with all anomalous 
        couplings  ($c_i^{\cal L}$) at $0.05$   
        as a function of $\sqrt{s}$  for  the SM analytic ({\it solid}/blue) and {\tt aTGC} analytic
        ({\it dashed} /green) with unpolarized beams. The {\it crossed} (black) points and {\it boxed} (red) 
        points with errorbars correspond to results from {\tt MadGraph5}.
        The errorbars are given for number of events of $10^4$. } 
\end{figure}
\begin{table}\caption{\label{tab:param_dependence}The dependence of observables 
        (numerators) on the  form factor  couplings in the form of  $c_i^{\cal L}$ (linear), 
        $(c_i^{\cal L})^2$ (quadratic) and $c_i^{\cal L}c_j^{\cal L},~i\ne j$ 
        (interference)   in the process  $e^+e^-\to W^+W^-$. Here, $V\in\{\gamma,Z\}$.
        The ``\checkmark" ({\it checkmark}) represents the presence  and ``---" ({\it big-dash})
        corresponds to  absence.}
    \renewcommand{\arraystretch}{1.70}
    \begin{scriptsize}
        \begin{tabular*}{\textwidth}{@{\extracolsep{\fill}}cccccccccccc@{}}\hline
            Parameters & $\sigma$ & $\sigma\times A_x$ & $\sigma\times A_y$ & $\sigma\times A_z$ &$\sigma\times A_{xy}$  
            &$\sigma\times A_{xz}$  & $\sigma\times A_{yz}$ & $\sigma\times A_{x^2-y^2}$ & $\sigma\times A_{zz}$ & $\sigma\times A_{fb}$ \\\hline
            $\Delta g_1^V$ & \checkmark & \checkmark & --- & \checkmark & --- & \checkmark & --- & \checkmark & \checkmark & \checkmark \\
            $g_4^V $& --- & --- & \checkmark & --- & \checkmark & --- & \checkmark & --- & --- & --- \\
            $g_5^V $& \checkmark & \checkmark & --- & \checkmark & --- & \checkmark & --- & \checkmark & \checkmark & \checkmark \\
            $\lambda^V $& \checkmark & \checkmark & --- & \checkmark & --- & \checkmark & --- & \checkmark & \checkmark & \checkmark \\
            $\wtil{\lambda^V}$ & --- & --- & \checkmark & --- & \checkmark & --- & \checkmark & --- & --- & --- \\
            $\Delta\kappa^V$ & \checkmark & \checkmark & --- & \checkmark & --- & \checkmark & --- & \checkmark & \checkmark & \checkmark \\
            $\wtil{\kappa^V}$ & --- & --- & \checkmark & --- & \checkmark & --- & \checkmark & --- & --- & --- \\
            $(\Delta g_1^V)^2 $& \checkmark & \checkmark & --- & --- & --- & --- & --- & \checkmark & \checkmark & --- \\
            $(g_4^V)^2$& \checkmark & --- & --- & --- & --- & --- & --- & \checkmark & \checkmark & --- \\
            $(g_5^V)^2 $& \checkmark & --- & --- & --- & --- & --- & --- & \checkmark & \checkmark & --- \\
            $(\lambda^V)^2$& \checkmark & \checkmark & --- & --- & --- & --- & --- & \checkmark & \checkmark & --- \\
            $(\wtil{\lambda^V})^2$& \checkmark & \checkmark & --- & --- & --- & --- & --- & \checkmark & \checkmark & --- \\
            $(\Delta\kappa^V)^2 $& \checkmark & \checkmark & --- & --- & --- & --- & --- & \checkmark & \checkmark & --- \\
            $(\wtil{\kappa^V})^2 $& \checkmark & \checkmark & --- & --- & --- & --- & --- & \checkmark & \checkmark & --- \\
            $\Delta g_1^V g_4^V $& --- & --- & --- & --- & --- & --- & \checkmark & --- & --- & --- \\
            $\Delta g_1^V g_5^V $& --- & --- & --- & \checkmark & --- & --- & --- & --- & --- & \checkmark \\
            $\Delta g_1^V \lambda^V $& \checkmark & \checkmark & --- & --- & --- & --- & --- & \checkmark & \checkmark & --- \\
            $\Delta g_1^V \wtil{\lambda^V} $& --- & --- & \checkmark & --- & \checkmark & --- & --- & --- & --- & --- \\
            $\Delta g_1^V \Delta\kappa^V $& \checkmark & \checkmark & --- & --- & --- & --- & --- & \checkmark & \checkmark & --- \\
            $\Delta g_1^V \wtil{\kappa^V} $& --- & --- & \checkmark & --- & \checkmark & --- & --- & --- & --- & --- \\
            $g_4^V g_5^V $& --- & --- & --- & --- & \checkmark & --- & --- & --- & --- & --- \\
            $g_4^V \lambda^V $& --- & --- & --- & --- & --- & --- & \checkmark & --- & --- & --- \\
            $g_4^V \wtil{\lambda^V} $& --- & --- & --- & \checkmark & --- & \checkmark & --- & --- & --- & \checkmark \\
            $g_4^V \Delta\kappa^V $& --- & --- & --- & --- & --- & --- & \checkmark & --- & --- & --- \\
            $g_4^V \wtil{\kappa^V} $& --- & --- & --- & \checkmark & --- & \checkmark & --- & --- & --- & \checkmark \\
            $g_5^V \lambda^V $& --- & --- & --- & \checkmark & --- & \checkmark & --- & --- & --- & \checkmark \\
            $g_5^V \wtil{\lambda^V} $& --- & --- & --- & --- & --- & --- & \checkmark & --- & --- & --- \\
            $g_5^V \Delta\kappa^V $& --- & --- & --- & \checkmark & --- & \checkmark & --- & --- & --- & \checkmark \\
            $g_5^V \wtil{\kappa^V} $& --- & --- & --- & --- & --- & --- & \checkmark & --- & --- & --- \\
            $\lambda^V \wtil{\lambda^V}$ & --- & --- & \checkmark & --- & \checkmark & --- & --- & --- & --- & --- \\
            $\lambda^V \Delta\kappa^V $& \checkmark & \checkmark & --- & --- & --- & --- & --- & \checkmark & \checkmark & --- \\
            $\lambda^V\wtil{\kappa^V}  $& --- & --- & \checkmark & --- & \checkmark & --- & --- & --- & --- & --- \\
            $\wtil{\lambda^V} \Delta\kappa^V $& --- & --- & \checkmark & --- & \checkmark & --- & --- & --- & --- & --- \\
            $\wtil{\lambda^V}\wtil{\kappa^V}  $& \checkmark & \checkmark & --- & --- & --- & --- & --- & \checkmark & \checkmark & --- \\
            $\Delta\kappa^V \wtil{\kappa^V} $& --- & --- & \checkmark & --- & \checkmark & --- & --- & --- & --- & --- \\
            \hline
        \end{tabular*}
    \end{scriptsize}
\end{table}
We use the vertex factors in Eq.~(\ref{eq:wwv_vertex}) for the analytical 
calculation of our observables and cross validate them  numerically
with {\tt MadGraph5}~\cite{Alwall:2014hca} implementation
of Eq.~(\ref{eq:WW-LagWWV}). As an example, we present two observables 
$\sigma_{W^+W^-}$ and $A_{zz}$ for the SM ($c_i^{\cal L}=0.0$) and for a 
chosen couplings point $c_i^{\cal L}=0.05$, in Fig.~\ref{fig:SanityCheck}. 
The agreement between the analytical and the numerical calculations over a range
of $\sqrt{s}$ indicates the validity of relations in Eq.~(\ref{eq:reltn_f_Lagrn}),
specially the $s$ dependence of $f_1^V$ and $f_6^V$.

Analytical  expressions of all the observables have been obtained and their
dependence on the anomalous couplings $c_i^{\cal L}$ are given  in
Table~\ref{tab:param_dependence}.
The $CP$-even couplings in $CP$-even observables $\sigma$, $A_x$, $A_z$, $A_{xz}$, 
$A_{x^2-y^2}$, and $A_{zz}$ appear in linear as well as in quadratic form but do not
appear  in the $CP$-odd observables $A_y$, $A_{xy}$, and $A_{yz}$. On the other hand, 
$CP$-odd couplings appears linearly in  $CP$-odd observables and quadratically
in $CP$-even observables. Thus the $CP$-even couplings may have
a double patch in their confidence intervals leading to asymmetric limits which will be discussed in subsection~\ref{sec:3.1}. The $CP$-odd
couplings, however, will have a single patch in their confidence intervals and will acquire 
symmetric limits. 
\subsection{Sensitivity of observables on anomalous couplings and their binning}
\label{sec:3.1}
\begin{figure}[h!]
    \centering
    \includegraphics[width=0.496\textwidth]{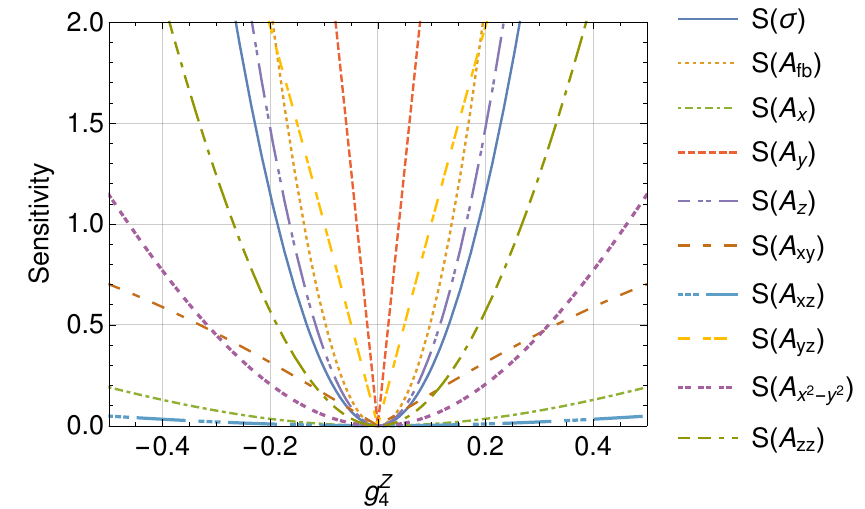}
    \includegraphics[width=0.496\textwidth]{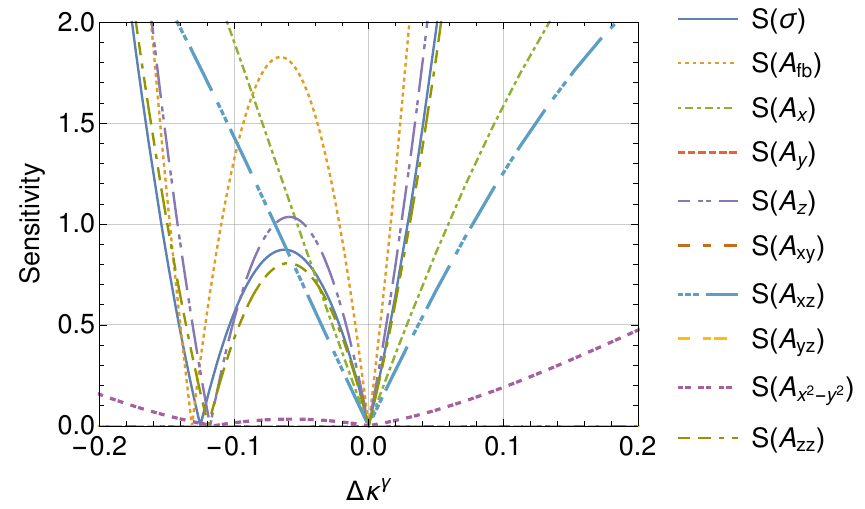}
    \caption{\label{fig:sensitivity} The one parameter sensitivities of the cross section $\sigma$, 
        $A_{fb}$ and $8$ polarization asymmetries ($A_i$) on $g_4^Z$ ({\em left-panel})
        and on $\Delta\kappa^\gamma$ ({\em right-panel})   for $\sqrt{s}=500$ GeV, ${\cal L}=100$ fb$^{-1}$ with unpolarized beams.} 
\end{figure}

We studied the sensitivities (see Eq.~(\ref{eq:epjc2-sensitivity-beam-pol}) for definition) of all $10$ observables to  all the $14$
couplings of the Lagrangian in Eq.~(\ref{eq:WW-LagWWV}).
We take ${\cal L}=50$ fb$^{-1}$ of integrated luminosity for each of the opposite beam polarizations and systematic uncertainties of  
$\epsilon_\sigma=2~\%$ for the cross section and $\epsilon_A=1~\%$ for the asymmetries as a benchmark scenario for the present
analyses. 
The sensitivities of all observables 
on $g_4^Z$ and $\Delta\kappa^\gamma$ are shown in Fig.~\ref{fig:sensitivity}
as representative. Being a $CP$-odd coupling (either only linear or only quadratic terms
present in the observables), $g_4^Z$ has a single patch in the confidence interval, 
while the  $\Delta\kappa^\gamma$ being a  $CP$-even 
(linear as well as quadratic terms present in the observables) 
 has two patches in the sensitivity curve, as noted earlier. The  $CP$-odd observable $A_y$
provides the tightest one parameter limit on $g_4^Z$. The tightest $1\sigma$ 
limit on $\Delta\kappa^\gamma$ is obtained using $A_{fb}$, while at $2\sigma$ 
level, a combination of $A_{fb}$ and $A_x$ provide the tightest limit.

\begin{figure}[h!]
    \centering
    \includegraphics[width=0.496\textwidth]{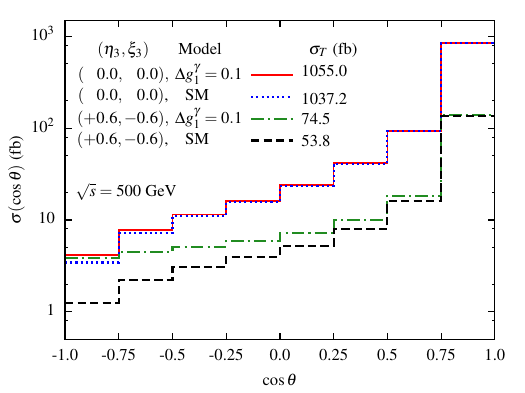}
    \includegraphics[width=0.496\textwidth]{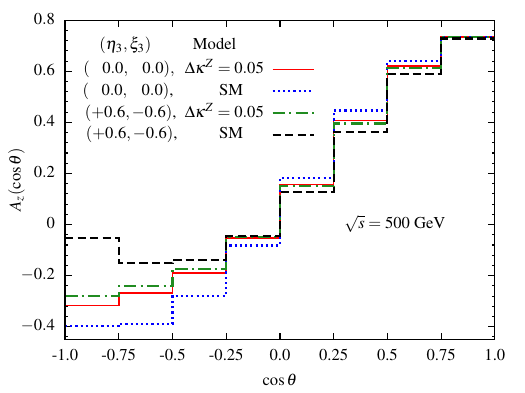}
    \includegraphics[width=0.496\textwidth]{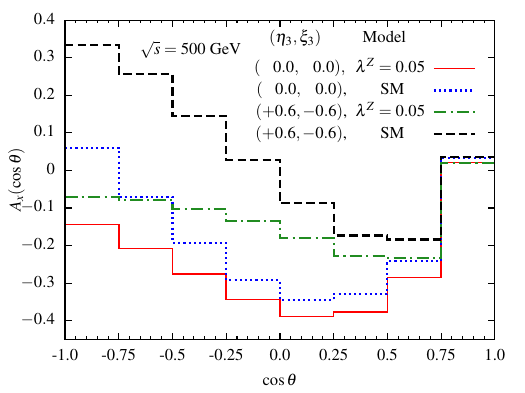}
    \includegraphics[width=0.496\textwidth]{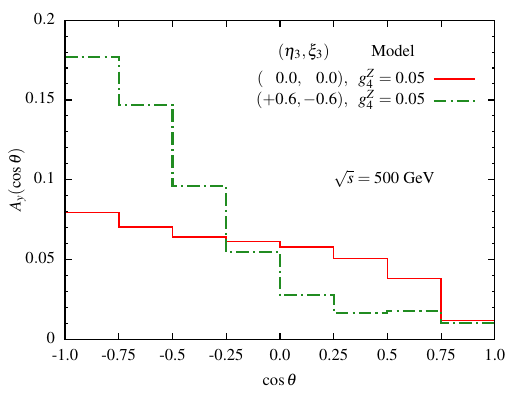}
    \caption{\label{fig:diffSigmaSM2} The cross section  $\sigma$  ({\em left-top}), $A_z$
        ({\em right-top}), $A_x$ ({\em left-bottom}) and $A_y$ ({\em right-bottom}) as a function of $\cos\theta$  
        of $W^-$ in $8$  bin  for $\sqrt{s}=500$ GeV. The {\it dotted} (blue) lines correspond to the SM unpolarized values, 
        {\it solid} (red) lines correspond to 
       the unpolarized {\tt aTGC} values, {\it dashed} (black) 
        lines represent the polarized SM values, and {\it dashed-dotted} (green) lines 
        represent polarized {\tt aTGC} values of 
        observables. For {\tt aTGC}, only one anomalous coupling  
        has been assumed non-zero and others kept at zero in each {\em panel}.} 
\end{figure}
Here, we have a total of $14$ different anomalous couplings to be measured, while 
we only have $10$ observables. A certain combination of large couplings may 
mimic the SM within the statistical errors. To avoid these we need more number 
of observables to be included in the analysis. We achieve this by dividing $\cos\theta_{W^-}$
 into eight bins and calculate the cross section and
polarization asymmetries in all of them.
In Fig.~\ref{fig:diffSigmaSM2} the cross section and  the polarization asymmetries $A_z$, $A_{x}$, and $A_y$ are shown  as 
a function of $\cos\theta_{W^-}$   for the SM and some  {\tt aTGC} couplings
for both polarized and unpolarized beams.
The sensitivities for unpolarized  SM cases are shown in {\it dotted} (blue) lines; 
SM with polarization of $(\eta_3,\xi_3)=(+0.6,-0.6)$ are shown in {\it dashed} (black) lines.
The {\it solid} (red) lines correspond to unpolarized {\tt aTGC} values, while   {\it dashed-dotted}
(green) lines represent polarized {\tt aTGC} values of observables. 
For the cross section ({\em left-top-panel}), we take $\Delta g_1^\gamma$  to be $0.1$  and all other 
couplings to zero  for both polarized and unpolarized beams.
We see that the fractional deviation from the SM value is larger in the
most backward bin ($\cos\theta_{W^-}\in(-1.0 , -0.75)$) and gradually reduces in the
forward direction.  The deviation is even larger in case
of beam polarization.
The sensitivity of the cross section on  $\Delta g_1^\gamma$  is thus expected to be high in the most 
backward bin. For the asymmetries $A_z$ ({\em right-top-panel}), $A_{xz}$ ({\em left-bottom-panel}) 
and $A_y$ ({\em right-bottom-panel}), the {\tt aTGC} are assumed to be $\Delta\kappa^Z=0.05$, 
$\lambda^Z=0.05$  and $g_4^Z=0.05$, respectively, while others are kept at 
zero. The changes in the asymmetries due to {\tt aTGC} are larger in the backward
bins for both  polarized and unpolarized beams. 
We note that the asymmetries may not have the highest sensitivity in the most backward bin, but 
in some other bins. We consider 
the cross section and eight polarization asymmetries  in all $8$ bins, i.e., we have  $72$ observables in our
analysis. 

\begin{figure}
    \centering
    \includegraphics[width=0.496\textwidth]{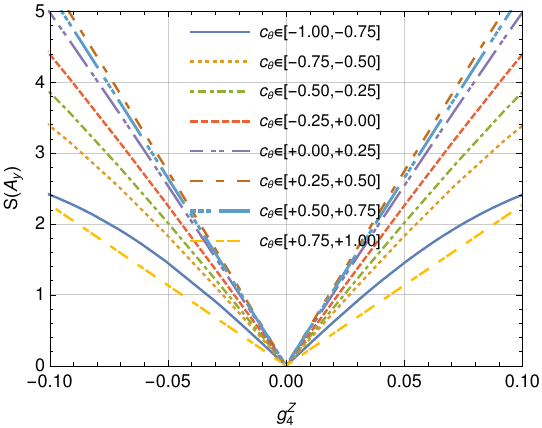}
    \includegraphics[width=0.496\textwidth]{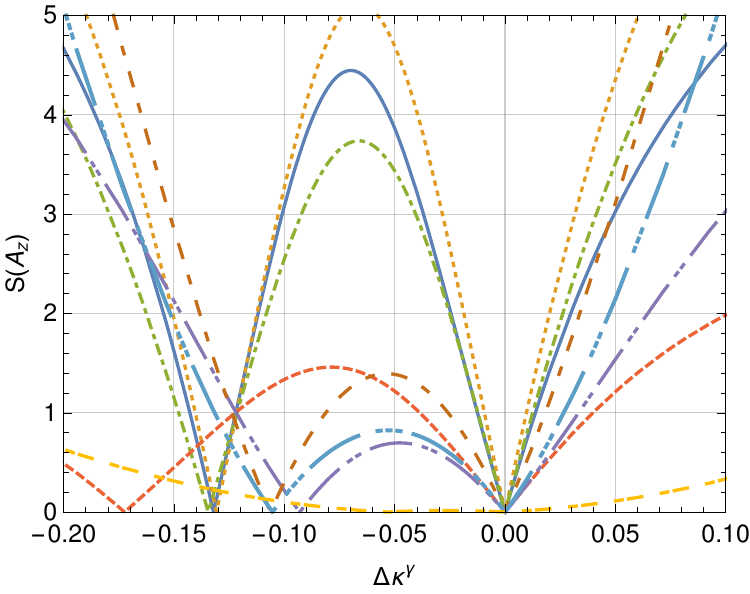}
    \caption{\label{fig:sensitivity_binned}The one parameter  sensitivities of $A_x$ 
        on $g_4^Z$ ({\em left-panel}) and of $A_z$ on $\Delta\kappa^\gamma$ ({\em right-panel}) in $8$ bins
         at $\sqrt{s}=500$ GeV,  ${\cal L}=100$ fb$^{-1}$ with $c_\theta = \cos\theta_{W^-}$  with unpolarized beams.}
\end{figure}
One parameter sensitivity of the set of $9$ observables in all  $8$ bins to all the couplings  have
been studied. We show sensitivity of $A_y$  on $g_4^Z$ and of $A_z$ on $\Delta\kappa^\gamma$ 
in the $8$ bin in Fig.~\ref{fig:sensitivity_binned} as representative. The tightest limits based on 
sensitivity (coming from one bin)  is roughly twice as tight as  compared 
to the unbin case in Fig.~\ref{fig:sensitivity}. Thus we expect simultaneous limits
on all the couplings to be tighter when using binned observables.
\begin{table}[!ht]\caption{\label{tab:obs-analysis-name} The list  of  
        analyses performed in the present work
        and  set of  observables used with  different kinematical cuts   to obtain 
        simultaneous limits on the anomalous couplings  
        at $\sqrt{s}=500$ GeV, ${\cal L}=100$ fb$^{-1}$ with unpolarized beams.
        The rectangular volumes of couplings at $95\%$ BCI are 
        shown in the last column for each analyses (see text
        for details).  }
    \renewcommand{\arraystretch}{1.50}
    \begin{tabular*}{\textwidth}{@{\extracolsep{\fill}}llll@{}}\hline
        Analysis name  & Set of observables & Kinematical cut on $\cos\theta_{W^-}$  & Volume of Limits \\\hline
        {\tt $\sigma$-ubinned} & $\sigma$ & $\cos\theta_{W^-}\in[-1.0,1.0]$& $4.4\times 10^{-11}$\\
        {\tt Unbinned} & 
        $\sigma$, $A_{fb}$, $A_i$
        & $\cos\theta_{W^-}\in[-1.0,1.0]$ & $3.1\times 10^{-12}$\\
        {\tt $\sigma$-binned} &  $\sigma$ & 
        $\cos\theta_{W^-}\in[\frac{m-5}{4},\frac{m-4}{4}]$, $m=1,2,\dots, 8$
        & $3.7 \times 10^{-12}$ \\
 {\tt Pol.-binned} &$A_i$ &" & $1.6\times 10^{-15} $ \\
        {\tt Binned} & $\sigma$,  $A_i$ &  "& $5.2 \times 10^{-17}$\\
        \hline
    \end{tabular*}
\end{table}
\begin{figure}[h!]
    \includegraphics[width=0.54\textwidth]{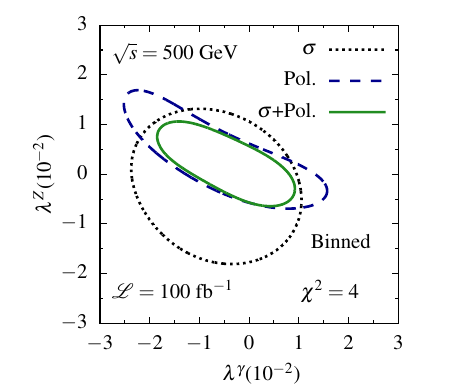}
    \includegraphics[width=0.46\textwidth]{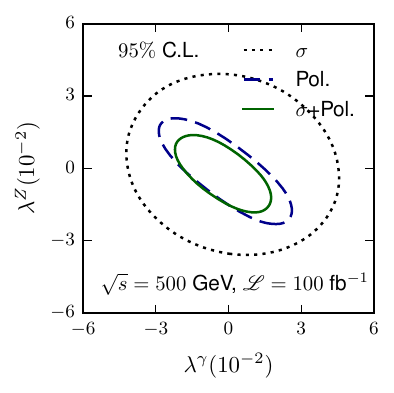}
    \caption{\label{fig:sig-vs-sigpol} The  $\chi^2=4$ contours in the {\em left-panel} and $95~\%$ C.L. contours from 
        simultaneous analysis in the  {\em right-panel} in the $\lambda^\gamma$--$\lambda^Z$ plane using the binned cross sections ($\sigma$)
        alone in {\it dotted} (black) lines, just binned polarizations asymmetries (Pol.) in {\it dashed} (blue) lines and the 
        bin cross section together with binned polarization 
        asymmetries ($\sigma$ + Pol.) in {\it solid} (green) lines for $\sqrt{s}=500$ GeV, ${\cal L}=100$ fb$^{-1}$. } 
\end{figure}
We perform a set of MCMC analyses with a different set of observables for different kinematical
cuts with unpolarized beams to understand their roles in providing limits on
the anomalous couplings. These analyses are listed in 
Table~\ref{tab:obs-analysis-name}. 
The corresponding $14$-dimensional rectangular 
volume\footnote[3]{This volume of limit  is the 
    the volume of a $14$-dimensional rectangular box bounding by the 
    $95\%$ BCI projection of simultaneous limits in each coupling, 
    which can be a measure of goodness of the benchmark beam
    polarization. We computed the cross section and other asymmetries keeping 
    term up to quadratic in couplings. In this case, even a single observable 
    can  give a finite volume of limit and constrain all $14$ couplings, which 
    would not be possible if only terms linear in couplings were present.}  made out 
of $95\%$ Bayesian confidence
interval  on the anomalous couplings are also listed in
Table~\ref{tab:obs-analysis-name} in the last column. 
The simplest analysis would be to consider only the cross section in the full $\cos\theta_{W^-}$ domain
and perform MCMC analysis which is named as  {\tt $\sigma$-ubinned}. The typical
$95\%$ limits on the parameters range from $\sim \pm0.04$ to $\pm0.25$ giving the volume of 
limits to be $4.4\times 10^{-11}$. As we have polarizations asymmetries, the straight
forward analysis would be to consider all the observables  
for the full domain of $\cos\theta_{W^-}$. This analysis is named  {\tt Unbinned} where  limits on 
anomalous couplings get constrained better reducing the volume of limits by a factor of $10$
compared to the {\tt $\sigma$-ubinned}.  
To see how binning improve the limits, we performed an analysis named {\tt $\sigma$-binned} 
using only the cross section in $8$ bins. We see that the analysis {\tt $\sigma$-binned} is better 
than the analysis {\tt $\sigma$-unbinned} and comparable to the analysis {\tt Unbinned}.
To see the strength of the polarization asymmetries, we performed an analysis named {\tt Pol.-binned} using
just the polarization asymmetries in $8$ bins. We see that this analysis is much better than the analysis  
{\tt $\sigma$-binned}.
The most natural and complete analysis would be to  consider all the  observables after binning.
The analysis is named as  {\tt Binned} which has  limits  much better than any analyses. 
The comparison between  the analyses {\tt $\sigma$-binned}, {\tt Pol.-binned} and {\tt Binned}
is shown in Fig.~\ref{fig:sig-vs-sigpol} in the panel $\lambda^\gamma$--$\lambda^Z$ in two-parameter
({\em left-panel}) as well as in multi-parameter ({\em right-panel}) analysis using MCMC as 
representative. The {\em right-panel} reflects the Table~\ref{tab:obs-analysis-name}. The behaviours are same   even in the two parameter 
analysis ({\em left-panel}) by keeping all other parameter to zero, i.e, the bounded region for $\chi^2=4$ is 
smaller in   {\tt Pol.-binned} (Pol.) than  {\tt $\sigma$-binned} ($\sigma$) and smallest for {\tt Binned} ($\sigma$+Pol.).

We also calculate one parameter limits on all the couplings at $95~\%$ C.L.  
considering all the binned observables with 
unpolarized beams in the effective vertex formalism as well as in the effective operator 
approach and list them in the last column of  Tables~\ref{tab:Limits-Lag} \&~\ref{tab:Limits-Op}, respectively for comparison.
In the next subsection, we study the effect of beam polarizations on the limits of the anomalous couplings. 
\subsection{Effect of beam polarizations to the limits on aTGC}\label{sec:WW-beampolEffect}
A suitable choice of beam polarizations can enhance the signal to background ratio tightening the constraints on the aTGC.
Below we discuss the comparison between various combinations of beam polarizations to better constrain the aTGC.
After that, we see the effect of beam polarizations in constraining the aTGC for both fixed choices and best combined choices.

\subsubsection{Combining  beam polarization with it's opposite values}\label{subsubsec:WW-Combinng-beampol}
\begin{figure}[h!]
    \includegraphics[width=0.505\textwidth]{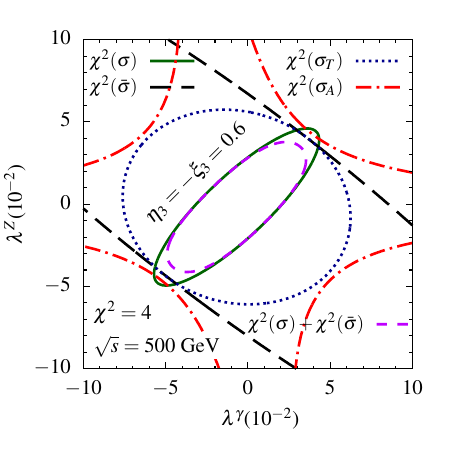}
    \includegraphics[width=0.488\textwidth]{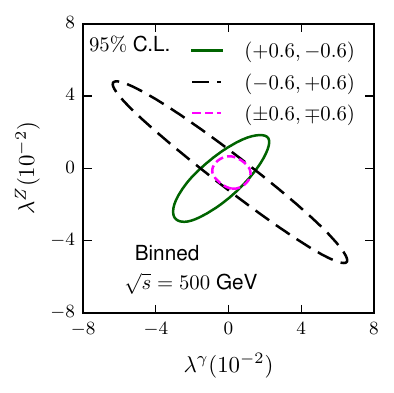}
    \caption{\label{fig:beampol-combine}  The $\chi^2=4$ contours of the unbinned cross sections   $\sigma=\sigma(+\eta_3,+\xi_3)$ in {\it solid}/green lines,
        $\bar{\sigma}=\sigma(-\eta_3,-\xi_3)$ in {\it big-dashed}/black lines, $\sigma_T=\sigma(+\eta_3,+\xi_3)+\sigma(-\eta_3,-\xi_3)$ in {\it dotted}/blue line,  $\sigma_A=\sigma(+\eta_3,+\xi_3)-\sigma(-\eta_3,-\xi_3)$ in {\it dash-dotted}/red line and the combined $\chi^2$ of $\sigma$ and
        $\bar{\sigma}$ in {\it dashed}/magenta lines   for polarization $(\eta_3,\xi_3)=(+0.6,-0.6)$ on  $\lambda^\gamma$--$\lambda^Z$ plane are shown in the {\em left-panel}. The $95~\%$ C.L. contours from simultaneous
        analysis in $\lambda^\gamma$--$\lambda^Z$ plane for the beam polarization $(+0.6,-0.6)$, $(-0.6,+0.6)$ and their combined one $(\pm0.6,\mp0.6)$
        are shown in the {\em right-panel} using all the binned observables, i.e., in {\tt Binned} case. The analyses are done for  $\sqrt{s}=500$ GeV and ${\cal L}=50$ fb$^{-1}$ luminosity to each beam polarization set.} 
\end{figure}
To reduce the systematic errors in an analysis due to luminosity, the beam polarizations 
are flipped between  two opposite choices frequently giving half the total luminosity to both
the polarization choices in an $e^+$--$e^-$ collider.
One can, in principle, use the observables, e.g., the total cross section ($\sigma_T$) or their difference ($\sigma_A$)
as in Eqs.~(\ref{eq:sigma_T}) \& (\ref{eq:sigma_A}), respectively or
for the two opposite polarization choices ($\sigma$ \& $\bar{\sigma}$) separately for a suitable analysis. 
In this work,  we do not  combine the 
beam polarization $(+\eta_3,+\xi_3)$ and it's opposite $(-\eta_3,-\xi_3)$ at the level of observables,  rather we combine them
 at the level of $\chi^2$ (as we did in chapter~\ref{chap:epjc2}, see Eq.~(\ref{eq:beampol-totChi2})) given by,
\begin{equation}\label{eq:beampol-combine}
\chi^2_{tot}(\pm\eta_3,\pm\xi_3)= \sum_{bin}  \sum_{N} \left(\chi^2\left[{\cal O}_N(+\eta_3,+\xi_3)\right] +\chi^2\left[{\cal O}_N(-\eta_3,-\xi_3)\right] \right) ,
\end{equation}
where $N$ runs over all the observables. This is because the later combination constrains the couplings better than any combinations and of-course the individuals.
To depict this, we present the $\chi^2=4$ contours 
of the unbinned cross sections in Fig.~\ref{fig:beampol-combine} ({\em left-panel}) for beam polarization $(+0.6,-0.6)$ ($\sigma$) and $(-0.6,+0.6)$ ($\bar{\sigma}$) and the
combinations $\sigma_T$ and $\sigma_A$   along with the combined $\chi^2$
in the $\lambda^\gamma$--$\lambda^Z$ plane     for  ${\cal L}=50$ 
fb$^{-1}$ luminosity to each polarization choice as representative.
A systematic error of $2\%$ is used as a benchmark in the cross section.  
The nature of the contours can be explain as follows: 
In the $WW$ production, the aTGC contributions appear only in the 
$s$-channel (see Fig.~\ref{fig:Feynman-ee-ww}), where initial state
$e^+e^-$ couples through $\gamma/Z$ boson and both left and right chiral electrons contribute
almost equally. The $t$-channel diagram, however, is pure background and  
receives contribution only from left chiral 
electrons. As a result the  $\bar{\sigma}$ ({\it big-dashed}/black) 
contains more background than $\sigma$ ({\it solid}/green)  leading to a weaker limit on the couplings.
Further,  inclusion of $\bar{\sigma}$ into 
$\sigma_T$ ({\it dotted}/blue) and  $\sigma_A$ ({\it dashed-dotted}/red) reduces the signal to background ratio
and hence they are less  sensitive to the couplings. The total $\chi^2$  for the combined
beam polarizations  shown in {\it dashed} (magenta) is, of course, the best to constrain the couplings.
This behaviour is reverified with the simultaneous analysis using the binned cross sections and 
polarization asymmetries ($72$ observables in the {\tt Binned} case) 
and shown in Fig.~\ref{fig:beampol-combine} ({\em right-panel}) in the same $\lambda^\gamma$--$\lambda^Z$ 
plane showing the  $95~\%$ C.L. contours for beam polarizations 
$(+0.6,-0.6)$, $(-0.6,+0.6)$, and their combinations $(\pm0.6,\mp0.6)$.
Thus we choose to combine the opposite beam polarization choices at the level 
of $\chi^2$ rather than combining them at the level of observables.

\subsubsection{Case of fixed beam polarizations}\label{subsubsec:WW-fixed-beampolEffect}
\begin{figure}[h!]
    \centering
    \includegraphics[width=0.9\textwidth]{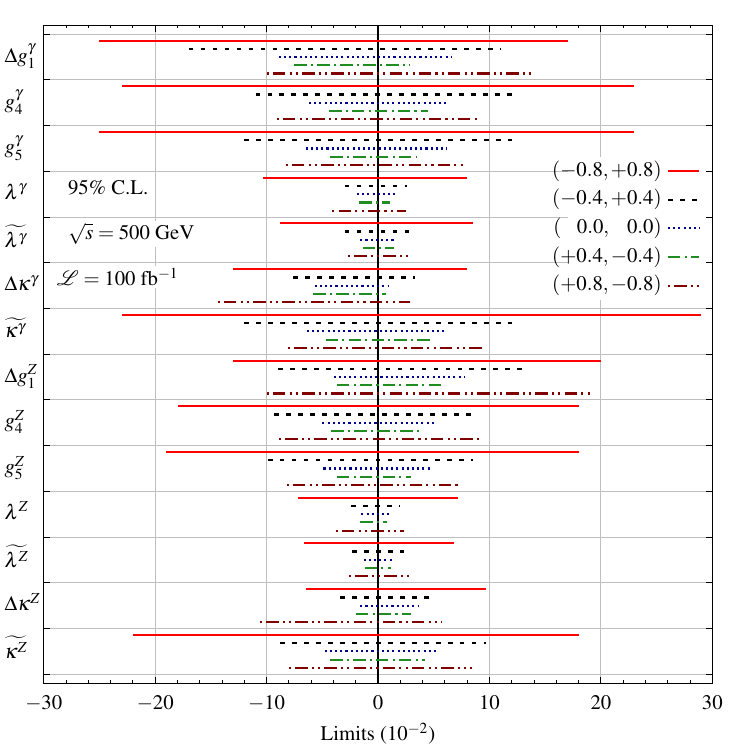}
    \caption{\label{fig:limit-beampol-fixed} The  $95~\%$ BCI limits   on the  anomalous couplings $c_i^{\cal L}$ for a set of fixed choices of beam polarizations
        for $\sqrt{s}=500$ GeV, ${\cal L}=100$ fb$^{-1}$ using the binned observables.} 
\end{figure}
\begin{figure}[h!]
    \centering
    \includegraphics[height=0.29\textheight]{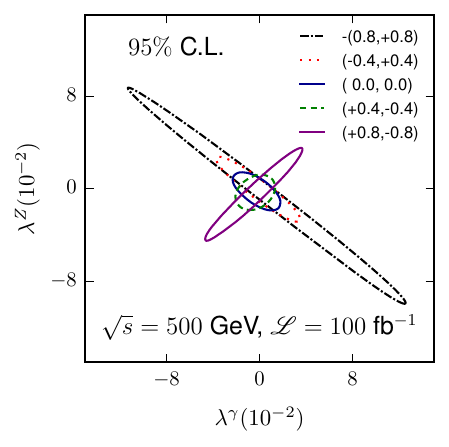}
    \includegraphics[height=0.29\textheight]{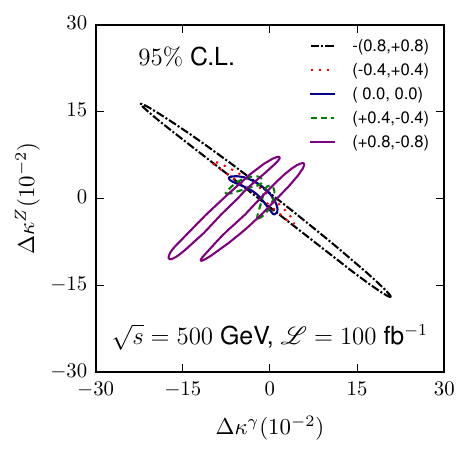}
    \caption{\label{fig:contours-beampol-fixed} The $95~\%$ C.L. contours from simultaneous
        analysis in  $\lambda^\gamma$--$\lambda^Z$ plane ({\em left-panel}) and $\lambda^\gamma$--$\lambda^Z$ plane ({\em right-panel}) for a set of fixed choices of beam polarizations for $\sqrt{s}=500$ GeV, ${\cal L}=100$ fb$^{-1}$ using the binned observables. } 
\end{figure}
Although, we will have data for two opposite choice of beam polarizations, we first 
investigate how the fixed beam polarizations of various amplitudes affect the limits.
We estimate simultaneous limits on all the $14$ (independent) anomalous couplings $c_i^{\cal L}$
using MCMC method in the {\tt Binned} case for  five different set of  fixed choices of beam polarizations 
($\eta_3,\xi_3$)  namely $(-0.8,+0.8)$, $(-0.4,+0.4)$, 
$(0.0,+0.0)$, $(+0.4,-0.4)$ and 
$(+0.8,-0.8)$. We choose the cross-diagonal choices as they provide optimal result for the cross section. 
The cross section depending on beam polarizations  can be expressed as,
    \begin{eqnarray}\label{eq:sigma_eta_xi}
    \sigma(\eta_3,\xi_3)
    &=&
    (1+\eta_3)(1-\xi_3) \frac{1}{4} \sigma_R
    +
    (1-\eta_3)(1+\xi_3)\frac{1}{4} \sigma_L \nonumber\\
    &=&
    (1 -\eta_3\xi_3)\frac{1}{4} (\sigma_R + \sigma_L) 
    +
    (\eta_3 - \xi_3)\frac{1}{4}  (\sigma_R - \sigma_L), 
    \end{eqnarray}
    where $\sigma_R$ denotes the $e_R$ annihilation cross
    section, while $\sigma_L$ is that for $e_L$ annihilation cross section.
Thus, the $\eta_3=-\xi_3$ polarizations will give optimal result for the cross section.
The $95~\%$ BCI limits on the couplings $c_i^{\cal L}$ are shown in Fig.~\ref{fig:limit-beampol-fixed} for the above choices of beam polarizations. 
We observe that the limits on anomalous couplings are tightest for the beam polarization 
$(+0.4,-0.4)$. We estimate simultaneous limits on the couplings on several other polarization points 
along $\eta_3=-\xi_3$ direction and find the $(+0.4,-0.4)$ polarization to be the best 
to provide tightest limits. The correlations among the parameters are also studied in this case. In Fig.~\ref{fig:contours-beampol-fixed}, we show the  $95~\%$ C.L. contours from simultaneous
analysis in  $\lambda^\gamma$--$\lambda^Z$ plane ({\em left-panel}) and $\Delta\kappa^\gamma$--$\Delta\kappa^Z$ plane ({\em right-panel}) for the set of fixed choices of beam polarizations. We see that, $(-0.8,+0.8)$ and $(+0.8,-0.8)$ polarizations give
orthogonal contours with maximal correlation and anti-correlation, respectively in both
planes much like seen in Fig.~\ref{fig:beampol-combine}. In the $\Delta\kappa^\gamma$--$\Delta\kappa^Z$ plane, we see an interesting case: 
An elliptical contour for beam polarization of $(-0.8,+0.8)$ 
({\it dotted}/black) breaks into two disconnected regions for $(+0.4,-0.4)$ 
({\it solid}/green) and then these disconnected regions grow in size for $(+0.8,-0.8)$ 
({\it dashed}/purple). The contours for beam polarization  $(+0.4,-0.4)$ are tighter and less
 correlated. The results and conclusions differs when two opposite choice of beam polarizations are considered, which are discussed below.

\subsubsection{Case of beam polarization combined with their flipped values}\label{subsubsec:WW-flipped-beampolEffect}
\begin{figure}[h]
    \begin{minipage}{0.5\textwidth}
        \includegraphics[width=\textwidth]{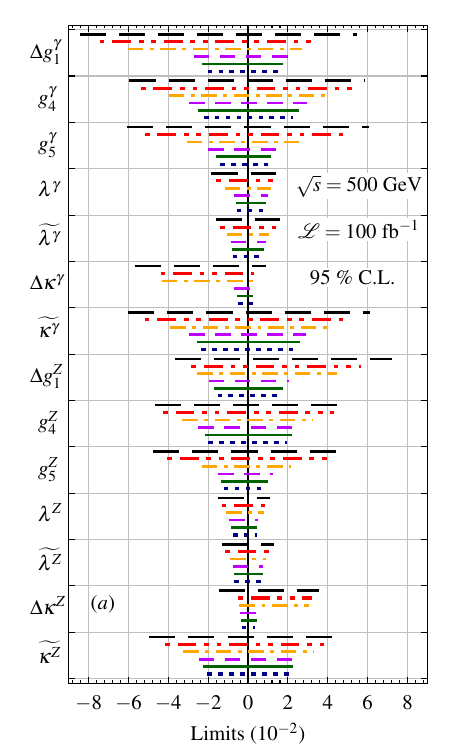}
    \end{minipage}
    \begin{minipage}{0.5\textwidth}
        \includegraphics[width=\textwidth]{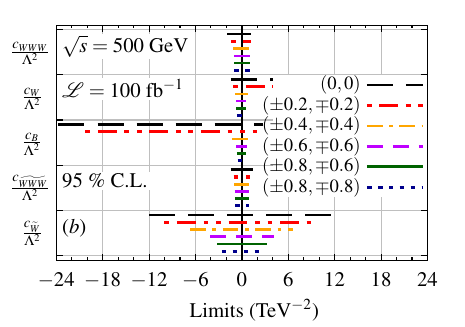}
        \includegraphics[width=\textwidth]{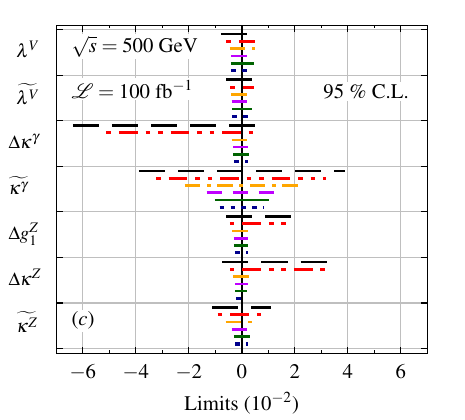}
    \end{minipage}
    \caption{\label{fig:Limits-combined} The pictorial visualisation of $95~\%$ BCI limits  $(a):$ on the  anomalous couplings $c_i^{\cal L}$ 
        in the {\em left-panel}, $(b):$ on  $c_i^{\cal O}$ in the {\em right-top-panel} and $(c):$ on $c_i^{{\cal L}_g}$ in the  {\em right-bottom-panel}
         for $\sqrt{s}=500$ GeV, ${\cal L}=100$ 
        fb$^{-1}$ using the binned observables. The numerical values of the 
        limits  can be read of  in Tables~\ref{tab:Limits-Lag} \&~\ref{tab:Limits-Op}.} 
\end{figure}

Here, we perform MCMC analysis to estimated simultaneous limits on the couplings of both form factors 
and effective operators for beam polarizations combined with their opposite values.
We perform the analysis for beam polarizations of $(\eta_3,\xi_3)$ to be
$(0,0)$, $(+0.2,-0.2)$, $(+0.4,-0.4)$, $(+0.6,-0.6)$, $(+0.8,-0.6)$, $(+0.8,-0.8)$ combined with their opposite values 
using the $\chi^2$  given in Eq.~(\ref{eq:beampol-combine}).
The $95~\%$ BCI simultaneous limits for the chosen set of 
beam polarizations combined according to Eq.~(\ref{eq:beampol-combine}) are shown in 
Table~\ref{tab:Limits-Lag} for effective vertex formalism ($c_i^{\cal L}$) 
and in Table~\ref{tab:Limits-Op} for  effective operator approach ($c_i^{\cal O}$).  
The corresponding translated limits to the vertex factor couplings  $c_i^{{\cal L}_g}$ 
are also shown  in the  Table~\ref{tab:Limits-Op}. While presenting limits, the  following notations are used:  
$$_{ low}^{ high}\equiv [ low,  high]$$ with $low$ being lower limit and $high$ being upper limit.
A pictorial visualization of the limits shown in Table~\ref{tab:Limits-Lag} \& and \ref{tab:Limits-Op}
is given in Fig.~\ref{fig:Limits-combined} for the easy comparisons. 
The limits on the couplings get tighter as the amplitude of beam polarizations are increased along $\eta_3=-\xi_3$ path
and become tightest at the extreme beam polarization 
$(\pm 0.8, \mp 0.8)$. However, the choice $(\pm 0.8, \mp 0.6)$ is best to put constraints on the couplings
within the technological reach~\cite{Vauth:2016pgg,MoortgatPick:2006qp}. 
\begin{sidewaystable}
    \centering
    \caption{\label{tab:Limits-Lag}List of  posterior $95~\%$ BCI
        of anomalous couplings $c_i^{\cal L}$ ($10^{-2}$) of the Lagrangian in Eq.~(\ref{eq:WW-LagWWV})   at $\sqrt{s}=500$ GeV, ${\cal L}=100$ fb$^{-1}$  for a chosen set of longitudinal beam polarizations $\eta_3$ and $\xi_3$ 
        from MCMC in {\tt Binned} case. The limits for the best choice of beam polarization within technological reach, i.e., $(\pm0.8,\mp0.6)$ are marked in \textbf{bold}. The pictorial visualisation for these $95~\%$ BCI of $c_i^{\cal L}$   
        is shown in Fig.~\ref{fig:Limits-combined} in the {\em left-panel}. 
        The one parameter ($1P$) limits ($10^{-2}$)   at $95~\%$ BCI  with unpolarized beams are given in the
        last column for comparison. The notations used here are $_{ low}^{ high}\equiv [ low,  high]$ with
        $low$ being lower limit and $high$ being upper limit.}
    \renewcommand{\arraystretch}{1.50}
\begin{tabular*}{\textheight}{@{\extracolsep{\fill}}cccccccc@{}}\hline
param              &$(0,0)          $&$(\pm 0.2,\mp 0.2)  $&$(\pm 0.4,\mp 0.4)  $&$(\pm 0.6,\mp 0.6)   $&          $ \mathbf{ (\pm 0.8,\mp 0.6)} $&$(\pm 0.8,\mp 0.8) $&$1P$$( 0, 0)   $\\ \hline
$ \Delta g_1^{\gamma}         $&$ _{ -8.5 }^{+5.5 }$&$ _{-7.4 }^{+3.3 }$&$ _{-6.0 }^{+ 2.7 }$&$ _{ -2.7 }^{+2.1 }$&$ \mathbf{_{-2.3 }^{+ 1.7 }} $&$ _{ -2.0 }^{+1.6 }$&$ _{-1.4 }^{+1.3 }$ \\  \hline 
$ g_4^{\gamma}                $&$ _{ -6.0 }^{+5.8 }$&$ _{-5.4 }^{+5.3 }$&$ _{-4.0 }^{+ 4.0 }$&$ _{ -3.0 }^{+3.0 }$&$ \mathbf{_{-2.5 }^{+ 2.5 }} $&$ _{ -2.2 }^{+2.2 }$&$ _{-1.9 }^{+1.9 }$ \\  \hline 
$ g_5^{\gamma}                $&$ _{ -6.1 }^{+6.1 }$&$ _{-5.2 }^{+5.1 }$&$ _{-3.1 }^{+ 2.6 }$&$ _{ -2.0 }^{+1.4 }$&$ \mathbf{_{-1.6 }^{+ 1.1 }} $&$ _{ -1.4 }^{+1.0 }$&$ _{-2.0 }^{+1.9 }$ \\  \hline 
$ \lambda^{\gamma}           $&$ _{ -1.8 }^{+1.4 }$&$ _{-1.6 }^{+1.2 }$&$ _{-1.2 }^{+ 1.2 }$&$ _{ -0.68}^{+1.0}$&$ \mathbf{_{-0.61}^{+ 0.89}} $&$ _{ -0.57}^{+0.81}$&$ _{-1.1 }^{+0.77}$ \\  \hline 
$ \widetilde{\lambda^{\gamma}}$&$ _{ -1.6 }^{+1.6 }$&$ _{-1.4 }^{+1.4 }$&$ _{-1.1 }^{+ 1.1 }$&$ _{ -0.88}^{+0.88}$&$ \mathbf{_{-0.82}^{+ 0.82}} $&$ _{ -0.78}^{+0.77}$&$ _{-1.0 }^{+1.0 }$ \\  \hline 
$  \Delta\kappa^{\gamma}      $&$ _{ -5.7 }^{+0.91}$&$ _{-4.4 }^{+0.32}$&$ _{-4.3 }^{+ 0.46}$&$ _{ -0.69}^{+0.28}$&$ \mathbf{_{-0.55}^{+ 0.27}} $&$ _{ -0.48}^{+0.25}$&$ _{-0.34}^{+0.33}$ \\  \hline 
$ \widetilde{\kappa^{\gamma}} $&$ _{ -6.0 }^{+6.1 }$&$ _{-5.2 }^{+5.2 }$&$ _{-3.9 }^{+ 4.0 }$&$ _{ -3.0 }^{+2.9 }$&$ \mathbf{_{-2.6 }^{+ 2.6 }} $&$ _{ -2.3 }^{+2.3 }$&$ _{-2.4 }^{+2.3 }$ \\  \hline 
$ \Delta g_1^Z                $&$ _{ -3.7 }^{+7.2 }$&$ _{-2.8 }^{+5.6 }$&$ _{-2.6 }^{+ 4.5 }$&$ _{ -2.0 }^{+2.1 }$&$ \mathbf{_{-1.7 }^{+ 1.8 }} $&$ _{ -1.5 }^{+1.6 }$&$ _{-1.3 }^{+1.3 }$ \\  \hline 
$ g_4^Z                       $&$ _{ -4.7 }^{+4.8 }$&$ _{-4.3 }^{+4.3 }$&$ _{-3.3 }^{+ 3.3 }$&$ _{ -2.5 }^{+2.5 }$&$ \mathbf{_{-2.2 }^{+ 2.2 }} $&$ _{ -2.0 }^{+2.0 }$&$ _{-1.4 }^{+1.4 }$ \\  \hline 
$ g_5^Z                       $&$ _{ -4.8 }^{+4.7 }$&$ _{-4.1 }^{+4.0 }$&$ _{-2.3 }^{+ 2.1 }$&$ _{ -1.5 }^{+1.3 }$&$ \mathbf{_{-1.3 }^{+ 1.0 }} $&$ _{ -1.2 }^{+0.86}$&$ _{-1.3 }^{+1.2 }$ \\  \hline 
$ \lambda^Z                   $&$ _{ -1.5 }^{+1.1 }$&$ _{-1.3 }^{+1.0 }$&$ _{-1.1 }^{+ 0.80}$&$ _{ -0.94}^{+0.49}$&$ \mathbf{_{-0.83}^{+ 0.47}} $&$ _{ -0.76}^{+0.44}$&$ _{-0.57}^{+0.56}$ \\  \hline 
$ \widetilde{\lambda^Z}       $&$ _{ -1.3 }^{+1.3 }$&$ _{-1.1 }^{+1.1 }$&$ _{-0.90}^{+ 0.90}$&$ _{ -0.77}^{+0.77}$&$ \mathbf{_{-0.73}^{+ 0.73}} $&$ _{ -0.68}^{+0.68}$&$ _{-0.56}^{+0.57}$ \\  \hline 
$ \Delta\kappa^Z              $&$ _{ -1.5 }^{+3.6 }$&$ _{-0.49}^{+3.2 }$&$ _{-0.44}^{+ 3.1 }$&$ _{ -0.38}^{+0.56}$&$ \mathbf{_{-0.35}^{+ 0.43}} $&$ _{ -0.32}^{+0.36}$&$ _{-0.48}^{+0.43}$ \\  \hline 
$ \widetilde{\kappa^Z}        $&$ _{ -5.0 }^{+4.7 }$&$ _{-4.2 }^{+4.2 }$&$ _{-3.3 }^{+ 3.3 }$&$ _{ -2.5 }^{+2.5 }$&$ \mathbf{_{-2.2 }^{+ 2.2 }} $&$ _{ -2.0 }^{+2.1 }$&$ _{-1.5 }^{+1.5 }$ \\  \hline
\end{tabular*}
\end{sidewaystable}
\begin{sidewaystable}
\centering
\caption{\label{tab:Limits-Op}The list of  posterior   $95~\%$ BCI
of anomalous couplings $c_i^{\cal O}$ (TeV$^{-2}$)  of effective operators  in Eq.~(\ref{eq:opertaors-dim6}) and their translated limits
on the couplings $c_i^{{\cal L}_g}$ ($10^{-2}$)
for $\sqrt{s}=500$ GeV, ${\cal L}=100$ fb$^{-1}$ in
{\tt Binned} case for a chosen set of longitudinal beam polarizations $\eta_3$ and $\xi_3$ from MCMC.
The pictorial visualisations for these $95~\%$ BCI of $c_i^{\cal O}$ and $c_i^{{\cal L}_g}$ are shown in   Fig.~\ref{fig:Limits-combined} in 
{\em right-top} and {\em right-bottom} panel, respectively.  Rest details are same 
as in Table~\ref{tab:Limits-Lag}. }
\renewcommand{\arraystretch}{1.50}
\begin{tabular*}{\textheight}{@{\extracolsep{\fill}}cccccccc@{}}\hline
param               &$(0,0)          $&$(\pm 0.2,\mp 0.2)  $&$(\pm 0.4,\mp 0.4)  $&$(\pm 0.6,\mp 0.6)   $                    &$\mathbf{ (\pm 0.8,\mp 0.6)  }$&$(\pm 0.8,\mp 0.8) $&$1P$$( 0, 0)   $\\ \hline
$\frac{c_{WWW}}{\Lambda^2}            $&$ _{ -1.9 }^{+1.3 }$&$ _{ -1.4 }^{+ 1.2 }$&$ _{ -1.1 }^{+1.2 }$&$ _{-0.96 }^{+1.1  }$&$\mathbf{ _{ -1.0  }^{+1.1  }}$&$ _{ -0.94 }^{+ 1.0  }$&$ _{-0.97}^{+ 0.84 }$ \\ \hline
$\frac{c_{W}}{\Lambda^2}              $&$ _{ -1.4 }^{+5.0 }$&$ _{ -1.1 }^{+ 4.6 }$&$ _{ -0.86}^{+0.83}$&$ _{-0.72 }^{+0.58 }$&$\mathbf{ _{ -0.73 }^{+0.60 }}$&$ _{ -0.63 }^{+ 0.55 }$&$ _{-0.58}^{+ 0.55 }$ \\ \hline
$\frac{c_{B}}{\Lambda^2}              $&$ _{-23.7 }^{+2.7 }$&$ _{-20.2 }^{+ 1.9 }$&$ _{ -1.3 }^{+0.98}$&$ _{-0.75 }^{+0.62 }$&$\mathbf{ _{ -0.64 }^{+0.56 }}$&$ _{ -0.53 }^{+ 0.47 }$&$ _{-1.3 }^{+ 1.2 }$ \\ \hline
$\frac{c_{\widetilde{WWW}}}{\Lambda^2}$&$ _{ -1.4 }^{+1.4 }$&$ _{ -1.1 }^{+ 1.1 }$&$ _{ -0.97}^{+0.97}$&$ _{-0.93 }^{+0.94 }$&$\mathbf{ _{ -0.90 }^{+0.91 }}$&$ _{ -0.87 }^{+ 0.87 }$&$ _{-0.98}^{+ 0.97 }$ \\ \hline
$\frac{c_{\widetilde{W}}}{\Lambda^2}  $&$ _{-12.0 }^{+2.1 }$&$ _{-10.0 }^{+ 9.8 }$&$ _{ -6.7 }^{+6.6 }$&$ _{-4.1  }^{+4.2  }$&$\mathbf{ _{ -3.2  }^{+3.2  }}$&$ _{ -2.6  }^{+ 2.6  }$&$ _{-9.9 }^{+ 10.1 }$ \\ \hline  \hline  
$\lambda^{V}                 $&$ _{-0.79 }^{+ 0.52 }$&$ _{-0.58 }^{+0.50 }$&$ _{-0.46 }^{+ 0.49 }$&$ _{-0.40 }^{+0.46 }$&$ \mathbf{_{-0.41 }^{+ 0.45 }}$&$ _{ -0.39 }^{+0.42 }$&$ _{-0.40 }^{+0.35 }$\\  \hline 
$\widetilde{\lambda^{V}}     $&$ _{-0.60 }^{+ 0.60 }$&$ _{-0.45 }^{+0.44 }$&$ _{-0.40 }^{+ 0.40 }$&$ _{-0.38 }^{+0.39 }$&$ \mathbf{_{-0.37 }^{+ 0.37 }}$&$ _{ -0.36 }^{+0.36 }$&$ _{-0.41 }^{+0.40 }$\\  \hline 
$\Delta\kappa^{\gamma}       $&$ _{-6.4  }^{+ 0.52 }$&$ _{-5.1  }^{+0.44 }$&$ _{-0.38 }^{+ 0.28 }$&$ _{-0.32 }^{+0.24 }$&$ \mathbf{_{-0.32 }^{+ 0.25 }}$&$ _{ -0.28 }^{+0.23 }$&$ _{-0.61 }^{+0.56 }$\\  \hline 
$\widetilde{\kappa^{\gamma}} $&$ _{-3.9  }^{+ 3.9  }$&$ _{-3.2  }^{+3.2  }$&$ _{-2.1  }^{+ 2.1  }$&$ _{-1.3  }^{+1.3  }$&$ \mathbf{_{-1.0  }^{+ 1.0  }}$&$ _{ -0.84 }^{+0.84 }$&$ _{-3.2}^{+3.2}$\\  \hline 
$\Delta g_1^Z                $&$ _{-0.59 }^{+ 2.1  }$&$ _{-0.45 }^{+1.9  }$&$ _{-0.36 }^{+ 0.34 }$&$ _{-0.30 }^{+0.24 }$&$ \mathbf{_{-0.30 }^{+ 0.25 }}$&$ _{ -0.26 }^{+0.23 }$&$ _{-0.24 }^{+0.23 }$\\  \hline 
$\Delta\kappa^Z              $&$ _{-0.73 }^{+ 3.6  }$&$ _{-0.45 }^{+3.2  }$&$ _{-0.33 }^{+ 0.34 }$&$ _{-0.24 }^{+0.21 }$&$ \mathbf{_{-0.24 }^{+ 0.21 }}$&$ _{ -0.20 }^{+0.19 }$&$ _{-0.30 }^{+0.30 }$\\  \hline 
$\widetilde{\kappa^Z}        $&$ _{-1.1  }^{+ 1.1  }$&$ _{-0.91 }^{+0.92 }$&$ _{-0.61 }^{+ 0.62 }$&$ _{-0.38 }^{+0.38 }$&$ \mathbf{_{-0.30 }^{+ 0.29 }}$&$ _{ -0.24 }^{+0.24 }$&$ _{-0.93 }^{+0.92 }$\\  \hline 
\end{tabular*}                                                                                                             
\end{sidewaystable}
\begin{figure}[h!]
    \centering
    \includegraphics[width=0.9\textwidth]{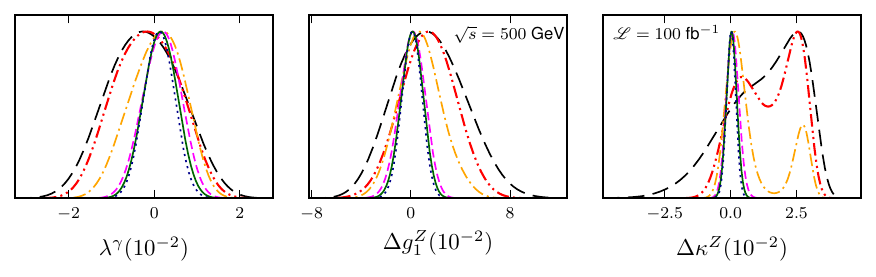}
    \includegraphics[width=0.325\textwidth]{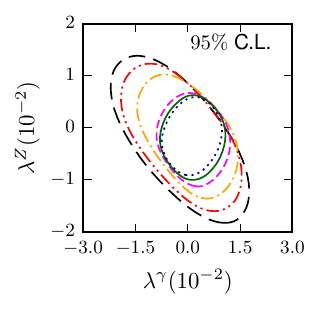}
    \includegraphics[width=0.325\textwidth]{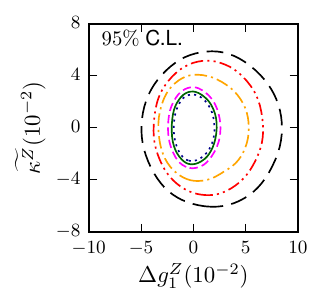}
    \includegraphics[width=0.325\textwidth]{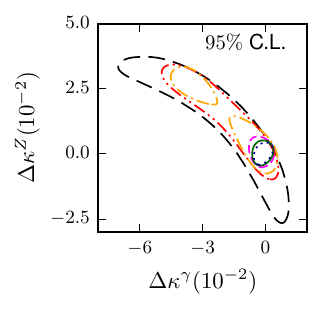}
    \caption{\label{fig:MCMC-beamPol-Lag} The marginalised $1D$ projections for the couplings
        $\lambda^\gamma$, $\Delta g_1^Z$ and $\Delta\kappa^Z$ in the {\em top-panel} 
        and $2D$ projection at $95~\%$ C.L. on 
        $\lambda^\gamma$--$\lambda^Z$, $\Delta g_1^Z$--$\wtil{\kappa^Z}$  and   
        $\Delta\kappa^\gamma$--$\Delta\kappa^Z$ planes in {\em bottom-panel}     
        from MCMC for a set of choice of beam polarizations are shown
    for $\sqrt{s}=500$ GeV, ${\cal L}=100$ fb$^{-1}$ using the binned observables in 
    the effective vertex formalism. The legend labels are same as in Figs.~\ref{fig:Limits-combined} \&~\ref{fig:MCMC-BeamPol-Op}. } 
\end{figure}
\begin{figure}[t]
    \includegraphics[width=\textwidth]{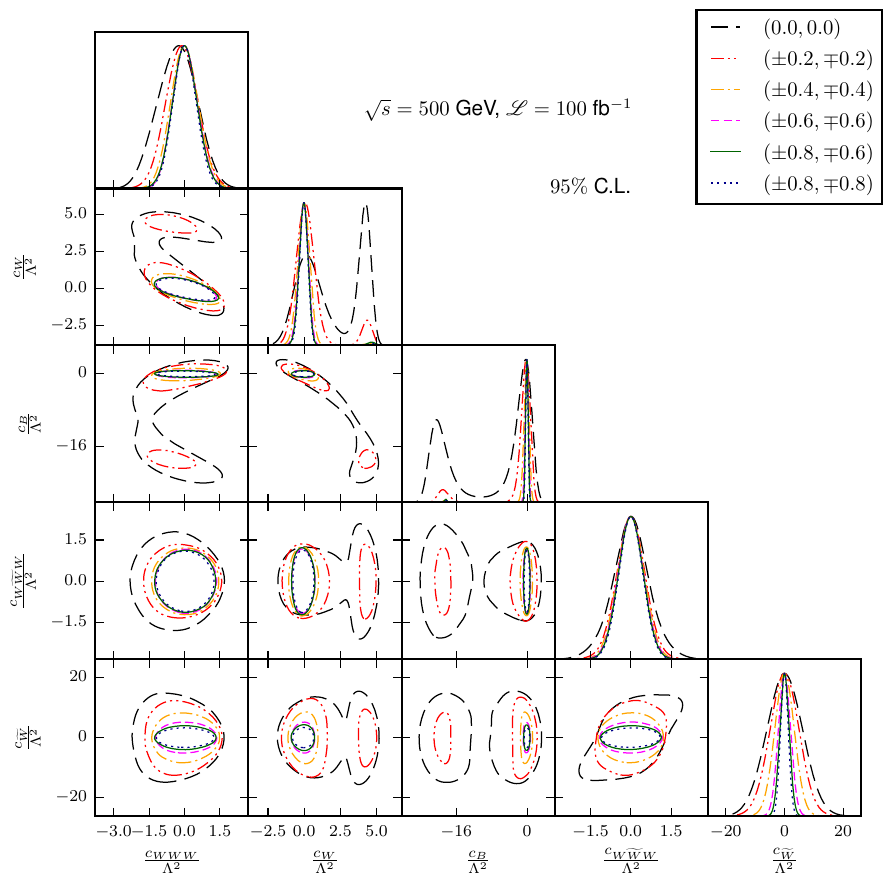}
    \caption{\label{fig:MCMC-BeamPol-Op} All the  marginalised $1D$ projections and $2D$ projections at $95~\%$ C.L. from MCMC in triangular array
    for the effective operators (TeV$^{-2}$) for a set of choice of beam polarizations
    for $\sqrt{s}=500$ GeV, ${\cal L}=100$ fb$^{-1}$ using the binned observables. } 
\end{figure}

To show the effect of beam polarizations, the marginalised $1D$ projections for the couplings
$\lambda^\gamma$, $\Delta g_1^Z$ and $\Delta\kappa^Z$ as well as $2D$ projections at $95~\%$ C.L. on 
$\lambda^\gamma$--$\lambda^Z$, $\Delta g_1^Z$--$\wtil{\kappa^Z}$ and   
$\Delta\kappa^\gamma$--$\Delta\kappa^Z$  planes are shown in Fig.~\ref{fig:MCMC-beamPol-Lag} for form factors ($c_i^{\cal L}$) as representative. 
We observe that as the amplitude of beam
polarizations are increased from $(0,0)$ to $(\pm 0.8, \mp 0.8)$, the contours get smaller centred 
around the SM values in the $2D$ projections which are reflected in the $1D$ projections as well.
In the $\Delta\kappa^\gamma$--$\Delta\kappa^Z$ plane, the contour gets divided into two parts 
at $(\pm 0.4, \mp 0.4)$ and then become one single contour later  centred around the SM values. 
In the case of effective operators ($c_i^{\cal O} $), all the $1D$ and  $2D$ ($95~\%$ C.L.)  projections 
after marginalization are shown in Fig.~\ref{fig:MCMC-BeamPol-Op}. In this case the couplings
$c_W$ and $c_B$ has two patches up-to beam polarization $(\pm 0.2, \mp 0.2)$  and become one single
patch starting at beam polarization $(\pm 0.3, \mp 0.3)$ centred around the SM values. As the amplitude of beam
polarizations are increased along the $\eta_3=-\xi_3$ line, the measurement of the anomalous couplings
gets improved. The set of beam polarizations chosen here are mostly  along the $\eta_3=-\xi_3$ line, but some
choices off to the line might provide the same results. A discussion on the choice of beam polarization is given in the next subsection.
\subsection{On the choice of beam polarizations}\label{sec:3.3}
\begin{figure}[h!]
    \centering
    \includegraphics[width=0.6\textwidth]{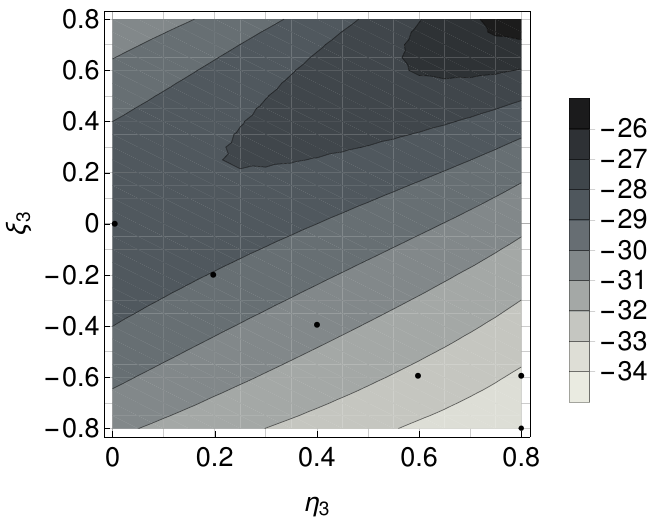}
    \caption{\label{fig:averageLikelihood} The  averaged likelihood 
        $L_{Av}=L(V_{\vec{f}}~;\eta_3,\xi_3)$ 
        in log scale as a function of $(\pm\eta_3 ,\pm\xi_3)$ in the
         effective vertex formalism for $\sqrt{s}=500$ GeV, ${\cal L}=100$ fb$^{-1}$.}
\end{figure}
In the previous subsection, we found that the beam polarization choice $(\pm\eta_3 ,\pm\xi_3)=(\pm 0.8, \pm 0.6)$ is the best choice of beam
polarizations to provide simultaneous limits on the anomalous couplings obtained by MCMC analysis.
Here, we discuss the average likelihood  or the
weighted volume of the parameter space defined as~\cite{Rahaman:2017qql},
\begin{eqnarray}\label{eq:average_likelihood}
L(V_{\vec{f}}~;\eta_3,\xi_3)&=&\int_{V_{\vec{f}}}  \exp{\left[-\frac{1}{2}\chi_{tot}^2(\vec{f},\eta_3,\xi_3)\right]} d\vec{f}
\end{eqnarray}
to cross-examine the beam polarization choices made in the previous section, as we did in section~\ref{sec:epjc2-average-like-best-choich} for $ZV$ production. 
Here $\vec{f}$ is the couplings vector and $V_{\vec{f}}$ is the volume 
of parameter space over which the average is done; 
$L(V_{\vec{f}}~;\eta_3,\xi_3)$ corresponds to  
 the volume of the parameter space that is statistically consistent with the SM . One naively expects the limits to be tightest when
$L(V_{\vec{f}}~;\eta_3,\xi_3)$ is minimum.
We calculate the above quantity as a function of $(\pm\eta_3,\pm\xi_3)$ 
for  {\tt Binned} case in the
effective vertex formalism given in Lagrangian in Eq.~(\ref{eq:WW-LagWWV}) and  present it in Fig.~\ref{fig:averageLikelihood}.
As the opposite beam polarizations are combined, only the half-portion
are shown in the $\eta_3$--$\xi_3$ plane. The dot ({\tiny$\bullet$}) points along the $\eta_3=-\xi_3$ are the 
chosen choices of beam polarizations for the MCMC analysis.
We see that the average likelihood decreases along $\eta_3=-\xi_3$ line while it increases
along $\eta_3=\xi_3$ line. The constant lines or contours of average likelihood in the figure imply that any beam polarizations
along the lines/contours will provide the similar shape of $1D$ and $2D$ projections of couplings and  their limits. For example, 
the point $(\pm 0.8,\mp 0.6)$ is equivalent to the point $(\pm 0.7,\mp 0.7)$ as well as 
$(\pm 0.6,\pm 0.8)$ roughly in providing simultaneous limits which are verified from the limits obtained by MCMC analysis.
It is clear that the polarization $(\pm 0.8,\mp 0.6)$ is indeed the best choice to provide simultaneous limits
on the anomalous couplings within the achievable range.


\section{Summary}\label{sec:WW-conclusion}
In summary, here, we studied the anomalous triple gauge boson couplings in 
$e^+e^-\to  W^+W^-$ with longitudinally polarized beams using    $W$ boson polarization observables
together with the total cross section and the forward-backward asymmetry
for $\sqrt{s}=500$ GeV and  luminosity of ${\cal L}=100$ fb$^{-1}$.
We estimated simultaneous limits on all the couplings 
for several chosen set of beam polarizations in both effective vertex formalism and
effective operator approach. The limits on the couplings are tighter when  $SU(2)\times U(1)$ symmetry is assumed. 
We showed the consistency between
the best choice of beam polarizations and minimum likelihood averaged over the anomalous
couplings. The extreme beam polarization $(\pm 0.8,\mp 0.8)$ appears to be the best to provide the tightest
constraint on the anomalous couplings in both approaches at the ILC (same as obtained for $e^+e^-\to ZV$ in Chapter~\ref{chap:epjc2}).
Our one parameter limits with unpolarized beams and simultaneous limits
for best polarization choice are much better
than the one parameter limits available from experiment, see Table~\ref{tab:Limits-Op}.

\chapter{The probe of aTGC in $W^\pm Z$ productions at the LHC  and the role of $Z/W$ boson  polarizations}\label{chap:WZatLHC}
\begingroup
\hypersetup{linkcolor=blue}
\minitoc
\endgroup
{\small\textit{\textbf{ The contents in this chapter are based on the published article in Ref.~\cite{Rahaman:2019lab}. }}}
\vspace{1cm}

In the previous chapter, the polarization asymmetries are shown to give  promising results
in probing the aTGC in the charge sector in a future $e^+$-$e^-$ collider. 
In this chapter,
we want to probe the aTGC in the charge sector at the current hadron collider LHC in the $W^\pm Z$ production processes in $3l+\cancel{E}_T$ channel.
The anomalous couplings appearing in the $W^\pm Z$ production at the LHC are 
\begin{eqnarray}
{\cal L}_{WWZ} &=&ig_{WWZ}\Big[\left(1+\Delta g_1^Z\right)(W_{\mu\nu}^+W^{-\mu}-
W^{+\mu}W_{\mu\nu}^-)Z^\nu
+\frac{\lambda^Z}{m_W^2}W_\mu^{+\nu}W_\nu^{-\rho}Z_\rho^{\mu}\nonumber\\
&+&\frac{\wtil{\lambda^Z}}{m_W^2}W_\mu^{+\nu}W_\nu^{-\rho}\wtil{Z}_\rho^{\mu}
+\left(1+\Delta \kappa^Z\right) W_\mu^+W_\nu^-Z^{\mu\nu}+\wtil{\kappa^Z}W_\mu^+W_\nu^-\wtil{Z}^{\mu\nu}
\Big] 
\label{eq:WZ-LagWWZ}
\end{eqnarray}
containing half ($7$) the couplings of the full $WWV$ Lagrangian in Eq.~(\ref{eq:WW-LagWWV}).
There has been a lot of studies of these aTGC at the LHC~\cite{Baur:1987mt,Dixon:1999di,Falkowski:2016cxu,Azatov:2017kzw,Azatov:2019xxn,Bian:2015zha,Campanario:2016jbu,Bian:2016umx,Butter:2016cvz,Baglio:2017bfe,Li:2017esm,Bhatia:2018ndx,Chiesa:2018lcs}
in different perspective. Direct measurement of these aTGC at the  LHC~\cite{Aaboud:2017cgf,
    Sirunyan:2017bey,Aaboud:2017fye,Khachatryan:2016poo,
    Aad:2016ett,Aad:2016wpd,Chatrchyan:2013yaa,
    RebelloTeles:2013kdy,ATLAS:2012mec,Chatrchyan:2012bd,Aad:2013izg,
    Chatrchyan:2013fya,Sirunyan:2017jej,Sirunyan:2019gkh,Sirunyan:2019dyi,Sirunyan:2019bez} are also available in different processes using the cross sections with various kinematical cuts.
Our aim, here, is to study these $WWZ$ anomalous couplings in $ZW^\pm$ production at the LHC at 
$\sqrt{s}=13$ TeV using the cross section, forward backward asymmetry and 
polarizations asymmetries of $Z$ and $W^\pm$ in the $3l+\cancel{E}_T$ channel.
In addition to the vertex form factor in Eq.~(\ref{eq:WZ-LagWWZ}), we will also
probe the effective dimension-$6$ operators given in Eq.~(\ref{eq:WW-LagWWV}) independently.
Similar to the study in the previous chapter, we see the modification of the form factors 
in Eq.~(\ref{eq:WZ-LagWWZ})  subjected to $SU(2)\times U(1)$ gauge invariance through relations given in Eq.~(\ref{eq:intro-Operator-to-Lagrangian}).
The 
polarizations of $W^\pm/Z$  has been estimated earlier in the same process $ZW^\pm$ production that we are looking at~\cite{Stirling:2012zt,Baglio:2018rcu,Baglio:2019nmc} 
and also has been measured recently at the LHC~\cite{Aaboud:2019gxl} in the SM.

The $W^\pm Z$ process in $3l+\cancel{E}_T$ channel has got quite a bit of attention recently
for having excess at the LHC~\cite{Sirunyan:2019bez}. This has been looked as an anomaly and has been addressed in terms of two BSM scalar~\cite{vonBuddenbrock:2019ajh}. 
This final state is also important for various BSM searches, including supersymmetry and dark matter.

\section{Signal cross sections and their sensitivity to anomalous couplings}\label{sec:signal-sigma}
\begin{figure}[t!]
    \centering
    \includegraphics[width=1.0\textwidth]{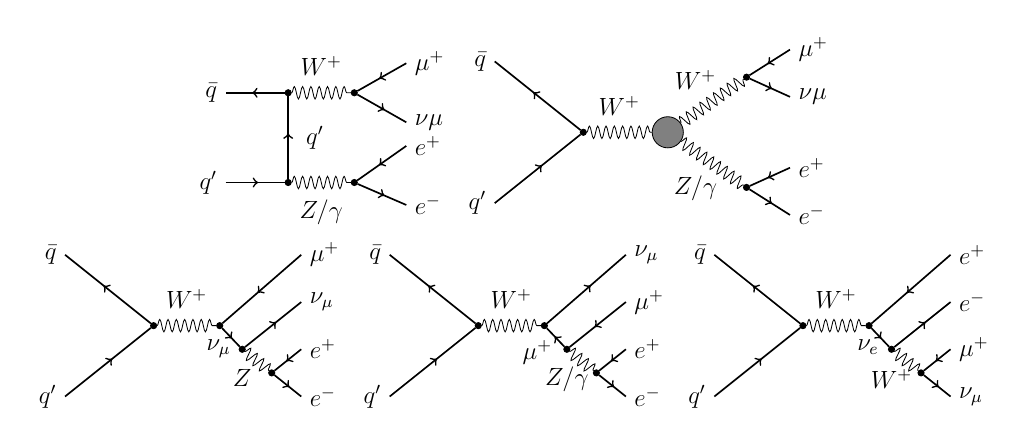}
    \caption{\label{fig:Feynman_WZ_LHC}
Sample of Born level Feynman diagrams for $ZW^+$ production in the $e^+e^-\mu^+\nu_\mu$ channel at the LHC. 
The diagrams for $ZW^-$ can be obtained by charge conjugation. The shaded blob represents the presence of
anomalous $WWV$ couplings on top of SM.}
\end{figure}

The process of interest is the $ZW^\pm$ production  in the $3l+\cancel{E}_T$ channel
at the LHC. The representative Feynman diagrams at Born level are displayed in Fig.~\ref{fig:Feynman_WZ_LHC}
containing doubly-resonant processes ({\em upper-row}) as well as singly-resonant processes ({\em lower-row}).
The presence of anomalous $WWZ$ couplings is shown by the shaded blob. While this may contain the $WW\gamma$
couplings due to the off-shell $\gamma$, this has been cut out by $Z$ selection cuts, described later.
The leading order result ($148.4$ fb estimated by {\tt MATRIX} in Ref.~\cite{Grazzini:2017ckn}) 
for the $3l+\cancel{E}_T$ cross section at the LHC is way below the 
measured cross section at the LHC ($258$ fb measured by CMS~\cite{Khachatryan:2016tgp}). 
Higher-order corrections are thus necessary to add to the tree level result. 
The NLO corrections in QCD appear in the vertices connected to the quarks 
(see, Fig.~\ref{fig:Feynman_WZ_LHC}) with either QCD loops or 
QCD radiations from the quarks.
The SM cross sections of $ZW^\pm$ production  in the $e^+e^-\mu^\pm$ channel  obtained
by {\tt MATRIX} and \MGvATNLO~v2.6.4 ({\tt mg5\_aMC}) for $\sqrt{s}=13$ TeV 
for the CMS fiducial phase-phase region are  presented in the Table~\ref{tab:WZ-sigma-SM}.
The CMS fiducial phase-phase region~\cite{Khachatryan:2016tgp} is given by,
\begin{eqnarray}\label{eq:CMS_fudicial_region}
p_T(l_{Z,1})>20~\text{GeV},~~p_T(l_{Z,2})>10~\text{GeV},~~p_T(l_{W})>20~\text{GeV} ,\nonumber\\
|\eta_l|<2.5,~~60~\text{GeV}<m_{l_Z^+l_Z^-}<120~\text{GeV},~~m_{l^+l^-}>4~\text{GeV} .
\end{eqnarray}
We use the values of the SM input parameters same as used  in Ref.~\cite{Grazzini:2017ckn} 
(default in {\tt MATRIX}). A fixed renormalization ($\mu_R$) and factorization ($\mu_F$) 
scale of $\mu_R=\mu_F=\mu_0=\frac{1}{2}\left (m_Z+m_W \right)$ is used and the uncertainties are estimated
by varying the $\mu_R$ and $\mu_F$ in the range of $0.5\mu_0 ≤ \mu_R , \mu_F ≤ 2\mu_0$ and shown in Table~\ref{tab:WZ-sigma-SM}. We use the NNPDF3.0 sets of parton distribution functions (PDFs) with $\alpha_s(m_Z)$ for our calculations.
\begin{table*}[!ht]\caption{\label{tab:WZ-sigma-SM} The theoretical estimates and experimental measurements of the cross sections of  
        $ZW^\pm$ productions in the $e^+e^-\mu^\pm\nu_\mu/\bar{\nu}_\mu$ channel  at $\sqrt{s}=13$ TeV 
        at the LHC for CMS fiducial phase-space. The uncertainties in the  theoretical
        estimates are due to scale variation.}
    \renewcommand{\arraystretch}{1.70}
    \begin{tabular*}{\columnwidth}{@{\extracolsep{\fill}}lllll@{}} \hline
        Process & Obtained at & $\sigma_{\text{LO}}$ (fb)& $\sigma_{\text{NLO}}$ (fb) & $\sigma_{\text{NNLO}}$ (fb) \\ \hline
\multirow{2}{*}{$pp\to e^+e^-\mu^+\nu_\mu$} &{\tt MATRIX} & $22.08_{-6.2\%}^{+5.2\%}$ & $43.95_{-4.3\%}^{+5.4\%}$ & $48.55_{-2.0\%}^{+2.2\%}$  \\ \cline{2-5}
& {\tt mg5\_aMC} & $22.02_{-7.2\%}^{+6.1\%}$ & $43.63_{-6.6\%}^{+6.6\%}$ & ------  \\ \hline
\multirow{2}{*}{$pp\to e^+e^-\mu^-\bar{\nu}_\mu$} & {\tt MATRIX}& $14.45_{-6.7\%}^{+5.6\%}$ & $30.04_{-4.5\%}^{+5.6\%}$ & $33.39_{-2.1\%}^{+2.3\%}$\\ \cline{2-5}
& {\tt mg5\_aMC}& $14.38_{-7.6\%}^{+6.4\%}$ & $29.85_{-6.8\%}^{+6.8\%}$ & ------\\ \hline
$pp \to 3l+\cancel{E}_T $& {\tt MATRIX}~\cite{Grazzini:2017ckn} &$148.4_{-6.4\%}^{+5.4\%}$ &$301.4_{-4.4\%}^{+5.1\%}$ &$334.3_{-2.1\%}^{+2.3\%}$ \\ \hline\hline
        \end{tabular*}
\begin{tabular*}{\columnwidth}{@{\extracolsep{\fill}}llllll@{}}
$pp \to 3l+\cancel{E}_T$&&~~~~~CMS~\cite{Khachatryan:2016tgp} & \hspace{1.2cm}$258.0\pm 8.1\%$ (stat)$^{+7:4\%}_{-7.7\%}$ (syst)$\pm 3.1$ (lumi)& \\ \hline
\end{tabular*}
\end{table*}
\begin{figure}[h!]
    \centering
    \includegraphics[width=0.495\textwidth]{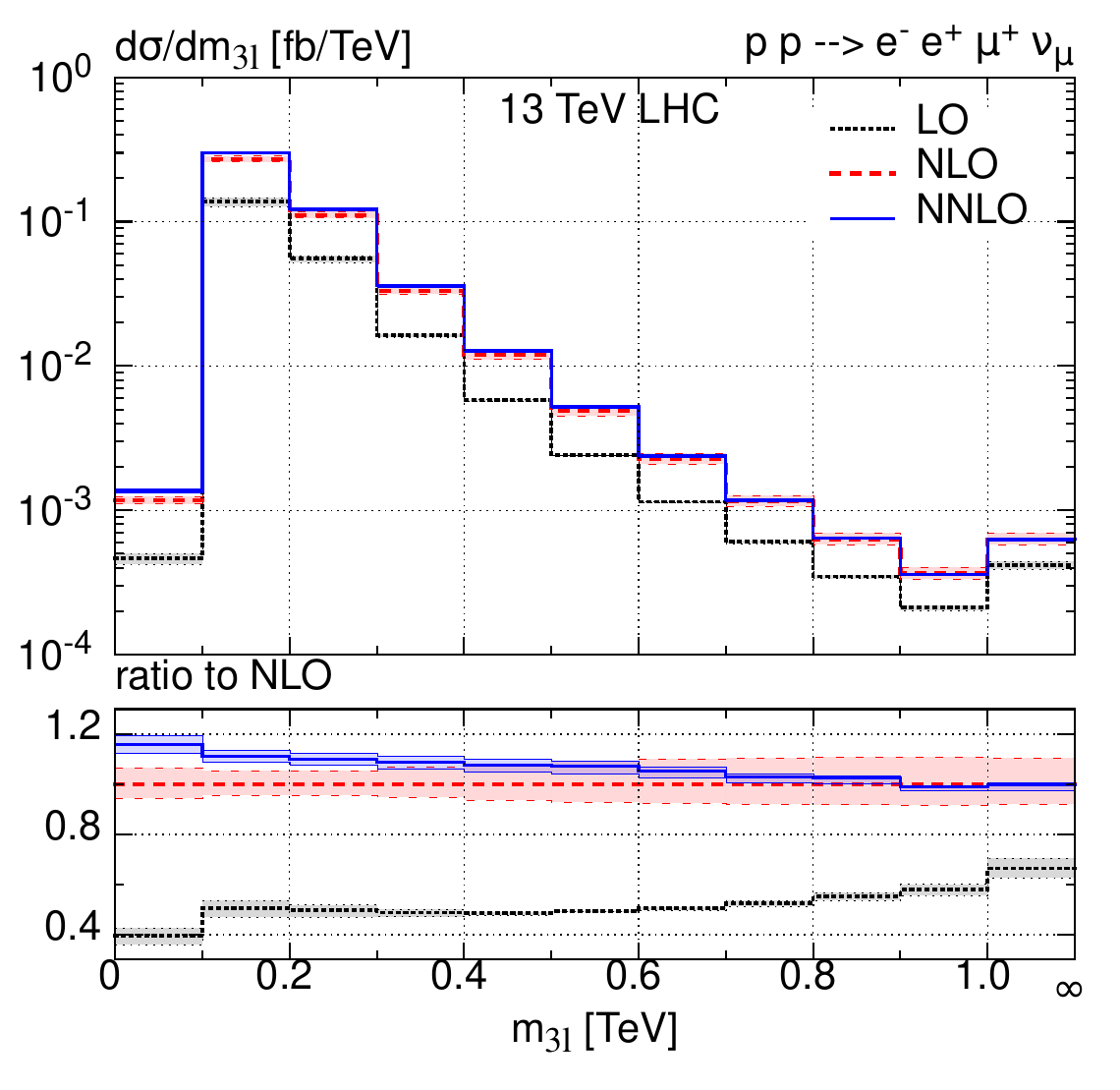}
    \includegraphics[width=0.495\textwidth]{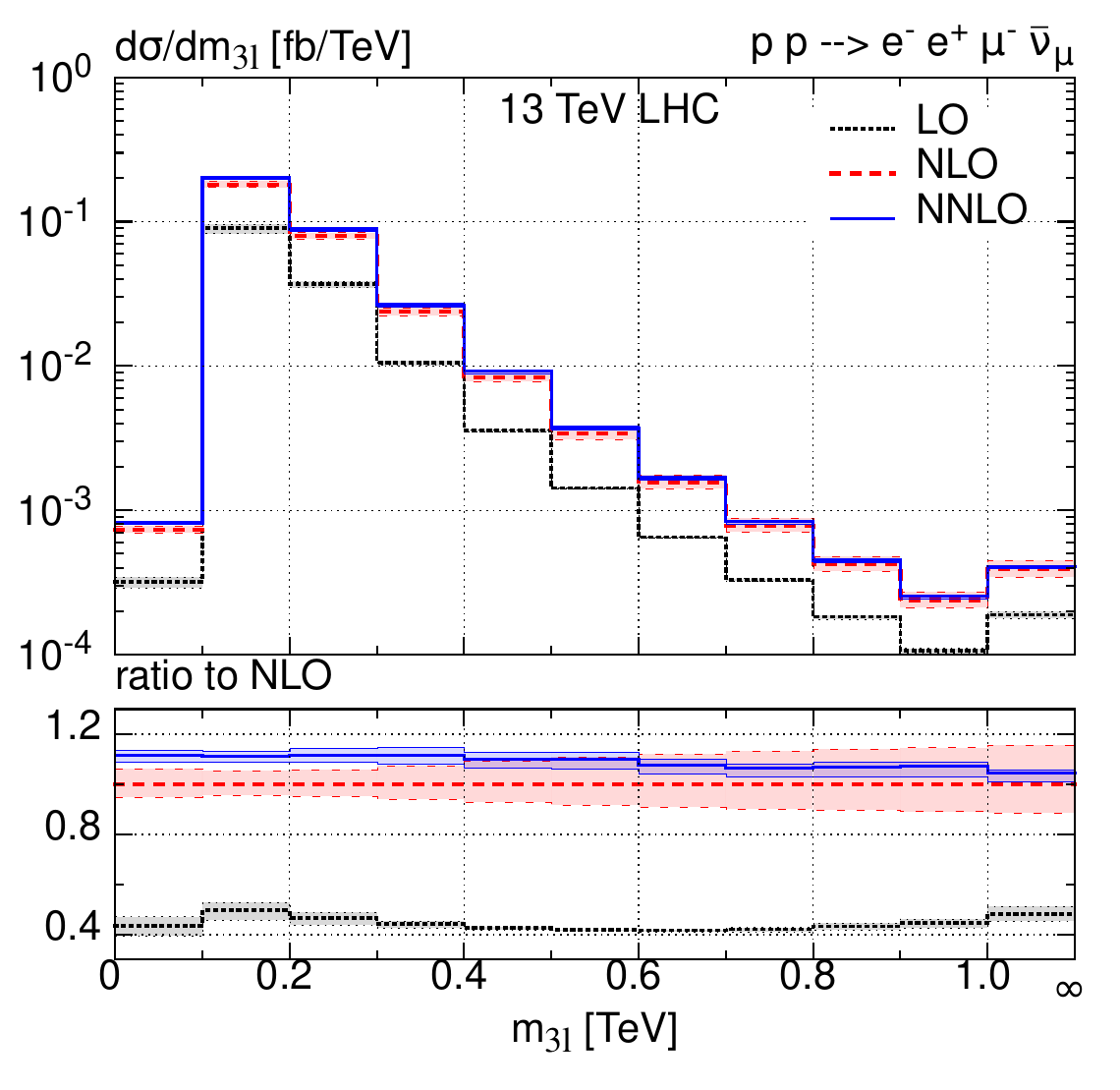}
    \includegraphics[width=0.495\textwidth]{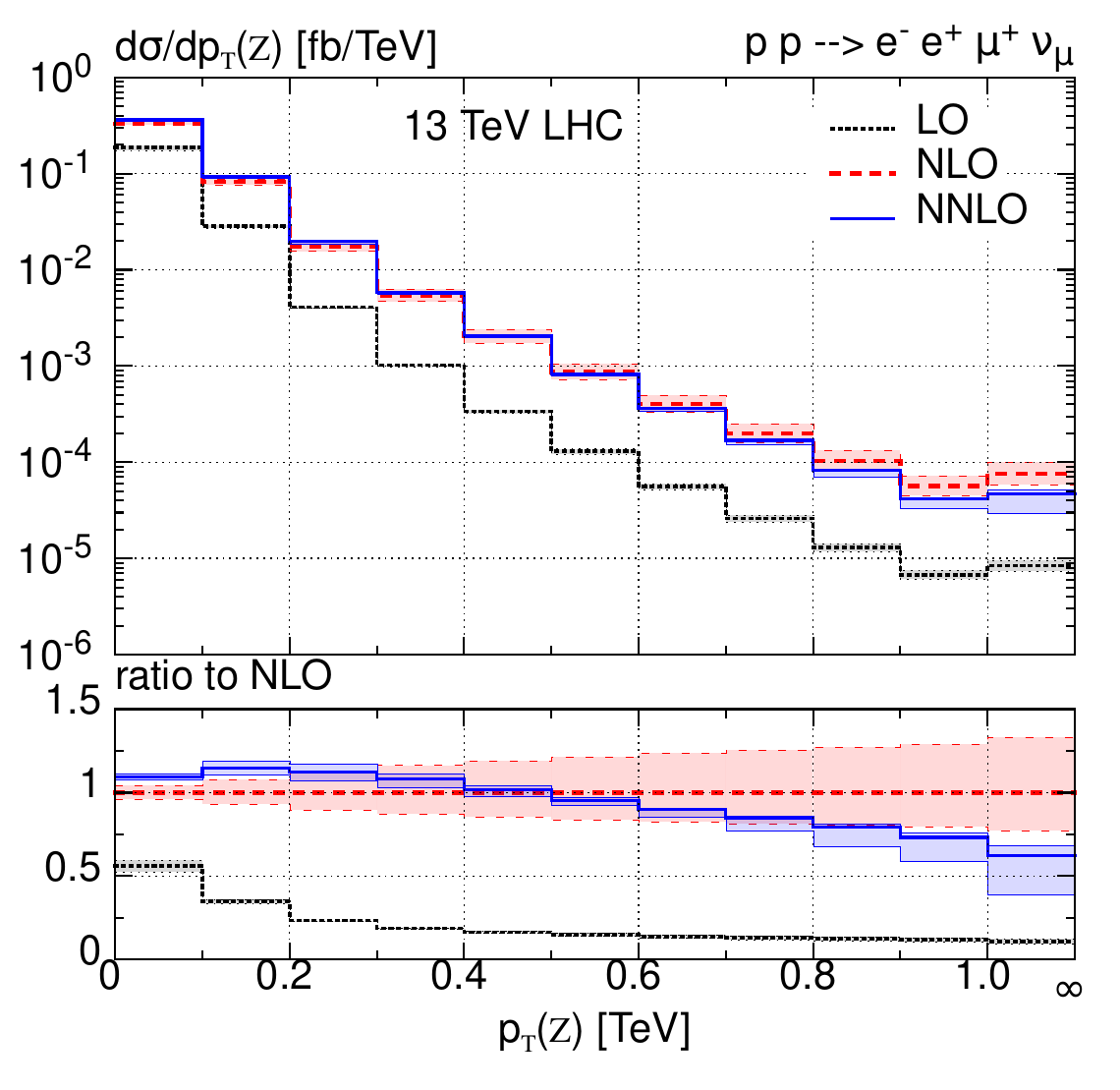}
    \includegraphics[width=0.495\textwidth]{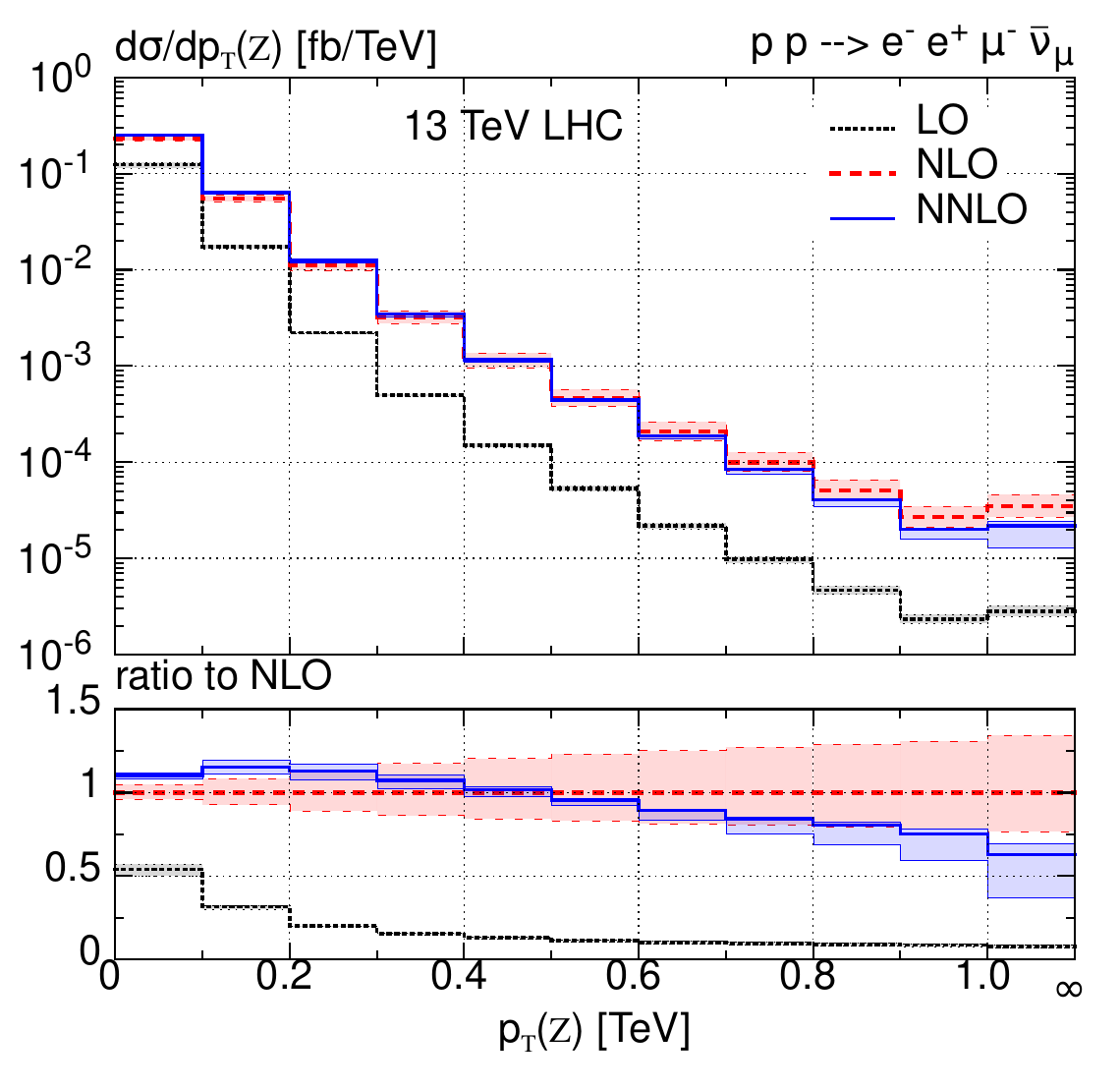}
    \caption{\label{fig:m3l-pTZ-matrx}
        The differential distributions of $m_{3l}$ ({\em top-row}) and $p_T(Z)$ ({\em bottom-row}) in the $ZW^+$ ({\em left-column}) and $ZW^-$ ({\em 
            right-column}) production in the $e^+e^-\mu^\pm+\cancel{E}_T$ channel at the LHC for $\sqrt{s}=13$ TeV at LO, NLO and NNLO  obtained using {\tt 
            MATRIX}~\cite{Grazzini:2016swo,Grazzini:2017ckn,Grazzini:2017mhc,Cascioli:2011va,Denner:2016kdg,Gehrmann:2015ora,Catani:2012qa,Catani:2007vq}
        for CMS fiducial phase-space.
    }
\end{figure}
The combined result for all leptonic channel given in Ref.~\cite{Grazzini:2017ckn} 
and the  measured cross section by CMS~\cite{Khachatryan:2016tgp} are also presented  in  the same table.
The uncertainties  in the  theoretical estimates are due to scale variation.
The result obtained by {\tt MATRIX} and {\tt mg5\_aMC} matches quite well at both LO and NLO level.
The NLO corrections have increased the LO cross section by up to $100~\%$, and the NNLO cross section is further
increased by $10~\%$ from the NLO value. It is thus necessary  to include QCD corrections to leading order result.
The higher order corrections to the cross section vary with kinematical variable like $m_{3l}$ and $p_T(Z)$, as shown
in Fig.~\ref{fig:m3l-pTZ-matrx}  obtained by  {\tt MATRIX}~\cite{Grazzini:2016swo,Grazzini:2017ckn,Grazzini:2017mhc,Cascioli:2011va,Denner:2016kdg,Gehrmann:2015ora,Catani:2012qa,Catani:2007vq}. 
 The lower panels display the respective bin-by-bin ratios to the NLO central predictions. 
 The NLO to LO ratio does not appear to be constant over the range of 
 $m_{3l}$ and $p_T(Z)$. Thus a simple  $k$-factor with LO events can not be used 
 as a proxy for NLO events. 
We use results from {\tt mg5\_aMC} with   NLO  QCD  corrections for our analysis in the rest of the paper. However, the SM values (Table~\ref{tab:WZ-sigma-SM}) and distributions (Fig.~\ref{fig:m3l-pTZ-matrx}) at LO and NLO in QCD are kept for completeness.

\begin{figure}[h!]
    \centering
    \includegraphics[width=0.495\textwidth]{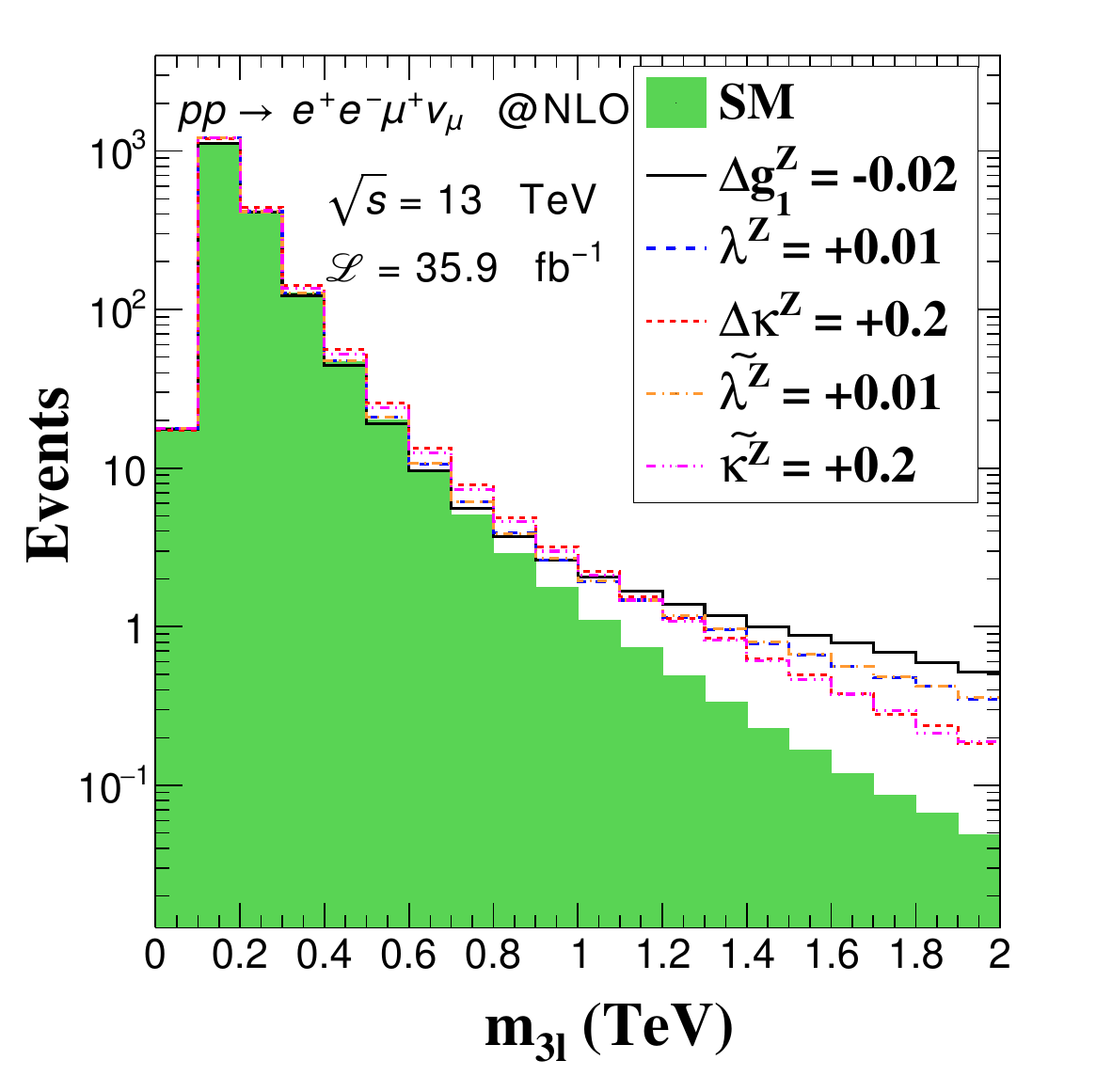}
    \includegraphics[width=0.495\textwidth]{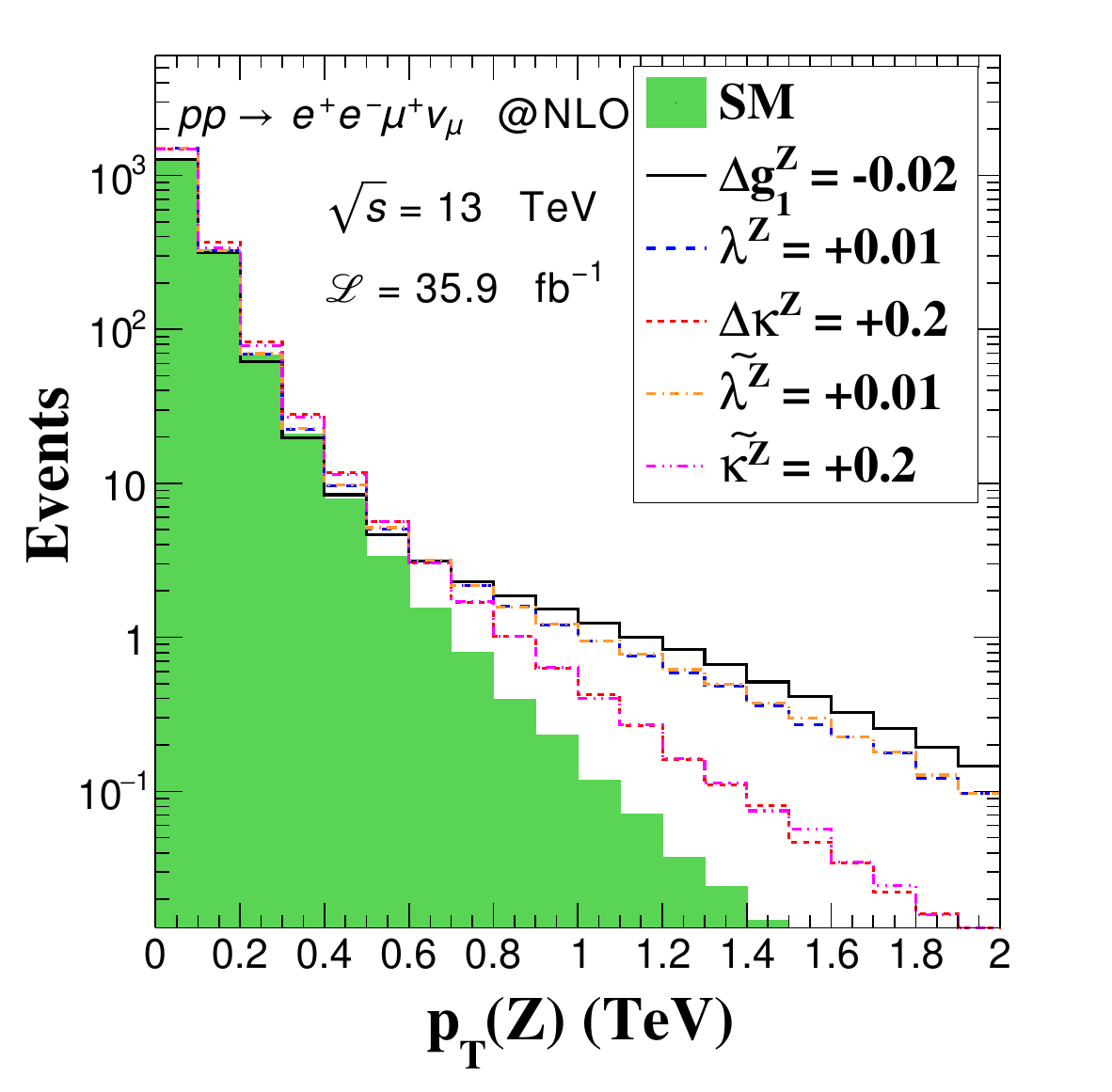}
    \caption{\label{fig:m3l-pTZ-ma5}
        The differential distributions of $m_{3l}$ and $p_T(Z)$ in the $W^+ Z$ production in the $e^+e^-\mu^+\nu_\mu$ channel
        at the LHC at $\sqrt{s}=13$ TeV and ${\cal L}=35.9$ fb$^{-1}$  at NLO in the SM and five benchmark anomalous couplings.
    }
\end{figure}
\begin{figure}[h!]
    \centering
    \includegraphics[width=0.495\textwidth]{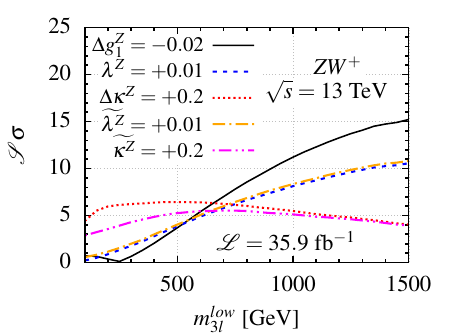}
    \includegraphics[width=0.495\textwidth]{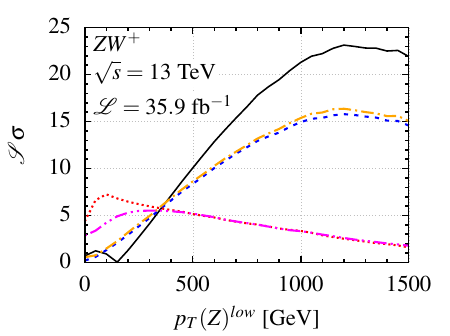}
    \includegraphics[width=0.495\textwidth]{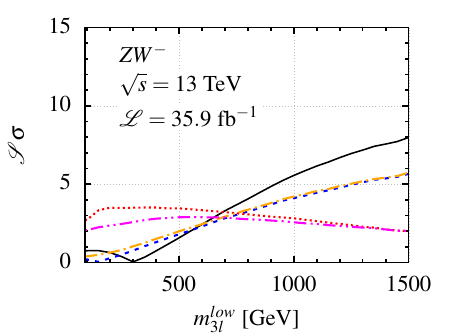}
    \includegraphics[width=0.495\textwidth]{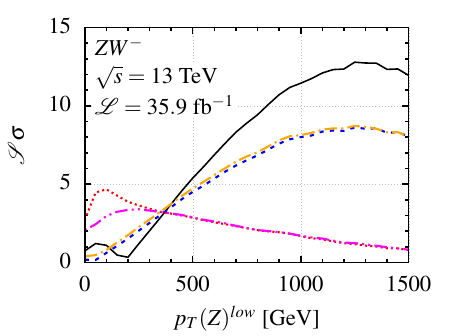}
    \caption{\label{fig:sen-sigma-bench}
        The sensitivity of cross sections  to the five benchmark aTGC as a function of 
        the lower cut on $m_{3l}$ and  $p_T(Z)$ in the $ZW^\pm$ production at the LHC at $\sqrt{s}=13$ TeV and ${\cal L}=35.9$ fb$^{-1}$.
    }
\end{figure}
The signal for the $e^+e^-\mu^+$ and $e^+e^-\mu^-$  are generated  
separately using {\tt mg5\_aMC}  at NLO in QCD for SM as well as  SM including aTGC.
We use the {\tt FeynRules}~\cite{Alloul:2013bka} to generate QCD NLO UFO model of the Lagrangian in Eq.~(\ref{eq:WZ-LagWWZ}) 
for {\tt mg5\_aMC}. These 
signal are then used as a proxy for the $3l+\cancel{E}_T$ final state upto a factor of four for the four channels. 
For these, the $p_T$  cut for $e^\pm$ and $\mu^\pm$ are  kept at the same value, i.e., $p_T(l)>10$ GeV. We use a threshold for the trilepton invariant mass ($m_{3l}$) of $100$ GeV to select the doubly resonant contribution of trilepton final state. We will see later that lower cuts of $m_{3l}$ higher than $100$ GeV are required  for best sensitivity to the anomalous couplings. 
The event selection cuts for this analysis are thus,
\begin{equation}\label{eq:selection-cuts}
p_T(l)>10~\text{GeV},~~|\eta_l|<2.5,~~60~\text{GeV}<m_{l_Z^+l_Z^-}<120~\text{GeV},~~m_{l^+l^-}>4~\text{GeV},~~m_{3l}>100~\text{GeV} .
\end{equation}

 We explore the  effect of aTGC in the distributions  of $m_{3l}$ and 
$p_{T}(Z)$ in both $ZW^+$ and $ZW^-$ production and show them in Fig.~\ref{fig:m3l-pTZ-ma5}.
The  distribution of  $m_{3l}$ in the {\em left-panel} and $p_{T}(Z)$ in the 
{\em right-panel} in the $e^+e^-\mu^+\nu_\mu$ channel are shown  for SM ({\it filled}/green) and five anomalous benchmark
couplings of $\Delta g_1^Z=-0.02$ ({\it solid}/black), $\lambda^Z=+0.01$ ({\it dashed}/blue), $\Delta\kappa^Z=+0.2$ ({\it dotted}/red),
$\wtil{\lambda^Z}=+0.01$ ({\it dash-dotted}/orange)  and $\wtil{\kappa^Z}=+0.2$ ({\em dashed-dotdotted}/ magenta)  
with events normalised to an integrated luminosity of ${\cal L}=35.9$ fb$^{-1}$. 
For each of the benchmark couplings, only one of the couplings get non-zero value at a time while others remain zero. More benchmark scenarios with more than one parameters getting non-zero values at a time are considered for the comparison of  reconstructed neutrino solutions in section~\ref{sec:Pol-Asym}.
The higher $m_{3l}$ 
and higher $p_T(Z)$ seems to have higher sensitivity to the anomalous couplings
which is due to higher momentum transfer at higher energies, for example see Ref.~\cite{Rahaman:2019mnz}. 
We studied the sensitivities (see Eq.~(\ref{eq:sensitivity}) of the cross sections to the  anomalous 
couplings by varying lower cuts on $m_{3l}$  and  $p_T(Z)$  for the above mentioned five benchmark scenarios. 
The sensitivity of the cross sections, ignoring the systematic uncertainty, for the five benchmark cases  (as used in Fig.~\ref{fig:m3l-pTZ-ma5}) are shown in Fig.~\ref{fig:sen-sigma-bench}
 for $ZW^+$ in the {\em upper-row} and for   $ZW^-$ in the {\em lower-row} as a function
 of lower cut  of $m_{3l}$ ({\em left-column}) and  $p_T(Z)$  ({\em right-column}) for luminosity of ${\cal L}=35.9$ fb$^{-1}$. 
 It is clear that the sensitivities increase as the cut  increases  for both 
$m_{3l}$ and  $p_T(Z)$ for couplings $\Delta g_1^Z$, $\lambda^Z$ and $\wtil{\lambda^Z}$,
 while they  decrease just after $\sim 150$ GeV of cuts for the couplings $\Delta\kappa^Z$ and $\wtil{\kappa^Z}$. This can also be seen in Fig.~\ref{fig:m3l-pTZ-ma5} where
 $\Delta\kappa^Z$ and $\wtil{\kappa^Z}$ contribute more than other three couplings for $m_{3l}<0.8$ TeV and $p_T(Z)<0.6$ TeV.
Taking hints from Fig.~\ref{fig:sen-sigma-bench}, we identify four bins in $m_{3l}$-$p_T(Z)$ plane
to maximize the sensitivity of all the couplings. These four bins are given by,
\begin{eqnarray}\label{eq:sigma-twobin}
Bin_{11} &:& 400~\text{GeV}<m_{3l}<1500~\text{GeV},~200~\text{GeV}<p_T(Z)<1200~\text{GeV} ,\nonumber\\
Bin_{12} &:& 400~\text{GeV}<m_{3l}<1500~\text{GeV},~p_T(Z)>1200~\text{GeV} ,\nonumber\\
Bin_{21} &:& m_{3l}>1500~\text{GeV},~200~\text{GeV}<p_T(Z)<1200~\text{GeV} ,\nonumber\\
Bin_{22} &:& m_{3l}>1500~\text{GeV},~p_T(Z)>1200~\text{GeV} .
\end{eqnarray}
The sensitivities of the cross sections to the benchmark anomalous couplings  are 
 calculated in the said four bins for luminosity of ${\cal L}=35.9$ 
 fb$^{-1}$ and they are shown  in Table~\ref{tab:sen-sigma-bench-twobin} in both 
$ZW^+$ and $ZW^-$ productions. 
As expected, we see that  $Bin_{22}$ has the higher sensitivity to couplings $\Delta g_1^Z$, $\lambda^Z$ and $\wtil{\lambda^Z}$, 
while $Bin_{11}$ has higher, but comparable sensitivity to couplings $\Delta\kappa^Z$ 
and $\wtil{\kappa^Z}$. 
The simultaneous cuts on both the variable have 
increased the sensitivity by a significant amount
as compared to the individual cuts. For example, the Fig.~\ref{fig:sen-sigma-bench} 
shows that cross section in $ZW^+$ 
has a maximum sensitivity of $15$ and $22$  on $\Delta g_1^Z = -0.02$  for individual $m_{3l}$  
and  $p_T(Z)$ lower cuts, respectively.  While imposing a simultaneous lower cuts on both the
variable,  the same sensitivity increases to $44.5$ (in $Bin_{22}$).
\begin{table}\caption{\label{tab:sen-sigma-bench-twobin} The sensitivity of the cross sections  on the five benchmark aTGC in 
the four bins (see Eq.~(\ref{eq:sigma-twobin})) of $m_{3l}$ and  $p_T(Z)$ in the $ZW^\pm$ 
productions at the LHC at $\sqrt{s}=13$ TeV and ${\cal L}=35.9$ fb$^{-1}$. }
\renewcommand{\arraystretch}{1.50}
\begin{tabular*}{\textwidth}{@{\extracolsep{\fill}}|c|cccc|cccc|@{}}\hline
& \multicolumn{4}{c|}{$ZW^+$}& \multicolumn{4}{c|}{$ZW^-$} \\ \hline
aTGC                           & $Bin_{11}$&$Bin_{12}$ &$Bin_{21}$&$Bin_{22}$ & $Bin_{11}$&$Bin_{12}$ &$Bin_{21}$ & $Bin_{22}$ \\ \hline
$\Delta g_1^Z = -0.02$         &$1.17 $&$1.14 $&$7.52 $ &$44.5 $ & $0.32 $ &$2.10 $&$3.95 $ & $23.19 $   \\ \hline
$\lambda^Z = 0.01$             &$3.08 $&$5.37 $&$6.08 $ &$26.2 $ & $ 1.58$ &$2.63 $&$3.32 $ & $13.68 $ \\ \hline
$\Delta\kappa^Z = 0.2$         &$8.52 $&$0.50 $&$3.28 $ &$4.87 $ & $5.01 $ &$0.15 $&$1.64 $ & $2.40 $ \\ \hline
$\widetilde{\lambda^Z} = 0.01$ &$3.20 $&$5.56 $&$6.18 $ &$27.2 $ & $1.70 $ &$2.69 $&$3.37 $ & $13.83 $ \\ \hline
$\widetilde{\kappa^Z} = 0.2$   &$6.50 $&$0.60 $&$3.15 $ &$4.89 $ & $3.86 $ &$0.22 $&$1.65 $ & $2.36 $ \\ \hline
\end{tabular*}
\end{table} 

At the LHC, the other contributions to the   $3l+\cancel{E}_T$ channel  come from the production of  $ZZ$, $Z\gamma$, $Z+j$, $t\bar{t}$, $Wt$, $WW+j$, 
$t\bar{t}+V$, $tZ$, $VVV$ as has been studied by CMS~\cite{Khachatryan:2016tgp,Sirunyan:2019bez} 
and ATLAS~\cite{Aaboud:2016yus,Aaboud:2019gxl}.
The total non-$WZ$ contributions listed above is about $40~\%$ of the $WZ$ contributions~\cite{Khachatryan:2016tgp}.
We include these extra contributions to the cross sections while estimating limits on the anomalous couplings in Sect.~\ref{sec:limits-and-bench}.

\section{The asymmetries}\label{sec:Pol-Asym}
\begin{figure}[h!]
    \centering
    \includegraphics[width=0.496\textwidth]{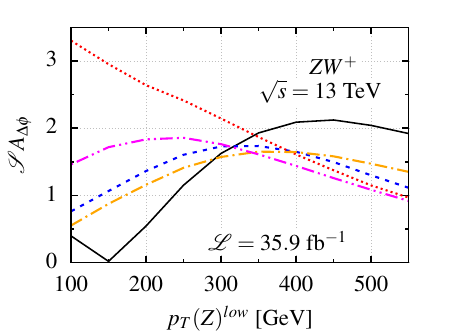}
    \includegraphics[width=0.496\textwidth]{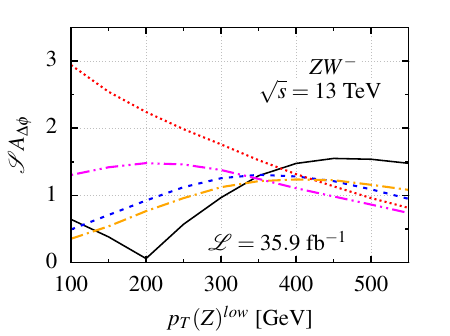}
    \caption{\label{fig:dist_deltaphi}
        The sensitivity of the asymmetry $A_{\Delta\phi}$  on the five benchmark aTGC as a function of 
        the lower cut on   $p_T(Z)$ in the $ZW^\pm$ productions at the LHC at $\sqrt{s}=13$ TeV and ${\cal L}=35.9$ fb$^{-1}$.
        The legend labels are same as in Fig.~\ref{fig:sen-sigma-bench}.}
\end{figure}
We use the polarization asymmetries of $Z$ and $W$ in each $W^+Z$ and $W^-Z$ processes, i.e, $4\times 8=32$ polarization asymmetries
along with the cross sections in four bins.
Beside these, the $Z$ and the $W^\pm$ boson produced in the $ZW^\pm$ production are not forward backward
symmetric owing to only a $t$-channel diagram and not having an $u$-channel diagram (see Fig.~\ref{fig:Feynman_WZ_LHC}). 
These provide an extra observable,  the forward-backward asymmetry defined as,
\begin{equation}\label{eq:def-Afb}
A_{fb}^V=\dfrac{\sigma(\cos\theta_V>0)-\sigma(\cos\theta_V<0)}{\sigma(\cos\theta_V>0)+\sigma(\cos\theta_V<0)} ,
\end{equation}
$\theta_V$ is the production angle of the $V$ w.r.t. the colliding quark-direction. 
One more angular variable
sensitive to aTGC is the angular separation of the lepton $l_W$ from $W^\pm$ and the $Z$ in the
transverse plane, i.e, 
\begin{equation}\label{eq:deltaphi_lw_Z}
\Delta\phi(l_W,Z) = \cos^{-1}\l(\dfrac{\vec{p}_T(l_W).\vec{p}_T(Z)}{p_T(l_W) p_T(Z)}\r) .
\end{equation}
One can construct an asymmetry based on the $\Delta\phi$  as,
\begin{equation}\label{eq:deltaphi_Asym}
A_{\Delta\phi}= \dfrac{\sigma\l(\cos\l( \Delta\phi(l_W,Z) \r)>0\r)-\sigma\l(\cos\l( \Delta\phi(l_W,Z) \r)<0\r)}{\sigma\l(\cos\l( \Delta\phi(l_W,Z) \r)>0\r)+\sigma\l(\cos\l( \Delta\phi(l_W,Z) \r)<0\r)} .
\end{equation}
The sensitivities of $A_{\Delta\phi}$ to the five benchmark 
aTGC are shown in Fig.~\ref{fig:dist_deltaphi} as a function of lower cuts on $p_T(Z)$ in both 
$ZW^\pm$ for luminosity of ${\cal L}=35.9$ fb$^{-1}$.  A choice of $p_T(Z)^{low}=300$ GeV appears to
be an optimal choice for sensitivity for all the couplings. The $m_{3l}$ cut, however, reduces the sensitivities
to all the aTGC.

To construct the asymmetries, we need to set a  reference frame and assign the leptons to the correct mother
spin-$1$ particle. For the present process with missing neutrino, we face a set of challenges in constructing the asymmetries. These are
discussed below.
\paragraph{Selecting $Z$ candidate leptons}
The $Z$ boson momentum
is required to be reconstructed to obtain all the asymmetries which require the right pairing of the $Z$
boson leptons $l_Z^+$ and  $l_Z^-$.
Although the opposite flavour channels $e^+e^-\mu^\pm/\mu^+\mu^-e^\pm$ are safe, the same flavour channels $e^+e^-e^\pm/\mu^+\mu^-\mu^\pm$ suffer ambiguity to select the right $Z$ boson
candidate leptons. The right paring of
leptons for the $Z$ boson in the same flavoured channel is possible with $\ge 96.5~\%$ accuracy for
$m_{3l}>100$ GeV and  $\ge 99.8~\%$ accuracy for $m_{3l}>550$ GeV in
both SM and benchmark aTGC by requiring a smaller value of $|m_Z-m_{l^+l^-}|$. This small  miss pairing is 
neglected to  use the $2e\mu\nu_\mu$ channel as a proxy 
for a $3l+\cancel{E}_T$ final state with good enough accuracy.

\paragraph{The  reconstruction of neutrino momentum}
The other major issue is to obtain the asymmetries related to $W^\pm$ bosons, which require to 
reconstruct their momenta. As the neutrino from $W^\pm$ goes missing, reconstruction
of $W^\pm$ boson momenta is possible with a two-fold ambiguity using the transverse missing 
energy $\cancel{p_T}/\cancel{E_T}$ and the on-shell $W$ mass ($m_W$) constrain. The two solutions for
the longitudinal momentum of the missing neutrino are given by,
\begin{equation}\label{eq:pznu-solution}
{p_z(\nu)}_\pm = \frac{-\beta p_z(l_W)\pm E(l_W)\sqrt[]{D}}{p_T^2(l_W)} 
\end{equation}
with 
\begin{equation}\label{eq:pznu-sol-Dbeta}
D = \beta^2 -p_T^2(\nu)p_T^2(l_W)~,~~~
\beta=m_W^2 + p_x(l_W) p_x(\nu) + p_y(l_W) p_y(\nu) .
\end{equation}

Because the $W$ is not produced on-shell all the time, among the two solutions of neutrino's longitudinal momenta, one of them will be closer to the true value, and another will be far from the true value.
There are no suitable selector or discriminator to select the correct solution from 
the two solutions. Even if we substitute the Monte-Carlo truth $m_W$ to 
solve for $p_z(\nu)$  we don't have any discriminator to distinguish between the two solutions $p_z(\nu)_\pm$. 
The smaller value of $|p_z(\nu)|$ corresponds to the correct solution only for $\approx 65\%$ times on average
in $ZW^+$ and little lower in $ZW^-$ production. One more discriminator which 
is  $||\beta_Z|-|\beta_W||$, the smaller value of this can choose the correct solution a little over
the boundary, i.e., $\approx 55\%$. We have tried machine-learning approaches (artificial neural network) to select the 
correct solutions, but the accuracy was not better than $ 65\%$. 
In some cases, we have $D<0$ with the on-shell $W$, for these cases
either one can throw those events (which affects the distribution and statistics) or one can 
vary the $m_W$ from its central value to have $D>0$. Here, we follow the later. 
 So, as the best available option, we choose the smaller value of  $|p_z(\nu)|$
to be the correct solution to reconstruct the $W$ boson momenta.
 At this point,
it becomes important to explore the effect of reconstruction on asymmetries and their
sensitivities to aTGC. To this end, we consider three scenarios: 
\begin{description}
    \item[\textbf{Abs.~True}] The first thing is to use the  Monte-Carlo truth events and estimate the asymmetries in the lab frame.
The observables in this scenario are directly related to the dynamics up to a rotation of frame~\cite{Bourrely:1980mr,V.:2016wba,Velusamy:2018ksp}.

    \item[\textbf{Reco.~True}] Using the pole mass of $W$ in Eq.~(\ref{eq:pznu-sol-Dbeta}) and choosing the solution closer to
    the Monte-Carlo true value is the best that one can do in  reconstruction. 
        The goal of any reconstruction algorithm would be to become as close to this scenario as possible. 
        
    \item[\textbf{Small}~$\mathbf{|p_z(\boldsymbol{\nu})|}$]   This choice is a best available realistic algorithm which we will be using for the analysis.
\end{description}
\begin{figure}[h!]
    \centering
    \includegraphics[width=0.496\textwidth]{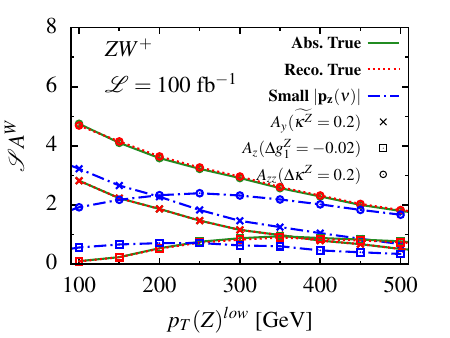}
    \includegraphics[width=0.496\textwidth]{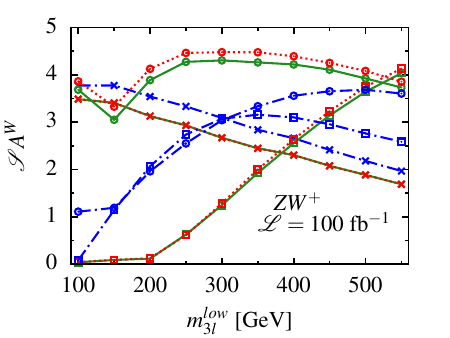}
    \includegraphics[width=0.496\textwidth]{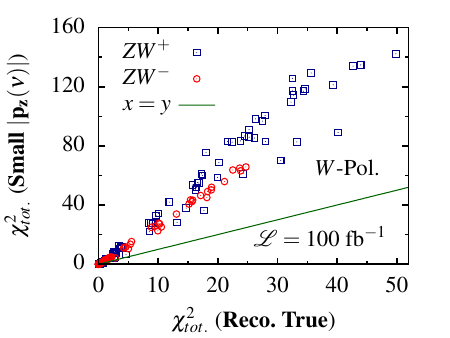}
    \caption{\label{fig:Sen-Wpol-Bench-TrueVsReco}
        The sensitivity of some polarization asymmetries of $W^+$ ($ZW^+$)  
        on some benchmark aTGC for three scenarios: with absolute truth (\textbf{Abs.~True}) information of 
        neutrino in {\em solid}/blue lines, with the close to true reconstructed solution of neutrino 
        (\textbf{Reco.~True}) in {\em dotted}/red lines and with the smaller $|p_z(\nu)|$ to be the true 
        solution (\textbf{Small}~$\mathbf{|p_z(\boldsymbol{\nu})|}$) in {\em dash-dotted}/blue lines as a function of 
        the lower cut on $p_T(Z)$ ({\em top-left-panel}) and $m_{3l}$ ({\em top-right-panel}) at $\sqrt{s}=13$ TeV and ${\cal L}=100$ fb$^{-1}$. The scatter plot of  the total $\chi^2$ for  about $100$ aTGC points
        using all the asymmetries of $W^\pm$ for \textbf{Reco.~True} in $x$-axis with \textbf{Small}~$\mathbf{|p_z(\boldsymbol{\nu})|}$  in $y$-axis is shown in the {\em bottom-panel}.
    }
\end{figure}

The values of reconstructed  asymmetries and hence polarizations   get shifted  from  \textbf{Abs.~True} case.
 In case of \textbf{Reco.~True}, the shifts are roughly constant, while
in case of \textbf{Small}~$\mathbf{|p_z(\boldsymbol{\nu})|}$, the shifts are not constant over
varying lower cut on $m_{3l}$ and $p_T(Z)$ due to the $35~\%$ wrong choice. It is, thus, expected that
the reconstructed sensitivities to aTGC remain same  in \textbf{Reco.~True} and change in \textbf{Small}~$\mathbf{|p_z(\boldsymbol{\nu})|}$ case when compared to the \textbf{Abs.~True} case. In the \textbf{Small}~$\mathbf{|p_z(\boldsymbol{\nu})|}$ reconstruction case, sensitivities 
of some asymmetries to aTGC are less than that of   the \textbf{Abs.~True} case, while they are higher for some other asymmetries.
This is illustrated
in Fig.~\ref{fig:Sen-Wpol-Bench-TrueVsReco} ({\it top-row}) comparing the sensitivity of some polarization
asymmetries, e.g., $A_y$ to $\wtil{\kappa^Z}=+0.2$ in cross ($\times$) points, $A_z$ to 
$\Delta g_1^Z=-0.02$ in square  ($\boxdot$) points,  and  $A_{zz}$ to  
$\Delta\kappa^Z=+0.2$ in circular ($\odot$) points for the three scenarios of \textbf{Abs.~True} ({\em solid}/blue line), \textbf{Reco.~True} ({\em dotted}/red) and \textbf{Small}~$\mathbf{|p_z(\boldsymbol{\nu})|}$ ({\em dash-dotted}/blue) for varying lower cuts on $p_T(Z)$ and $m_{3l}$ in $ZW^+$ production with a luminosity of ${\cal L}=100$ 
fb$^{-1}$. The sensitivities are roughly same for \textbf{Abs.~True} and 
\textbf{Reco.~True} reconstruction in all asymmetries for both $p_T(Z)$ and $m_{3l}$ cuts.
In the \textbf{Small}~$\mathbf{|p_z(\boldsymbol{\nu})|}$ reconstruction case, sensitivity is smaller for $A_{zz}$;  higher for $A_y$;  and it depends on cut for $A_z$ when compared to the \textbf{Abs.~True} case.  When all the $W$ asymmetries are combined, the total 
 $\chi^2$  is higher in the     \textbf{Small}~$\mathbf{|p_z(\boldsymbol{\nu})|}$ case compared to  the \textbf{Reco.~True} case  for about $100$ chosen
 benchmark point, see Fig.~\ref{fig:Sen-Wpol-Bench-TrueVsReco} ({\em bottom-panel}).   
Here, a  total $\chi^2$  of all the  asymmetries of $W$ ($A_i^W$) for a benchmark point ($\{c_i\}$) is given by,
\begin{equation}
\chi^2(A_i^W)(\{c_i\})=\sum_{j}^{N=9} \l({\cal S}A_j^W(\{c_i\}) \r)^2.
\end{equation}
The said increment of $\chi^2$ is observed in both $W^+Z$ ($\boxdot$/blue) and $W^-Z$ ($\odot$/red)  production processes. 
So even if we are not able to reconstruct the $W$ and hence it's polarization observables correctly, realistic effects end up enhancing the overall sensitivity 
of the observables to the aTGC.

\paragraph{Reference $z$-axis for polarizations}
The other challenge to obtain the polarization of $V$ is that one needs a reference axis ($z$-axis) 
to get the momentum direction of $V$, which is not possible at the LHC as it is a
symmetric collider. Thus, for the asymmetries related to $Z$ boson, we consider the direction of total visible longitudinal momenta as an unambiguous choice for
positive $z$-axis. 
For the case of $W$,  the direction of reconstructed boost is used as a proxy for the positive $z$-axis. The latter choice is inspired by the fact that
in $q^\prime\bar{q}$ fusion the quark is supposed to have larger momentum
than the anti-quark at the LHC, thus the above proxy could  stand statistically for the direction of 
the quark direction.

\paragraph{List of observables}\label{para:list-of-obs}
The set of observables used in this analysis are,
\begin{itemize}
\item[$\sigma_i$]: The cross sections in four bins ($4$),
\item[$A_{pol}^Z$]: Eight polarization asymmetries of $Z$ ($8$), 
\item[$A_{fb}^Z$]: Forward backward asymmetry of $Z$ ($1$), 
\item[$A_{\Delta\phi}$]: Azimuthal asymmetry ($1$),
\item[$A_{pol}^W$]: Eight polarization asymmetries of reconstructed $W$ ($8$),
\item[$A_{fb}^W$]: Forward backward asymmetry of reconstructed $W$\footnote{We note that the forward backward asymmetry of $Z$ and $W$ are ideally the same in the CM frame. However, since we measure the $Z$ and $W$ $\cos\theta$  w.r.t. different quantity, i.e., visible $p_z$ for $Z$ and reconstructed boost for $W$, they are practically different and we use them as two independent observables.} ($1$),
\end{itemize}
which make a total of $N({\cal O})=(4+8+1+1+8+1)\times 2=46$ observables including both processes.
All the asymmetry from $Z$ side and all the asymmetries from $W$ side are termed as $A_i^Z$ and 
$A_i^W$, respectively for latter uses.
The total $\chi^2$ for all observables would be the quadratic sum of 
sensitivities (Eq.~(\ref{eq:sensitivity}))  given by,
\begin{equation}\label{eq:tot-chi2}
\chi^2_{tot}(c_i) = \sum_{j}^{N=46} \l({\cal S}{\cal O}_j(c_i) \r)^2 .
\end{equation}
We use these set of observables in some chosen kinematical region to obtain limits on aTGC in the next section.

\section{Probe of the anomalous couplings}\label{sec:limits-and-bench}
\begin{table}[h!]\caption{\label{tab:m3lpTZcut-on-Asym}  The list of  optimized lower cuts ({\tt opt.cut}) on  ($m_{3l}$,$p_T(Z)$)  for various asymmetries to maximize the sensitivity to the anomalous couplings.}
    \renewcommand{\arraystretch}{1.50}
    \begin{tabular*}{\textwidth}{@{\extracolsep{\fill}}|cccc|@{}}\hline
        ${\cal O}$      & $Z$ in $ZW^+$ & $Z$ in $ZW^-$ & $W^\pm$ in $ZW^\pm$  \\ \hline
        $A_x$           & $(200,100)$   & $(100,150)$   &  $(250,0)$         \\ 
        $A_y$           & $(150,100)$   & $(100,100)$   &  ''                  \\
        $A_z$           & $(550,50)$    & $(100,250)$   &  ''                  \\
        $A_{xy}$        & $(150,100)$   & $(150,100)$   &  ''                  \\
        $A_{xz}$        & $(150,0)$     & $(200,50)$    &  ''                  \\
        $A_{yz}$        & $(100,50)$    & $(100,0)$     &  ''                  \\
        $A_{x^2-y^2}$   & $(400,150)$   & $(300,100)$   &  ''                  \\
        $A_{zz}$        & $(550,0)$     & $(300,400)$   &  ''                  \\
        $A_{fb}$        & $(300,0)$     & $(550,0)$     &  ''                \\ \hline
    \end{tabular*}            
    \begin{tabular*}{\textwidth}{@{\extracolsep{\fill}}|ccc|@{}}\hline
        & $ZW^+$       &   $ZW^-$                            \\\hline
        $A_{\Delta\phi}$& $(100,300)$   & $(100,300)$                      \\ \hline
    \end{tabular*}
\end{table}
\begin{figure}[h!]
    \centering
    \includegraphics[width=0.496\textwidth]{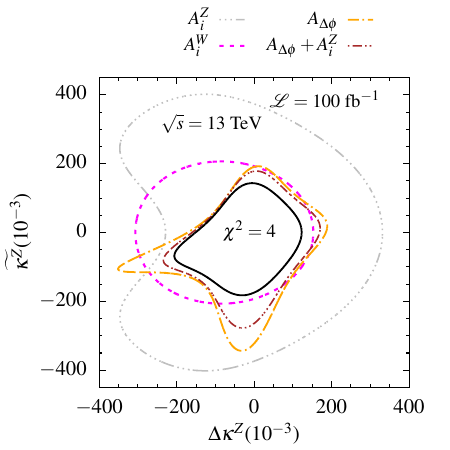}
    \includegraphics[width=0.496\textwidth]{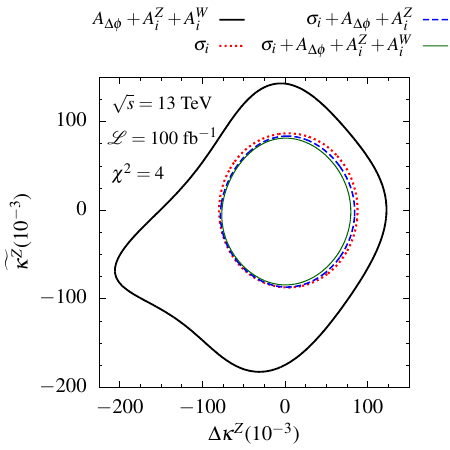}
    \caption{\label{fig:chi2-sigZW-contour-Lum100fb}
        The $\chi^2=4$ contours are shown in the $\Delta\kappa^Z$--$\wtil{\kappa^Z}$ plane with different asymmetries and their combinations in the {\em left-panel},  various combinations of the  cross sections  and asymmetries in the {\em right-panel}   
        for $\sqrt{s}=13$ TeV and  ${\cal L}=100$ fb$^{-1}$. The contour for $A_{\Delta\phi}+A_i^Z+A_i^W$ ({\em thick-solid}/black line ) is repeated in both {\em panel} for comparison.
    }
\end{figure}
We studied the sensitivities of all the ($N({\cal O})=46$) observables   for varying lower cuts on $m_{3l}$ and $p_T(Z)$
separately as well as simultaneously (grid scan in the step of $50$ GeV in each direction) for the chosen benchmark anomalous couplings.
The maximum sensitivities are observed for simultaneous lower cuts on $m_{3l}$ and $p_T(Z)$
given in  Table~\ref{tab:m3lpTZcut-on-Asym} for all the asymmetries in both $ZW^\pm$ processes. Some of these
cuts can be realised from Fig.~\ref{fig:dist_deltaphi} \& \ref{fig:Sen-Wpol-Bench-TrueVsReco}.
The SM values of the asymmetries of $Z$ and $W$ and their corresponding polarizations
for the selection  cuts ({\tt sel.cut} in Eq.~(\ref{eq:selection-cuts})) and for the optimized  cuts ({\tt opt.cut} in Table~\ref{tab:m3lpTZcut-on-Asym}) are listed in Table~\ref{tab:SM-values-Asym-Pol}
in appendix~\ref{app:SM-values-Asym} for completeness. 
We use the cross sections in the four bins and  all asymmetries with the optimized cuts to obtain limits on the anomalous 
couplings for both effective vertices and  effective operators. We  use the semi-analytical expressions for the observables fitted with the simulated data from {\tt mg5\_aMC}. The details of the fitting procedures  are described in appendix~\ref{app:fitting}.
The uncertainty on the cross sections and asymmetries are taken as $\epsilon_\sigma=20~\%$ and $\epsilon_A=2~\%$, respectively consistent with the analysis by CMS~\cite{Khachatryan:2016tgp} and ATLAS~\cite{Aaboud:2019gxl}. We note that these uncertainties are not considered
in the previous sections for qualitative analysis and optimization of cuts.

The sensitivities of all the observables to the aTGC are studied by varying one-parameter, two-parameter and all-parameter at a time in the optimized cut region.
We look at the $\chi^2=4$ contours in the $\Delta\kappa^Z$-$\wtil{\kappa^Z}$ plane  for a luminosity of ${\cal L}=100$ fb$^{-1}$ for various combinations
of asymmetries and cross sections and show them in Fig.~\ref{fig:chi2-sigZW-contour-Lum100fb}.
We observe that the $Z$-asymmetries ($A_i^Z$) are weaker than the $W$-asymmetries ($A_i^W$);  $A_i^W$ provides
very symmetric limits, while $A_i^Z$ has a sense of directionality.
The  $A_{\Delta\phi}$ is better than both $A_i^Z$ and $A_i^W$ in most of the directions in $\Delta\kappa^Z$-$\wtil{\kappa^Z}$ plane.
After combining  $A_i^Z$, $A_i^W$ and $A_{\Delta\phi}$, we get a tighter contours; but the shape is dictated by $A_{\Delta\phi}$.
We see (Fig.~\ref{fig:chi2-sigZW-contour-Lum100fb} {\em right-panel})  that the cross sections have  higher sensitivities compared to
the asymmetries to the aTGC.  
The cross sections dominate constraining  the couplings, while the contribution from  the asymmetries  remain sub-dominant at best. 
Although the directional constraints provided by the asymmetries  get washed way when combined with the cross sections, they are expected to remain prominent to extract non-zero couplings   should a deviation from the SM  be observed. This possibility is discussed in   the  subsection~\ref{subsec:rol-of-asym}.


\subsection{Limits on the couplings}
\begin{table}[h!]\caption{\label{tab:simul-limits-Lag-Op-OpLag} The list of simultaneous limits from MCMC
at $95~\%$ BCI on  the effective vertex couplings $c_i^{\cal L}$  and the effective operator 
couplings $c_i^{\cal O}$  along with translated limits on effective vertices $c_i^{{\cal L}_g}$ for
various luminosities. The notations for $c_i^{\cal L}$, $c_i^{\cal O}$, and $c_i^{{\cal L}_g}$ are given in Eqs.~(\ref{eq:ciL}),~(\ref{eq:ciO}), and~(\ref{eq:ciLg}), respectively.}
\renewcommand{\arraystretch}{1.50}
\begin{tabular*}{\textwidth}{@{\extracolsep{\fill}}ccccc@{}}\hline
$c_i^{\cal L}$ $(10^{-3})$                  &$35.9$ fb$^{-1}$       & $100$ fb$^{-1}$        & $300$ fb$^{-1}$        & $1000$ fb$^{-1}$      \\ \hline
$\Delta g_1^Z$                  &$[  -4.20 ,+   2.15 ]$ &$[   -3.47 ,+   1.50 ]$ &$[   -2.92 ,+   0.963]$ &$[   -2.48 ,+  0.565]$ \\\hline
$\lambda^Z $                    &$[  -2.24 ,+   2.11 ]$ &$[   -1.78 ,+   1.66 ]$ &$[   -1.42 ,+   1.30 ]$ &$[   -1.14 ,+   1.01]$ \\\hline
$\Delta\kappa^Z$                &$[ -83.0 ,+  83.5   ]$ &$[  -64.1  ,+  66.6  ]$ &$[  -47.9  ,+  52.8  ]$ &$[  -34.2  ,+  42.1 ]$ \\\hline
$\widetilde{\lambda^Z}$         &$[  -2.19 ,+   2.19 ]$ &$[   -1.74 ,+   1.72 ]$ &$[   -1.38 ,+   1.36 ]$ &$[   -1.09 ,+   1.09]$ \\\hline
$\widetilde{\kappa^Z}$          &$[ -88.4 ,+  86.2   ]$ &$[  -70.4  ,+  67.5  ]$ &$[  -54.9  ,+  51.8  ]$ &$[  -43.2  ,+  40.1 ]$ \\\hline
$c_i^{\cal O}$  (TeV$^{-2}$) &&&                                                                                                                  \\\hline
$\frac{c_{WWW}}{\Lambda^2}$               &$[ -0.565 ,+  0.540 ]$ &$[  -0.445 ,+  0.426 ]$ &$[  -0.365 ,+  0.327 ]$ &$[  -0.258 ,+  0.257]$ \\\hline
$\frac{c_{W}}{\Lambda^2}$                 &$[ -0.747 ,+  0.504 ]$ &$[  -0.683 ,+  0.397 ]$ &$[  -0.624 ,+  0.274 ]$ &$[  -0.390 ,+  0.196]$ \\\hline
$\frac{c_{B}}{\Lambda^2}$                 &$[-67.1   ,+ 67.8   ]$ &$[ -59.2   ,+ 60.1   ]$ &$[ -52.6   ,+ 47.6   ]$ &$[ -33.3   ,+ 30.9  ]$ \\\hline
$\frac{c_{\wtil{WWW}}}{\Lambda^2}$        &$[ -0.514 ,+  0.516 ]$ &$[  -0.430 ,+  0.415 ]$ &$[  -0.342 ,+  0.339 ]$ &$[  -0.244 ,+  0.252]$ \\\hline
$\frac{c_{\wtil{W}}}{\Lambda^2}$          &$[-68.5   ,+ 69.2   ]$ &$[ -60.4   ,+ 61.2   ]$ &$[ -52.0   ,+ 52.7   ]$ &$[ -32.7   ,+ 34.2  ]$ \\\hline
 $c_i^{{\cal L}_g}$ $(10^{-3})$ &&&                                                                                                          \\\hline
 $\Delta g_1^Z$                  &$[  -3.10 ,+   2.10 ]$ &$[   -2.84 ,+   1.65 ]$ &$[   -2.59 ,+   1.14 ]$ &$[   -1.62 ,+  0.814]$ \\\hline
 $\lambda^Z $                    &$[  -2.31 ,+   2.21 ]$ &$[   -1.82 ,+   1.74 ]$ &$[   -1.49 ,+   1.34 ]$ &$[   -1.06 ,+   1.05]$ \\\hline
 $\Delta\kappa^Z$                &$[ -63.4  ,+  62.1  ]$ &$[  -56.4  ,+  54.6  ]$ &$[  -44.8  ,+  48.3  ]$ &$[  -29.1  ,+  30.6 ]$ \\\hline
 $\widetilde{\lambda^Z}$         &$[  -2.10 ,+   2.11 ]$ &$[   -1.76 ,+   1.70 ]$ &$[   -1.40 ,+   1.39 ]$ &$[   -1.00 ,+   1.03]$ \\\hline
 $\widetilde{\kappa^Z}$          &$[ -64.5  ,+  63.8  ]$ &$[  -57.1  ,+  56.3  ]$ &$[  -49.1  ,+  48.4  ]$ &$[  -31.9  ,+  30.5 ]$ \\\hline
  \end{tabular*} 
\end{table}
\begin{figure}[h!]
    \centering
    \includegraphics[width=1.0\textwidth]{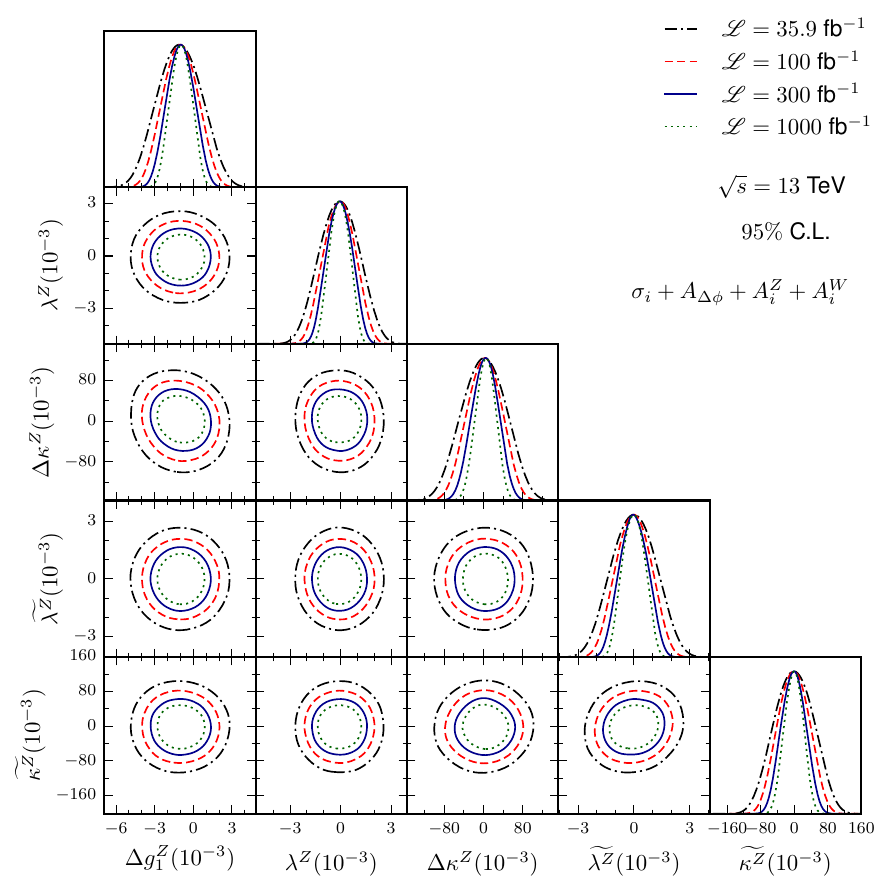}
    \caption{\label{fig:mcmc_Limit_Lag}
        All the  marginalised $1D$ projections and $2D$ projections at $95~\%$ BCI from MCMC in triangular array
        for the effective vertices ($c_i^{\cal L}$)  for  various luminosities at $\sqrt{s}=13$ TeV using all the observables.}
\end{figure}
\begin{figure}[h!]
    \centering
    \includegraphics[width=1.0\textwidth]{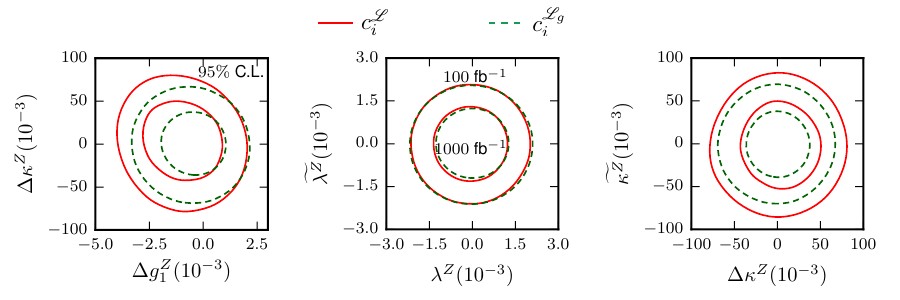}
    \caption{\label{fig:mcmc_Lag_vs_OpLag} The marginalised $2D$ projections at $95~\%$ BCI from MCMC
        in the $\Delta g_1^Z$-$\Delta\kappa^Z$, $\lambda^Z$-$\wtil{\lambda^Z}$, and 
        $\Delta\kappa^Z$-$\wtil{\kappa^Z}$ planes are shown in {\em solid}/red when the effective vertex factors
        ($c_i^{\cal L}$) are treated independent, while shown in {\em dashed}/green
        when the operators are treated independent ($c_i^{{\cal L}_g}$) for luminosities ${\cal L}=1000$ fb$^{-1}$
        (two inner contours) and ${\cal L}=100$ fb$^{-1}$ (two outer contours)
        at $\sqrt{s}=13$ TeV using all the observables.}
\end{figure}

\begin{figure}[h!]
    \centering
    \includegraphics[width=0.49\textwidth]{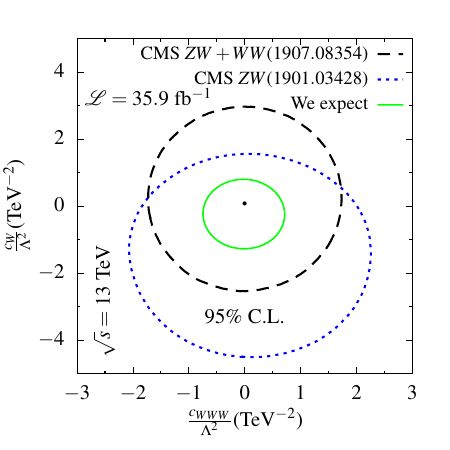}
    \includegraphics[width=0.5\textwidth]{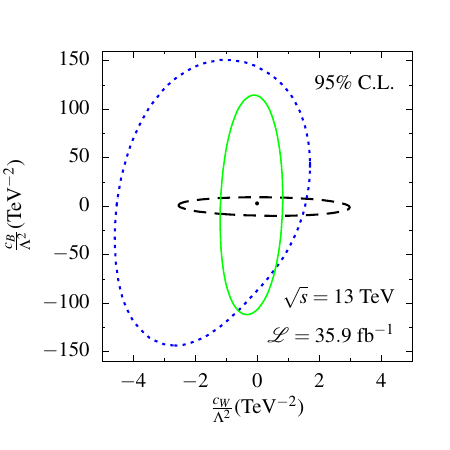}
    \caption{\label{fig:comparision_with_CMS_Op}
        The two parameter $95~\%$ C.L. contours in the $c_{WWW}/\Lambda^2$--$c_W/\Lambda^2$ plane ({\em left-panel}) and $c_{W}/\Lambda^2$--$c_B/\Lambda^2$ plane ({\em right-panel}) for our estimate in {\em solid}/green lines, for CMS $ZW+WW$ in {\em dashed}/black lines  and for  CMS $ZW$ in {\em dotted}/blue lines
        at $\sqrt{s}=13$ TeV and  ${\cal L}=35.9$ fb$^{-1}$ using all the observables.
    } 
\end{figure}
We extract simultaneous limits  on all the anomalous couplings using all the  observables
using MCMC method. We perform this analysis in two ways:  ($i$) vary effective vertex factors 
couplings ($c_i^{\cal L}$)  and
($ii$) vary effective operators couplings ($c_i^{\cal O}$) and translate them in to effective vertex  factors couplings ($c_i^{{\cal L}_g}$). The definitions for $c_i^{\cal L}$, $c_i^{\cal O}$, and $c_i^{{\cal L}_g}$ can be found in Eqs.~(\ref{eq:ciL}),~(\ref{eq:ciO}), and~(\ref{eq:ciLg}), respectively.
The   $95~\%$ BCI (Bayesian confidence interval) obtained on aTGC are listed  in Table~\ref{tab:simul-limits-Lag-Op-OpLag}  for four choices of integrated  luminosities:  ${\cal L}=35.9$ fb$^{-1}$, 
${\cal L}=100$ fb$^{-1}$, ${\cal L}=300$ fb$^{-1}$ and ${\cal L}=1000$ fb$^{-1}$.
The correlation among the parameters are studied  (using {\tt GetDist}~\cite{Antony:GetDist}) and 
they  are shown in Fig.~\ref{fig:mcmc_Limit_Lag} along with $1D$ projections for effective vertex  factors.
The limits on the couplings get tighter as the luminosity is increased, as it should be. The shape of 
the contours are very circular in all two-parameter projections as the cross sections dominate
in constraining the aTGC. The same conclusions are drawn when effective operators are
varied as independent parameters. 
The limits on $c_i^{{\cal L}_g}$ are tighter compared to the limits on $c_i^{\cal L}$ (see Table~\ref{tab:simul-limits-Lag-Op-OpLag});
the comparison between them are shown in the two-parameter marginalised plane  in Fig.~\ref{fig:mcmc_Lag_vs_OpLag}
in $\Delta g_1^Z$-$\kappa^Z$,  $\lambda^Z$-$\wtil{\lambda^Z}$ and $\kappa^Z$-$\wtil{\kappa^Z}$ planes as representative
for luminosity ${\cal L}=100$ fb$^{-1}$ (outer contours) and ${\cal L}=1000$ fb$^{-1}$ (inner contours). 
The limits and the contours are roughly same in  $\lambda^Z$-$\wtil{\lambda^Z}$  plane.  The contours are more symmetric
around the SM for $c_i^{{\cal L}_g}$ compared to $c_i^{\cal L}$, e.g., see $\Delta g_1^Z$-$\kappa^Z$ plane.
The  limits obtained here for luminosity $35.9$ fb$^{-1}$ 
are better than the experimentally observed limits at the LHC given in 
Table~\ref{tab:aTGC_constrain_form_collider}  except on $c_B$ and hence 
on $\Delta\kappa^Z$. This  is due to the fact that
the LHC analysis~\cite{Sirunyan:2019gkh} uses $WW$ production on top of $WZ$ production
whereas we only use $WZ$ production process. But our  limits on the couplings are better when compared
with the $WZ$ production process alone at the LHC~\cite{Sirunyan:2019bez}. 
In Fig.~\ref{fig:comparision_with_CMS_Op}, we present the comparison of limits obtained by the CMS analyses with 
$ZW+WW$~\cite{Sirunyan:2019gkh} process and $ZW$~\cite{Sirunyan:2019bez}
with our estimate with two parameter $95~\%$ BCI contours  in the $c_{WWW}/\Lambda^2$--$c_W/\Lambda^2$ plane ({\em left-panel}) and $c_{W}/\Lambda^2$--$c_B/\Lambda^2$ plane ({\em right-panel}). The contour in the plane $c_{WWW}/\Lambda^2$--$c_W/\Lambda^2$
in our estimate (We~expect) ({\em solid}/green line) is tighter compared to both  
CMS $ZW+WW$ ({\em dashed}/black line) and CMS $ZW$ analyses ({\em dotted}/blue line). 
This is because we use binned cross sections  in the analysis.
The limit on the couplings $c_B/\Lambda^2$ ({\em right-panel}) on the other hand  
is tighter, yet comparable, with CMS $ZW$ and weaker than the CMS $ZW+WW$ analysis because
the $ZW$ process itself is less sensitive to $c_W$.

\subsection{The role of asymmetries in parameter extraction}\label{subsec:rol-of-asym}
\begin{figure}[h!]
    \centering
    \includegraphics[width=1\textwidth]{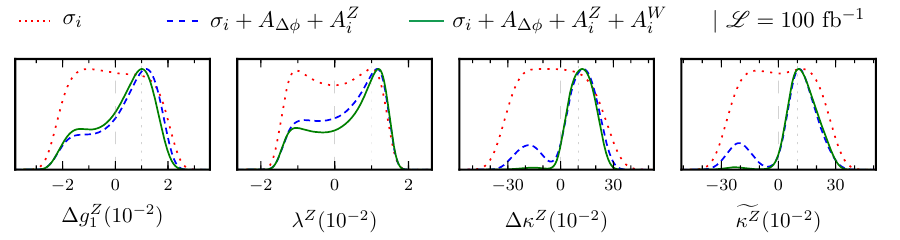}
    \includegraphics[height=0.225\textwidth]{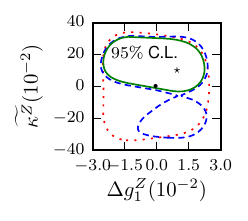}
    \includegraphics[height=0.225\textwidth]{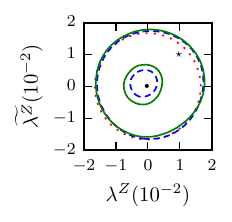}
    \includegraphics[height=0.225\textwidth]{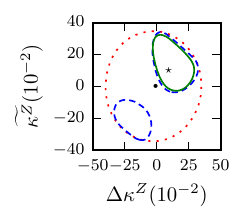}
    \includegraphics[height=0.225\textwidth]{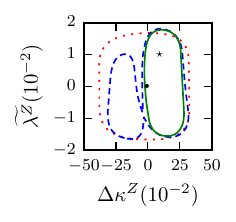}
    \caption{\label{fig:mcmcBench-Lum100fb}
        The marginalised $1D$ projections for the couplings
        $\Delta g_1^Z$, $\lambda^Z$, $\Delta\kappa^Z$, and $\wtil{\kappa^Z}$ in {\em top-panel} and $2D$ projections at $95~\%$ BCI on 
        $\Delta g_1^Z$--$\wtil{\kappa^Z}$, $\lambda^Z$--$\wtil{\lambda^Z}$,   
        $\Delta\kappa^Z$--$\wtil{\kappa^Z}$, and $\Delta\kappa^Z$--$\wtil{\lambda^Z}$ planes in {\em bottom-panel}     from MCMC with observables $\sigma_i$ ({\em dotted}/red line), $\sigma_i + A_{\Delta\phi}+A_i^Z$ 
        ({\em dashed}/blue line) and $\sigma_i + A_{\Delta\phi}+A_i^Z+A_i^W$ ({\em solid}/green line)
        for {\tt aTGC-Bench} couplings $\{\Delta g_1^Z,\lambda^Z,\Delta\kappa^Z,\wtil{\lambda^Z},\wtil{\kappa^Z}\}=\{0.01,0.01,0.1,0.01,0.1\}$
        at $\sqrt{s}=13$ TeV and  ${\cal L}=100$ fb$^{-1}$.}
\end{figure}
The asymmetries are subdominant in constraining the couplings much like seen in Ref.~\cite{Rahaman:2018ujg}
for $pp\to ZZ$ case. 
But the asymmetries help significantly giving directional constraint
in the parameter space. To see this,  
we perform a toy analysis to extract non zero anomalous couplings 
with pseudo data generated by the set of anomalous couplings of
\begin{equation}\label{eq:bench-aTGC}
 \text{{\tt aTGC-Bench}}:~ \{\Delta g_1^Z,\lambda^Z,\Delta\kappa^Z,\wtil{\lambda^Z},\wtil{\kappa^Z}\}=\{0.01,0.01,0.1,0.01,0.1\}
\end{equation}
using MCMC method.
We choose these benchmark couplings well above the current limits so as to mimic a situation of deviation from the SM is observed, as we have considered for $ZZ$ production in chapter~\ref{chap:ZZatLHC}.
 In Fig.~\ref{fig:mcmcBench-Lum100fb}, 
we show the posterior marginalised $1D$ projections for the couplings
$\Delta g_1^Z$, $\lambda^Z$, $\Delta\kappa^Z$, and $\wtil{\kappa^Z}$ in {\em top-panel} and $2D$ 
projections at $95~\%$ BCI on $\Delta g_1^Z$--$\wtil{\kappa^Z}$, $\lambda^Z$--$\wtil{\lambda^Z}$,   
$\Delta\kappa^Z$--$\wtil{\kappa^Z}$, and $\Delta\kappa^Z$--$\wtil{\lambda^Z}$ planes in {\em 
bottom-panel}.  We extract the limits using   $\sigma_i$ only  ({\em dotted}/red line),  using $\sigma_i$ along with  
$A_{\Delta\phi}+A_i^Z$ ({\em dashed}/blue line) and all observables $\sigma_i + A_{\Delta\phi}+A_i^Z+A_i^W$ 
({\em solid}/green line)  for integrated luminosity of  ${\cal L}=100$ fb$^{-1}$.  The {\em dashed} lines 
in $1D$ projections and dot ($\bullet$) in the $2D$ contours represent the SM point, while  the 
{\em dotted} lines in $1D$ projections and star-mark ($\star$) in the contours represent the 
  couplings from {\tt aTGC-Bench}. As the asymmetries $A_{\Delta\phi}$ and  asymmetries of $Z$ ($A_i^Z$) 
  are added on top of the cross sections, the measurement gets better and it  
  further gets better when the asymmetries 
of $W$ ($A_i^W$) are added, which can be seen from both $1D$ projections and $2D$ contours. The 
cross sections are blind to the orientation  of  {\tt aTGC-Bench} couplings and sensitive only to the magnitude 
of deviation from the SM. 
 The asymmetries, however, give direction to the measurement, e.g., in 
$\Delta\kappa^Z$--$\wtil{\kappa^Z}$ plane $\sigma_i + A_{\Delta\phi}+A_i^Z$ provide two patches 
(excluding the SM), and we get  one single (true) region when using all the asymmetries along with the cross sections. 
In the $\lambda^Z$-$\wtil{\lambda^Z}$ plane the asymmetries could not provide a direction, however, they 
shrink the $95~\%$ contours from simply connected patch to an annular region 
(excluding the SM).   For the other couplings  the asymmetries favour the regions of the correct 
solution of {\tt aTGC-Bench} couplings.
For higher luminosities (not shown here), the 
contours become tighter, and the $1D$ curves become sharper centred around the {\tt aTGC-Bench} couplings when 
using $\sigma_i+A_i$, while  $\sigma_i$ alone  remain blind to the {\tt aTGC-Bench}. Thus the asymmetries help 
in the measurement of anomalous couplings provided an   excess of events of aTGC kind are observed. 
 
We note  that the   $3l+\cancel{E}_T$  excess   in the lower $p_T(Z)$ region at the LHC~\cite{Sirunyan:2019bez}     
interpreted by two extra scaler\cite{vonBuddenbrock:2019ajh} may be fitted by aTGC, which is beyond the scope of this present work.

\section{Summary}\label{sec:wz-conclusion}
To summarize this chapter, we studied the $WWZ$ anomalous  couplings in the $ZW^\pm$ production at the LHC and 
examined the role of polarization asymmetries together with $\Delta\phi(l_W,Z)$ asymmetry and forward-backward asymmetry
on the estimation of limits on the anomalous couplings.
We reconstructed the missing neutrino momentum by choosing the small $|p_z(\nu)|$ from the two-fold solutions and estimated the $W$ polarization asymmetries, while the 
$Z$ polarization asymmetries are kept free from any reconstruction ambiguity.
We generated NLO events at {\tt mg5\_aMC} for about $100$ sets of anomalous couplings and used them for the numerical fitting
of semi-analytic expressions of all the observables as a function of couplings. We estimated simultaneous limits on the
anomalous couplings using MCMC method  for both effective vertex formalism and effective operator approach
for luminosities $35.9$ fb$^{-1}$, $100$ fb$^{-1}$, $300$ fb$^{-1}$ and $1000$ fb$^{-1}$. The limits obtained  for ${\cal L}=35.9$ fb$^{-1}$   are tighter
than the available limits obtained at the LHC (see Table~\ref{tab:aTGC_constrain_form_collider} \&~\ref{tab:simul-limits-Lag-Op-OpLag}) except on $c_W$ (and $\Delta\kappa^Z$).  
The asymmetries are helpful in extracting the values of anomalous
couplings if a deviation from the SM is observed at the LHC. We performed a toy analysis of parameter extraction with some benchmark
aTGC couplings and observed that the inclusion of  asymmetries to  the cross sections improves the parameter extraction significantly.


\chapter{\label{chap:conclusion} Conclusions and outlooks}


It is for sure that new physics beyond the SM is needed to address many open questions of the SM.
So far, there is no hint of new physics at the LHC. The new physics scale could be standing
at very high energy not to be directly observed at the current energy range. 
These high scale new physics may leave footprints
in the present energy through higher dimensional effective operators, which often predict new vertices or modify some of the SM vertices.
These new or modified vertices are called anomalous couplings. This thesis deals in probing the anomalous triple gauge boson couplings, i.e., new $VVZ$ ($V=Z/\gamma$) vertices as well as new and modified $WWV$ vertices at various colliders. Precise measurements of these gauge boson self couplings can help us to
understand the electroweak symmetry breaking in detail.
As the $Z$ boson and $W$ boson has polarized couplings to the electrons and quarks, they are produced polarized in any process. The inclusion of aTGC further modifies the values of the polarizations of the gauge bosons. Thus, the polarizations, along with the cross sections, can be used to probe the aTGC in a collider.

A spin-$1$ particle has a total of eight polarizations which are three degrees of vector polarizations and five degrees of tensor polarizations, see chapter~\ref{chap:polarization}. The polarizations are related to the various combinations of the elements of the production density matrix of the particle. The information for the polarizations is transferred to the angular distributions of the decay products of the particle.  
So, the polarizations can be measured from the production density matrix as well as from the angular information of their decayed products by constructing some asymmetries in a  Monte-Carlo event generator or a real collider.  
The density matrix method and the method of asymmetry from decay distributions    of calculating polarizations of a spin-$1$ particle are 
shown to be consistent through the  examples of the processes $ZZ$, $Z\gamma$ and $W^+W^-$ productions at an $e^+e^-$
collider in the SM. 
These polarization asymmetries, along with the cross sections and other asymmetries, are used to probe the 
aTGC in the di-boson productions processes ($ZZ$, $Z\gamma$, and $W^+W^-$)  at an $e^+$-$e^-$ collider as well as  ($ZZ$ and $W^\pm Z$) at the LHC.

In the $e^+$-$e^-$ collider (chapters~\ref{chap:epjc1},~\ref{chap:epjc2} \&~\ref{chap:eeWW}), we calculated the polarization asymmetries from the production density matrix by analytical helicity amplitude method, while we estimated them from the decay angular distributions at the LHC (chapters~\ref{chap:ZZatLHC} \&~\ref{chap:WZatLHC}). 
Some of the polarization asymmetries are $CP$-even while some others are $CP$-odd.  
The $CP$-odd nature of the couplings can be identified by the $CP$-odd polarization asymmetries $A_y$, $A_{xy}$, and $A_{yz}$.
A $CP$-even polarization asymmetry can also identify a $CP$-odd couplings in a suitable kinematic region, e.g., the $CP$-even asymmetry $A_{x^2-y^2}$ probes $CP$-odd couplings in the $ZZ$ production, discussed in chapter~\ref{chap:ZZatLHC}. 
Although $A_z$, $A_{xz}$, and $A_{yz}$ are zero in $ZZ$ and $Z\gamma$ production (see chapter~\ref{chap:epjc1}), they can be made non-zero by little modification using the direction of $Z$,  see chapter~\ref{chap:epjc2}.
The spin of the gauge bosons has different orientation for different polar angle orientation in a general two-body reaction. Thus the gauge boson poses different values of polarizations for different polar orientations.
We used this information to make bins w.r.t. the polar angle of $W$ and calculated it's polarizations in each bin in the $WW$ production process, discussed in chapter~\ref{chap:eeWW}.

For the aTGC, we restricted ourselves up to the dimension-$6$ terms for both the form factors (respecting the Lorentz invariance and $U(1)_{EM}$) as well as the effective operators (respecting the SM gauge symmetry).
As the effective operators for neutral aTGC appear  at dimension-$8$ onward, we studied the neutral aTGC using the form factor approach (chapters~\ref{chap:epjc1},~\ref{chap:epjc2}, \& \ref{chap:ZZatLHC}).
For the charged aTGC, however, we studied both the form factors as well as effective operators  (chapter~\ref{chap:eeWW} \&~\ref{chap:WZatLHC}).
We considered the dimension-$6$ operator up to their quadratic contribution as the linear approximation is not valid, see appendix~\ref{app:zzlhc-b}.

We studied the sensitivity of the polarization asymmetries along with the cross sections to the anomalous couplings and found their one parameter limits.
There is a particular polarization asymmetry which is highly correlated to
a particular aTGC parameter  giving us the best suitable observable for
each parameter, e.g., see chapter~\ref{chap:epjc1}.
We estimated the simultaneous limits on the aTGC using the MCMC method for different luminosities. The polarization asymmetries give significant improvement on the constraints of the aTGC at $e^+$-$e^-$ collider, but not at the LHC. 
However, the polarization asymmetries show directional limits and help to pinpoint non-zero aTGC values at the LHC. To demonstrate  this,  we performed a toy analysis of parameter extraction with some benchmark
aTGC and observed that the inclusion of the asymmetries to the cross sections improves the parameter extraction significantly in both $ZZ$ (chapter~\ref{chap:ZZatLHC}) and $ZW^\pm$ (chapter~\ref{chap:WZatLHC}) productions at the LHC. The limits on the couplings, for the charged aTGC, are tighter when  $SU(2)\times U(1)$ symmetry is assumed, see  chapter~\ref{chap:eeWW} \&~\ref{chap:WZatLHC}. 
In the $ZW^\pm$ production process (chapter~\ref{chap:WZatLHC}), the $\Delta\phi(l_W,Z)$ asymmetry is used, along with the forward-backward asymmetry, polarization asymmetries and cross sections. This azimuthal asymmetry is more sensitive to aTGC than the polarization asymmetries. 
We reconstructed the missing neutrino momenta by choosing small $|p_z(\nu)|$ from the two-fold solutions and estimated the $W$  asymmetries, while the 
$Z$  asymmetries are  free from any such reconstruction ambiguity.

The beam polarization at $e^+$-$e^-$ collider plays a vital role in probing the aTGC. 
The limits on the aTGC get improved by suitably tuning the beam polarizations.  
We estimated the best choice of beam polarization based on average likelihood to put the tightest constraints on the anomalous couplings. The best choice of beam polarization changes as the values of the anomalous couplings change. Thus, 
we estimated the best choice of beam polarizations using the likelihood, averaged over the couplings.
The best choices are found to be  in the extreme corner ($\pm 0.8,\mp 0.8$ for $e^-$ and $e^+$)  when beam polarizations are combined with their opposite choices  for both $ZV$ (chapter~\ref{chap:epjc2}) and $WW$ (chapter~\ref{chap:eeWW}) production processes. For fixed beam polarizations, however, the best choices 
are $(\sim 0.1,\sim -0.1)$ and $(0.4,-0.4)$ for $ZV$ and  $WW$ production, respectively.

The polarization asymmetries of the $Z$ and $W$ have shown promising results in probing the anomalous gauge boson couplings. They provide a large set
of observables to obtain simultaneous limits on a large set of anomalous couplings, $ZZ/Z\gamma$ has $4$ couplings, and $W^+W^-$ has $14$ couplings. 
In all the analyses, we consider certain simplifying assumptions,
such as the absence of initial-state/final-state radiation, hadronization, and detector effects. 
The limits on the aTGC are expected to get dilute if we consider those realistic effects, but the qualitative features of the observables will remain the same.

In conclusion, we proposed some novel techniques to probe new physics beyond the SM at colliders. 
Application of these techniques to the data from the LHC and future colliders may 
reveal the underlying mechanism of the electroweak symmetry breaking and help to discover new physics.

\subsubsection*{Outlooks}
The spins of the gauge bosons in the di-boson production processes are correlated, providing  $64$ potential correlators, see chapter~\ref{chap:polarization}. Their spins could be correlated even if they are produced unpolarized individually, e.g., $t\bar{t}$ production case at the LHC.
In our analyses, we did not use the spin-spin correlations of the gauge bosons for simplicity; 
they would improve the results if used.
Apart from the anomalous triple gauge boson couplings, the polarization observables can be used to study
any new physics associated with them. For example, one can study the anomalous quartic gauge boson couplings and  Higgs to gauge boson  couplings in processes like
triple gauge boson production (VVV)~\cite{Senol:2016axw,Wen:2014mha}, vector boson scattering~\cite{Perez:2018kav}, $t\bar{t}Z$ production, $VHH$ production~\cite{Kumar:2019bmk}, $VVH$ production.  
The polarization observables are also helpful in probing the FCNC interactions~\cite{Behera:2018ryv}, e.g.,  $tqZ$ couplings. Besides these,  polarizations of gauge bosons can help in probing SUSY, extra dimension, and dark matter by looking at their SM backgrounds. 
The effective operators that we consider for the aTGC may also provide quartic couplings which appear in different  processes~\cite{Senol:2016axw,Wen:2014mha,Perez:2018kav,Almeida:2018cld}.
One can consider all the processes containing the contribution of a  given
operator and derive a limit on it at a given collider for a complete study.

%
\begin{appendices}
\chapter{The Standard Model Feynman rules in the electroweak theory}\label{appendix:intro}
The Pauli sigma matrices are
\begin{equation}\label{eq:app:pauli-sigma}
\sigma_1=\sigma_x=\begin{pmatrix} 0 & 1 \\ 1 & 0  \end{pmatrix},~~
\sigma_2=\sigma_y=\begin{pmatrix} 0 & -i \\ i & 0  \end{pmatrix},~~
\sigma_3=\sigma_z=\begin{pmatrix} 1 & 0 \\ 0 & -1  \end{pmatrix}.
\end{equation}
The $\gamma$ matrices in Dirac basis are given by:
\begin{equation}\label{eq:app:gamma-dirac}
\gamma^0 = \begin{pmatrix} \mathbb{I}_{2\times 2} & 0 \\ 0 & -\mathbb{I}_{2\times 2} \end{pmatrix},~~
\gamma^k = \begin{pmatrix} 0 & \sigma_k \\ -\sigma_k & 0 \end{pmatrix},~~
\gamma_5 = i\gamma^0\gamma^1\gamma^2\gamma^3= \begin{pmatrix} 0 & \mathbb{I}_{2\times 2}  \\  \mathbb{I}_{2\times 2} & 0 \end{pmatrix},~~
k=1,2,3.
\end{equation}

\vspace{2cm}
The Feynman rules in the electroweak theory, which are used in this thesis, are given below.
\begin{figure}[H]
	\qquad
	\begin{minipage}{2in}
\vspace{0.5cm}
Dirac propagator :
	\end{minipage}
	\begin{minipage}{1in}
		\begin{tikzpicture}[line width=0.6 pt, scale=0.7]
		\draw[fermion] (-2,-2) -- (2,-2);
		\draw[->] (-1,-1.5) -- (1,-1.5);
		\node at (0,-1.0) {$p$};
		\end{tikzpicture}
	\end{minipage}
	\begin{minipage}{2in}
		\begin{align*}
=~~\frac{\iu}{{\s p}-m+\iu \varepsilon} 
		\end{align*}
	\end{minipage}
	
		\vspace{1cm}
	\qquad
	\begin{minipage}{2in}
		\vspace{0.5cm}
		Photon  propagator :
	\end{minipage}
	\begin{minipage}{1in}
		\begin{tikzpicture}[line width=0.6 pt, scale=0.7]
		\draw[photon] (-2,-2) -- (2,-2);
		\draw[->] (-1,-1.5) -- (1,-1.5);
\node at (0,-1.0) {$\gamma(p)$};
		\node at (-2.5,-2) {$\mu$};
		\node at (2.5,-2) {$\nu$};
		\end{tikzpicture}
	\end{minipage}
	\hspace{0.5cm}
	\begin{minipage}{2in}
		\begin{align*}
		=~~\frac{-\iu g_{\mu\nu}  }{p^2+\iu \varepsilon} 
		\end{align*}
	\end{minipage}
	
	\vspace{1cm}
	\qquad
	\begin{minipage}{2in}
		\vspace{0.5cm}
		Massive boson propagator :
	\end{minipage}
	\begin{minipage}{1in}
		\begin{tikzpicture}[line width=0.6 pt, scale=0.7]
		\draw[photon] (-2,-2) -- (2,-2);
		\draw[->] (-1,-1.5) -- (1,-1.5);
\node at (0,-1.0) {$p$};
		\node at (-2.5,-2) {$\mu$};
		\node at (2.5,-2) {$\nu$};
		\end{tikzpicture}
	\end{minipage}
\hspace{0.5cm}
	\begin{minipage}{2in}
		\begin{align*}
		=~~\frac{\iu \l(-g_{\mu\nu} + \frac{p_\mu p_\nu}{m^2}  \r) }{p^2-m^2+\iu \varepsilon} 
		\end{align*}
	\end{minipage}

	\end{figure}

\begin{figure}[t!]
\begin{minipage}{0.5\textwidth}
	\begin{tikzpicture}[line width=0.6 pt, scale=0.7]
	\draw[fermion] (-2,3) -- (2,0);
	\draw[fermion] (2,0) -- (-2,-3);
	\draw[photon] (2,0) -- (6,0);
	\coordinate [vertex] () at (2,0);
	\node at (6.5,0.0) {$A_\mu$};
	\node at (-2.5,3) {$f$};
	\node at (-2.5,-3) {$\bar{f}$};
	\end{tikzpicture}
\end{minipage}
\begin{minipage}{0.5\textwidth}
	\begin{align*}
		=~~ \iu Q_f e \gamma^\mu
	\end{align*}
\end{minipage}

\qquad
\begin{minipage}{0.5\textwidth}
	\begin{tikzpicture}[line width=0.6 pt, scale=0.7]
	\draw[fermion] (-2,3) -- (2,0);
	\draw[fermion] (2,0) -- (-2,-3);
	\draw[photon] (2,0) -- (6,0);
	\coordinate [vertex] () at (2,0);
	\node at (6.8,0.0) {$W_\mu^\pm$};
	\node at (-2.8,3) {$f(l^-)$};
	\node at (-2.8,-3) {$\bar{f}(\bar{\nu_l})$};
	\end{tikzpicture}
\end{minipage}
\begin{minipage}{0.5\textwidth}
	\begin{align*}
		=~~ -\frac{ \iu g}{2\sqrt{2}} \gamma^\mu(1-\gamma_5)
	\end{align*}
\end{minipage}

\qquad
\begin{minipage}{0.5\textwidth}
	\begin{tikzpicture}[line width=0.6 pt, scale=0.7]
	\draw[fermion] (-2,3) -- (2,0);
	\draw[fermion] (2,0) -- (-2,-3);
	\draw[photon] (2,0) -- (6,0);
	\coordinate [vertex] () at (2,0);
	\node at (6.8,0.0) {$W_\mu^\pm$};
	\node at (-2.8,3) {$q_u,A$};
	\node at (-2.8,-3) {$q_d,B$};
	\end{tikzpicture}
\end{minipage}
\begin{minipage}{0.5\textwidth}
	\begin{align*}
	=~~ -\frac{ \iu g}{2\sqrt{2}} \gamma^\mu(1-\gamma_5) V_{AB}
	\end{align*}
\end{minipage}
\qquad
\begin{minipage}{0.5\textwidth}
	\begin{tikzpicture}[line width=0.6 pt, scale=0.7]
	\draw[fermion] (-2,3) -- (2,0);
	\draw[fermion] (2,0) -- (-2,-3);
	\draw[photon] (2,0) -- (6,0);
	\coordinate [vertex] () at (2,0);
	\node at (6.5,0.0) {$Z_\mu$};
	\node at (-2.5,3) {$f$};
	\node at (-2.5,-3) {$\bar{f}$};
	\end{tikzpicture}
\end{minipage}
\begin{minipage}{0.5\textwidth}
	\begin{align*}
	=~~ -\frac{\iu g}{\cos\theta_W} \gamma^\mu \l(\underbrace{T_3\frac{1}{2} (1-\gamma_5)-Q\sin^2\theta_W}_{C_L\frac{1}{2} (1-\gamma_5)+C_R\frac{1}{2} (1+\gamma_5)\equiv \left({\sf v}_f-a_f\gamma_5\right)}\r)
	\end{align*}
\end{minipage}
\qquad
\begin{minipage}{0.5\textwidth}
	\begin{tikzpicture}[line width=0.6 pt, scale=0.7]
	\draw[photon] (-2,3) -- (2,0);
	\draw[photon] (2,0) -- (-2,-3);
	\draw[photon] (2,0) -- (6,0);
	\coordinate [vertex] () at (2,0);
	\node at (6.5,0.0) {$V_\rho$};
	\node at (-2.8,3) {$W_\mu^+$};
	\node at (-2.8,-3) {$W_\nu^-$};
	\draw[->] (-1,3.0) -- (1,1.5);
\node at (0.5,2.50) {$p_1$};	
\draw[->] (-1,-3.0) -- (1,-1.5);	
\node at (0.5,-2.50) {$p_2$};	
\draw[->] (5,0.50) -- (3,0.5);	
\node at (4.0,1.0) {$p_3$};	
	\end{tikzpicture}
\end{minipage}
\begin{minipage}{0.5\textwidth}
	\begin{align*}
	=~~ -\iu g_{WWV} \Big[g^{\mu\nu}(p_1-p_2)^\rho +g^{\nu\rho}(p_2-p_3)^\mu\\ + g^{\rho\mu}(p_3-p_1)^\mu  \Big]\\g_{WWZ}=-e\cot\theta_W,~g_{WW\gamma}=-e
	\end{align*}
\end{minipage}
\end{figure}

\chapter{Helicity amplitudes and polarization observables in $e^+e^-\to ZZ/Z\gamma$}\label{appendix:epjc1}
\begin{figure}
	\centering
	\includegraphics[width=0.6\textwidth]{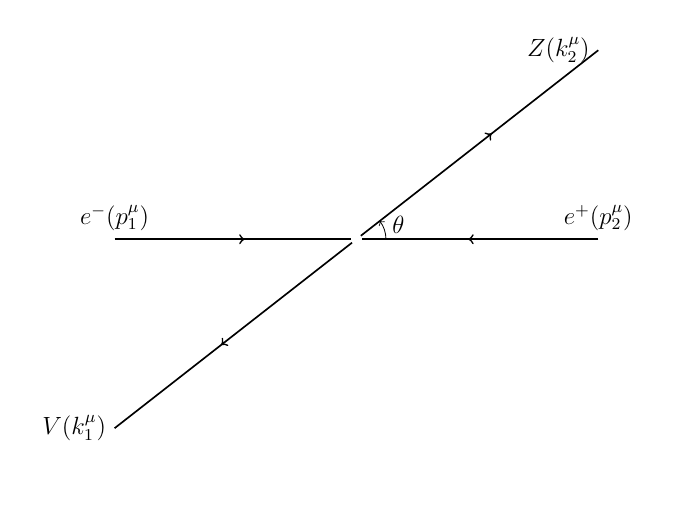}
	\caption{\label{fig:eezv-kinematic} Schematic diagram for $e^-e^+\to ZV$ kinematics.   }
\end{figure}
\section{Helicity amplitudes}\label{appendix:helicity_amplitude}

Vertices in SM are taken as
\begin{eqnarray}
e^- e^+ Z^\mu \Rightarrow && -i\frac{g_Z}{2} \gamma^\mu \left(C_L P_L + C_R P_R\right),\nonumber \\
e^- e^+ \gamma^\mu \Rightarrow && i g_e \gamma^\mu ,
\end{eqnarray}
where $C_L=-1+2\sin^2\theta_W$, $C_R=2\sin^2\theta_W$, 
with $\sin^2\theta_W=1-\left(\frac{m_W}{m_Z}\right)^2$. 
Here $\theta_W$ is the Weinberg mixing angle.
The couplings $g_e$, $g_Z$ are given by,
\begin{equation}
g_e=e=\sqrt{4\pi\alpha_{EM}} \hspace{1cm}\text{and}\hspace{1cm} 
g_Z=\frac{g_e}{\cos\theta_W \sin\theta_W}.
\end{equation}
$P_L=\frac{1-\gamma_5}{2}$, $P_R=\frac{1+\gamma_5}{2}$ are the left and right chiral operators.
Here $\sin\theta$ and $\cos\theta$ are written as $s_\theta$ and $c_\theta$ respectively.
\subsection{For the process $e^+e^-\to ZZ$}
The four-momentum of the particles in this process are (see Fig.~\ref{fig:eezv-kinematic})
\begin{eqnarray}
p_1^\mu=\frac{\sqrt{\hat{s}}}{2}\left\{1,0,0,1\right \},
&&p_2^\mu=\frac{\sqrt{\hat{s}}}{2}\left\{1,0,0,-1 \right\}\nonumber\\
k_1^\mu=\frac{\sqrt{\hat{s}}}{2}\left\{1,-\beta\sin\theta,0,-\beta\cos\theta \right\},
&&k_2^\mu=\frac{\sqrt{\hat{s}}}{2}\left\{1,\beta\sin\theta,0,\beta\cos\theta \right\},
\end{eqnarray}
$\sqrt{\hat{s}}$ being the centre-of-mass energy of the colliding beams.
The on-shell condition of $Z$, $k_1^2=k_2^2=m_Z^2$ gives $\beta$ as
\begin{equation}
\beta=\sqrt{1-\frac{4m_Z^2}{\hat{s}}}.
\end{equation}
We also define
$f^Z=f_4^Z+if_5^Z\beta$, 
$f^\gamma=f_4^\gamma+if_5^\gamma\beta$.
\begin{eqnarray}
\mathcal{M}_{\lambda_{e^-},\lambda_{e^+},\lambda_{z(1)},\lambda_{z(2)}}&=& \mathcal{M}_{SM}+\mathcal{M}_{aTGC} \nonumber\\
\mathcal{M}_{+,-,+,+}  &=&-\frac{ig_Z^2C_R^2 \left(1-\beta ^2\right) c_{\theta } s_{\theta }  }{1+2\beta ^2\left(1-2 c_{\theta }^2\right)+\beta ^4}\nonumber
\end{eqnarray}
\begin{eqnarray}
\mathcal{M}_{+,-,+,0} &=&
- \frac{i g_Z^2C_R^2 \sqrt{1-\beta ^2 }  \left(1-2 c_{\theta }+\beta ^2\right)(1+c_{\theta })  }{\sqrt{2}\left(  1+2\beta ^2\left(1-2 c_{\theta }^2\right)+\beta ^4\right)} 
-\frac{\sqrt{2} \left(1+c_{\theta }\right) g_e \beta  \left(2 g_e  f^\gamma-C_R g_Z f^Z\right)}{\left(1-\beta ^2\right)^{3/2}}\hspace{10cm} \nonumber\\
\mathcal{M}_{+,-,+,-}  &=&
\frac{ig_Z^2C_R^2 \left(1+\beta ^2\right) (1+c_{\theta })  s_{\theta }  }{1+2\beta ^2\left(1-2 c_{\theta }^2\right)+\beta ^4} \nonumber\\
\mathcal{M}_{+,-,0,+}  &=&
\frac{i g_Z^2C_R^2 \sqrt{1-\beta ^2 }  \left(1+2 c_{\theta }+\beta ^2\right)(1-c_{\theta })  }{\sqrt{2}\left(  1+2\beta ^2\left(1-2 c_{\theta }^2\right)+\beta ^4\right)}
+\frac{\sqrt{2} \left(1-c_{\theta }\right) g_e \beta  \left(2 g_e  f^\gamma-C_R g_Z f^Z\right)}{\left(1-\beta ^2\right)^{3/2}}\nonumber\\
\mathcal{M}_{+,-,0,0} &=&
-\frac{2ig_Z^2C_R^2 \left(1-\beta ^2\right) c_{\theta } s_{\theta }  }{1+2\beta ^2\left(1-2 c_{\theta }^2\right)+\beta ^4}\nonumber\\
%
\mathcal{M}_{+,-,0,-} & =&
- \frac{i g_Z^2C_R^2 \sqrt{1-\beta ^2 }  \left(1-2 c_{\theta }+\beta ^2\right)(1+c_{\theta })  }{\sqrt{2}\left(  1+2\beta ^2\left(1-2 c_{\theta }^2\right)+\beta ^4\right)}\nonumber\\
&+&\frac{\sqrt{2} \left(1+c_{\theta }\right) g_e \beta  \left(2 g_e  {f^\gamma}^\star-C_R g_Z {f^Z}^\star \right)}{\left(1-\beta ^2\right)^{3/2}}\nonumber\\
\mathcal{M}_{+,-,-,+} & =&
-\frac{ig_Z^2C_R^2 \left(1+\beta ^2\right) (1-c_{\theta })  s_{\theta }  }{1+2\beta ^2\left(1-2 c_{\theta }^2\right)+\beta ^4}\nonumber\\
\mathcal{M}_{+,-,-,0} & =&
\frac{i g_Z^2C_R^2 \sqrt{1-\beta ^2 }  \left(1+2 c_{\theta }+\beta ^2\right)(1-c_{\theta })  }{\sqrt{2}\left(  1+2\beta ^2\left(1-2 c_{\theta }^2\right)+\beta ^4\right)}
-\frac{\sqrt{2} \left(1-c_{\theta }\right) g_e \beta  \left(2 g_e  {f^\gamma}^\star-C_R g_Z {f^Z}^\star \right)}{\left(1-\beta ^2\right)^{3/2}}\nonumber\\
\mathcal{M}_{+,-,-,-} & =&
-\frac{ig_Z^2C_R^2 \left(1-\beta ^2\right) c_{\theta } s_{\theta }  }{1+2\beta ^2\left(1-2 c_{\theta }^2\right)+\beta ^4}\nonumber\\
\mathcal{M}_{-,+,+,+} & =&
-\frac{ig_Z^2C_L^2 \left(1-\beta ^2\right) c_{\theta } s_{\theta }  }{1+2\beta ^2\left(1-2 c_{\theta }^2\right)+\beta ^4}\nonumber\\
\mathcal{M}_{-,+,+,0} & =&
-\frac{i g_Z^2C_L^2 \sqrt{1-\beta ^2 }  \left(1+2 c_{\theta }+\beta ^2\right)(1-c_{\theta })  }{\sqrt{2}\left(  1+2\beta ^2\left(1-2 c_{\theta }^2\right)+\beta ^4\right)} 
+\frac{\sqrt{2} \left(1-c_{\theta }\right) g_e \beta  \left(2 g_e  f^\gamma- C_L g_Z f^Z \right)}{\left(1-\beta ^2\right)^{3/2}}  \nonumber\\
\mathcal{M}_{-,+,+,-} & =&
\frac{ig_Z^2C_L^2 \left(1+\beta ^2\right) (1-c_{\theta })  s_{\theta }  }{1+2\beta ^2\left(1-2 c_{\theta }^2\right)+\beta ^4} \nonumber\\
\mathcal{M}_{-,+,0,+} & =&
\frac{i g_Z^2 C_L^2 \sqrt{1-\beta ^2 }  \left(1-2 c_{\theta }+\beta ^2\right)(1+c_{\theta })  }{\sqrt{2}\left(  1+2\beta ^2\left(1-2 c_{\theta }^2\right)+\beta ^4\right)}
-\frac{\sqrt{2} \left(1+c_{\theta }\right) g_e \beta  \left(2 g_e  f^\gamma-C_L g_Z f^Z \right)}{\left(1-\beta ^2\right)^{3/2}}\nonumber\\
\mathcal{M}_{-,+,0,0} & =&
-\frac{2ig_Z^2C_L^2 \left(1-\beta ^2\right) c_{\theta } s_{\theta }  }{1+2\beta ^2\left(1-2 c_{\theta }^2\right)+\beta ^4}\nonumber\\
\mathcal{M}_{-,+,0,-} & =&
- \frac{i g_Z^2C_L^2 \sqrt{1-\beta ^2 }  \left(1+2 c_{\theta }+\beta ^2\right)(1-c_{\theta })  }{\sqrt{2}\left(  1+2\beta ^2\left(1-2 c_{\theta }^2\right)+\beta ^4\right)}\nonumber\\
&-&\frac{\sqrt{2} \left(1-c_{\theta }\right) g_e \beta  \left(2 g_e  {f^\gamma}^\star - C_L g_Z {f^Z}^\star \right)}{\left(1-\beta ^2\right)^{3/2}}\nonumber\\
\mathcal{M}_{-,+,-,+} & =&
\frac{ig_Z^2C_L^2 \left(1+\beta ^2\right) (1+c_{\theta })  s_{\theta }  }{1+2\beta ^2\left(1-2 c_{\theta }^2\right)+\beta ^4} \nonumber\\
\mathcal{M}_{-,+,-,0} & =&
\frac{i g_Z^2C_L^2 \sqrt{1-\beta ^2 }  \left(1-2 c_{\theta }+\beta ^2\right)(1+c_{\theta })  }{\sqrt{2}\left(  1+2\beta ^2\left(1-2 c_{\theta }^2\right)+\beta ^4\right)}
+\frac{\sqrt{2} \left(1+c_{\theta }\right) g_e \beta  \left(2 g_e  {f^\gamma}^\star -C_L g_Z {f^Z}^\star \right)}{\left(1-\beta ^2\right)^{3/2}}\nonumber\\
\mathcal{M}_{-,+,-,-} & =&
-\frac{i g_Z^2C_L^2 \left(1-\beta ^2\right) c_{\theta } s_{\theta }  }{1+2\beta ^2\left(1-2 c_{\theta }^2\right)+\beta ^4}
\end{eqnarray}

\subsection{For the process $e^+e^-\to Z\gamma$}
The four momentum of the particles in this process  are 
\begin{eqnarray}
p_1^\mu=\frac{\sqrt{\hat{s}}}{2}\left\{1,0,0,1\right \},
&&p_2^\mu=\frac{\sqrt{\hat{s}}}{2}\left\{1,0,0,-1 \right\}\nonumber\\
k_1^\mu=\frac{\sqrt{\hat{s}}}{2}\left\{\beta,-\beta\sin\theta,0,-\beta\cos\theta \right\},
&&k_2^\mu=\frac{\sqrt{\hat{s}}}{2}\left\{2-\beta,\beta\sin\theta,0,\beta\cos\theta \right\}.
\end{eqnarray}
The four momentum of $\gamma$  satisfy $k_1^2=0$.  The onn-shell condition of $Z$,
$k_2^2=m_Z^2$    provides $\beta$ as
\begin{equation}
\beta=1-\frac{M_Z^2}{\hat{s}}.
\end{equation}

We define $h^\gamma=h_1^\gamma + i h_3^\gamma$ and $h^Z=h_1^Z+ih_3^Z$.

\begin{eqnarray}
\mathcal{M}_{ \lambda_{e^-},\lambda_{e^+},\lambda_{z},\lambda_{\gamma}} &= & \mathcal{M}_{SM}+\mathcal{M}_{aTGC} \nonumber\\
\mathcal{M}_{+,-,+,+}&= &
\frac{-i  g_e g_Z C_Rs_{\theta } (1-\beta )}{\beta (1-c_{\theta }) } 
-\frac{g_e \left( 2g_e h^\gamma - C_Rg_Z h^Z \right) s_{\theta } \beta }{4 (1-\beta
	)}\nonumber\\
\mathcal{M}_{+,-,+,-}&=&
\frac{i g_e g_ZC_R  s_{\theta }}{\beta(1 -  c_{\theta })}\nonumber\\
\mathcal{M}_{+,-,0,+}&=&
\frac{i \sqrt{2}g_e g_ZC_R\sqrt{1- \beta }  }{\beta }
 +\frac{\left(1-c_{\theta }\right)g_e\left(2g_e h^\gamma - C_Rg_Z h^Z \right)
	\beta }{4 \sqrt{2} (1-\beta )^{3/2}}\nonumber\\
\mathcal{M}_{+,-,0,-}&=&
\frac{-i \sqrt{2}g_e g_ZC_R\sqrt{1- \beta }  }{\beta }
+\frac{\left(1+c_{\theta }\right)g_e\left(  2g_e {h^\gamma}^\star -C_Rg_Z {h^Z}^\star \right)
	\beta }{4 \sqrt{2} (1-\beta )^{3/2}} \nonumber\\
\mathcal{M}_{+,-,-,+}&=&
\frac{-i g_e g_Z C_R s_{\theta }}{\beta(1 +  c_{\theta })}\nonumber\\
\mathcal{M}_{+,-,-,-}&=&
\frac{i g_e g_Z (1-\beta )  s_{\theta }}{\beta  (1+c_{\theta })}
-\frac{g_e \left(  2g_e {h^\gamma}^\star -C_Rg_Z {h^Z}^\star \right) s_{\theta } \beta }{4 (1-\beta
	)}\nonumber\\
\mathcal{M}_{-,+,+,+}&=&
\frac{i  g_e g_Z C_Ls_{\theta } (1-\beta )}{ \beta ({1+\text{c}_{\theta} })}
-\frac{g_e \left( 2g_e h^\gamma - C_L g_Z h^Z \right) s_{\theta } \beta }{4
	(1-\beta )}\nonumber\\
\mathcal{M}_{-,+,+,-}&=&
\frac{-i g_e g_ZC_Ls_{\theta }}{\beta(1 +c_{\theta })}\nonumber\\
\mathcal{M}_{-,+,0,+}&=&
\frac{i \sqrt{2}g_e g_ZC_L\sqrt{1- \beta }  }{\beta }
-\frac{\left(1+c_{\theta }\right)g_e\left(2g_e h^\gamma - C_L g_Z h^Z \right)
	\beta }{4 \sqrt{2} (1-\beta )^{3/2}} \nonumber\\
\mathcal{M}_{-,+,0,-}&=&
\frac{-i \sqrt{2}g_e g_ZC_L\sqrt{1- \beta }  }{\beta }
 -\frac{\left(1-c_{\theta }\right)g_e\left(2g_e{h^\gamma}^\star -C_L g_Z {h^Z}^\star \right)
	\beta }{4 \sqrt{2} (1-\beta )^{3/2}}\nonumber\\
\mathcal{M}_{-,+,-,+}&=&
\frac{i g_e g_Z C_L s_{\theta }}{\beta(1 - c_{\theta })}\nonumber\\
\mathcal{M}_{-,+,-,-}&=&
\frac{-i g_e g_ZC_L (1-\beta )  s_{\theta }}{\beta  (1-c_{\theta })}
 -\frac{g_e \left(2g_e{h^\gamma}^\star -C_L g_Z {h^Z}^\star \right) s_{\theta } \beta }{4
	(1-\beta )}
\end{eqnarray}

\section{Polarization observables}\label{appendix:Expresson_observables}
\subsection{For the process $e^+e^-\to ZZ$}
\begin{equation}
\sigma(e^+e^-\to Z Z)=\frac{1}{2} \dfrac{1}{32\pi\beta \hat{s}} \ 
\tilde{\sigma}_{ZZ}
\end{equation}

\begin{eqnarray}
\tilde{\sigma}_{ZZ}&=&\frac{1}{2} g_Z^4\left(C_L^4+C_R^4\right)\left(\frac{\left(5-2 \beta ^2+\beta ^4\right) \log\left(\frac{1+\beta
	}{1-\beta }\right)}{2 \left(\beta +\beta ^3\right)}-1\right)\nonumber\\
&+& g_Z^3 g_e \left(C_L^3-C_R^3\right) f_5^Z\frac{ \left(3-\beta ^2\right)\left(2 \beta  +\left(1+\beta ^2\right) \log\left(\frac{1+\beta
	}{1-\beta }\right)\right)}{2 \beta  \left(1-\beta ^2\right)}\nonumber\\
&-& g_Z^2 g_e^2\left(C_L^2-C_R^2\right) f_5^{\gamma } \frac{ \left(3-\beta ^2\right)\left(2 \beta  -\left(1+ \beta ^2\right) \log\left(\frac{1+\beta
	}{1-\beta }\right)\right)}{\beta  \left(1-\beta ^2\right)}\nonumber\\
&+& 16 g_Z^2 g_e^2\left(C_L^2+C_R^2\right)\frac{\beta ^2 \left((f_4^Z)^2+(f_5^Z)^2 \beta ^2\right)}{3 \left(1-\beta ^2\right)^3}\nonumber\\
&-& 64 g_Z g_e^3 \left(C_L+C_R\right)\frac{ \beta ^2 \left(f_4^{\gamma } f_4^Z+f_5^{\gamma } f_5^Z \beta ^2\right)}{3 \left(1-\beta ^2\right)^3}\nonumber\\
&+& 128 g_e^4\frac{ \beta ^2 \left((f_4^{\gamma })^2+(f_5^{\gamma })^2 \beta ^2\right)}{3 \left(1-\beta ^2\right)^3}
\end{eqnarray}
\begin{eqnarray}
\tilde{\sigma}_{ZZ}\times P_x &=& g_Z^4 \left(C_L^4-C_R^4\right)\frac{\pi(1-\beta^2)^{3/2}}{4(1+\beta^2)}
+ g_Z^3 g_e \left(C_L^3+C_R^3\right)\frac{f_5^Z\pi\beta^2(1+2\beta^2)}{4(1-\beta^2)^{3/2}}\nonumber\\
&-& g_Z^2 g_e^2 \left(C_L^2+C_R^2\right)\frac{f_5^\gamma\pi\beta^2(1+2\beta^2)}{2(1-\beta^2)^{3/2}}
\end{eqnarray}
\begin{eqnarray}
\tilde{\sigma}_{ZZ}\times P_y &=& 
- g_Z^3 g_e \left(C_L^3+C_R^3\right)\frac{f_4^Z\pi\beta(2+\beta^2)}{4(1-\beta^2)^{3/2}}
+ g_Z^2 g_e^2 \left(C_L^2+C_R^2\right)\frac{f_4^\gamma\pi\beta(2+\beta^2)}{2(1-\beta^2)^{3/2}}\nonumber\\
\end{eqnarray}
\begin{eqnarray}
\tilde{\sigma}_{ZZ}\times T_{xy}&=&
g_Z^3g_e\left(C_L^3-C_R^3\right)\frac{\sqrt{3}}{8}f_4^Z f_{xy}(\beta)\nonumber\\
&&-g_Z^2g_e^2 \left(C_L^2-C_R^2\right)\frac{\sqrt{3}}{4}f_4^{\gamma }f_{xy}(\beta)
+g_Z^2 g_e^2 \left(C_L^2+C_R^2\right)\frac{4}{\sqrt{3}}\frac{ f_4^Z f_5^Z\beta ^3}{\left(1-\beta ^2\right)^3}\nonumber\\
&&+g_Z g_e^3\left(C_L+C_R\right)\frac{8}{\sqrt{3}}\frac{ \left(f_4^Z f_5^{\gamma }+f_4^{\gamma } f_5^Z\right)\beta ^3}{\left(1-\beta
	^2\right)^3}
+ g_e^4\frac{32f_4^{\gamma } f_5^{\gamma }\beta ^3}{\sqrt{3} \left(1-\beta ^2\right)^3},\nonumber\\
&&f_{xy}(\beta)=\frac{\left(2 \left(\beta +\beta ^3\right)-\left(1-\beta
	^2\right)^2 \log\left(\frac{1+\beta }{1-\beta }\right)\right)}{ \beta ^2 \left(1-\beta ^2\right)}
\end{eqnarray}
\begin{eqnarray}
\tilde{\sigma}_{ZZ}\times(T_{xx}-T_{yy})&=&
g_Z^4 \left(C_L^4+C_R^4\right) \sqrt{\frac{3}{2}}\frac{ \left(1-\beta ^2\right) \left(2 \beta -\left(1+\beta ^2\right) \log\left(\frac{1+\beta
	}{1-\beta }\right)\right)}{8 \beta ^3}\nonumber\\
&-&\bigg( g_Z^3g_e \left(C_L^3-C_R^3\right) f_5^Z
+ g_Z^2g_e^2\left(C_L^2-C_R^2\right) 2f_5^{\gamma }\bigg) f_{x^2-y^2}(\beta)  \nonumber\\
&+& g_Z^2g_e^2 \left(C_L^2+C_R^2\right)2 \sqrt{\frac{2}{3}}\frac{\beta ^2 \left((f_4^Z)^2-(f_5^Z)^2 \beta ^2\right)}{\left(1-\beta
	^2\right)^3}\nonumber\\
&-& g_Z g_e^3\left(C_L+C_R\right)8 \sqrt{\frac{2}{3}}\frac{\beta ^2 \left(f_4^{\gamma } f_4^Z-f_5^{\gamma } f_5^Z \beta ^2\right)}{\left(1-\beta
	^2\right)^3}\nonumber\\
&+& g_e^416 \sqrt{\frac{2}{3}}\frac{\beta ^2 \left((f_4^{\gamma })^2-(f_5^{\gamma })^2 \beta ^2\right)}{\left(1-\beta ^2\right)^3},\nonumber\\
f_{x^2-y^2}(\beta)&=&\sqrt{\frac{3}{2}}\frac{\left(2 \left(\beta +\beta ^3\right)-\left(1-\beta ^2\right)^2
	\log\left(\frac{1+\beta
	}{1-\beta }\right)\right)}{4 \beta  \left(1-\beta ^2\right)}
\end{eqnarray}
\begin{eqnarray}
\tilde{\sigma}_{ZZ}\times T_{ZZ}&=&
g_Z^4 \left(C_L^4+C_R^4\right)\frac{f_{zz}(\beta) }{8 \sqrt{6} \beta ^3 \left(1+\beta ^2\right)}
+ g_Z^3g_e \left(C_L^3-C_R^3\right) f_5^Z\frac{f_{zz}(\beta) }{4 \sqrt{6} \beta  \left(1-\beta ^2\right)}\nonumber\\
&-&g_Z^2g_e^2 \left(C_L^2-C_R^2\right)  \frac{f_{zz}(\beta)f_5^{\gamma } }{2 \sqrt{6} \beta  \left(1-\beta ^2\right)}\nonumber\\
&-&g_Z^2 g_e^2  \left(C_L^2+C_R^2\right) \frac{4}{3} \sqrt{\frac{2}{3}}\frac{\beta ^2 \left((f_4^Z)^2+(f_5^Z)^2 \beta ^2\right)}{
	\left(1-\beta ^2\right)^3}\nonumber\\
&+&g_Zg_e^3\left(C_L+C_R\right)\frac{16 }{3}\sqrt{\frac{2}{3}}\frac{\beta ^2 \left(f_4^{\gamma } f_4^Z+f_5^{\gamma }
	f_5^Z \beta ^2\right)}{ \left(1-\beta ^2\right)^3}\nonumber\\
&-&g_e^4\frac{32}{3} \sqrt{\frac{2}{3}}\frac{\beta ^2 \left((f_4^{\gamma })^2+(f_5^{\gamma })^2 \beta ^2\right)}{ \left(1-\beta
	^2\right)^3},\nonumber\\
f_{zz}(\beta)&=&\bigg(\left(3+\beta ^2+5 \beta ^4-\beta ^6\right) \log\left(\frac{1+\beta }{1-\beta}\right)
 -2 \beta  \left(3-\beta ^2\right) \left(1+\beta ^2\right)\bigg)
\end{eqnarray}
\subsection{For the process $e^+e^-\to Z\gamma$}
\begin{equation}
\sigma(e^+e^-\to Z \gamma)=\dfrac{1}{32\pi\beta \hat{s}} \ 
\tilde{\sigma}_{Z\gamma} ,\ \ \ c_\theta\in[-c_{\theta_0},c_{\theta_0}]
\end{equation}
\begin{eqnarray}
\tilde{\sigma}_{Z\gamma}&=&
- g_e^2 g_Z^2\left(C_L^2+C_R^2\right)\frac{1}{\beta ^2}\bigg(c_{\theta _0} \beta ^2- 2 \left(2-2 \beta +\beta ^2\right) \tanh ^{-1}(c_{\theta_0})\bigg)\nonumber \\
&+& g_e^2 g_Z^2\left(C_L^2-C_R^2\right)\frac{c_{\theta _0}h_3^Z (2-\beta )}{2 (1-\beta )} 
- g_e^3 g_Z \left(C_L-C_R\right)\frac{c_{\theta _0}h_3^{\gamma } (2-\beta )}{(1-\beta) } \nonumber\\
&+& g_e^2 g_Z^2\left(C_L^2+C_R^2\right)c_{\theta _0}\left((h_1^Z)^2+(h_3^Z)^2\right) h_{\sigma}(\beta) \nonumber\\
&-& g_e^3 g_Z\left(C_L+C_R\right) c_{\theta _0}\left(h_1^{\gamma } h_1^Z+h_3^{\gamma } h_3^Z\right) 4 h_{\sigma}(\beta)\nonumber\\
&+& g_e^4 c_{\theta _0}\left((h_1^{\gamma })^2+(h_3^{\gamma })^2\right) 8h_{\sigma}(\beta),\nonumber \\
h_{\sigma}(\beta)&=&\frac{\beta ^2 \left(9-6 \beta -c_{\theta
		_0}^2 (1-2 \beta )\right)}{96 (1-\beta )^3}
\end{eqnarray}
\begin{eqnarray}
\tilde{\sigma}_{Z\gamma}\times P_x &=&
  g_e^2 g_Z^2\left(C_L^2-C_R^2\right)2\frac{\sqrt{1-\beta }}{\beta ^2} (2-\beta ) \sinh ^{-1}(c_{\theta_0})\nonumber\\
&+& g_e^2 g_Z^2 \left(C_L^2+C_R^2\right)h_3^Z  h_x^1(\beta)
- g_e^3 g_Z\left(C_L+C_R\right)h_3^{\gamma }  2 h_x^1(\beta)\nonumber\\
&+& g_e^2 g_Z^2 \left(C_L^2-C_R^2\right) \left((h_1^Z)^2+(h_3^Z)^2\right)h_x^2(\beta) 
- g_e^3 g_Z\left(C_L-C_R\right)\left(h_1^{\gamma } h_1^Z+h_3^{\gamma } h_3^Z\right) 4h_x^2(\beta),\nonumber\\
&&h_x^1(\beta)=\frac{ \left(c_{\theta _0} s_{\theta _0} (2-3 \beta )-3 (2-\beta ) \sinh ^{-1}(c_{\theta_0})\right)}{8 (1-\beta )^{3/2}},\nonumber\\
&&h_x^2(\beta)=\frac{ \beta ^2 \left(c_{\theta _0} s_{\theta
		_0}+\sinh ^{-1}(c_{\theta_0})\right)}{32 (1-\beta )^{5/2}}
\end{eqnarray}
\begin{eqnarray}
\tilde{\sigma}_{Z\gamma}\times P_y&=&
g_e^2 g_Z^2\left(C_L^2+C_R^2\right) \frac{h_1^Z\bigg(c_{\theta _0}s_{\theta _0}\beta-(4-\beta)\sin^{-1}(c_{\theta _0})\bigg)}{8(1-\beta)^{3/2}}\nonumber\\
&&- g_e^3 g_Z\left(C_L+C_R\right) \frac{h_1^\gamma\bigg(c_{\theta _0}s_{\theta _0}\beta-(4-\beta)\sin^{-1}(c_{\theta _0})\bigg)}{4(1-\beta)^{3/2}}
\end{eqnarray}
\begin{eqnarray}
\tilde{\sigma}_{Z\gamma}\times T_{xy}=
g_e^3 g_Z \left(C_L-C_R\right)\frac{\sqrt{3}}{2}\frac{h_1^\gamma c_{\theta _0}}{2(1-\beta)}
- g_e^2 g_Z^2 \left(C_L^2-C_R^2\right)\frac{\sqrt{3}}{4}\frac{h_1^Z c_{\theta _0}}{4(1-\beta)}
\end{eqnarray}
\begin{eqnarray}
\tilde{\sigma}_{Z\gamma}\times(T_{xx}-T_{yy})=
g_e^2 g_Z^2\left(C_L^2+C_R^2\right)\frac{\sqrt{6} c_{\theta _0}(1-\beta)}{\beta^2}
- g_e^3 g_Z \left(C_L-C_R\right)\sqrt{\frac{3}{2}} \frac{h_3^\gamma c_{\theta _0}}{(1-\beta)} 
\nonumber\\
\end{eqnarray}
\begin{eqnarray}
\tilde{\sigma}_{Z\gamma}\times T_{zz}&=&
- g_e^2 g_Z^2\left(C_L^2+C_R^2\right)\frac{1}{\sqrt{6} \beta ^2}\bigg(c_{\theta _0} \left(6-6 \beta +\beta ^2\right)
-2 \left(2-2 \beta
+ \beta ^2\right) \tanh ^{-1}(c_{\theta_0})\bigg)\nonumber\\
&-& g_e^2 g_Z^2\left(C_L^2-C_R^2\right)\frac{c_{\theta _0}h_3^Z (1+\beta )}{2 \sqrt{6} (1-\beta )}
+ g_e^3 g_Z\left(C_L-C_R\right)\frac{c_{\theta _0}h_3^{\gamma } (1+\beta )}{\sqrt{6} (1-\beta )}\nonumber\\
&-& g_e^2 g_Z^2\left(C_L^2+C_R^2\right) c_{\theta _0}\left((h_1^Z)^2+(h_3^Z)^2\right) h_{zz}(\beta) \nonumber\\
&+& g_e^3 g_Z \left(C_L+C_R\right) c_{\theta _0}\left(h_1^{\gamma } h_1^Z+h_3^{\gamma } h_3^Z\right) 4h_{zz}(\beta) 
- g_e^4 c_{\theta _0}\left((h_1^{\gamma })^2+(h_3^{\gamma })^2\right) 8h_{zz}(\beta) ,\nonumber\\
h_{zz}(\beta)&=&\frac{ \left(c_{\theta _0}^2 (2-\beta )+3 \beta \right)
	\beta ^2}{48 \sqrt{6} (1-\beta )^3}
\end{eqnarray}


\chapter{Polarization observables of $Z$ boson in $ZZ$ production at the LHC}\label{appendix:ZZatLHC}
The production density matrix at the LHC will be
\begin{eqnarray}
\rho\l(\lambda,\lambda^\prime\r) = \int dx_1 f_1\l(x_1,Q^2\r)  \int dx_2 f_2\l(x_2,Q^2\r) ~~ \hat{\rho}\l(\lambda,\lambda^\prime\r), 
\end{eqnarray}
where $\hat{\rho}$ is the parton level density matrix, $f_1\l(x_1,Q^2\r) $ and $f_2\l(x_2,Q^2\r)$ are the parton distribution functions
for two parton from the two colliding protons  with energy fraction $x_1$ and $x_2$, respectively.  The $Q$ is the scale factor assumed.
The observables are calculated using the distribution of the decay products of the $Z$ boson, not from the production density matrix.
Events were generated in \MGvATNLO~ for a set of couplings values,  functional form of the observables were obtained by numerical fitting
the data. The observables are listed below.		
\section{Expressions of observables}\label{app:zzlhc-a}
\begin{eqnarray}\label{eq:sigma-300GeV}
\sigma (M_{4l} > 0.3~\text{TeV}) 
&=& 7.9503 + f_5^\gamma \times 16.886 + f_5^Z \times 4.0609 
+ f_4^\gamma f_4^Z\times 58561 \nonumber\\
&+& f_5^\gamma f_5^Z\times 54131 
+ (f_4^\gamma)^2 \times 58771 + (f_4^Z)^2 \times 81647 \nonumber\\
&+& (f_5^\gamma)^2 \times 55210 + (f_5^Z)^2 \times 78325 
\hspace{0.5cm}\text{fb}
\end{eqnarray}
\begin{eqnarray}\label{eq:sigma-700GeV}
\sigma (M_{4l} > 0.7~\text{TeV}) 
&=& 0.37616 + f_5^\gamma \times 3.8161 + f_5^Z \times 2.9704 
+ f_4^\gamma f_4^Z\times 55005 \nonumber\\
&+& f_5^\gamma f_5^Z\times 52706 
+ (f_4^\gamma)^2 \times 57982 + (f_4^Z)^2 \times 80035 \nonumber\\
&+& (f_5^\gamma)^2 \times 57131 + (f_5^Z)^2 \times 78515 \hspace{0.5cm}\text{fb}
\end{eqnarray}
\begin{eqnarray}\label{eq:sigma-1TeV}
\sigma (M_{4l} > 1~\text{TeV}) 
&=& 0.096685 + f_5^\gamma \times 1.9492 + f_5^Z \times 1.7106 
+ f_4^\gamma f_4^Z\times 51788 \nonumber\\
&+& f_5^\gamma f_5^Z\times 51204 
+ (f_4^\gamma)^2 \times 54933 + (f_4^Z)^2 \times 75432 \nonumber\\
&+& (f_5^\gamma)^2 \times 54507 + (f_5^Z)^2 \times 74466 \hspace{0.5cm}\text{fb}
\end{eqnarray}
\begin{eqnarray}\label{eq:axz-300GeV}
A_{xz}^{\text{num.}} (M_{4l} > 0.3~\text{TeV}) 
&=& -0.77152 + f_5^\gamma \times 6.1912+ f_5^Z \times 7.8270 
+ f_4^\gamma f_4^Z\times 2869.5\nonumber\\
& +& f_5^\gamma f_5^Z\times 396.94 
+ (f_4^\gamma)^2 \times 1029.7 + (f_4^Z)^2 \times 2298.7 \nonumber\\
&-& (f_5^\gamma)^2 \times 274.02 - (f_5^Z)^2 \times 1495.5\hspace{0.5cm}\text{fb}
\end{eqnarray}
\begin{eqnarray}\label{eq:axxyy-300GeV}
	A_{x^2-y^2}^{\text{num.}} (M_{4l} > 0.3~\text{TeV}) 
	&=& -0.94583 + f_5^\gamma \times 2.4091 - f_5^Z \times 0.17878 
	+ f_4^\gamma f_4^Z\times 2700.4 \nonumber\\
	&-& f_5^\gamma f_5^Z\times 5491.1 
	+ (f_4^\gamma)^2 \times 4298.7 + (f_4^Z)^2 \times 5835.1 \nonumber\\
	&-& (f_5^\gamma)^2 \times 6576.7 - (f_5^Z)^2 \times 8467.8 \hspace{0.5cm}\text{fb}
\end{eqnarray}
\begin{eqnarray}\label{eq:axxyy-700GeV}
A_{x^2-y^2}^{\text{num.}} (M_{4l} > 0.7~\text{TeV}) 
&=& -0.04295 + f_5^\gamma \times 1.5563+ f_5^Z \times 0.37094
+ f_4^\gamma f_4^Z\times 3299.8\nonumber\\
 &-& f_5^\gamma f_5^Z\times 5853.9
+ (f_4^\gamma)^2 \times 4241.8 + (f_4^Z)^2 \times 5679.3 \nonumber\\
&-& (f_5^\gamma)^2 \times 6520.1 - (f_5^Z)^2 \times 8559.3 \hspace{0.5cm}\text{fb}
\end{eqnarray}
\begin{eqnarray}\label{eq:azz-700GeV}
A_{zz}^{\text{num.}} (M_{4l} > 0.7~\text{TeV}) 
&=& 0.048175 - f_5^\gamma \times 0.12125 - f_5^Z \times 1.5339
- f_4^\gamma f_4^Z\times 6449.2 \nonumber\\
&-& f_5^\gamma f_5^Z\times 5860.4 
- (f_4^\gamma)^2 \times 6344.7 - (f_4^Z)^2 \times 8907.4 \nonumber\\
&-& (f_5^\gamma)^2 \times 6457.7 - (f_5^Z)^2 \times 8346.8 
\hspace{0.5cm}\text{fb}
\end{eqnarray}
The asymmetries will be given as,
\begin{eqnarray}
\wtil{A_{xz}}=\dfrac{A_{xz}^{\text{num.}} (M_{4l} > 0.3~\text{TeV})}{\sigma (M_{4l} > 0.3~\text{TeV}) }\nonumber\\
A_{x^2-y^2}=\dfrac{A_{x^2-y^2}^{\text{num.}} (M_{4l} > 0.7~\text{TeV}) }{\sigma (M_{4l} > 0.7~\text{TeV}) }\nonumber\\
A_{zz}=\dfrac{A_{zz}^{\text{num.}} (M_{4l} > 0.7~\text{TeV})}{\sigma (M_{4l} > 0.7~\text{TeV}) }
\end{eqnarray}
\section{Note on linear approximation}\label{app:zzlhc-b}
We note that, the linear approximation of considering anomalous couplings will be valid if the quadratic 
contribution on the cross section will be much smaller than the linear contribution, i.e.,
\begin{equation}
|f_i \times \sigma_i |\gg |f_{i}^2\times \sigma_{ii}|,~~~\text{or}~~ |f_i|\ll \frac{\sigma_i}{\sigma_{ii}},
\end{equation}
where $\sigma_{i}$ and $\sigma_{ii}$ are the linear and quadratic
coefficient of the coupling $f_i$ in the cross section.
Based on $\sigma (M_{4l} > 1~\text{TeV})$ in Eq.~(\ref{eq:sigma-1TeV}) the linear approximation constrain $f_5^V$ as
\begin{eqnarray}
|f_5^Z|\ll 2.2\times 10^{-5},~~ |f_5^\gamma|\ll 3.5\times 10^{-5},
\end{eqnarray}
which are much much smaller than the limit (see Eq.~(\ref{eq:CMS-limit})) 
observed at the LHC~\cite{Sirunyan:2017zjc}. To this end we
keep terms upto quadratic in couplings in our analysis. 
\chapter{The helicity amplitudes in $e^+e^-\to W^+W^-$ in SM+aTGC}\label{appendix:eeWW}

\begin{figure}
	\centering
	\includegraphics[width=0.6\textwidth]{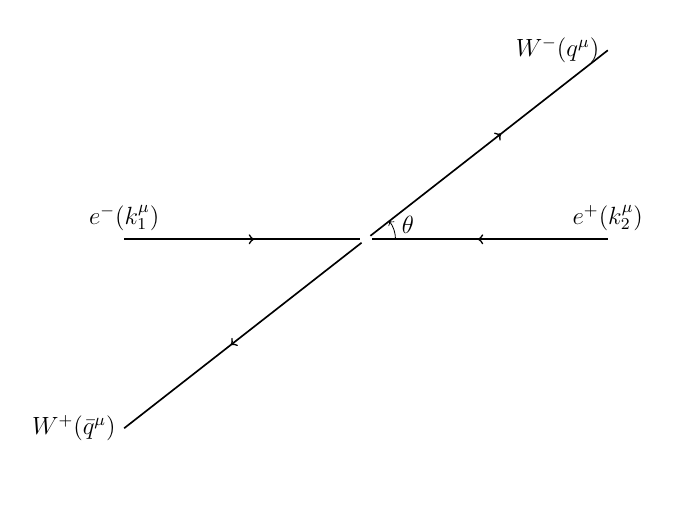}
	\caption{\label{fig:eeww-kinematic} Schematic diagram for $e^-e^+\to W^-W^+$ kinematics.   }
\end{figure}
We compute the  processes 
\begin{equation}\label{eq:eeWW-process}
e^-(k_1,\lambda_{e^-}) + e^+(k_2,\lambda_{e^+}) \to   W_\alpha^-(q,\lambda_{W^-}) +  W_\beta^+(\bar{q},\lambda_{W^+}).
\end{equation}
The helicity amplitudes for this process in SM in the t-channel (${\cal M}_T$), SM+aTGC in  $Z$-mediated s-channel (${\cal M}_{sZ}$) and $\gamma$-mediated s-channel (${\cal M}_{s\gamma}$) 
 are given by,
\begin{eqnarray}\label{eq:matrixelement-WW}
i{\cal M}_T(\lambda_{e^-},\lambda_{e^+},\lambda_{W^-},\lambda_{W^+} )&=&
i \bar{v}(k_2,\lambda_{e^+})\left( \frac{-ig_W}{2\sqrt{2}}\gamma^\beta(1-\gamma_5) \right)\left(\frac{i}{(\slashed{k}_1-\slashed{q})} \right) \left( \frac{-ig_W}{2\sqrt{2}}\gamma^\alpha(1-\gamma_5) \right) \nonumber\\
&&u(k_1,\lambda_{e^-}) \epsilon_\beta^\star(\bar{q},\lambda_{W^+})\epsilon_\alpha(q,\lambda_{W^-}),
\end{eqnarray}
\begin{eqnarray}
i{\cal M}_{sZ}(\lambda_{e^-},\lambda_{e^+},\lambda_{W^-},\lambda_{W^+} )&=&
i \bar{v}(k_2,\lambda_{e^+})\left(\frac{-ig_Z}{2}\gamma^\rho(v_e-a_e\gamma_5)  \right)u(k_1,\lambda_{e^-})\nonumber\\
&& \left( \frac{-i \bigg(g_{\rho\mu}-\frac{P_\rho P_\mu}{m_Z^2}\bigg)}{P^2-m_Z^2} \right)
\left(ig_{WWZ} \Gamma_{Z^\star}^{\mu\alpha\beta}\left(P,q,\bar{q}\right) \right) \nonumber\\
&&\epsilon_\beta^\star(\bar{q},\lambda_{W^+})\epsilon_\alpha(q,\lambda_{W^-}),
\end{eqnarray}
\begin{eqnarray}
i{\cal M}_{s\gamma}(\lambda_{e^-},\lambda_{e^+},\lambda_{W^-},\lambda_{W^+} )&=&
i \bar{v}(k_2,\lambda_{e^+})\bigg(ig_e\gamma^\rho \bigg)u(k_1,\lambda_{e^-}) \left( \frac{-ig_{\rho\mu}}{P^2} \right)\nonumber\\
&&
\left(ig_{WW\gamma} \Gamma_{\gamma^\star}^{\mu\alpha\beta}\left(P,q,\bar{q}\right) \right) \epsilon_\beta^\star(\bar{q},\lambda_{W^+})\epsilon_\alpha(q,\lambda_{W^-}).
\end{eqnarray}
The total helicity matrix element is thus
\begin{eqnarray}\label{eq:M-tot-WW}
{\cal M}_{tot}(\lambda_{e^-},\lambda_{e^+},\lambda_{W^-},\lambda_{W^+} ) &=& 
{\cal M}_T(\lambda_{e^-},\lambda_{e^+},\lambda_{W^-},\lambda_{W^+} ) +
{\cal M}_{sZ}(\lambda_{e^-},\lambda_{e^+},\lambda_{W^-},\lambda_{W^+} )\nonumber\\
&+&{\cal M}_{s\gamma}(\lambda_{e^-},\lambda_{e^+},\lambda_{W^-},\lambda_{W^+} )
\end{eqnarray}
The $WWV$ vertex factor $\left(ig_{WW\gamma} \Gamma_{\gamma^\star}^{\mu\alpha\beta}\left(P,q,\bar{q}\right)\right)$ contains both SM and
aTGC contribution and they are given in Eq.~(\ref{eq:wwv_vertex}) with the relation in Eq.~(\ref{eq:reltn_f_Lagrn}) to 
the Lagrangian in Eq.~(\ref{eq:WW-LagWWV}).  Here, the coupling constants are
\begin{eqnarray}
&g_W=\frac{g_e}{s_w},~g_Z=\frac{g_W}{c_w},~c_w=\cos\theta_W=\frac{m_W}{m_Z},~s_w=\sqrt{1-c_w^2},~
\nonumber\\
&g_{WW\gamma}=-g_e,~g_{WWZ}=-g_e\frac{c_w}{s_w},~g_e=e=\sqrt{4\pi\alpha_{EM}}.
\end{eqnarray}

The four-momentum of the particles in this process are (see Fig.~\ref{fig:eeww-kinematic})
\begin{eqnarray}
k_1^\mu=\frac{\sqrt{\hat{s}}}{2}\left\{1,0,0,1\right \},
&&k_2^\mu=\frac{\sqrt{\hat{s}}}{2}\left\{1,0,0,-1 \right\}\nonumber\\
q^\mu=\frac{\sqrt{\hat{s}}}{2}\left\{1,-\beta\sin\theta,0,-\beta\cos\theta \right\},
&&\bar{q}^\mu=\frac{\sqrt{\hat{s}}}{2}\left\{1,\beta\sin\theta,0,\beta\cos\theta \right\},
\end{eqnarray}
$\sqrt{\hat{s}}$ being the centre-of-mass energy of the colliding beams.
The on-shell condition of $W^\pm$, $q^2=\bar{q}^2=m_W^2$ gives $\beta$ as
\begin{equation}
\beta=\sqrt{1-\frac{4m_W^2}{\hat{s}}}.
\end{equation}
The polarization vector for the $W$'s are similar as in Eq.~(\ref{eq:polvec-Z}) and they are
\begin{eqnarray}\label{eq:polvec-W}
\epsilon^\mu(q,\pm)&=&\frac{1}{\sqrt{2}}\left\{0,\mp \cos\theta, -i, \pm  \cos\theta\right\},\nonumber\\
\epsilon^\mu(q,0)&=&\frac{1}{m_W}\left\{|\vec{q}|, q_0\sin\theta, 0, q_0\cos\theta\right\}.
\end{eqnarray}
The density matrix is calculated in the similar way as in given  Eq.~(\ref{eq:ZV-rho-modM}) with the symmetry factor $S=1$ and replacing $Z$ and $V$ with $W^-$ and $W^+$ respectively.  

The non-zero helicity amplitudes in SM including aTGC for this process are given below with the following notations,
$$c_\theta=\cos\theta,~s_\theta=\sin\theta .$$
\begin{eqnarray}
&&{\cal M}_T(\lambda_{e^-},\lambda_{e^+},\lambda_{W^+},\lambda_{W^-} )={\cal M}_{SM} + {\cal M}_{aTGC}\nonumber\\
{\cal M}_{-,+,-,-}&=&\frac{(1+c_\theta) g_W^2 s_\theta}{1-2 c_\theta \beta +\beta ^2}\nonumber\\
{\cal M}_{-,+,-,0}&=&\frac{(1+c_\theta) \left(2 g_{WW\gamma} g_e \beta -\frac{\hat{s} g_Z ~g_{WWZ} (a_f+v_f)
		\beta }{\hat{s}-m_Z^2}-\frac{g_W^2 \bigg(1-2 c_\theta+2 \beta -\beta ^2\bigg)}{1-2 c_\theta \beta +\beta ^2}\right)}{\sqrt{2-2
		\beta ^2}}\nonumber\\
&+&\frac{1}{\sqrt{2} \bigg(1-\beta
	^2\bigg)^{3/2}}(1+c_\theta) \bigg(
g_{WW\gamma} g_e 
\bigg(i f_6^\gamma \bigg(1-\beta ^2\bigg)-\beta  \bigg(f_3^\gamma-f_3^\gamma \beta ^2\nonumber\\
&-&if_4^\gamma \bigg(1-\beta ^2\bigg)
-\beta  \bigg(2 i \widetilde{f_7^\gamma}+f_5^\gamma \bigg(1-\beta ^2\bigg)\bigg)\bigg)\bigg)
-\frac{\frac{1}{2}\hat{s} g_Z
	g_{WWZ} (a_f+v_f) }{\hat{s}-m_Z^2} \nonumber\\
&&\bigg(i f_6^Z \bigg(1-\beta ^2\bigg)
-\beta  \bigg(f_3^Z-f_3^Z \beta ^2+i f_4^Z \bigg(1-\beta
^2\bigg)\nonumber\\
&+&\beta  \bigg(2 i \widetilde{f_7^Z}+f_5^Z \bigg(1-\beta ^2\bigg)\bigg)\bigg)\bigg)\bigg)\nonumber\\
{\cal M}_{-,+,-,+}&=& -s_\theta \left(g_{WW\gamma} g_e \beta -\frac{\frac{1}{2}\hat{s} g_Z ~g_{WWZ} (a_f+v_f)
	\beta }{\hat{s}-m_Z^2}+\frac{g_W^2 (c_\theta-\beta )}{1-2 c_\theta \beta +\beta ^2}\right)\nonumber\\
&+&s_\theta
\left(g_{WW\gamma} g_e (f_1^\gamma \beta-i f_6^\gamma )-\frac{\frac{1}{2}\hat{s} g_Z ~g_{WWZ} (a_f+v_f) ( f_1^Z
	\beta -i f_6^Z )}{\hat{s}-m_Z^2}
\right)\nonumber\\
{\cal M}_{-,+,0,-}&=&-\frac{(1+c_\theta) \left(2 g_{WW\gamma} g_e \beta -\frac{\hat{s} g_Z ~g_{WWZ} (a_f+v_f)
		\beta }{\hat{s}-m_Z^2}-\frac{g_W^2 \bigg(1-2 c_\theta+2 \beta -\beta ^2\bigg)}{1-2 c_\theta \beta +\beta ^2}\right)}{\sqrt{2-2
		\beta ^2}} \nonumber\\
&+&\frac{(1+c_\theta) }{\sqrt{2} \bigg(1-\beta
	^2\bigg)^{3/2}}\Bigg(
g_{WW\gamma} g_e \bigg(i f_6^\gamma \bigg(1-\beta ^2\bigg)+\beta  \bigg(f_3^\gamma-f_3^\gamma \beta ^2+i
f_4^\gamma \bigg(1-\beta ^2\bigg)\nonumber\\
&+&\beta  \bigg(2 i \widetilde{f_7^\gamma}+f_5^\gamma \bigg(-1+\beta ^2\bigg)\bigg)\bigg)\bigg)
-\frac{\frac{1}{2}\hat{s} g_Z
	g_{WWZ} (a_f+v_f)}{\hat{s}-m_Z^2} \bigg(i f_6^Z \bigg(1-\beta ^2\bigg)\nonumber\\
&+&\beta  \bigg(f_3^Z-f_3^Z \beta ^2+i f_4^Z \bigg(1-\beta
^2\bigg)+\beta  \bigg(2 i \widetilde{f_7^Z}-f_5^Z \bigg(1-\beta ^2\bigg)\bigg)\bigg)\bigg)\Bigg)\nonumber\\
{\cal M}_{-,+,0,0}&=&\frac{s_\theta \left(g_{WW\gamma} g_e \beta  \bigg(3-\beta ^2\bigg)-\frac{\frac{1}{2}\hat{s} g_Z
		g_{WWZ} (a_f+v_f) \beta  \bigg(3-\beta ^2\bigg)}{\hat{s}-m_Z^2}+\frac{g_W^2 \bigg(2 c_\theta-3 \beta
		+\beta ^3\bigg)}{1-2 c_\theta \beta +\beta ^2}\right)}{\left(1-\beta ^2\right)}\nonumber\\
&+&\frac{s_\theta }{\left(1-\beta
	^2\right)^2}\Bigg(g_{WW\gamma} g_e \beta  \bigg(f_1^\gamma-2 f_3^\gamma-4
f_2^\gamma \beta ^2+2 f_3^\gamma \beta ^2-f_1^\gamma \beta ^4\bigg)\nonumber\\
&+&\frac{\frac{1}{2}\hat{s} g_Z ~g_{WWZ} (a_f+v_f) \beta  \bigg(2
	\bigg(f_3^Z+2 f_2^Z \beta ^2-f_3^Z \beta ^2\bigg)-f_1^Z \bigg(1-\beta ^4\bigg)\bigg)}{\hat{s}-m_Z^2}\Bigg)\nonumber\\ 
{\cal M}_{-,+,0,+}&=&\frac{(1-c_\theta) }{\bigg(\hat{s}-m_Z^2\bigg) \sqrt{2-2 \beta ^2} \bigg(1-2 c_\theta \beta +\beta ^2\bigg)}
\bigg(m_Z^2 \bigg(g_W^2 \bigg(1+2 c_\theta-2 \beta -\beta ^2\bigg)\nonumber\\
&+&2g_{WW\gamma} g_e \beta  \bigg(1-2 c_\theta \beta +\beta ^2\bigg)\bigg)
-\hat{s} \bigg(g_W^2 \bigg(1+2 c_\theta-2
\beta -\beta ^2\bigg)\nonumber\\
&+&(2 g_{WW\gamma} g_e-g_Z ~g_{WWZ} (a_f+v_f)) \beta  \bigg(1-2 c_\theta \beta +\beta ^2\bigg)\bigg)\bigg)\nonumber\\
%
%
&+&\frac{(1-c_\theta) }{\sqrt{2} \bigg(1-\beta ^2\bigg)^{3/2}}
\bigg(-g_{WW\gamma} g_e \bigg(-i
f_6^\gamma \bigg(-1+\beta ^2\bigg)+\beta  \bigg(f_3^\gamma \bigg(-1+\beta ^2\bigg)\nonumber\\
&+&i f_4^\gamma \bigg(-1+\beta ^2\bigg)
+\beta  \bigg(2 i \widetilde{f_7^\gamma}+f_5^\gamma
\bigg(-1+\beta ^2\bigg)\bigg)\bigg)\bigg)+\frac{\frac{1}{2}\hat{s} g_Z ~g_{WWZ} (a_f+v_f) }{\hat{s}-m_Z^2}
\nonumber\\&&
\bigg(-i f_6^Z \bigg(-1+\beta
^2\bigg)
+\beta  \bigg(f_3^Z \bigg(-1+\beta ^2\bigg)+i f_4^Z \bigg(-1+\beta ^2\bigg)\nonumber\\
&+&\beta  \bigg(2 i \widetilde{f_7^Z}+f_5^Z \bigg(-1+\beta
^2\bigg)\bigg)\bigg)\bigg)\bigg)\nonumber\\
%
%
%
%
{\cal M}_{-,+,+,-}&=&s_\theta \left(-g_{WW\gamma} g_e \beta +\frac{\frac{1}{2}\hat{s} g_Z ~g_{WWZ} (a_f+v_f)
	\beta }{\hat{s}-m_Z^2}-\frac{g_W^2 (c_\theta-\beta )}{1-2 c_\theta \beta +\beta ^2}\right)\nonumber\\
&+&s_\theta
\left(g_{WW\gamma} g_e (i f_6^\gamma+f_1^\gamma \beta )-\frac{\frac{1}{2}\hat{s} g_Z ~g_{WWZ} (a_f+v_f) (i f_6^Z+f_1^Z
	\beta )}{\hat{s}-m_Z^2}\right)\nonumber\\
{\cal M}_{-,+,+,0}&=&\frac{(1-c_\theta) \left(2 g_{WW\gamma} g_e \beta -\frac{\hat{s} g_Z ~g_{WWZ} (a_f+v_f)
		\beta }{\hat{s}-m_Z^2}+\frac{g_W^2 \left(1+2 c_\theta-2 \beta -\beta ^2\right)}{1-2 c_\theta \beta +\beta ^2}\right)}{\sqrt{2-2
		\beta ^2}}\nonumber\\
&+&\frac{(1-c_\theta)}{\sqrt{2} \bigg(1-\beta
	^2\bigg)^{3/2}}\Bigg(-g_{WW\gamma} g_e \bigg(-i f_6^\gamma \bigg(-1+\beta ^2\bigg)+\beta  \bigg(f_3^\gamma-f_3^\gamma \beta ^2\nonumber\\
&+&if_4^\gamma \bigg(-1+\beta ^2\bigg)+\beta  \bigg(f_5^\gamma+2 i \widetilde{f_7^\gamma}-f_5^\gamma \beta ^2\bigg)\bigg)\bigg)+\frac{\frac{1}{2}\hat{s} g_Z
	g_{WWZ} (a_f+v_f) }{\hat{s}-m_Z^2}\nonumber\\&&
\bigg(-i f_6^Z \bigg(-1+\beta ^2\bigg)+\beta  \bigg(f_3^Z-f_3^Z \beta ^2+i f_4^Z \bigg(-1+\beta
^2\bigg)\nonumber\\
&+&\beta  \bigg(f_5^Z+2 i \widetilde{f_7^Z}-f_5^Z \beta ^2\bigg)\bigg)\bigg)\Bigg)\nonumber\\
%
{\cal M}_{-,+,+,+}&=&-\frac{(1-c_\theta) g_W^2 s_\theta}{1-2 c_\theta \beta +\beta ^2}\nonumber\\
{\cal M}_{+,-,-,0}&=&-\frac{(1-c_\theta)  \beta  \sqrt{2-2 \beta ^2}}{\bigg(\hat{s}-m_Z^2\bigg) \bigg(1-\beta ^2\bigg)}
\bigg(-g_{WW\gamma} g_e m_Z^2+\frac{1}{2}\hat{s} \Big(2 g_{WW\gamma} g_e\nonumber\\
&+&g_Z
g_{WWZ} (a_f-v_f)\Big)\bigg)
\nonumber\\
&+&\frac{(1-c_\theta)}{\sqrt{2} \bigg(1-\beta ^2\bigg)^{3/2}}\bigg(-g_{WW\gamma} g_e \bigg(-i f_6^\gamma \bigg(-1+\beta ^2\bigg)+\beta  \bigg(f_3^\gamma \bigg(-1+\beta ^2\bigg)\nonumber\\
&-&i f_4^\gamma \bigg(-1+\beta
^2\bigg)+\beta  \bigg(f_5^\gamma+2 i \widetilde{f_7^\gamma}-f_5^\gamma \beta ^2\bigg)\bigg)\bigg)+\frac{\frac{1}{2}\hat{s} g_Z ~g_{WWZ} (a_f-v_f)
}{\hat{s}-m_Z^2}\nonumber\\
&&\bigg(i f_6^Z \bigg(-1+\beta ^2\bigg)+\beta  \bigg(f_3^Z-f_3^Z \beta ^2+i f_4^Z \bigg(-1+\beta ^2\bigg)\nonumber\\
&+&\beta  \bigg(-2 i \widetilde{f_7^Z}+f_5^Z
\bigg(-1+\beta ^2\bigg)\bigg)\bigg)\bigg)\bigg)\nonumber\\
%
%
%
{\cal M}_{+,-,-,+}&=&s_\theta \left(-g_{WW\gamma} g_e+\frac{\frac{1}{2}\hat{s} g_Z ~g_{WWZ} (v_f-a_f)}{\hat{s}-m_Z^2}\right) \beta \nonumber\\
&+&g_{WW\gamma} g_e s_\theta (-i f_6^\gamma+f_1^\gamma \beta )+\frac{\frac{1}{2}\hat{s} g_Z ~g_{WWZ}
	s_\theta (a_f-v_f) (-i f_6^Z+f_1^Z \beta )}{\hat{s}-m_Z^2}\nonumber\\
{\cal M}_{+,-,0,-}&=&\frac{\sqrt{2} (1-c_\theta) \bigg(-g_{WW\gamma} g_e m_Z^2+\frac{1}{2}\hat{s} (2 g_{WW\gamma} g_e+g_Z
	g_{WWZ} (a_f-v_f))\bigg) \beta }{\bigg(\hat{s}-m_Z^2\bigg) \sqrt{1-\beta ^2}}\nonumber\\
&+&\frac{(1-c_\theta) }{\sqrt{2} \bigg(1-\beta ^2\bigg)^{3/2}}\bigg(-g_{WW\gamma}
g_e \bigg(-i f_6^\gamma \bigg(-1+\beta ^2\bigg)+\beta  \bigg(f_3^\gamma-f_3^\gamma \beta ^2\nonumber\\
&-&i f_4^\gamma \bigg(-1+\beta ^2\bigg)+\beta  \bigg(2
i \widetilde{f_7^\gamma}+f_5^\gamma \bigg(-1+\beta ^2\bigg)\bigg)\bigg)\bigg)\nonumber
+\frac{\frac{1}{2}\hat{s} g_Z ~g_{WWZ} (a_f-v_f) }{\hat{s}-m_Z^2}\nonumber\\
&&\bigg(i f_6^Z
\bigg(-1+\beta ^2\bigg)-\beta  \bigg(f_3^Z-f_3^Z \beta ^2-i f_4^Z \bigg(-1+\beta ^2\bigg)\nonumber\\
&+&\beta  \bigg(2 i \widetilde{f_7^Z}+f_5^Z
\bigg(-1+\beta ^2\bigg)\bigg)\bigg)\bigg)\bigg)\nonumber\\
{\cal M}_{+,-,0,0}&=&\frac{s_\theta \bigg(-g_{WW\gamma} g_e m_Z^2+\frac{1}{2}\hat{s} (2 g_{WW\gamma} g_e+g_Z
	g_{WWZ} (a_f-v_f))\bigg) \beta  \bigg(3-\beta ^2\bigg)}{\bigg(\hat{s}-m_Z^2\bigg) \bigg(1-\beta ^2\bigg)}\nonumber\\
&+&\frac{s_\theta \beta  }{\bigg(1-\beta ^2\bigg)^2}
\bigg(g_{WW\gamma} g_e \bigg(f_1^\gamma-2 f_3^\gamma-4 f_2^\gamma \beta ^2+2 f_3^\gamma \beta ^2-f_1^\gamma \beta ^4\bigg)\nonumber\\
&+&\frac{\frac{1}{2}\hat{s} g_Z ~g_{WWZ} (a_f-v_f) \bigg(f_1^Z-2 f_3^Z-4 f_2^Z \beta ^2+2 f_3^Z \beta ^2-f_1^Z \beta
	^4\bigg)}{\hat{s}-m_Z^2}\bigg)\nonumber\\
{\cal M}_{+,-,0,+}&=&\frac{\sqrt{2} (1+c_\theta) \bigg(-g_{WW\gamma} g_e m_Z^2+\frac{1}{2}\hat{s} (2 g_{WW\gamma} g_e+g_Z
	g_{WWZ} (a_f-v_f))\bigg) \beta }{\bigg(\hat{s}-m_Z^2\bigg) \sqrt{1-\beta ^2}}\nonumber\\
&+&\frac{(1+c_\theta) }{\sqrt{2} \bigg(1-\beta ^2\bigg)^{3/2}}\bigg(-g_{WW\gamma}
g_e \bigg(i f_6^\gamma \bigg(-1+\beta ^2\bigg)+\beta  \bigg(f_3^\gamma-f_3^\gamma \beta ^2\nonumber\\
&-&i f_4^\gamma \bigg(-1+\beta ^2\bigg)+\beta  \bigg(f_5^\gamma-2
i \widetilde{f_7^\gamma}-f_5^\gamma \beta ^2\bigg)\bigg)\bigg)+\frac{\frac{1}{2}\hat{s} g_Z ~g_{WWZ} (a_f-v_f) }{\hat{s}-m_Z^2}\nonumber\\
&&\bigg(-i f_6^Z \bigg(-1+\beta
^2\bigg)+\beta  \bigg(f_3^Z \bigg(-1+\beta ^2\bigg)+i f_4^Z \bigg(-1+\beta ^2\bigg)\nonumber\\
&+&\beta  \bigg(2 i \widetilde{f_7^Z}+f_5^Z \bigg(-1+\beta
^2\bigg)\bigg)\bigg)\bigg)\bigg)\nonumber\\
{\cal M}_{+,-,+,-}&=&s_\theta \left(-g_{WW\gamma} g_e+\frac{\frac{1}{2}\hat{s} g_Z ~g_{WWZ} (v_f-a_f)}{\hat{s}-m_Z^2}\right) \beta \nonumber\\
&+&g_{WW\gamma} g_e s_\theta (i f_6^\gamma+f_1^\gamma \beta )+\frac{\frac{1}{2}\hat{s} g_Z ~g_{WWZ}
	s_\theta (a_f-v_f) (i f_6^Z+f_1^Z \beta )}{\hat{s}-m_Z^2}\nonumber\\
%
%
%
{\cal M}_{+,-,+,0}&=&-\frac{\sqrt{2} (1+c_\theta) \bigg(-g_{WW\gamma} g_e m_Z^2+\frac{1}{2}\hat{s} (2 g_{WW\gamma} g_e+g_Z
	g_{WWZ} (a_f-v_f))\bigg) \beta }{\bigg(\hat{s}-m_Z^2\bigg) \sqrt{1-\beta ^2}}\nonumber\\
&+&\frac{(1+c_\theta) }{\sqrt{2} \bigg(1-\beta ^2\bigg)^{3/2}}\bigg(\frac{\frac{1}{2}\hat{s} g_Z ~g_{WWZ} (a_f-v_f) }{\hat{s}-m_Z^2}\bigg(i f_6^Z \bigg(1-\beta ^2\bigg)\nonumber\\
&+&\beta  \bigg(f_3^Z-f_3^Z \beta ^2+i
f_4^Z \bigg(-1+\beta ^2\bigg)+\beta  \bigg(f_5^Z+2 i \widetilde{f_7^Z}-f_5^Z \beta ^2\bigg)\bigg)\bigg)\nonumber\\
&-&g_{WW\gamma}
g_e \bigg(i f_6^\gamma \bigg(-1+\beta ^2\bigg)-\beta  \bigg(f_3^\gamma \bigg(1-\beta ^2\bigg)-i f_4^\gamma \bigg(1-\beta ^2\bigg)\nonumber\\
&+&\beta
\bigg(2 i \widetilde{f_7^\gamma}+f_5^\gamma \bigg(1-\beta ^2\bigg)\bigg)\bigg)\bigg)\bigg)\nonumber\\
\end{eqnarray}

\chapter{The fitting procedures of the observables and their SM values in $WZ^\pm$ productions at the LHC }\label{appendix:WZLHC}
\section{The SM values of the asymmetries and the corresponding polarizations}\label{app:SM-values-Asym}
In Table~\ref{tab:SM-values-Asym-Pol}, we show the SM estimates (with $1\sigma$ MC error) of the polarization 
asymmetries of $Z$ and $W$ and their corresponding polarizations along with the other 
asymmetries for our selection cuts ({\tt sel.cut}) given in Eq.~(\ref{eq:selection-cuts}) 
and optimized cuts ({\tt opt.cut}) given Table~\ref{tab:m3lpTZcut-on-Asym}.  
A  number of events  of $N\simeq9.9\times 10^6$ satisfy our selection cuts   which give the same error
($\delta A_i=1/\sqrt{N}$)  for all asymmetries, and hence they are given in the top row. As the
optimized cuts for $W$ are same for all asymmetries, the errors for them are also given in the top row. For the
optimized cuts of $Z$ observables, however, the number of events vary and hence the MC error are given to each asymmetry.  
The $CP$-odd polarizations $p_y$, $T_{xy}$, $T_{yz}$
and their corresponding asymmetries are consistent with zero in the SM within MC error.
\afterpage{\clearpage
	\begin{sidewaystable}
		\caption{\label{tab:SM-values-Asym-Pol} The SM values with MC error of the polarization 
			asymmetries of $Z$ and $W$ and their corresponding polarizations along with the other 
			asymmetries in  $ZW^\pm$ production in the $e^+e^-\mu^\pm+\cancel{E}_T$ channel are shown 
			for event selection cuts ({\tt sel.cut}) given in 
			Eq.~(\ref{eq:selection-cuts}) and optimized cuts ({\tt opt.cut}) given Table~\ref{tab:m3lpTZcut-on-Asym}. 
		}
		\renewcommand{\arraystretch}{1.50}
		\centering
		\begin{scriptsize}
			\begin{tabular*}{\textwidth}{@{\extracolsep{\fill}}|l|l|l|l|l|l|l|l|l|@{}}\hline
				& \multicolumn{4}{c|}{$ZW^+$}& \multicolumn{4}{c|}{$ZW^-$} \\ \hline
				&  \multicolumn{2}{c|}{$Z$}& \multicolumn{2}{c|}{$W^+$} &  \multicolumn{2}{c|}{$Z$}& \multicolumn{2}{c|}{$W^-$}\\ \hline
				${\cal O}$& {\tt sel.cut} & {\tt opt.cut}& {\tt sel.cut} & {\tt opt.cut}& {\tt sel.cut} & {\tt opt.cut}& {\tt sel.cut} & {\tt opt.cut}\\ 
				$\delta A_i$ &$\pm 0.0003$&          &$\pm 0.0003$&$\pm 0.0007$ &$\pm 0.0003$ &  &$\pm 0.0003$ &$\pm 0.0007$ \\ \hline
				$A_x$          & $-0.0196$&$-0.0150\pm 0.0008$ &$-0.2303$ &$-0.0550$&$+0.0074$&$-0.0046\pm 0.0010$&$-0.0826$&$-0.0001$  \\ 
				$p_x$          & $+0.1192\pm0.0018$&$+0.0912\pm0.0049$ &$+0.3071\pm0.0004$ &$0.0733\pm0.0009$&$-0.0450\pm0.0018$&$+0.0280\pm0.0061$&$+0.110\pm0.00041$&$+0.00013\pm0.0009$  \\ \hline
				$A_y$          &$+0.0003$ &$+0.0004\pm 0.0007$ &$-0.0007 $ &$-0.0005$&$-0.0013$&$-0.0021\pm0.0007$&$0.0$&$+0.0007$  \\ 
				$p_y$          &$-0.0018\pm0.0018$ &$-0.0024\pm0.0146$ &$+0.0009\pm0.0004$ &$+0.0006\pm0.0009$&$+0.0079\pm0.0018$&$+0.0127\pm0.0042$&$0.0\pm0.0004$&$-0.0009\pm0.0009$  \\ \hline
				$A_z$          &$-0.0040$ &$+0.0502\pm 0.0025$ &$ +0.1337$ &$+0.6615$&$+0.0316$&$+0.0482\pm0.0019$&$+0.1954$&$+0.7381$  \\ 
				$p_z$          &$+0.0243\pm0.0018$ &$-0.3051\pm0.0152$ &$-0.1783\pm0.0004$ &$-0.8820\pm0.0009$&$-0.1921\pm0.0018$&$-0.2930\pm0.0115$&$-0.2605\pm0.0004$&$-0.9841\pm0.0009$  \\ \hline
				$A_{xy}$       &$-0.0017$ &$+0.0005\pm 0.0007$ &$-0.0011$ &$-0.0006$&$+0.0008$&$+0.0014\pm0.0007$&$+0.0013$ &$-0.0003$ \\ 
				$T_{xy}$       &$-0.0033\pm0.0006$ &$+0.00096\pm0.0013$ &$-0.0021\pm0.0006$ &$-0.0012\pm0.0013$&$+0.0015\pm0.0006$&$+0.0027\pm0.0013$&$+0.0025\pm0.0006$ &$-0.0006\pm0.0013$ \\ \hline
				$A_{xz}$       &$+0.0196$ &$+0.0914\pm 0.0004$ &$+0.0048$ &$-0.0063$&$+0.0961$&$+0.0547\pm0.0006$&$+0.0010$ &$-0.0136$ \\ 
				$T_{xz}$       &$+0.0377\pm0.0006$ &$+0.1758\pm0.0008$ &$+0.0092\pm0.0006$ &$-0.0121\pm0.0013$&$+0.1849\pm0.0006$&$+0.1052\pm0.0011$&$+0.0019\pm0.0006$ &$-0.0262\pm0.0013$ \\ \hline
				$A_{yz}$       &$+0.0002$ &$-0.0001\pm 0.0004$ &$+0.0003$ &$-0.0005$&$-0.0017$&$-0.0016\pm0.0003$&$+0.0001$&$-0.0001$  \\ 
				$T_{yz}$       &$+0.0004\pm0.0006$ &$-0.0002\pm0.0008$ &$+0.0006\pm0.0006$ &$-0.0009\pm0.0013$&$-0.0033\pm0.0006$&$-0.0031\pm0.0006$&$+0.0002\pm0.0006$&$-0.0002\pm0.0013$  \\ \hline
				$A_{x^2-y^2}$  &$-0.0878$ &$-0.0925\pm 0.0019$ &$-0.0266$ &$-0.1326$&$-0.0935$&$-0.0899\pm0.0012$&$-0.0923$&$-0.1588$  \\ 
				$T_{xx}-T_{yy}$&$-0.3378\pm0.0011$ &$-0.3559\pm0.0073$ &$-0.1023\pm0.0011$ &$-0.5102\pm0.0027$&$-0.3597\pm0.0011$&$-0.3459\pm0.0046$&$-0.3551\pm0.0011$&$-0.6110\pm0.0027$  \\ \hline
				$A_{zz}$       & $-0.0137$&$+0.0982\pm 0.0024$ &$+0.0519$ &$+0.1406$&$+0.0030$&$+0.0863\pm0.0048$&$+0.1046$&$+0.2547$  \\ 
				$T_{zz}$       & $-0.0298\pm0.0006$&$+0.2138\pm.0052$ &$+0.1130\pm0.0006$ &$+0.3061\pm0.0015 $&$+0.0065\pm0.0006$&$+0.1879\pm0.0104$&$+0.2277\pm0.0006$&$+0.5546\pm0.0015$  \\ \hline
				$A_{fb}$       &$+0.6829$ &$+0.4475\pm 0.0009$  &  $+0.4699$&$+0.2627$ &$+0.6696$&$+0.2791\pm0.0025$&$+0.2060$&$+0.3174$ \\ \hline\hline
				&  \multicolumn{2}{c|}{{\tt sel.cut} }& \multicolumn{2}{c|}{{\tt opt.cut}} &  \multicolumn{2}{c|}{{\tt sel.cut} }& \multicolumn{2}{c|}{{\tt opt.cut}}\\ \hline
				$A_{\Delta\phi}$&  \multicolumn{2}{c|}{$-0.3756\pm 0.0003$}& \multicolumn{2}{c|}{$-0.4151\pm 0.0022$} &  \multicolumn{2}{c|}{$-0.3880\pm 0.0003$}& \multicolumn{2}{c|}{$-0.4208\pm 0.0025$}\\ \hline
			\end{tabular*}
		\end{scriptsize}
	\end{sidewaystable}
	\clearpage }

\newpage
\section{Fitting procedure for obtaining observables as a function of couplings}\label{app:fitting}
The SM+aTGC  events are generated for about $100$ set  of couplings
$$\{c_i\}=\{\Delta g_1^Z,\lambda^Z,\Delta\kappa^Z,\wtil{\lambda^Z},\wtil{\kappa^Z}\}$$
in both processes.  The values of all the observables are obtained for  the set couplings in the optimized
cuts (Table~\ref{tab:m3lpTZcut-on-Asym})  and then those are used for numerical fitting to obtain the semi-analytical expression of all the observables as a function of the couplings.
For the cross sections the following $CP$-even expression is used to fit the data:
\begin{equation}\label{eq:sigma-fit}
\sigma(\{c_i\}) =\sigma_{SM} + \sum_{i=1}^{3} c_i \times \sigma_i +\sum_{i=1}^{5} (c_i)^2 \times \sigma_{ii}  + \frac{1}{2}\sum_{i=1}^{3}\sum_{j(\ne i)=1}^{3} c_i c_j \times \sigma_{ij}
+ c_4 c_5 \times \sigma_{45} .
\end{equation}
For asymmetries, the numerator and the denominator  are fitted separately and then used as
\begin{equation}
A_j(\{c_i\})=\dfrac{\Delta\sigma_{A_j}(\{c_i\})}{\sigma_{A_j}(\{c_i\})} \ \  .
\end{equation}
The numerator ($\Delta\sigma_A$) of $CP$-odd asymmetries are fitted with  the $CP$-odd expression
\begin{equation}\label{eq:CP-odd-fit}
\Delta\sigma_A(\{c_i\}) =\sum_{i=4}^{5} c_i \times \sigma_i + \sum_{i=1}^{3} \left( c_i c_4 \times\sigma_{i4}
+  c_i c_5 \times\sigma_{i5}\right) .
\end{equation}
The denominator ($\sigma_{A_j}$) of all the asymmetries and the numerator  ($\Delta\sigma_A$) of $CP$-even
asymmetries are fitted with the $CP$-even expression given in Eq.~(\ref{eq:sigma-fit}).

\begin{figure}[h!]
	\centering
	\includegraphics[width=0.496\textwidth]{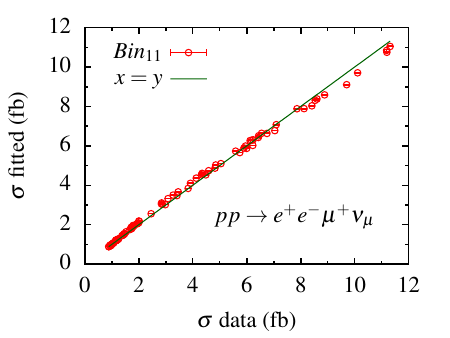}
	\includegraphics[width=0.496\textwidth]{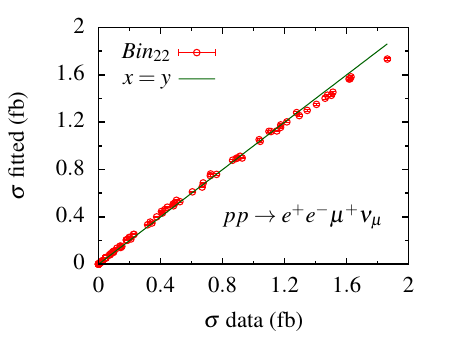}
	\includegraphics[width=0.496\textwidth]{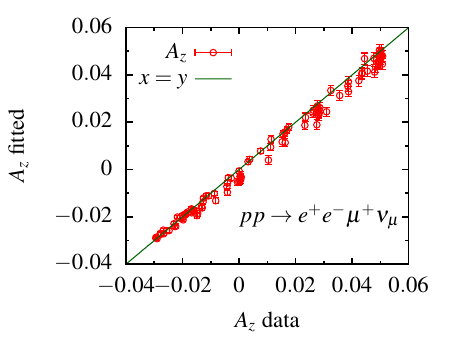}
	\includegraphics[width=0.496\textwidth]{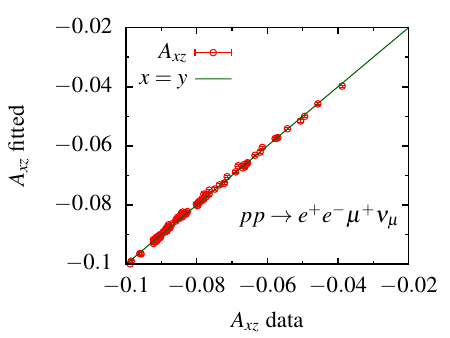}
	\caption{\label{fig:Fitt-Test-sigma-ZWp}
		The simulated data (in $x$-axis) vs. fitted values (in $y$-axis) for the cross section in the two diagonal bins ({\em top-panel})
		and the polarization asymmetries $A_z$ and $A_{xz}$ ({\em bottom-panel}) in
		in $ZW^+$ production in $e^+e^-\mu^+\nu_\mu$ channel at the LHC at $\sqrt{s}=13$ TeV. 
	}
\end{figure}
We use MCMC method to
fit the coefficients of the cross sections with positivity demand, i.e., $\sigma(\{c_i\})\ge 0$.
We use $80~\%$ data to fit the coefficients of the cross sections, and then the fitted expressions are
validated against the rest   $20~\%$ of the  data  and found to be matching within $2\sigma$ MC error. 
We generated  $10^7$ events to keep the MC  error as small as possible  even in the tightest
optimized cuts. For example, the $A_{zz}$ in $ZW^+$ has the tightest cut on $m_{3l}$ (see Table~\ref{tab:m3lpTZcut-on-Asym}) and yet have very small ($0.2~\%$) MC error (see Table~\ref{tab:SM-values-Asym-Pol}).  
In Fig.~\ref{fig:Fitt-Test-sigma-ZWp} fitted values of observables are compared against the simulated data for
the cross section in two diagonal bins ({\em top-panel})
and the polarization asymmetries $A_z$ and $A_{xz}$ ({\em bottom-panel}) 
in $ZW^+$ production in $e^+e^-\mu^+\nu_\mu$ channel as representative. The fitted values seem
to agree with the simulated data used within the MC error.
\chapter{HEP packages that are used in this thesis}\label{appendix:HEP-packages}
We have used various high energy physics (HEP) packages for modelling the
anomalous couplings, implementing to event generators, and for analysing events. Below, we give
brief descriptions of them and their role in my thesis.

\paragraph{FeynRules}
The {\tt FeynRules}~\cite{Alloul:2013bka} is a Mathematica based  HEP package, where one generates model files for various event generators and get the Feynman rules 
by implementing the SM or BSM Lagrangian. 
One has to provide the {\tt FeynRules} with the required information  
such as the gauge symmetry, particles, parameters, etc.,
to describe the QFT 
model, contained in the model file with `fr' extension.
The {\tt FeynRules} than calculate the set of Feynman rules in momentum space associated with the given Lagrangian. The {\tt FeynRules} can provide output for various other packages
such as, {\tt CalcHep}~\cite{Pukhov:2004ca}, {\tt FeynArts}~\cite{Hahn:2000kx}, {\tt Sherpa}~\cite{Hoche:2014kca}, UFO (Universal FeynRules output), {\tt Whizard}~\cite{Kilian:2007gr} etc for physics analysis. 

\paragraph{MadGRaph5}
The {\tt MadGRaph5}~\cite{Alwall:2014hca}  package named as \MGvATNLO~ is a Monte-Carlo event generator for scattering and decay processes in a QFT model. One imports the UFO model files in a given model in the {\tt MadGRaph5} interface and generates events of a given process
with/without kinematical cuts on the final state particles. This can calculate 
the cross section and generates events at up to NLO in QCD.  

\paragraph{MATRIX}
The {\tt MATRIX}~\cite{Grazzini:2017mhc} can calculate cross section up to NNLO in QCD for some specific processes
already implemented. One can get distributions of various kinematical variables
of the final state particles at LO, NLO and NNLO in QCD and compare the QCD corrections
over the ranges of the variables. This package has no features of generating events of a process.

\paragraph{GetDist} The package {\tt GetDist}~\cite{Antony:GetDist,Lewis:2019xzd}
 is a Python based package which performs Bayesian statistics of Monte-Carlo samples. It produces $1D$, $2D$, even $3D$ marginalised plots on the parameter space of interest from the   Monte-Carlo samples, e.g., MCMC correlated samples, 
 and gives simultaneous confidence  limits on the parameters.   

\paragraph{MadAnalysis5}
The {\tt MadAnalysis5}~\cite{Conte:2012fm} is a package for analysing events generated by a Monte-Carlo event generator ({\tt MadGRaph5}). The {\tt MadAnalysis5} which uses the package {\tt ROOT}/python,
can give a cut flow analysis and produces nice plots containing several
event files. The package has an expert mode, where one can perform some 
cut based analysis for casting some analysis performed at the LHC to derive confidence level exclusions on the signals. It can perform detector simulations using hadronized events generated by {\tt MadGRaph5} and Pythia.

\paragraph{Uses of the packages in the thesis}
We calculated Feynman rules and produced UFO model files for {\tt MadGRaph5}
using the {\tt FeynRules} package with aTGC Lagrangian in all the processes
from chapters~\ref{chap:epjc1}-\ref{chap:WZatLHC}. We generated LO model files
for processes in $e^+$-$e^-$ collider used in chapters~\ref{chap:epjc1},~\ref{chap:epjc2} \&~\ref{chap:eeWW}, while we generated NLO in QCD model files for processes at the LHC as used in chapters~\ref{chap:ZZatLHC} \&~\ref{chap:WZatLHC}. 
Using the UFO models files generated in {\tt FeynRules},
we generated events in {\tt MadGRaph5} for $e^+$-$e^-$ collider and the  LHC over the range of aTGC and beam energy, and obtained different angular observables from the events. We used these events to get differential distributions using the package {\tt MadAnalysis5} and used in chapters~\ref{chap:ZZatLHC} \&~\ref{chap:WZatLHC}. 
The package {\tt MATRIX} is used to calculate NLO and NNLO cross sections 
and distributions of kinematical variables in chapters~\ref{chap:ZZatLHC} \&~\ref{chap:WZatLHC}. The package  {\tt GetDist} is used in all  chapters~\ref{chap:epjc1}-\ref{chap:WZatLHC} to obtain posterior $1D$ and $2D$
marginalised plots and simultaneous confidence limits on the aTGC.
 
\end{appendices}

\cleardoublepage
\phantomsection
\addcontentsline{toc}{chapter}{Bibliography}
\adjustmtc

\bibliography{References/main}
\bibliographystyle{utphysM}


\end{document}